\documentclass[11pt,a4paper]{report}

\usepackage{latexsym,amsmath,amsthm,amssymb,amsfonts,amsbsy,calrsfs,multibox}
\usepackage{graphicx}

\newcommand\fverb{\setbox\fverbbox=\hbox\bgroup\verb}
\newcommand\fverbdo{\egroup\medskip\noindent%
			\fbox{\unhbox\fverbbox}\ }
\newcommand\fverbit{\egroup\item[\fbox{\unhbox\fverbbox}]}
\newbox\fverbbox

\newtheorem{theorem}{Theorem}[chapter]
\newtheorem{lemma}{Lemma}[chapter]

\newtheorem{prop}{Proposition}[chapter]

\def\a{\alpha}
\def\b{\beta}
\def\g{\gamma}
\def\c{\gamma}
\def\d{\delta}
\def\h{\eta}
\def\l{\lambda}
\def\L{\Lambda}
\def\G{\Gamma}
\def\D{\Delta}
\def\m{\mu}
\def\n{\nu}
\def\N{\nabla}
\def\r{\rho}
\def\o{\omega}
\def\s{\sigma}

\def\t{\tau}
\def\p{\pi}
\def\ve{\varepsilon}

\def\x{\xi}
\def\z{\zeta}
\def\e{\varepsilon}
\def\pa{\partial}
\def\6{\partial}

\def\be{\begin{equation}}
\def\ee{\end{equation}}
\def\bea{\begin{eqnarray}}
\def\eea{\end{eqnarray}}
\def\beq{\begin{eqnarray}}
\def\eeq{\end{eqnarray}}

\def\bs{{\textbf{s}}}
\def\ca{{\cal A}}
\def\cb{{\cal B}}

\def\cd{{\cal D}}

\def\cg{{\cal G}}
\def\ch{{\cal H}}

\def\cl{{\cal L}}
\def\cm{{\cal M}}

\def\co{{\cal O}}

\def\car{{\cal R}}

\def\ct{{\cal T}}

\def\cy{{\cal Y}}

\def\mathun{\mbox{l\hspace{-0.55em}1}}

\def\ba{\begin{array}}
\def\ba{\end{array}}

\begin{document} 

\begin{titlepage}
\begin{center}
{\large Universit\'e de Mons-Hainaut}

Acad\'emie universitaire Wallonie-Bruxelles

Facult\'e des Sciences

Service de m\'ecanique et gravitation

\vspace{4cm}
{\Huge{Higher spin interactions: cubic deformations on Minkowski and (Anti)de Sitter backgrounds}}

\vspace{5cm}
{\LARGE{Serge Leclercq}}
\vspace{5mm}

Th\`ese pr\'esent\'ee en vue de l'obtention 

du grade l\'egal de Docteur en Sciences

\vspace{2cm}

\includegraphics[width=7cm]{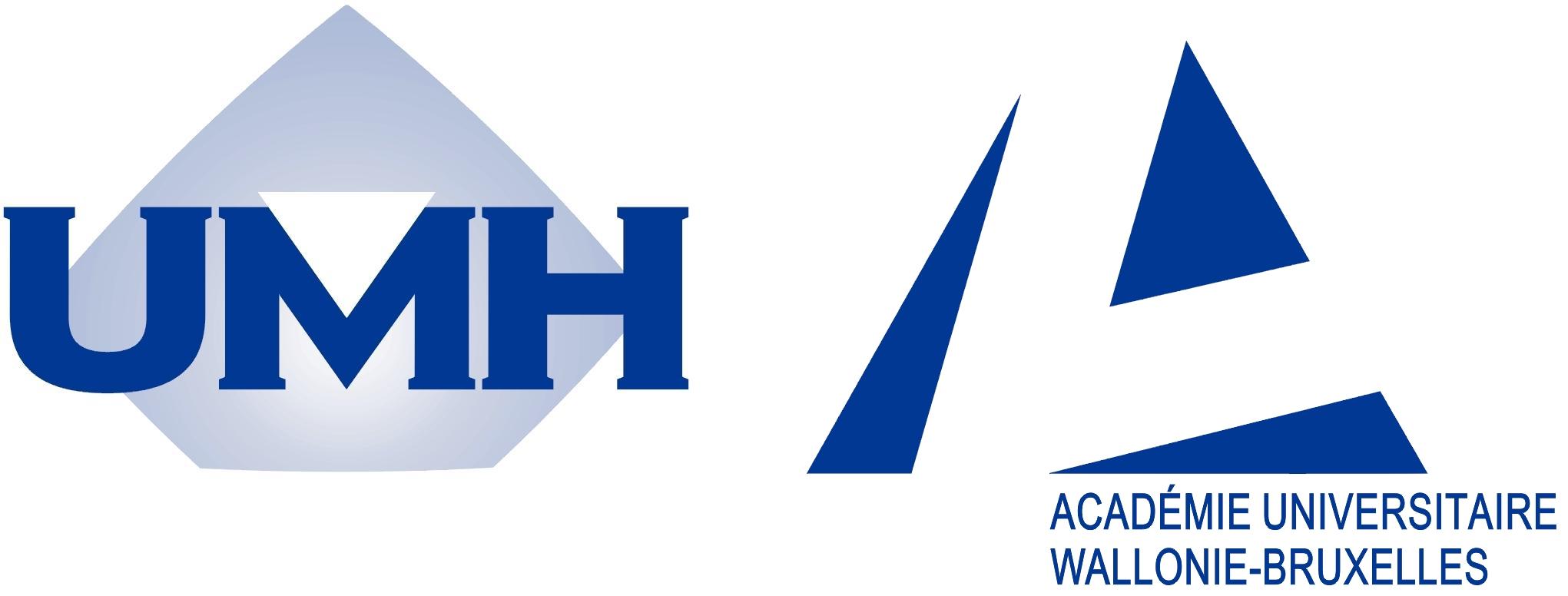}

\vspace{2cm}

Ann\'ee acad\'emique 2008-2009

\end{center}
\end{titlepage}

\thispagestyle{empty}

\newpage

${}$

\newpage

\begin{center}
{\large Universit\'e de Mons-Hainaut}

Acad\'emie universitaire Wallonie-Bruxelles

Facult\'e des Sciences

Service de m\'ecanique et gravitation

\vspace{3cm}
{\Huge{Higher spin interactions: cubic deformations on Minkowski and (Anti)de Sitter backgrounds}}

\vspace{3cm}
{\LARGE{Serge Leclercq}}
\vspace{5mm}

Th\`ese pr\'esent\'ee en vue de l'obtention 

du grade l\'egal de Docteur en Sciences

\vspace{5mm}

{\bf{Directeurs de th\`ese :}}

Dr Nicolas Boulanger

Prof. Philippe Spindel

{\bf{Membres du jury :}}

Prof. Glenn Barnich

Prof. Yves Brihaye

Prof. Marc Henneaux

Prof. Christian Michaux

Prof. Claude Semay

\vspace{1cm}

\includegraphics[width=7cm]{logo.jpg}

Ann\'ee acad\'emique 2008-2009

\end{center}

\thispagestyle{empty}

\newpage

${}$
\thispagestyle{empty}

\newpage

\section*{Remerciements}

A l'heure de terminer le travail difficile qu'est la r\'edaction d'une th\`ese de doctorat, je me dois naturellement de remercier les diff\'erentes personnes qui ont eu une importance, grande ou petite, dans cette r\'ealisation.

En premier lieu, je remercie Philippe Spindel, qui m'a soutenu et pouss\'e en avant, tout en faisant preuve de gentillesse, de patience et de compr\'ehension, depuis la fin de ma licence o\`u j'ai d\'ecid\'e de tenter ma chance dans la recherche. De plus, bien que n'ayant finalement r\'ealis\'e que peu de projets ensemble, il a toujours \'et\'e de bon conseil en tant que chef de service et directeur de th\`ese.

Ensuite, je remercie mon coll\`egue principal, Nicolas Boulanger, dont la culture en physique ne laisse pas de m'\'etonner. Depuis son arriv\'ee \`a Mons, il m'a guid\'e dans mon travail, les travaux r\'ealis\'es dans cette th\`ese n'auraient pas \'et\'e possibles sans lui. Il m'a \'egalement ouvert les yeux sur certaines choses, qu'elles concernent la physique, mon comportement... ou m\^eme la musique.

Je tiens maintenant \`a remercier les autres personnes avec qui j'ai travaill\'e. Tout d'abord Sandrine Cnockaert, avec qui j'ai eu l'occasion de collaborer \`a Bruxelles, lors d'un projet conjoint avec Nicolas. Ensuite, Xavier Bekaert, avec qui j'ai eu d'int\'eressantes discussions lors des deux premi\`eres \'editions de l'\'ecole de Modave. Enfin, Per Sundell, qui m'a tr\`es gentillement accueilli chez lui lors d'un s\'ejour \`a Pise o\`u nous avons travaill\'e avec Nicolas.

Un travail est plus gratifiant quand on est entour\'e de coll\`egues sympathiques. Je remercie tout particuli\`erement Claude Semay, Fabien Buisseret, Vincent Mathieu, Jean Nuyts, Fernand Grard, Martine Dumont, Yves Brihaye et Terence Delsate, dont la pr\'esence et le soutien dans ces derniers mois m'ont \'et\'e tr\`es b\'en\'efiques. Je remercie \'egalement Francis Michel et St\'ephane Detournay qui ont \'et\'e pr\'esents \`a Mons auparavant.

Enfin, bien s\^ur, je remercie ma famille au grand complet. En particulier mon p\`ere Daniel, mon fr\`ere Dominique et sa compagne Nad\`ege Capouillez, qui est \'egalement mon amie depuis que nous avons fait ensemble nos \'etudes de physique. Bien que n'\'etant pas en mesure de m'aider dans ma rédaction, leur pr\'esence dans les moments difficiles que l'on vit parfois en r\'ealisant un travail de cette envergure a \'et\'e essentielle.

\thispagestyle{empty}

\newpage

${}$

\thispagestyle{empty}

\setcounter{page}{0}

\tableofcontents
\newpage
${}$
\thispagestyle{empty}
\newpage

%%%%%%%%%%%%%%%%%%%%%%%%%%%%%%%%%%%%%%%%%%%%%%%%
%\chapter*{Introduction}
{\renewcommand{\thechapter}{}\renewcommand{\chaptername}{}
\addtocounter{chapter}{-1}
\chapter{Introduction}
{\markboth{Introduction}{\bf Introduction}}}

%%%%%%%%%%%%%%%%%%%%%%%%%%%%%%%%%%%%%%%%%%%%%%%%

\section*{Higher spin theories in Minkowski and Anti de Sitter backgrounds}

The subject of this thesis is to investigate the structure of higher spin theories. The concept of higher spin is well defined now, but it has evolved since its first appearance. The fundamental interactions that are considered nowadays only involve values of the spin lower than two. The three forces described in quantum field theory are Yang-Mills-like models, that involve some spin-1 gauge bosons that transmit the interactions and spin-$\frac{1}{2}$ massive fermions that constitute matter: the quarks and leptons. The first definition of higher spin was therefore ``spin higher than one''. However, some consistent, interacting theories can be defined for spin-$\frac{3}{2}$ and spin-2 fields, while only free theories are known for the higher values, with good reasons that we explain below. As a matter of fact, gravitation can be considered as a spin-2 theory. Einstein's theory describes the dynamics of the metric in relation with the matter content of spacetime. The metric is a tensor bearing two symmetric indices. By considering this theory as a perturbation around the Minkowski spacetime, the first order in the metric clearly behaves like a spin-2 field. As for the spin-$\frac{3}{2}$ field, it is part of supergravity models (where it is the gravitino). This is a first motivation to set the separation between lower and higher spin at two. Moreover, no mathematical model describing a {\em finite} set of interacting fields with spin greater than two is known, contrary to the spin-$\frac{3}{2}$ case, though a free theory is known, for bosons as well as for fermions.

Before we present our results, let us make a brief historical review of higher spin theories. Fields with an arbitrary spin were first introduced by Majorana in \cite{Majorana}, but it is Wigner \cite{Wigner:1939} who described them precisely in terms of unitary irreducible representations of the Poincar\'e group, the symmetry group of the four-dimensional Minkowski spacetime $\cm_4$. Each physical field is characterized by its mass and its spin, which are related to the eigenvalues of the Casimir operators. Two main sectors have to be considered: the massive sector, with $m^2>0$ and the massless sector, with $m=0$. The concept of spin is slightly different in the massless case, where it should rather be called helicity and corresponds to the projection of the spin angular momentum in the direction of the quadri-momentum of the particle. At the same time, Fierz and Pauli \cite{Fierz:1939ix} built free covariant actions describing spin-$\frac{3}{2}$ and spin-2 particles, then Rarita and Schwinger provided the spinor-tensor description of the free spin-$\frac{3}{2}$ theory \cite{Rarita:1941mf}. Let us emphasize that, although the spin is defined quantum mechanically, the fields and their spin have a meaning at the classical level, where equations of motion and Lagrangians can be built independently of any operator considerations. The first major step in the description of higher spin fields was the completion of Wigner's programme by Bargmann and Wigner \cite{Bargmann:1948ck}, which consisted in finding linear differential equations for each unitary representation of the Poincar\'e group. Later on, quadratic Lagrangians describing free massive fields were obtained by Singh and Hagen, for both bosonic and fermionic fields \cite{Singh:1974qz,Singh:1974rc}. Then, Fang and Fronsdal found a massless limit of these results \cite{Fronsdal:1978rb,Fang:1978wz}. In Fronsdal's theory, the spin-$s$ fields appear as tensors, totally symmetric and double traceless in their spacetime indices, bearing $s$ indices in the bosonic case, and $s-\frac{1}{2}$ spacetime indices plus a spinorial index in the fermionic case. Fronsdal's theory is a gauge theory. In dimension 4, the fixing of all the gauge freedom leaves only two components of the spin-$s$ fields on-shell, which are the desired helicity degrees of freedom. Furthermore, in dimension 4, the symmetric fields are the only ones needed to describe spin-$s$ free particles. 

In higher dimension $n>4$, the concept of spin is a bit less clearly defined, since the Poincar\'e group in dimension $n$ is larger and the number of its Casimir operators grows. From the tensorial point of view, the totally symmetric fields are not exhaustive, tensors with ``mixed'' symmetries must be considered. These symmetries are related to the representations of the permutation group, which are well described by Young tableaux. The tensors which have a maximum of $s$ symmetrizable indices are said to be spin-$s$ tensor fields. Equivalently, they can be visualized as Young tableaux whose first row is of length $s$. Spin-$s$ fields in arbitrary dimensions have been studied in \cite{deWit:1979pe,Curtright:1979uz}. A covariant description and a free Lagrangian have been given for bosonic mixed symmetry fields in \cite{Labastida:1987ft,Labastida:1989kw}. The interest of studying fields with an arbitrary spin in an arbitrary dimension arose in string field theory \cite{Siegel:1984wx,Siegel:1984xd,Banks:1986ff,Neveu:1986sh,Siegel:1986tw,Witten:1986cc}, in which a massive spectrum of modes with arbitrarily high spin appears, whose tensorial expressions show various symmetries. Though this theory is massive, the limit when the string tension vanishes leads to a massless higher spin theory, where all values of the spin should be present. This has been presented in \cite{Ouvry:1986dv,Koh:1986vg,Bengtsson:1986ys,Henneaux:1987cp}, and articles \cite{Labastida:1987ft,Labastida:1989kw} are presented in the context of string field theories as well. Furthermore, string field theory could as well possess a highly symmetric massless phase, giving back the usual theory by spontaneously breaking some symmetries. Let us notice that some other problems have been investigated more recently relating higher spin theory to some tensionless string theories \cite{Sundborg:2000wp,Sezgin:2001zs,Sezgin:2002rt}.

The next step in the investigation about higher spin fields was to consider the fields in a curved background. Fang and Fronsdal have extended their Lagrangians for totally symmetric fields to de Sitter ($dS$) and Anti de Sitter ($AdS$) spacetimes \cite{Fronsdal:1979vb,Fang:1979hq}. The major difference of such a background compared with Minkowski spacetime is the absence of translations in their symmetry group. Thus, the notion of mass itself becomes slightly less intuitive. A correspondence can be established between massless totally symmetric tensor fields in Minkowski spacetime and a category of totally symmetric tensor fields in $(A)dS$, which are called massless too by analogy. However, some ``mass terms'' are present, depending on the $(A)dS$ radius, or equivalently on the cosmological constant. These fields are similar to the Minkowski massless fields by virtue of their identical gauge nature. The higher spin massless fields in $(A)dS$ are very promising, thanks to the works of Vasiliev who first gave a new formulation of Fronsdal's theory in terms of generalized vielbeins and connections, which is called the frame formulation \cite{Vasiliev:1980as,Lopatin:1987hz,Vasiliev:1987td,Vasiliev:2001wa}. The fields in the frame formulation are equivalent to the double traceless totally symmetric tensor fields. Fradkin and Vasiliev managed to build an action, together with gauge transformations, that are consistent at cubic order \cite{Fradkin:1987ks,Fradkin:1987qy}. This describes for example the coupling of a spin-2 field with any spin-$s$ field. The key feature of this construction is the presence of negative powers of the cosmological constant in the vertex. It is thus not defined in Minkowski spacetime. Later on, Vasiliev constructed a set of equations of motion describing consistent interactions among an infinite set of fields, with any value of the spin \cite{Vasiliev:1990en,Vasiliev:1992av,Vasiliev:2003ev,Vasiliev:2004qz,Vasiliev:2004cp}. As a matter of fact, some conditions on the algebra that is used to describe the fields show that a consistent theory involving fields with spin $s>2$ have to involve an infinity of fields with an unbounded value of the spin. Unfortunately, no Lagrangian formulation has been built yet. Both the Fradkin--Vasiliev action and Vasiliev's equations of motion admit Fronsdal's theory as a free limit. On the other hand, it has not been showed yet that Vasiliev's equations are related to the cubic Fradkin--Vasiliev action alluded to before. In fact, the requirement of a non zero cosmological constant does not appear at first sight in Vasiliev's equations, since these equations are background independent, however $(A)dS$ spacetime is the most natural solution of Vasiliev's equation around which perturbation expansion can be done -- and has been done. The high spin fields together with the metric fluctuations around $(A)dS$ are taken as weak fields, see \cite{Vasiliev:1989yr,Sezgin:2000hr,Sezgin:2002ru}. Let us emphasize that Vasiliev's equations are proved to be non-trivial, several exact solutions have been studied, for example in \cite{Sezgin:2005pv,Sezgin:2005hf,Iazeolla:2007wt,Didenko:2006zd,Didenko:2008va,Didenko:2009tc}. Furthermore, the basic Vasiliev construction is a bosonic theory, but it admits supersymmetric extensions, that have been studied in dimension 4 (see \cite{Konstein:1989ij,Engquist:2002vr,Engquist:2002gy} and references therein, see also \cite{Sezgin:2001yf,Sezgin:2002rt} about dimension 5 and 7 cases).

To finish this review, let us present more particularly consistent deformations and cubic vertices. Given Fronsdal's free actions for both bosons and fermions, it is natural to try to deform them in powers of a coupling constant, in order to build interacting actions. This is called the Fronsdal programme. Many results have been obtained about this problem in Minkowski spacetime, using several different approaches. First, some cubic first order couplings have been obtained in the light-cone gauge in \cite{Bengtsson:1983pd,Bengtsson:1983pg,Bengtsson:1986kh,Fradkin:1991iy}. More recently, some more complete studies have been achieved in \cite{Metsaev:2005ar,Metsaev:2007rn}. However, since the gauge is fixed in that approach, the distinction cannot be made between Abelian and nonabelian deformations (i.e. deformations of the gauge transformations that still commute or not). On the other hand, the problem of gauge independent computations is slightly more complicated since one has to simultaneously deform the Lagrangian and the gauge transformations \cite{Berends:1984rq}. This was made for spin-3 fields by Berends, Burgers and van Dam, who obtained a first order cubic vertex in \cite{Berends:1984wp}. However, it was soon showed that this deformation cannot be continued to higher orders in the coupling parameter \cite{Berends:1984rq}. A conjecture was then made, about the need to introduce other fields with a higher value of the spin in order to cure the obstructions. Some couplings between gravity and higher spin fields were also considered, but once again, this is only consistent at first order \cite{Berends:1979wu,Aragone:1981yn,Deser:1990bk}. This indicates that consistent higher spin theories, compatible with the Einstein--Hilbert spin-2 theory, cannot be built in Minkowski spacetime, unless maybe every spin is present in the same theory. Some more results about cubic covariant couplings have been presented in \cite{Bengtsson:1985iw,Bengtsson:1983bp,Bengtsson:1986bz}. As we said above, the $(A)dS$ cubic deformation problem has been addressed by Fradkin and Vasiliev, who have provided their cubic action in dimension 4 \cite{Fradkin:1987ks,Fradkin:1987qy} and 5 \cite{Vasiliev:2001wa,Alkalaev:2002rq}. Furthermore, a construction of Abelian cubic vertices in operator formalism, in relation with string field theory, is presented in \cite{Buchbinder:2006eq, Fotopoulos:2008ka} and other related articles, in both Minkoswki and $AdS$ spacetime.

\section*{Problems addressed}

This thesis is a continuation of all the earlier works about consistent deformations and cubic vertices in Minkowski spacetime and in $(A)dS$ spacetime. Let us emphasize that our work concerns bosons, in an arbitrary dimension $n$. To be more precise about consistent deformations, let us say that they are expansions of initial theories in powers of a parameter $g$, that preserve the number and type of gauge transformations. We require the Minkowski deformations to be Poincar\'e invariant and local: a finite number of derivatives must be present at each order in the deformation parameter. This is not very restrictive, since the fields are defined in terms of representations of the Poincar\'e group and since the theories considered usually are local. However, even a local theory deformed in a local way can prove to be nonlocal in the end, which seems to be the case for higher spin theory, in which an unbounded number of derivatives appears, because of the presence of an infinite set of fields and because the number of derivatives increases with the spin. A powerful way to deal with gauge independent deformations is to use the antifield formalism for Lagrangian gauge theories. This is a cohomological formalism, similar to the Hamiltonian BRST formalism \cite{Becchi:1974md,Becchi:1975nq,Tyutin:1975qk}, first presented by Batalin and Vilkovisky \cite{Batalin:1981jr,Batalin:1983}. Many features that we used have been developed later in various publications \cite{Henneaux:1992ig,Barnich:1993vg,Barnich:1994,Henneaux:1998i,Barnich:1995db,Barnich:1995mt,Barnich:2000zw,Boulanger:2000rq}. In this formalism, the information about the theory is contained in a single functional, which is thus the only object to deform. This allows to determine exhaustively the possible consistent local deformations of the Lagrangian and gauge transformations that do not preserve the Abelian nature of the free gauge algebra. Several encouraging results have been obtained a few years ago. First, the antifield formulation allows one to recover Yang--Mills theory \cite{Barnich:1994} and Einstein--Hilbert theory \cite{Boulanger:2000rq} as the only consistent deformations for a set of spin-1 or spin-2 fields. Furthermore, it has been found that there is no multi-graviton theory (each spin-2 field deforms independently of the others). Then, a study of the spin-3 consistent deformations in arbitrary dimension has been achieved \cite{Bekaert:2006jf}, in which the BBvD vertex found in \cite{Berends:1984wp} has been rederived, as well as another vertex involving more derivatives. The obstruction about the BBvD vertex has been confirmed while the other vertex still needs some study. Let us also refer to \cite{Boulanger:2001vr,Bekaert:2006us} for reviews of the spin-2 and spin-3 deformation problems in Minkowski spacetime.

Various computations about cubic vertices in dimension-$n$ Minkowski spacetime are achieved in this thesis. First, the complete determination of consistent cubic vertices involving spin-2 and spin-3 fields is realized, for both configurations $2-2-3$ and $2-3-3$ (which means that the vertex is quadratic in the spin-2 or the spin-3 field). Then, we complete the spin-3 problem, by computing the parity-breaking cubic deformations. More generally, we compute the generic $1-s-s$ vertices and we provide the gauge algebra and gauge transformations of the only possible $2-s-s$ nonabelian cubic deformations. We also provide general rules to find the list of possible deformations of the gauge algebra for any $s-s'-s''$ cubic configuration. These considerations involving generic spin-$s$ fields highlight that the number of derivatives in the deformation increases with the spin. Finally, we make some computations at second order in the coupling constant. Some vertices appear to be strongly obstructed. The BBvD vertex cannot be saved by introducing spin-4 and spin-5 fields. Furthermore, some deformations that are not obstructed appear to be incompatible with Einstein--Hilbert theory. Nothing shows that a full Lagrangian theory cannot be built in Minkowski spacetime, but there seems to be serious restrictions on its existence. It remains to be showed if either it must involve every spin and an arbitrary number of derivatives, or the deformation vanishes.

We then consider Lagrangian deformations in $(A)dS$ backgrounds, which cannot be studied with currently the same success by using directly the antifield formalism. Fortunately, the complete determination of nonabelian Minkowski deformations can be related to the nonabelian $(A)dS$ deformations. We first study a quasi-minimal construction between spin-2 and spin-$s$ that has been suggested by Fradkin and Vasiliev and we establish a relation between those vertices and the possible first order deformations that we found in Minkowski spacetime. This relation, as well as results obtained in the light-cone gauge \cite{Metsaev:2005ar,Metsaev:2007rn}, can be used to show the uniqueness of the quasi-minimal deformation, which is thus the only $2-s-s$ $(A)dS$ deformation. In the same way, any $(A)dS$ cubic deformation must admit a particular Minkowski limit. However, since it is known that higher spin theories in $(A)dS$ contain fields with any spin and thus an arbitrary number of derivatives \cite{Vasiliev:2004cp}, it is known that such a theory never admits a flat limit. We have clarified this point at the level of the cubic action. Let us note that the presence of an arbitrary number of derivatives indicates the nonlocal nature of higher spin theories. The precise nature of these nonlocal interactions is still to be made more precise.

\section*{Overview of the thesis}

Most of our results in Minkowski spacetime are obtained thanks to the power of the antifield formalism. Therefore, we first present a review of gauge theories and the construction of the antifield formalism in {\bf Chapter \ref{ch:anti}}, as well as some general results.

The concept of massless spin-$s$ gauge field, in arbitrary dimension, and the Fronsdal action for totally symmetric tensor fields are recalled in {\bf Chapter \ref{ch:Frons}}, in Minkowski spacetime and in (Anti)de Sitter spacetime. We provide the antifield formulation of Fronsdal's theory in Minkowski spacetime, for any set of fields with different spins, as well as some particular results, very useful for the computation of consistent deformations.

In {\bf Chapter \ref{ch:antidef}}, we discuss the problem of consistent deformations of a gauge theory. In the antifield treatment of the problem, one equation, called the {\em master equation}, must be deformed. This allows us to establish general rules to build Poincar\'e invariant cubic deformations of a Fronsdal theory, containing a finite number of derivatives. If all the fields in presence have a spin lower than four, we also show that cubic deformations are the only possible first order deformations of a Fronsdal theory. 

In {\bf Chapter \ref{ch:FV}}, we present the construction of a quasi-minimal deformation of a sum of Fronsdal Lagrangians describing spin-2 and spin-$s$ fields in (Anti)de Sitter spacetime. It consists in completing an inconsistent minimal deformation of the spin-$s$ Lagrangian by adding terms containing more than two derivatives in the action and the gauge transformations. We then briefly present the Fradkin-Vasiliev action in $(A)dS$. Finally, we address the problem of taking a limit when the cosmological constant goes to zero, which can be done for any triplet of spin, and that we used to demonstrate the uniqueness of (Anti)de Sitter deformations when knowing the uniqueness of their equivalent in Minkowski spacetime. However, no limit can be taken on a infinite sum of terms involving terms with increasing spin and number of derivatives, thus we show that a full $(A)dS$ theory is not related to a full Minkowski theory.

The next chapters are devoted to the computation of cubic vertices in Minkowski spacetime. The general problem of combining spin-2 and spin-3 fields is addressed in {\bf Chapter \ref{ch:int23}}. Under the sole assumptions of locality, Poincar\'e invariance and nonabelian consistent deformation, we have found a unique vertex for a $2-2-3$ configuration, and a unique vertex for a $2-3-3$ configuration. The latter is the flat limit of the quasi-minimal $2-3-3$ deformation in $(A)dS$, which is thus proved to be unique as well.

In {\bf Chapter \ref{ch:exo3}}, we briefly recall results about the spin-3 cubic vertices obtained in \cite{Bekaert:2006jf}, and then address the determination of parity-breaking spin-3 deformations. There are two parity-breaking nonabelian cubic vertices, in dimension 3 and 5, involving respectively two and four derivatives.

We then make general considerations about $1-s-s$ and $2-s-s$ cubic vertices in Minkowski spacetime in {\bf Chapter \ref{ch:intmisc}}. We show that there is only one nonabelian cubic deformation in those cases. We build explicitly the $1-s-s$ deformation, that involves $2s-1$ derivatives. We also provide the $2-4-4$ deformation and relate it to the corresponding quasi-minimal deformation in $(A)dS$. Finally, we establish some explicit rules to build the possible cubic deformations of the gauge algebra for an arbitrary $s-s'-s''$ configuration, that are the only ways of building nonabelian vertices.

Finally, in {\bf Chapter \ref{ch:socomp}}, we compute a component of the second order of the master equation for the first order deformations that we have obtained in the previous chapters. This must allow the existence of a second order deformation in order for a full Minkowski theory to exist. We find that some first order deformations cannot be completed in an arbitrary dimension. In particular, we show that the BBvD spin-3 deformation is obstructed if $n\geqslant 4$. The unique $2-2-3$ vertex also leads to an inconsistency. We also demonstrate that the unique $2-3-3$ deformation cannot coexist with the spin-2 Einstein--Hilbert deformation.

After the {\bf Conclusions}, we present an {\bf Appendix} about Young tableaux.

\chapter{The antifield formalism\label{ch:anti}}

This first chapter consists of an introduction to the antifield formalism for local Lagrangian gauge theories. This formalism, also called the Batalin--Vilkovisky formalism, is the Lagrangian version of the BRST formalism. It provides a rigid structure instead of the usual arbitrary functions, and thus allows one to achieve computations without fixing the gauge, and under very few assumptions. We use the notation and the theoretical developments of references \cite{Henneaux:1992ig,Henneaux:1991rx,Henneaux:1998i,Barnich:1995db}. We first make some recalls about Lagrangian gauge theories and graded algebras. The fields that are considered throughout the thesis are assumed to be bosonic, but the formalism holds for fermions as well. The third section consists of the BRST-BV construction itself, while the last section is devoted to some results that we need in the next chapters.

\section{Local Lagrangian gauge theories}

\subsubsection*{The local action}

A local Lagrangian action is a functional, defined as the integral over a spacetime domain of a function called the Lagrangian density. This function depends on some fields $\phi^i$ and their derivatives up to finite order $l$: \begin{eqnarray}S[\phi]=\int_{\mathcal{D}\subset\mathbb{R}^n}\mathcal{L}(x^\mu,\phi^i,\partial_\mu\phi^i,...,\partial_{\mu_1...\mu_l}\phi^i)d^n x\quad.\end{eqnarray} 
The action is defined over the history space $I$, the functional space of the values of the fields. More precisely, a history is a section, relating the spacetime coordinates $x^\mu$ defined on $\cd$ and the element of the fiber space $\left(x^\m, \phi^i(x^\m)\right)$.
In the local case, the fiber space that can be considered is the jet space of order $l$, denoted $J_l$, and whose elements take the form $(x^\mu,\phi^i,\partial_\mu\phi^i,...,\partial_{\mu_1...\mu_l}\phi^i)$. We see that local Lagrangians are indeed functions over jet spaces. The interesting point is that the fiber is a vectorial space of finite dimension defined at each point $x^\m$ and not a set of functions. Though most of the definitions that will be given make sense in the general case, they are much easier to deal with in the local case.

First of all, let us define the total derivative with respect to a spacetime variable. By ``total'', we mean that, at a given point, the variables of the jet space are independent, but we must take into account the fact that the fields vary from one point to another. We adopt the notation $\6_\m$ for such a derivative, which must of course still be considered as being a partial derivative. Let us insist on the difference between $\6_\m$ and $\frac{\6}{\6 x^\m}$, which is the ``local partial'' derivative with respect to $x^\m$. The definition is as follows: \begin{eqnarray}\partial_\mu := \frac{\partial}{\partial x^\mu} + \partial_\mu\phi^i\frac{\partial}{\partial \phi^i}+...+\partial_{\nu_1...\nu_{l}\mu}\phi^i\frac{\partial}{\partial (\partial_{\nu_1...\nu_{l}} \phi^i)}\quad.\end{eqnarray}

\subsubsection*{Equations of motion and stationary surface}

The functional derivatives of the action provide the dynamical equations that select the histories that extremize $S$. In the case of a local theory, the functional derivatives of a functional coincide with the Euler-Lagrange variational derivatives $\frac{\d \cl}{\d \phi}$ of the integrand. The equations of motion are thus denoted: 
\begin{eqnarray}\frac{\d \mathcal{L}}{\d \phi^i}=\sum_{j=0}^l (-1)^j \partial^j_{\mu_1...\mu_j}\frac{\partial\mathcal{L}}{\partial(\partial^j_{\mu_1...\mu_j}\phi^i)}\ \approx 0\quad,\end{eqnarray} where the weak equality $\approx$ is a convenient way to distinguish the on-shell and the off-shell cases, which is needed because we will mostly work off-shell. The equations determine a surface in $I$ called the stationary surface: \begin{eqnarray}\Sigma\equiv\left\{\phi^i(x^\mu)\in I|\frac{\d \cl}{\d \phi^i}=0\right\}\quad.\end{eqnarray}

\noindent In $J_l$, the stationary surface is determined by the dynamical equations and their derivatives up to a finite order $m$: \begin{eqnarray}\Sigma_J\equiv\left\{(x_\mu,\phi^i,...,\6_{\mu_1...\mu_l}\phi^i)\in J_l|\forall j<m<+\infty : \partial_{\mu_1...\mu_j}\frac{\d \mathcal{L}}{\d \phi^i}=0\right\}\quad.\label{smoothsigma}\end{eqnarray} The weak equality can be extended to more general functions:
\begin{eqnarray}f\approx 0 \Leftrightarrow f =\sum_a \l_a\left(x^\m,\phi^i(x^\m)\right)f^a\left(\frac{\d \mathcal{L}}{\d \phi^i},...,\partial_{\mu_1...\mu_m}\frac{\d \mathcal{L}}{\d \phi^i}\right)\quad|\quad f^a(0,...,0)=0\quad,\end{eqnarray}and of course: $f\approx g \Leftrightarrow f-g\approx 0$.

\subsubsection*{Gauge transformations and Noether identities}

Gauge transformations are transformations of the fields that depend on arbitrary functions while leaving the action invariant. They take the generic form: \begin{eqnarray}\d_\varepsilon \phi^i = R^i_\a \varepsilon^\a\quad, \end{eqnarray} where $R^i_\a$ is basically a set of differential operators. Most of the times, the De Witt notation will be used for such operators. It consists in defining the operators as expansions in the distribution $\d$ and its derivatives and including an integration over $\cd$ in the summations over the indices of the operator. The operator is thus defined as follows:
 \begin{eqnarray}R^i_\a := \mathcal{R}^{i(0)}_\a \d^n(x^n-x'{}^n) + \mathcal{R}^{i\mu}_\a \partial_\mu\d^n(x^n-x'{}^n) + ... + \mathcal{R}^{i\mu_1...\mu_j}_\a \partial_{\mu_1...\mu_j}\d^n(x^n-x'{}^n)\quad.\end{eqnarray} The summation over $\a$ implicitly involves an integration over $x^\n$ while the summation over $i$ involves an integration over $x'{}^\n$. However, we will use the differential operator notation $\car^i_\a=\car^{i(0)}_\a+\car^{i\m}_\a \6_\m+...$ in some theorems, in that case, no integration is made. The coefficients depend on the coordinates of $J_l$, and the number $j$ of derivatives in $R^i_\a$ is once again finite. The gauge parameters $\varepsilon^\a$ are arbitrary functions defined on the spacetime domain $\mathcal{D}$.
The above transformations are gauge transformations if they satisfy: \begin{eqnarray}\forall \{\varepsilon^\a\}\ :\ \d_\varepsilon S = 0\quad. \end{eqnarray}
A complete set $R^i_\a$, that generates all the gauge symmetries of a given action, is called a generating set.
Let us detail what the variation of the action looks like: 
\begin{eqnarray}\d_\varepsilon S &=& \int_\mathcal{D} \d_\varepsilon \mathcal{L}\ d^n x = \frac{\d \mathcal{L}}{\d \phi^i} R_\a^i \varepsilon^\a\ \nonumber\\ &=& \int_\mathcal{D} \partial_\mu j^\mu\ d^n x +  (R^i_\a \frac{\d \mathcal{L}}{\d \phi^i})\varepsilon^\a = 0\quad\nonumber.\end{eqnarray}
The last step is obtained by integration by parts. In the operator notation, the second term could be written $\int(\car^{\dagger i}_\a\frac{\d \cl}{\d\phi^i})\e^\a d^n x$ where $\car^{\dagger i}_\a$ is the adjoint operator of $\car^i_\a$. The first term is the integral of a divergence and thus depends only on the values of the fields, the gauge parameters and their derivatives up to a finite order on the boundary of $\mathcal{D}$. We need the requirement that the fields have definite values on the boundary and thus that the gauge parameters and their derivatives vanish on the boundary while being totally arbitrary inside the domain. This implies that the coefficients of $\ve^\a$ in the second term identically vanish everywhere in $\mathcal{D}$. These relations among the equations of motion are called the Noether identities : \begin{eqnarray}R^{i}_\a \frac{\d \mathcal{L}}{\d \phi^i}\equiv 0\quad.\end{eqnarray} There is one and only one Noether identity for each gauge transformation. Conversely, studying the algebraic structure of the equations of motion is a good way to find every gauge transformation of a Lagrangian theory.

Let us remark that every theory possesses some trivial gauge transformations: \begin{eqnarray}\forall S : \d_\mu \phi^i = \mu^{[ij]}\frac{\d S}{\d \phi^j} \Rightarrow \d_\mu S = \frac{\d S}{\d \phi^i}\mu^{[ij]}\frac{\d S}{\d \phi^j} = 0\quad,\end{eqnarray}
where $\mu^{[ij]}$ is any antisymmetric operator\footnote{The fields have been considered here as commuting, which is the reason why the integrand of the last integral is identically zero. In the case where anticommuting fields would be present (for example spinorial fields), the corresponding components of the operator should be ``graded'' antisymmetric, in the sense that we recall in the next section.}(the simplest example is an arbitrary set of functions, but, with the same requirement of vanishing on the boundary, any operator antisymmetric modulo a divergence will cause the vanishing of this expression). Such transformations have no physical meaning and can be discarded.

\subsubsection*{Reducibility, gauge algebra}

The generating set can sometimes be overcomplete, for example to ensure Lorentz invariance. 
In that case, differential operators $Z_A^\a$ can be constructed such that the gauge transformations with gauge parameter $\varepsilon^\a = Z_A^\a \eta^A$ are trivial ($\eta^A$ is a shorter set of arbitrary functions): \begin{eqnarray}\exists\ \mu^{[ij]}\ |\ \d_\eta\phi^i=R^i_\a Z^\a_A\eta^A=\mu^{[ij]}\frac{\d S}{\d\phi^i}\quad.\end{eqnarray}
Such theories are called reducible of order (at least) 1.
It is possible to have further reducibilities if the $Z_A^\a$ operators form themselves an overcomplete set of reducibility conditions. 

\noindent Similarly to the Noether identities associated with gauge transformations, there are relations among the gauge generators associated with reducibilities.
 Indeed, by integrating by parts in the variation of the fields,
the identities $Z_A^{\a}R^{i}_\a = C^{[ij]}_A \frac{\d S}{\d\phi^j}$ are obtained, where $C^{[ij]}_A(\phi^i)$ is a new set of operators (let us notice that the operators $Z$ and $C$ are also considered here with the De Witt notation). 

Let us now define the gauge algebra. The commutator of two gauge transformations is required to be a gauge transformation: 
\begin {eqnarray}[\d_\e,\d_\eta]\phi^i=R^i_\a \zeta^\a+\mu^{[ij]}\frac{\d \mathcal{L}}{\d \phi^j}\quad,\label{galg}\end{eqnarray} where $\zeta^\a=C^\a_{\b\g}\e^\b\eta^\g$ and $\mu^{[ij]}=M^{[ij]}_{\b\g}\e^\b\eta^\g$. The double De Witt summation of the operators $C^\a_{\b\g}$ and $M^{[ij]}_{\b\g}$ means that these operators act independently on the two sets of gauge parameters. The first one is called the structure operator while the second tells to what extent the algebra is open off-shell. The Jacobi identity gives rise to on-shell conditions for these coefficients. For example: $C^\a_{\beta[\mu}C^\beta_{\nu\rho]}\approx 0$. These are consistency conditions of a theory.

\section{Gradings, differentials and homologies}

\subsubsection*{Grassmann parity, bosonic and fermionic fields}

Even when the fields are commuting, some anticommuting fields will have to be introduced. Therefore, we recall some general tools to handle commuting and anticommuting fields and operators.

Let us consider a set of fields \{$\phi^i$\} = $\{q^a,\theta^\a\}$ such that the $q^a$ commute with any $\phi^i$ and the $\theta^\a$ anticommute: \begin{eqnarray}q^a q^b = q^b q^a\ ,\ q^a \theta^\a = \theta^\a q^a\ and\ \theta^\a\theta^\beta= - \theta^\beta\theta^\a\quad.\end{eqnarray} It is natural to introduce the Grassmann parity: $\epsilon_a\equiv\epsilon(q^a) = 0$ and $\epsilon_\a\equiv\epsilon(\theta^\a)=1$. Commuting fields are also said to be even or bosonic. Anticommuting fields are odd or fermionic. With the above notation, the commutation rule of two fields $\phi^i$ can be written: \begin{eqnarray}\phi^i\phi^j=(-1)^{\epsilon_i\epsilon_j}\phi^j\phi^i\quad.\end{eqnarray}
Then, the set of all polynomials in those fields can be considered (with real or complex coefficients). Given the product and the commutation rule of the basic fields, it is clearly an associative algebra $A$. The Grassmann parity is also called a $\mathbb{Z}_2$-grading, because the value of the parity is 0 or 1. The algebra is said to be $\mathbb{Z}_2$-graded-commutative, or supercommutative. It is natural to consider that polynomials of degree zero in the fields are in the algebra, so that the number 1 is the unit for the product in this algebra. When the polynomials are composed only of terms of odd/even powers of the $\theta^\a$, they have a definite Grassmann parity. The following property holds: \begin{eqnarray}&\forall x,y\in A\ {\textrm{, with definite parities }}\epsilon_x{\textrm{ and }}\epsilon_y\ :&\nonumber\\&\epsilon_{xy}=\epsilon_x+\epsilon_y\quad.&\end{eqnarray}
Even elements define a subset $A_0\subset A$, odd ones define $A_1\subset A$. Any element of $A$ can be written as the sum of an odd and an even elements, that is why: $A=A_0\oplus A_1$. This remains consistent if the algebra is extended to any function of the bosonic fields. Let us remark that because of the finite number, say $k$, of the anticommuting $\theta^\a$, there are no polynomials of degree higher than $k$ in the fermionic fields.

\subsubsection*{Operators, differentials and gradings}

The set $End(A)$ is the set of endomorphisms of $A$ (i.e. the set of linear maps from $A$ to $A$). Given the composition of operators, $End(A)$ is an associative algebra with unit. Some operators can have a definite Grassmann parity : $\tau\in End(A)$ is said to be even if $\forall x\in A\ |\ \exists\epsilon_x\ :\ \epsilon_{\tau(x)}=\epsilon_x$, it is said to be odd if $\forall x\in A\ |\ \exists\epsilon_x\ :\ $ $\epsilon_{\tau(x)}=1-\epsilon_x$. $End(A)$ is the direct sum of its subsets of definite parity. The following property holds: \begin{eqnarray}&\forall \tau,\sigma\in End(A){\textrm{ , with definite parities }}\epsilon_\tau{\textrm{ and }}\epsilon_\sigma\ :&\nonumber\\&\epsilon_{\tau\sigma}=\epsilon_\tau+\epsilon_\sigma\quad.&\end{eqnarray} In general, $End(A)$ is not  supercommutative. It is thus useful to introduce the graded commutator of two operators $\tau$, $\sigma$ with definite parities: \begin{eqnarray}[\tau,\sigma]=\tau\sigma - (-1)^{\epsilon_\tau\epsilon_\sigma}\sigma\tau\quad.\end{eqnarray} 
This commutator satisfies a graded Jacobi identity: 
\begin{eqnarray}[[\rho,\sigma],\tau]+(-1)^{\varepsilon_\rho(\varepsilon_\sigma+\varepsilon_\tau)}[[\sigma,\tau],\rho]+(-1)^{\varepsilon_\tau(\varepsilon_\rho+\varepsilon_\sigma)}[[\tau,\rho],\sigma]=0\quad.\end{eqnarray}

In the same way, it is often possible to introduce $\mathbb{N}$-gradings or $\mathbb{Z}$-gradings in an algebra. Generally, $A$ is then the direct sum of a set of classes labeled by a natural or integer number: 
\begin{eqnarray}A = \bigoplus_{n\in \mathbb{N}\,or\,\mathbb{Z}} A_n\quad{\textrm{such that}}\quad \forall\ n,m : A_n A_m\subset A_{n+m}\quad.\end{eqnarray}
The label is called a grading or a degree, and is denoted : $x\in A_n\Leftrightarrow deg\ x=n$. 
A common example is the polynomial degree in some of the generators $\phi^i$ of $A$ (in particular, the form degree related to the exterior differential is such a grading). More generally, any sum of integer multiples of such polynomial degrees is a grading. The $\mathbb{N}$($\mathbb{Z}$)-grading can be extended to operators: \begin{eqnarray}deg\ \tau = n \Leftrightarrow \forall m, \forall x\in A_m : \tau(x)\in A_{n+m}\quad.\end{eqnarray} $End(A)$ is the direct sum of its subsets of definite degree.

\noindent \textbf{Remark:} $0$ has no definite degree, as it belongs to every class $A_n$ (and of course, the same holds for the 0 operator). 

Derivatives are operators with a definite parity, that satisfy a $\mathbb{Z}_2$-graded Leibniz rule. We mostly use derivatives acting from the left, the definition of a right derivative is similar:
\begin{eqnarray}& d\in End(A) \textrm{ is a left derivative}&\nonumber\\&\nonumber\Leftrightarrow &\\&\forall x,y\in A\textrm{, with definite parity: }\ d(xy)= (dx)y + (-1)^{\epsilon_x\epsilon_d}xdy\quad.\nonumber\end{eqnarray} Derivatives form a subalgebra of $End(A)$, that will be denoted $Der(A)$, and which also decomposes into classes of definite degree. Let us note that the commutator is internal to $Der(A)$.

A differential $D$ is a nilpotent odd derivative: \begin{eqnarray}D\in Der(A) \textrm{ is a differential}\Leftrightarrow D^2 = 0 \textrm{ and } \epsilon_D = 1\quad.\end{eqnarray} We assume that there is a $\mathbb{N}$ or $\mathbb{Z}$-grading such that $deg\ D = 1$ or $-1$ (in the case of a $\mathbb{Z}$-grading, there is a freedom on the sign of the label and it can always be chosen such that $deg\ D=1$).

\subsubsection*{Homology and cohomology}

The kernel of $D$ is the set \begin{eqnarray}Ker\ D=\left\{x\in A\ |Dx=0\right\}\end{eqnarray} and the image of $D$ is the set \begin{eqnarray}Im\ D=\left\{x\in A\ |\ \exists y\in A\ |\ x=Dy\right\}\quad.\end{eqnarray} Elements of $Ker\ D$ are called $D$-closed objects ; elements of $Im\ D$ are called $D$-exact objects. Furthermore, an equation of the type $Dx=0$ is called a cocycle and an equation of the type $x=Dy$ is called a coboundary.

Because of the nilpotency of $D$, it is obvious that $Im\ D\subset Ker\ D$. Furthermore, $Im\ D$ is an ideal of $Ker\ D$ 
: $\forall x\in Ker\ D, y\in Im\ D\ :\ \exists z\ |\ yx=(Dz)x=D(zx)\in Im D$. We can thus define the coset space of $Ker\ D$ modulo $Im\ D$: \begin{eqnarray}H(D)= \frac{Ker\ D}{Im\ D}=\left\{[a]\ |\ Da=0\ ,\ a'\in[a]\Leftrightarrow \exists b| a'=a+Db\right\}\quad.\end{eqnarray}
If $deg\ D=-1\;$, this space is called the homology of $D$ and is denoted $H_*(D)\,;$ if $deg\ D=1$, it is called the cohomology of $D$ and is denoted $H^*(D)$. (Co)homologies are the direct sum of their classes of definite degree, which will often be considered individually: \begin{eqnarray}H_*(D)=\bigoplus_{n\in\mathbb{N}} H_n(D)\quad \textrm{or}\quad H^*(D)=\bigoplus_{n\in\mathbb{N}{\textrm{ or }}\mathbb{Z}} H^n(D)\quad.\end{eqnarray}

It is also possible to define the (co)homology of a differential $D$ in $Der(A)$. When a derivative $d$ supercommutes with $D$ : $[d,D]=0$, $d$ is said to be $D$-closed. When there is a derivative $\Delta$ such that $d = [\Delta,D]$, $d$ is said to be $D$-exact. The set of $D$-closed derivatives is a subalgebra of $Der(A)$ for the commutator. The Jacobi identity provides that: \begin{eqnarray}\nonumber\begin{array}{l}\forall\ d\ :\ [D,[D,d]]=\pm\frac{1}{2}[d,[D,D]]=0\\  \forall\,d\,|\,[d,D]=0\ ,\ \forall\,\Delta\ :\ [d,[D,\Delta]]=\pm[D,[\Delta,d]]\quad.\end{array}\end{eqnarray}This means that the $D$-exact derivatives form an ideal of the $D$-closed ones. The (co)homology of $D$ in $Der(A)$ is the coset space and is denoted $\mathcal{H}_*(D)$ or $\mathcal{H}^*(D)$.

Some derivatives are differentials only in a subspace $B\subset A$. A derivative $\g$ is called a differential in $B$ if the projection of its square on $B$ vanishes. For example, we consider in the sequel the algebra of functions of the fields $C^\infty(I)$ and the subset of functions on the stationary surface $C^\infty(\Sigma)$, both in tensor product with a polynomial space in some fermionic fields $C^\a$. Any derivative $\g$ such that $\g^2\approx 0$ will then be called a differential in $B=C^\infty(\Sigma)\otimes\mathbb{R}[C^\a]$ (remember that $\approx$ means ``equal on the stationary surface''). The cohomology of $\g$ in $B$ can also be defined, which is denoted $H^*(\g,B)$.

Finally, let us define a differential modulo a differential. Let $D$ be a differential in $A$, of negative grading. If $\g$ is an odd derivative such that $[D,\g]=0$ and $\exists \Delta\ :\ \g^2=-[D,\Delta]$, then $\g$ is called a differential modulo $D$. This implies that $\g$ is a differential in a space of representatives of $H_*(D)$ : if $Da=0$ then $\g^2 a=D(\Delta a)\in[0]\in H_*(D)$. In the sense of $\mathcal{H}_*(D)$, $\g$ is $D$-closed and $\g^2$ is $D$-exact. The cohomology of $\g$ modulo $D$ is denoted $H(\g|D)$ or $H(\g,H_*(D))$. \begin{eqnarray}H(\g|D)=\left\{[a]\ |\ Da=0\ ,\ \exists b\;|\;\g a+Db=0\ {\textrm{ and }}\ a'\in [a]\Leftrightarrow \exists c,e\;|\;a'=a+\g c+De\right\}\quad.\end{eqnarray}

\noindent Remark : in the case where $\g$ is a true differential (i.e. $\Delta=0$), the cohomology of $\gamma$ is of course defined in $A$, the condition $D a=0$ is removed and the cohomology of $\g$ modulo $D$ is then $$H(\g|D)=\left\{[a]\ |\ \exists b\;|\;\g a+Db=0\ {\textrm{ and }}\ a'\in [a]\Leftrightarrow \exists c,e|a'=a+\g c+De\right\}\quad.$$

\section{Construction of the formalism}

\subsection{Longitudinal derivative $\g$}

Two sets of values of the fields differing by a gauge transformation, say $\phi^i(x^\mu)$ and $\phi^i(x^\mu)+R_\a^i\varepsilon^\a$, give the same value of the action, by definition of the gauge transformations. This is true in particular for histories extremizing the action: if $\phi^i(x^\mu)\in\Sigma$, then $\forall \varepsilon^\a(x^\mu) : \phi^i(x^\mu)+R_\a^i\varepsilon^\a \in\Sigma$. The gauge transformations thus generate submanifolds on the stationary surface, which are called gauge orbits, with a dimension equal to the number $M$ of gauge transformations. This dimension has to be seen as the number of functions $\varepsilon^\a$, which is not very satisfying, because of the functional nature of the history space. In the local case, in which the jet space $J_l$ is considered instead, the dimension of the gauge orbits is the number of $\varepsilon^\a$'s and their derivatives up to the appropriate finite order. 

The gauge orbits are subspaces of the fiber, we can formally consider a basis of those, denoted $X_\a$. In terms of a natural basis $E_i$ of the fiber, these generators are such that $E_i R^i_\a \ve^\a=X_\a\ve^\a$. In the jet space, one may consider a basis $E_{i\mu_1...\mu_j}\ j\leqslant l$ of the fiber and a basis $X_{\a\mu_1...\mu_i}$ of the gauge orbits. The Lie bracket of the $X_\a$ reproduces the coefficients of the gauge algebra: $[X_\a,X_\b]\approx C^\g_{\a\b}X_\g$. A weak equality must be considered, because the gauge algebra is only closed on-shell. One may now consider the dual space of the gauge orbits, which is generated by fermionic objects $C^\a$, that are called ghosts. In the jet space, the dual space is generated by the ghosts and their derivatives up to a finite order. 

Then, a derivative $\g$ can be introduced, which acts along the gauge orbits, and is thus called the longitudinal derivative. The action of such a derivative on the fields is similar to considering an arbitrary move along the gauge orbits, i.e. a gauge transformation. Its action on the fields thus takes the form \begin{eqnarray}\g \phi^i = R_\a^i C^\a\quad.\end{eqnarray} The dual version of the Lie bracket of the $X_\a$ generators is the action of $\g$ on the ghosts : \begin{eqnarray}\g C^\a=-\frac{1}{2}C_{\beta\g}^\a C^\b C^\g\quad.\end{eqnarray} 
Let us now study $\g^2$. On $\phi^i$, we get: \begin{eqnarray}\g^2\phi^i=\g\left(R^i_\a C^\a\right)=\frac{\d R^i_\a}{\d\phi^j}R^j_\b C^\b C^\a -\frac{1}{2}R^i_\a C^\a_{\b\g}C^\b C^\g\quad.\end{eqnarray} The coefficient of the commutators of two gauge transformations is given by: \begin{eqnarray}[\d_\ve,\d_\eta]\phi^i=\left[\frac{\d R^i_\b}{\d\phi^j}R^j_\a-\frac{\d R^i_\a}{\d\phi^j}R^j_\b\right]\ve^\a \eta^\b\quad.\end{eqnarray} Thus, thanks to Eq.(\ref{galg}) and the anticommuting nature of the ghosts, we finally get: \begin{eqnarray}\g^2\phi^i=\frac{1}{2}\frac{\d \cl}{\d\phi^j}M^{ij}_{\a\b}C^\a C^\b \approx 0\quad.\end{eqnarray}
The action of $\g^2$ on the ghosts yields: \begin{eqnarray}\g^2 C^\a=C^\a_{\beta\g}\g C^\g C^\beta=\frac{1}{2}C^\a_{\beta\g}C^\g_{\mu\nu}C^\beta C^\mu C^\nu\approx 0\quad,\end{eqnarray} thanks to the weak Jacobi identity of the gauge algebra, the antisymmetry being due to the product of fermionic ghosts. We have thus proved that $\g$ is a differential on the stationary surface but not off-shell. The grading related to $\g$ is the polynomial degree in the ghosts, which is called the pure ghost number: \begin{eqnarray}pgh\ C^\a = 1\ ,\ pgh\ \phi^i = 0\ ,\ pgh\ \g=1\ .\end{eqnarray} The cohomology of $\g$ in $\Sigma$ can be defined, and is denoted $H^*(\g,C^\infty(\Sigma)\otimes\mathbb{R}[C^\a])$. We have considered an algebra of functions on $\Sigma$ and polynomial in the ghosts, which are the most general arguments of $\g$ for the moment. The cohomology in $pgh$ 0 is the set of functions closed under $\g$, which are none other than the gauge invariant functions. There are no $\g$-exact objects in $pgh$ 0, because the pure ghost number is a natural degree raised by $\g$.

The ghosts are defined at any point of the manifold, they can thus be considered as fields of an extended history space. In the local case, the ghosts and their derivatives up to a finite order are added to build an extended jet space, where $\g$ is an algebraic operator. The longitudinal derivative is assumed to commute with partial derivatives. In terms of the exterior derivative of the manifold, which is an odd operator, the relation $\g d+d\g=0$ is satisfied.

\subsection{Koszul-Tate differential $\d$}

Up to this point, we have managed to define a symmetry on the stationary surface that replaces the arbitrary gauge transformations. The second step is to relate the fact of being on-shell to a rigid transformation that is a differential off-shell. To do so we must build a homological resolution of the algebra where $\g$ is a differential.

A homological resolution of an algebra $A$ is realized when there is a differential $\d$, acting in an algebra $A'\supset A$, related to a grading labeled by a natural number $k$ with $deg\ \d=-1$, such that: \begin{eqnarray}\forall k>0 : H_k(\d)=0{\textrm{ and }}H_0(\d)\cong A\quad.\end{eqnarray} The second equation is an isomorphism, the elements of $A$ being representatives of the homology cosets.

In our case, the algebra $A$ will be $C^\infty(\Sigma)\otimes\mathbb{R}[C^\a]$ and the algebra $A'$ will be an extension of $C^\infty(I)\otimes\mathbb{R}[C^\a]$.
The grading of $\d$ is called the antifield number, and denoted $antigh$. 
By definition, if $C^\infty(\Sigma)\otimes\mathbb{R}[C^\a]$ is to be in the $antigh$ $0$ class, it is required that: \begin{eqnarray}antigh\ \phi^i=0 \textrm{ and } antigh\ C^\a=0\quad.\end{eqnarray}
Since $antigh\ \d=-1$ and the antifield number is a natural grading, it is required that: \begin{eqnarray}\d\phi^i=0\textrm{ and } \d C^\a=0\quad.\end{eqnarray} If there were no further fields, there would not be any $\d$-exact combinations of the $\phi^i$, and the homological class in degree zero would be $C^\infty(I)\otimes\mathbb{R}[C^\a]$. To restrict  it to $C^\infty(\Sigma)$, we must somehow make the equations of motion  $\d$-exact. The solution is to extend the space so as to include a new set of fields with $antigh$ $1$, called the ``$antigh$ 1 antifields'' and denoted $\phi^*_i$. There is the same number of antifields $\phi^*_i$ than fields $\phi^i$. The nature of the antifields does not matter, they are just considered through the differential $\d$. The action of $\delta$ on the antifields is defined as follows:
\begin{eqnarray}\d\phi^*_i:=\frac{\d S}{\d\phi^i}\ ,\ antigh\ \phi^*_i=1\quad.\end{eqnarray}This implies that the parity of an antifield is the opposite of that of the corresponding field: \begin{eqnarray}\varepsilon(\phi^*_i)=1-\varepsilon(\phi^i)\quad.\end{eqnarray}Furthermore, we find that $\d^2\phi^*_i=0$ automatically. 

Then, we have to make sure that the other homology classes are zero. For the moment, this is not the case: the Noether identities imply that $\d (R^i_\a\phi^*_i)=0$. The combinations $R^i_\a\phi^*_i$ are $\d$-closed but not $\delta$-exact, hence they would appear in $H_1(\d)$. It is thus required to introduce another set of antifields $C^*_\a$, called the ``$antigh$ 2 antifields'', such that \begin{eqnarray}\d C^*_\a :=R^i_\a\phi^*_i\ ,\ antigh\ C^*_\a=2\quad.\end{eqnarray}By construction, $\d^2 C^*_\a=0$. If the theory is irreducible, there are no relations among the gauge generators, no combinations of the $C^*_\a$ can be $\d$-closed and the construction stops. 

If the theory is reducible of order one, it is needed to introduce a further set of antifields, of $antigh$ $3$, $C^*_A$, such that $\d C^*_A=-Z^\a_A C^*_\a-\frac{1}{2}C^{[ij]}_A \phi^*_i\phi^*_j$ in order to compensate for the redundancy of the gauge generators. Unfortunately, when the theory is reducible, the longitudinal derivative cannot be easily extended to the whole space. It has to be replaced by another derivative, $\tilde{\g}$, called a ``model'' for $\g$, the cohomology of which is isomorphic to $H(\g)$ but whose action on the different families of fields is a bit more complicated. Furthermore, new families of ghosts, corresponding to the different generations of antifields, have to be introduced. For example, at first order of reducibility, a family of bosonic fields $C^A$ of $pgh$ 2 are introduced and one has $\tilde{\g}C^\a=\frac{1}{2}C^\a_{\beta\g}C^\beta\ C^\g+Z^\a_A\ C^A+...$ We will not give more details about this, since we have only worked on irreducible theories.

As usual, this was a bit formal but everything is very well defined in $J_l$. The jet space is just extended to the different families of antifields and their derivatives up to a finite order, then $\d$ is a simple derivation commuting with $\partial_\mu$ (or anticommuting with the exterior derivative $d$), namely: 
\begin{eqnarray}\displaystyle\d=\sum_{j=0}^{j_{max}}\partial_{\mu_1...\mu_j}(R_\a^i \phi^*_i)\frac{\partial}{\partial(\partial_{\mu_1...\mu_j} C^*_\a)}+\sum_{m=0}^{m_{max}}\partial_{\nu_1...\nu_m}\frac{\d\mathcal{L}}{\d\phi^i}\frac{\partial}{\partial(\partial_{\nu_1...\nu_m} \phi^*_i)}\quad.\end{eqnarray}
This definition ensures that $\forall k>0 : H_k(\d)=0$ in $J_l$. An important fact is that this construction not only ensures that $\d$ provides a resolution of $C^\infty(\Sigma)\otimes\mathbb{R}[C^\a]$ in $\mathbb{R}[C^*_\a,\phi^*_i]\otimes C^\infty(I)\otimes\mathbb{R}[C^\a]$, but it also provides a resolution of $Der(C^\infty(\Sigma)\otimes\mathbb{R}[C^\a])$ in $Der(\mathbb{R}[C^*_\a,\phi^*_i]\otimes C^\infty(I)\otimes\mathbb{R}[C^\a])$. This means that $\forall k>0\ :\ \mathcal{H}_k(\d)=0$ and $\mathcal{H}_0(\d)=Der(C^\infty(\Sigma)\otimes\mathbb{R}[C^\a])$. The proof can be found in \cite{Henneaux:1992ig}. This property of the homology of $\d$ in the derivative space is required for the consistency of the antifield formalism, as we will see in a subsequent theorem.

\subsection{The differential $\bs$}

First, we have to define the action of $\g$ on the antifields. In fact, this is quite arbitrary, and it is possible to choose $\g\phi^*_i$ and $\g C^*_\a$ in order for $\g$ and $\d$ to anticommute. It is obvious that the latter property is already true for the fields and the ghosts, because $[\g,\d]$ is an $antigh$ $-1$ operator. For the $antigh$ 1 antifields, by taking: \begin{eqnarray}\g\phi^*_i=\frac{\d[\phi^*_jR^j_\a C^\a]}{\d \phi^i}\quad,\end{eqnarray} it is found that $(\d\g+\g\d)\phi^*_i=0$ thanks to the Noether identities and to the fact that the variational derivatives of a divergence are identically zero. The value of $\g C^*_\a$ is chosen similarly (we do not write it explicitly but it will be straightforward to recover it given the differential $\bs$ a bit later). 

Since $\gamma$ anticommutes with $\delta$, it is $\d$-closed. In addition to that, $\g^2$ is $\d$-exact: \begin{eqnarray}\exists\ \D\in Der(A)\ |\ \g^2=-\d\Delta-\Delta\d\ ,\ antigh\ \D=1\quad.\end{eqnarray}The action of $\Delta$ can be built, on objects of increasing $antigh$, using the property that $\g^2 a\approx 0$ if $\d a=0$.
 Given an $antigh$ $0$ object $a_0$, which trivially satisfies $\d a_0=0$ :  $\exists\ b\ |\ \g^2 a_0=\d b\,.$ We can simply define $\Delta a_0= -b$ and the relation $\g^2=-\d\Delta-\Delta\d$ holds (when acting on $antigh$ 0 quantities).
Then, if $a_1$ is an $antigh$ $1$ object, $\delta a_1$ is of $antigh$ 0, so it satisfies $\g^2\d a_1=-\d\Delta\d a_1$. This can be rewritten as $\d[\g^2 a_1+\Delta\d a_1]=0$. The vanishing of $H_1(\d)$ now implies that $\exists c|\g^2 a_1=-\Delta\d a_1-\d c$, and the action of $\Delta$ in $antigh$ $ 1$ is defined as $\Delta a_1= c$. The same kind of argument gives the value of $\Delta$ for higher $antigh$.  Putting the two properties of $\gamma$ with respect to $\delta$ together, we have shown that $\g$ is a differential modulo $\d$.

Now, the conditions $\d^2=0$, $\g\d+\d\g=0$ and $\g^2+\d\Delta+\Delta\d=0$ can be seen as the first three  terms of the $antigh$ expansion of the equation $\bs^2$=0 with $\bs=\d+\g+\Delta+\textrm{higher}\ antigh\ \textrm{terms}$. We will prove that, given that $\d$ provides a resolution in the derivative space, such a differential $\bs$ actually exists. This differential, called the BRST-BV differential, defined on the whole functional space and which encodes every characteristic of the gauge theory, is the central object of the formalism.

\begin{theorem}\label{homper} Homological perturbations: Let us show that if:

\begin{itemize}
\item $\d$ is a differential related to the $\mathbb{N}$-grading $antigh$ noted $k$,\\ with $antigh\ \d=-1$, such that $\forall k>0\ :\ H_k(\d)=0\textrm{ and }\mathcal{H}_k(\d)=0$ ;
\item$\g$ is a differential modulo $\d$ (of $antigh$ $0$), i.e. $\g\d+\d\g=0$\\ and $\exists\stackrel{(1)}{s}\ |\ \g^2=-[\d,\stackrel{(1)}{s}]$ ;

$\g$ is associated with a $\mathbb{N}$-grading $pgh$, with $pgh\ \g=1$ and $pgh\ \d=0$ ; 
\end{itemize}
Then there exists a differential $\bs$ associated with the $\mathbb{Z}$-grading $gh=pgh-antigh$ with $gh\ \bs=1$, and such that $\displaystyle \bs=\d+\g+\sum_{k\geqslant 1}\stackrel{(k)}{\bs}$ with $antigh\ \stackrel{(k)}{\bs}=k$.

\noindent Furthermore, the cohomology classes of $\bs$ in positive $gh$ are isomorphic to those of $\g$ in $antigh$ $0$ : \begin{eqnarray}\nonumber &\forall i\geqslant 0\ :\ H^i(\bs)\cong H^i(\g,H_0(\d))&\\&\forall i<0\ :\ H^i(\bs)=0&\end{eqnarray} where $i$ is the $gh$ number for $\bs$ and is the $pgh$ number for $\g$.
In particular, gauge-invariant functionals correspond to cosets of the group $H^0(\bs)\cong H^0(\gamma, H_0(\delta))$.\end{theorem}

\vspace{2mm}
\noindent\underline{Proof}

\begin{enumerate}
\item
Let us consider $\bs_n=\d+\g+\stackrel{(1)}{\bs}+...+\stackrel{(n)}{\bs}$ and let us assume that its square has no terms of $antigh<n$ : $\bs_n^2=\stackrel{(n)}{\rho}+\stackrel{(n+1)}{\rho}+\stackrel{(n+2)}{\rho}+...$

The hypothesis tells us that it is true for $n=1$. It is sufficient to show that for any $n$, there exists $\stackrel{(n+1)}{\bs}$ such that $\bs_{n+1}^2$ begins at $antigh$ $(n+1)$.

It is trivial that $[\bs_n^2,\bs_n]=\bs_n^3-\bs_n^3=0$. The term of lowest $antigh$ of this expression is $[\stackrel{(n)}{\rho},\d]=0$, so $\stackrel{(n)}{\rho}$ is $\d$-closed. 

But we know that $\mathcal{H}_n(\d)=0$, thus $\exists \stackrel{(n+1)}{\bs}\ |\ \stackrel{(n)}{\rho}=-[\stackrel{(n+1)}{\bs},\d]$.

If $\bs_{n+1}=\bs_n+\stackrel{(n+1)}{\bs}$,\\ then $\bs_{n+1}^2=\stackrel{(n)}{\rho}+\d\stackrel{(n+1)}{\bs}+\stackrel{(n+1)}{\bs}\d+\stackrel{(n+1)}{\rho'}+...=\stackrel{(n+1)}{\rho'}+...$.

This proves the existence of the full $\bs$.

\item

Let us consider any element $x$ of the algebra, it can be expanded according to the $antigh$ number: $\displaystyle x=\sum_{k\geqslant 0} \stackrel{(k)}{x}$. The isomorphism needed to prove the second statement is simply given by the map $\pi$ that applies $x$ on $\stackrel{(0)}{x}$. The reason is that $\bs x=\g \stackrel{(0)}{x}+\d \stackrel{(1)}{x}+...$ so $\pi \bs x=\g\pi x$ in $H_0(\d)$.
It is clearly a morphism: $\forall x,y\ :\ \pi(xy)=\stackrel{(0)}{x}\stackrel{(0)}{y}=\pi(x)\pi(y)$.
\begin{enumerate}
\item \underline{$\pi$ is surjective:}
We need to prove that any $antigh$ $0$ object $x_0$ that is $\g$-closed in $H_0(\d)$ can be deformed into an $\bs$-closed $x$.
By assumption, $\g \stackrel{(0)}{x}\approx 0\Rightarrow\exists \stackrel{(1)}{x}\ |\ \g \stackrel{(0)}{x}+\d \stackrel{(1)}{x}=0$.\\ This tells us that $\bs(\stackrel{(0)}{x}+\stackrel{(1)}{x})=\stackrel{(1)}{t}+\stackrel{(2)}{t}+...$.

Now, it is sufficient to prove that if $x_n=\stackrel{(0)}{x}+\stackrel{(1)}{x}+...+\stackrel{(n)}{x}$ is such that $\bs x_n=\stackrel{(n)}{t}+\stackrel{(n+1)}{t}+...$ then there exists a $\stackrel{(n+1)}{x}$ such that $\bs x_{n+1}$ begins at $antigh$ $n+1$. 

Since $\bs^2 x_n=0$, its lowest $antigh$ term vanishes too: $\d \stackrel{(n)}{t}=0$, and as $H_n(\d)=0$:\\ $\exists\ \stackrel{(n+1)}{x}\ |\ \stackrel{(n)}{t}=-\d \stackrel{(n+1)}{x}$. 

If $x_{n+1}=x_n+\stackrel{(n+1)}{x}$, then $\bs x_{n+1}=\stackrel{(n)}{t}+\d\stackrel{(n+1)}{x}+\stackrel{(n+1)}{t'}+...=\stackrel{(n+1)}{t'}+...$

By induction, the full $x$ can be constructed, thus $\pi$ is surjective.

\item\underline{$\pi$ is injective:}
We have to prove that: \begin{eqnarray}\bs x=0,\ \pi x\in [0]\subset H^i(\g,H_0(\d))\Rightarrow x\in[0]\subset H^i(\bs)\quad.\nonumber\end{eqnarray}
More explicitly, the left-hand side means that $\exists\ \stackrel{(0)}{z},\stackrel{(1)}{z}\ |\ \stackrel{(0)}{x}=\g\stackrel{(0)}{z}+\d\stackrel{(1)}{z}$. 

If $x'=x-\bs(\stackrel{(0)}{z}+\stackrel{(1)}{z})$, then $\bs x'=0$ and $x'$ begins at $antigh$ $1$, so $\d\stackrel{(1)}{x'}=0$. Since $H_1(\d)=0$, we find that $\exists \stackrel{(2)}{z}\ | \stackrel{(1)}{x'}=\d \stackrel{(2)}{z}$. Then $x''=x-\bs(\stackrel{(0)}{z}+\stackrel{(1)}{z}+\stackrel{(2)}{z})$ is such that $\bs x''=0$ and $x''$ begins at $antigh$ $2$. Going on like this recursively, the different $antigh$ components of $x$ are removed one by one. It is finally found that $x$ is $\bs$-exact, which proves that $\pi$ is injective.
\end{enumerate}
Thus, $\pi$ is bijective and the isomorphism is established.
\end{enumerate}

\subsection{The antibracket and the generator $W$}

Even more interesting is the fact that the differential $\bs$ admits a generating functional. Let us introduce the antibracket. It is very similar to a Poisson bracket, but with pairs of variables of opposite parity. For an irreducible theory, its action on two functionals is defined as follows: \begin{eqnarray}(A,B)=\frac{\d^R A}{\d \phi^i}\frac{\d^L B}{\d \phi^*_i}-\frac{\d^R A}{\d \phi^*_i}\frac{\d^L B}{\d \phi^i}+\frac{\d^R A}{\d C^\a}\frac{\d^L B}{\d C^*_\a}-\frac{\d^R A}{\d C^*_\a}\frac{\d^L B}{\d C^\a}\quad,\end{eqnarray} where the summation over $i$ and $\a$ implicitly contains an integration over spacetime. The indices $L$ and $R$ just indicate whether the derivatives act from the left or from the right, which is not equivalent for fermionic fields. The antibracket is also well-defined on local functions, for which the functional derivatives are replaced by the variational derivatives (and no integration is made). The antibracket raises the $gh$ number by one and is fermionic, the first two terms lower the $antigh$ by one, the others lower the $pgh$ by one and the $antigh$ by two. It satisfies the following graded rules:\\
Symmetry\quad~~~~: $(A,B)=-(-1)^{(\varepsilon_A+1)(\varepsilon_B+1)}(B,A)$\\ 
Jacobi identity~: $(-1)^{(\varepsilon_A+1)(\varepsilon_C+1)}(A,(B,C))+\textrm{ cyclic permutations }=0$ \\
``Leibniz rule''~~: $(A,BC)=(A,B)C+(-1)^{\varepsilon_B(\varepsilon_A+1)}B(A,C)$

It can be shown that there exists a definite local functional $W$ such that for any functional or local function $A$: \begin{eqnarray}\bs A=(W,A)\quad.\end{eqnarray}The functional $W$ is of $gh$ 0 and bosonic. It is called the BRST-BV generator (or simply generator). It can be seen as an extended action, because its $antigh$ $0$ component is none other than the action $S$: \begin{eqnarray}W=S+\displaystyle\stackrel{(1)}{W}+\displaystyle\stackrel{(2)}{W}+\ldots\quad.\end{eqnarray}
Indeed, it reproduces the $antigh$ -1 part of the action of $\bs$ (i.e. the differential $\delta$) on the antifields $\phi^*_i$: $(S,\phi^*_i)=\d\phi^*_i=\frac{\d \cl}{\d\phi^i}$.
The second term of $W$, in $antigh \ 1$, can also be easily written, it reads: \begin{eqnarray}\displaystyle\stackrel{(1)}{W}=\int_\mathcal{D} \phi^*_i R^i_\a C^\a d^nx\quad,\end{eqnarray} and it generates the relations $\g \phi^i = R^i_\a C^\a$ and $\d C^*_\a=R^i_\a\phi^*_i$. 

An important feature of $W$ is provided by nilpotency of the differential $\bs$. Using the Jacobi identity, it is obtained that $\forall A:\ 0=\bs^2A=(W,(W,A))=\pm\frac{1}{2}(A,(W,W))$. This is true  if and only if \begin{eqnarray}(W,W)=0\quad.\end{eqnarray} This very important condition is called the master equation. 

The master equation is the key constraint that the generating functional $W$ must satisfy.
We will prove recursively that if the $antigh$ expansion of $W$ starts with the first two terms written above, then the further orders can always be constructed one by one in such a way that the master equation is satisfied.
To do so, it is enough to show that, if 
$\displaystyle\stackrel{(n-1)}{R}=\stackrel{(0)}{W}+\stackrel{(1)}{W}+\ldots +\stackrel{(n-1)}{W}$ satisfies the master equation up to its $antigh$ component $n-2$ , then one can build $\displaystyle\stackrel{(n)}{R}$ that satisfies it up to $antigh$ $n-1$. Indeed, $\displaystyle\stackrel{(1)}{R}=S+\stackrel{(1)}{W} $
satisfies the $antigh\  0$ component of the master equation, which is equivalent to the Noether identities.\\
Let us define  $\stackrel{(n-1)}{D}$ as the component of $antigh$ $n-1$ of $(\stackrel{(n-1)}{R},\stackrel{(n-1)}{R})$, which is its first non zero component. The $antigh$ $n-1$ term of $(W,W)=0$ is $2\d \stackrel{(n)}{W}+\stackrel{(n-1)}{D}=0$. On the other hand, the Jacobi identity implies that $(\stackrel{(n-1)}{R},(\stackrel{(n-1)}{R},\stackrel{(n-1)}{R}))=0$, the lower $antigh$ term of which is  $\d\stackrel{(n-1)}{D}=0$. Thanks to the vanishing of $H_{n-1}(\d)$, it is now obvious that $\stackrel{(n)}{W}$ exists. 

For an irreducible theory, the $antigh\ 2$ component of $W$ is $$\displaystyle\stackrel{(2)}{W}=\int_\mathcal{D} (\frac{1}{2} C^*_\a C^\a_{\g\beta}C^\beta C^\g - \frac{1}{4}\phi^*_i\phi^*_j M^{ij}_{\a\beta}C^\a C^\beta)d^n x \quad.$$ When the theory is reducible, the reducibility constants appear in terms like $C^*_\a Z^\a_A C^A$ or $\phi^*_i\phi^*_j C^{ij}_A C^A$, and so forth.
We see that the first terms of the generator involve all the various coefficients characterizing the theory: the action, the generators of the gauge transformations, the structure functions, etc. These coefficients are very easily singled out, as the terms are all linearly independent. In fact, $W$ is a single functional which contains all the information about the theory in a very simple way. The consistency of the whole is secured by a single constraint: the master equation. This is a fundamental advantage, which makes it possible for example to study deformations of an action with very few hypotheses (see Chapter \ref{ch:antidef}). 

\subsection{Locality, (co)homologies modulo $d$}

We must now worry a bit more about the local nature of the objects. If we want to make computations that are strictly local, we can only allow exact objects for a given differential that are the image of a local functional. We will perform computations with the integrands of those functionals, that are well defined over the jet space $J_l$. Since we make the assumption that boundary terms in the integrals vanish, the integrands of the functionals are defined modulo a divergence. The equivalent of (co)homologies for functionals are thus (co)homologies modulo divergences. A convenient way of taking this into account is to rather consider functionals as integrals of $n$-forms and to use the exterior differential $d$ of the spacetime manifold.

\subsubsection*{Exterior differential $d$}

The usual exterior differential $d=dx^\mu{\partial_\mu}$ can be introduced on the spacetime manifold. It is easily extended to the jet space thanks to the definition of the partial derivative on $J_k$. The form degree is the $\mathbb{N}$-grading of $d$ and is given by the number of anticommuting $1$-forms $dx^\mu$. This grading is bounded from above by $n$ because there are only $n$ independent $dx^\mu$. The Poincar\'{e} lemma states that, in a contractible domain $\cd$, the only $d$-closed object that are not $d$-exact are the constant zero forms (i.e. the numbers). In other words: \begin{eqnarray}\forall\ i>0\ :\ H^i(d,C^{\infty}(\cd))=0\textrm{ and }H^0(d,C^{\infty}(\cd))=\mathbb{R}\quad.\end{eqnarray}
The same result holds in a jet space, except in form degree $n$, for bosonic and/or fermionic fields and is called the algebraic Poincar\'{e} lemma. We use the following dual notation: \begin{eqnarray}d^n x=\frac{1}{n!}\varepsilon_{\mu_1...\mu_n}dx^{\mu_1}\wedge....\wedge dx^{\mu_n}\ ,\ d^{n-1}x_\mu=\frac{1}{(n-1)!}\varepsilon_{\mu\mu_1...\mu_{n-1}}dx^{\mu_1}\wedge...\wedge dx^{\mu_{n-1}}\quad.\end{eqnarray} Any $n$-form is obviously closed. A $(n-1)$-form can be written $v=V^\mu d^{n-1}x_\mu$ and its exterior derivative is the $d$-exact $n$-form : $dv=\partial_\mu V^\mu d^n x$. It is well-known that the Euler-Lagrange derivatives of a divergence are identically zero, which is equivalent to saying that divergences are trivial terms in a Lagrangian. The cohomology of $d$ in form degree $n$ in $J_k$ is thus isomorphic to the set of functions which have the same variational derivatives: \begin{eqnarray}&\displaystyle H^n(d)=\left\{[a]\ |\ deg\ a=n\;,\;a'\in[a]\Leftrightarrow \forall \phi^i\ :\ \frac{\d a'}{\d\phi^i}=\frac{\d a}{\d\phi^i}\right\}&\nonumber\\&\forall 0<i<n\ :\ H^i(d)=0&\nonumber\\& H^0(d)=\mathbb{R}\quad.&\end{eqnarray}
The local Lagrangian and any function defined on $J_l$ can be seen as the coefficient of a $n$-form. Let us note that if the metric is not flat (as for the Einstein--Hilbert theory of course), coefficients of $n$-forms must be tensorial densities (for example, $\sqrt|g|$, which appears in the Einstein--Hilbert action). So, in that case, the Lagrangian, the equations of motion and the antifields all have that behaviour. Trivial terms in the Lagrangian are just $d$-exact $n$-forms, the integral of which are of course boundary terms in the action, which are assumed to vanish. 

\subsubsection*{(Co)homologies modulo $d$}

The fact that we consider only vanishing boundary terms means that two functionals are equal if they are the integral of $n$-forms differing by a $d$-exact term: \begin{eqnarray} {\textrm{If }}A=\int a {\textrm{ and }}B=\int b{\textrm{ , then }}A=B\Leftrightarrow \exists v\ |\ a=b+dv\quad.\end{eqnarray}
This implies that a $\bs$-closed functional is related to an $n$-form $\bs$-closed modulo $d$ : \begin{eqnarray}\bs A=0=s\int a\Rightarrow\exists b\ | \bs a+db=0\quad.\end{eqnarray} The same holds for $\bs$-exact local functionals: \begin{eqnarray}A=\bs B\Rightarrow \exists c|a=\bs b+dc\quad.\end{eqnarray} The differentials $\d$ and $\bs$ are constructed in such a way as to commute with $d$ in $J_k$, so $\d$ and $\bs$ are differentials modulo $d$ and it is very natural to define $H^i_*(\d|d)$ and $H^{*,i}(s|d)$, where $i$ is the form degree. Under the locality assumption, one must rather compute the cohomology of $\bs$ modulo $d$ in $J_k$ than the cohomology of $\bs$.  

\section{Various results}

We will now review some important general theorems about (co)homologies modulo $d$ and make some considerations about linear theories.

\subsection{Results about (co)homologies modulo $d$}

The homology of $\d$ modulo $d$ does not vanish in general. However, since $\d$ and $d$ have vanishing (co)homologies for most values of their degrees, some isomorphisms can be established between classes $H_k^p(\d|d)$.

\begin{theorem} The homology of $\d$ modulo $d$ has no sector at strictly positive $pgh$ and $antigh$ numbers: 
\begin{eqnarray}\forall k>0\ :\ [a]\in H_k^p(\d|d) \Rightarrow pgh\ a=0\nonumber\end{eqnarray} 
\label{deltad1}\end{theorem}
\noindent Proof: Many details about this theorem can be found in \cite{Henneaux:1991rx}. It is in fact equivalent to proving that every component of the generator $W$ is local. Some natural assumptions for local gauge theories are required, namely, the stationary surface in $J_l$ is smooth (which we assumed in Eq.(\ref{smoothsigma})) and the generating set of the gauge transformations is locally complete, i.e. the Noether identities are algebraic relations between the coefficients of the operators in the jet space. A powerful tool to prove the vanishing of an homology is to find a contracting homotopy: a derivative whose anticommutation with the differential yields a counter operator of the related grading. For example, in the case of $\d$, such an homotopy can be built: $\s\d+\d\s=K\ ,\ K a_k=k a_k$ where $k=antigh\ a_k>0$. The construction of $\d$ ensures both the existence of $\s$ and the vanishing of the homology. Now, the above requirements allow one to consider the contracting homotopy as acting on functions of the jet space. The local homology of $\d$ thus vanishes in strictly positive antifield number, because: $\d a=0\Rightarrow a=\frac{1}{k}Ka=\d\left(\frac{1}{k}\s a\right)$. 

Let us now decompose the exterior derivative into a $pgh\ 0$ sector $d_0$ and a $pgh\ 1$ sector $d_1$: $d=d_0+d_1$. Both part are nilpotent derivatives. The interest of doing this is that, since $\d$ and $\s$ have nothing to do with the ghosts, they both anticommute with $d_1$. Thanks to this property, we find that, in strictly positive antifield number $k$: \begin{eqnarray}\d a=d_1 b\Rightarrow a=\frac{1}{k}Ka=\frac{1}{k}(\d\s a+\s\d a)=\d\left(\frac{1}{k}\s a\right)-d_1\left(\frac{1}{k}\s b\right)\quad.\label{deld1}\end{eqnarray} The next step is to introduce a new grading: the number of derivatives of the ghosts. In strictly positive $pgh$ number, the equation $\d a=d b$ can then be decomposed into components with a definite value of this grading: $\d \stackrel{(i)}{a}=d_0 \stackrel{(i)}{b}+d_1\stackrel{(i-1)}{b}$. The locality ensures that the grading is bounded, and we can assume that the expansion of $b$ stops at degree $t-1$ if the expansion of $a$ stops at degree $t$. If it were not the case, the top component would be of the form $d_1 \stackrel{(m)}{b}=0\Rightarrow \stackrel{(m)}{b}=d_1 \stackrel{(m-1)}{v}$. Then, by redefining $b'=b-d\stackrel{(m-1)}{v}$, the equation $\d a=d b'$ holds and $b'$ is of degree $m-1$. When all those trivial components of $b$ have been removed, the top equation becomes: $\d \stackrel{(t)}{a}=d_1\stackrel{(t-1)}{b}$. Thanks to Eq.(\ref{deld1}), we find that $\stackrel{(t)}{a}=\d \stackrel{(t)}{c}+d_1\stackrel{(t-1)}{e}$. We can now redefine $a$ into $a'=a-\d\stackrel{(t)}{c}-d\stackrel{(t-1)}{e}$, which reduces the maximum degree of $a$ to $t-1$. Going on in the same way, all the components of definite degree are found to be trivial and it is finally obtained that $a=\d c+ de$ $\Box$.

\begin{theorem} The following isomorphisms can be established: if $p\geqslant 1$ and $k>1$: $$H_k^p(\d|d)\cong H_{k-1}^{p-1}(\d|d)\quad.$$
\label{desctheo}\end{theorem}\noindent The proof is quite simple: Let us consider a cocycle of $H_k^p(\d|d)$: $\d a_k^p+d a_{k-1}^{p-1}=0$. By applying $\d$ to this equation and thanks to the algebraic Poincar\'e lemma, it is found that: $\exists\ a_{k-2}^{p-2}\ |\ \d a_{k-1}^{p-1}+d a_{k-2}^{p-2}=0$. In the $p=1$ case, it is just found that $\d a_{k-1}^0=0$. The map between $a_k^p$ and $a_{k-1}^{p-1}$ is injective and surjective thanks to the vanishing of the homology of $\d$ $\Box$. 

The $p=1$ case of the above theorem is quite interesting, since it emphasizes that a $\d$ modulo $d$ cocycle in form degree 0 is just a $\d$ cocycle : $\d a_k^0=0$. In other words : $\forall k>0\ :\ H_k^0(\d|d)=H_k(\d)=0$. Thanks to the isomorphisms between the different classes, it is now obvious that every class with $k>p$ is vanishing: $$\forall k>p\ :\ H^p_k(\d|d)\cong H^0_{k-p}(\d|d)=0\quad.$$ The next theorem involves the cohomology of $d$ modulo $\d$. The $antigh$ 0 classes of this cohomology can be seen as the set of nontrivial objects $d$-closed on shell and are called the characteristic cohomology.
\begin{theorem}If $p,k\geqslant 1$ and $(p,k)\neq (1,1)$: $$H_k^p(\d|d)\cong H_{k-1}^{p-1}(d|\d)\quad.$$ Furthermore: $$H_1^1(\d|d)\otimes\mathbb{R}\cong H_0^0(d|\d)\quad.$$\label{ddeltheo}\end{theorem}\noindent It is the same kind of proof as the previous one. Let us consider the cocycle $\d a_k^p + d b_{k-1}^{p-1}=0$. It tells simultaneously that $[a_k^p]\in H_k^p(\d|d)$ and $[b_{k-1}^{p-1}]\in H_{k-1}^{p-1}(d|\d)$. An element $a'\phantom{}_k^{p}$ belongs to $[a_k^p]$ if: $a'\phantom{}_k^{p}=a_k^p+\d m_{k+1}^p+ d n_k^{p-1}$. By applying $\d$, we find $\d a'\phantom{}_k^{p}=-d(b_{k-1}^{p-1}+\d n_k^{p-1})$. If $p,k>0$ and $(p,k)>(1,1)$, there is no nontrivial solution of $d c_{k-1}^{p-1}=0$ and we find that $\exists\ e_{k-1}^{p-2}\ |\ \d a'\phantom{}_k^p+d b'\phantom{}_{k-1}^{p-1}=0$ with $b'\phantom{}_{k-1}^{p-1}=b_{k-1}^{p-1}+\d n_k^{p-1}+ d e_{k-1}^{p-2}$. This establishes a surjective map between $[a_k^p]$ and $[b_{k-1}^{p-1}]$. This map is injective because $H_k(\d)=0$. If $(p,k)=(1,1)$, the map is no longer surjective: different elements $b_0^0$ differing by a constant correspond to the same $a_1^1$. The correspondence is thus established between $b_0^0$ and a couple $(a_1^1,C),\ C\in\mathbb{R}$ $\Box$.

The cohomology of $\g$ modulo $d$ in $H_0(\d)$ can be defined and the following theorem can be established: \begin{theorem}\begin{eqnarray}&\forall k\geqslant 0\ :\ H^{k,n}(\bs|d)\cong H^{k,n}(\g|d,H_0(\d))&\\&\forall k<0\ : H^{k,n}(\bs|d)\cong H_{-k}^n(\d|d)\quad.&\end{eqnarray}\end{theorem}\noindent It is the equivalent of Theorem \ref{homper} in $J_k$. Let us consider a cocycle of $\bs$ modulo $d$: $\bs a+db=0$. In the case of a positive ghost number, $a$ is the sum of various terms with $antigh\leqslant pgh$: $a=a_0+a_1+...$ where $antigh\ a_i=i$ and $pgh\ a_i=k+i$. As an $\mathbb{N}$-grading divides an algebra into independent subspaces, the cocycle of $\bs$ modulo $d$ can be decomposed into its components of different $antigh$. The bottom equation is thus: $\g a_0+\d a_1+d b_0=0$. This is a $\g$ modulo $d$ cocycle in $H_0(\d)$ and the wanted isomorphism is the one that applies $[a]$ on $[a_0]$. It is injective and surjective thanks to Theorem \ref{deltad1}, the argument is similar to that of Theorem \ref{homper}. In the $k<0$ case, the expansion of $a$ begins at $antigh=-k$: $a=a_{-k}+a_{-k+1}+...$ and the bottom equation is $\d a_{-k}+d b_{-k-1}=0$, which defines an element $[a_{-k}]$ of $H_{-k}^n(\d|d)$. Once again, the map between $[a]$ and $[a_{-k}]$ is injective and surjective, thanks to Theorem \ref{deltad1}.

\subsection{Linear theories\label{linth}}

What we call here a linear theory is a theory where the Lagrangian is quadratic in the fields and their derivatives, where the equations of motion are linear in them and the gauge transformations are independent of them. The free Fronsdal theories that we present in the next chapter are such theories. In a linear theory, the gauge transformations take the form: \begin{eqnarray}\displaystyle\d_\ve \phi^i=\car^i_\a\phi^i=\sum_{m=1}^{l}\mathcal{R}^{i\mu_1...\mu_m}_\a \6^m_{\mu_1...\mu_m}\ve^\a=\mathcal{R}^{i(\mu)}_\a \6^{|\m|}_{(\mu)} \ve^\a\quad.\label{multiind}\end{eqnarray} We have introduced the multiindex notation $(\mu)$, that is very convenient when making summations over different numbers of spacetime indices. The current number of derivatives, that we call the length of the multiindex, is denoted $|\m|$. The generating set $R^i_\a$ depends only on $x^\m$. The equations of motion are linear in the fields: \begin{eqnarray}\frac{\d \mathcal{L}}{\d \phi^i}=\cd_{ij}\phi^j=\mathcal{D}^{(\m)}_{ij}\6_{(\m)}^{|\m|}\phi^j\quad,\end{eqnarray} where the operator $D_{ij}$ depends only on $x_\m$. The Noether identities read: \begin{eqnarray}\car^{\dagger i}_\a \cd_{ij}\phi^j=(-1)^{|\r|}\6^{|\r|}_{(\r)}\Big(\mathcal{R}^{i(\r)}_\a \mathcal{D}^{(\s)}_{ij}\6^{|\s|}_{(\s)}\phi^j\Big)\equiv 0\quad.\end{eqnarray} As this is true off-shell, i.e. for any history of the fields, the identities are purely algebraical relations between the sets of coefficients : $\mathcal{R}^{i(\r)}_\a \mathcal{D}^{(\s)}_{ij}=0$. It is obvious that $[\d_\e,\d_\eta]\phi^i=0$, a linear theory is thus automatically Abelian. In an Abelian theory, the longitudinal derivative is a differential off-shell. For example, if the theory is also irreducible : $\g C^\a=0\Rightarrow \g^2\phi^i=\g(R^i_\a C^\a)=0$. Since $\d^2=0$ and $\d\g+\g\d=0$ by construction, the differential $\bs$ has a very simple, finite expansion : $\bs=\g+\d$. Furthermore, the generator $W$ consists of only two terms: \begin{eqnarray}W=S+\int_\mathcal{D}\phi^*_i \car^i_\a C^\a d^n x\quad.\end{eqnarray} Since $\g$ is a true differential in this case, its cohomology $H^*(\g)$ can be defined and will be very important in various problems. Its computation is a very general study that depends on the gauge structure of the theory. An important result about $H(\d|d)$ can also be established: 
\begin{theorem} In a linear theory of reducibility order $r$: $$H_k^n(\d|d)=0\ if\ k>r+2\quad.\label{hkndeltd}$$ \end{theorem} The reducibility order of a theory is the number of generations of reducibility relations. Let us sketch the proof in the irreducible case ($r=0$) with bosonic fields $\phi^i$. Let us consider a cocycle $\d a_k^n+d b_{k-1}^{n-1}=0$. If $a_k^n=a_k d^n x$, we can use the equivalent writing $\d a_k+\6_{\m}j_{k-1}^m=0$. 
The idea is to take variational derivatives of this cocycle, with respect to the fields and the antifields (let us remind that nontrivial elements have $pgh=0$). The divergence vanishes, but we have to take care of the commutation of $\d$ with the variational derivatives. Then, $a_k$ can be built back from its derivatives with an homotopy formula. The following relations can be established: \begin{eqnarray}\d\frac{\d^L a_k}{\d C^*_\a}&=&0\\ \d\frac{\d^L a_k}{\d \phi^*_i}&=&\car^i_\a\frac{\d^L a_k}{\d C^*_\a}\\ \d\frac{\d^L a_k}{\d\phi^i}&=&-\cd^{\dagger}_{ji}\frac{\d^L a_k}{\d \phi^*_j}\quad,\end{eqnarray} where $\cd^{\dagger}_{ij}=(-1)^{|\mu|}\6^{|\mu|}_{(\m)}(\mathcal{D}^{(\mu)}_{ij}\ .\,)$ is the adjoint of $\cd_{ij}$. The first relation is at $antigh = k-2>0$, thus, thanks to the vanishing of $H_k(\d)$ in strictly positive $antigh$ and thanks to the Noether identities, the system can be solved: \begin{eqnarray}\frac{\d^L a_k}{\d C^*_\a}&=&\d f^\a_{k-1}\\ \frac{\d^L a_k}{\d\phi^*_i}&=&\car^i_\a f^\a_{k-1}+\d f^i_k \\ \frac{\d^L a_k}{\d\phi^i}&=&-\cd^{\dagger}_{ji} f_k^j+\d f_{k+1,i}\quad.\end{eqnarray} Finally, let us write the following formula: \begin{eqnarray}a_k&=&\int^1_0\left[C^*_\a\frac{\d^L a_k}{\d C^*_\a}+\phi^*_i\frac{\d^L a_k}{\d \phi^*_i}+\phi^i\frac{\d^L a_k}{\d\phi^i}\right](t)dt\ +\ \textrm{div}\\&=&\d\int_0^1\left[C^*_\a f^\a_{k-1}-\phi^*_i f^i_k+\phi^j f_{k-1,j}\right](t)dt\ +\ \textrm{div}\quad.\end{eqnarray} The dependence in $t$ is of the form $F(t)=F(tC^*_\a,t\phi^*_i,t\phi^j)$. The last equality, which ensures that $a_k$ is trivial, is true because $\d(t)=\d$ and thus commutes with the integral.

The results obtained in this section are needed to study the deformations of consistent deformations of Fronsdal theories, that we discuss in the next two chapters.

%%%%%%%%%%%%%%%%%%%%%%%%%%%%%%%%%%%%%%%%%%%%%%%%
\chapter{Fronsdal theory for free spin-$s$ fields in Minkowski, de Sitter or Anti de Sitter spacetime}\label{ch:Frons}
%%%%%%%%%%%%%%%%%%%%%%%%%%%%%%%%%%%%%%%%%%%%%%%%

In this chapter, we first recall the concept of massless spin-$s$ field, in the extended sense of an arbitrary dimension $n$, in Minkowski or (Anti)de Sitter spacetime. Then, we recall Fronsdal's theory describing totally symmetric massless fields in Minkowski spacetime and provide its antifield formulation as well as some results that will prove to be useful for the computation of consistent deformations. Finally, we recall the $(A)dS$ version of Fronsdal's theory.

\noindent {\bf Notation:} Throughout the thesis, we use brackets to indicate a strength-one symmetrization and square brackets to indicate an strength-one antisymmetrization: for example $S_{(\m_1...\m_k)}$ is totally symmetrized in its $k$ indices and $A_{[\m_1...\m_k]}$ is totally antisymmetric. Furthermore, we use vertical bars to indicate that some indices are not involved in a (anti)symmetrization. For example: $A_{(\m|\r|\n)}$ is only symmetric in $\m\n$. Vertical bars are also used to separate groups of antisymmetric indices in the antisymmetric notation of the Young components of tensors.

\section{Massless spin-$s$ fields}

The concepts presented in this section are adapted from the book \cite{Weinberg:1995mt}, as well as the articles \cite{Bekaert:2006py,Boulanger:2008up}, where more details can be found.

\subsection*{Spin-$s$ fields in ${\cal{M}}_4$}

In the four-dimensional Minkowski spacetime, the spin and the quantum-mechanical expression of the mass are eigenvalues of some operators. In order for a theory to be relativistic, one requires the Poincar\'e covariance: vectors of an Hilbert space are required to transform under unitary representations of the Poincar\'e group: if $x'{}^\m=\L^\m_\n x^\n + a^\n$, then $|\psi'\rangle=U(\L,a)|\psi\rangle$ where the representation $U$ is such that $U(\L_1,a_1)U(\L_2,a_2)=U(\L_1\L_2,a_1+\L_1a_2)$. Infinitesimal representations can be written $U(\;\mathun+\o,\e)=\mathun+\frac{1}{2}i\o_{\m\n}J^{\m\n}-i\e_\r P^\r$ where the generators $J^{\m\n}$ and $P^\r$ are hermitian. These generators form a basis of the Lie algebra and satisfy the relations:
\begin{eqnarray}
\left[J^{\m\n},J^{\r\s}\right]&=&-2i\left[\eta^{\m[\r}J^{\s]\n}-\eta^{\n[\r}J^{\s]\m}\right]\nonumber\\
\left[P^\m,J^{\n\r}\right]&=&-2i\eta^{\m[\n}P^{\r]}\nonumber\\
\left[P^\m,P^\n\right]&=&0\quad.
\end{eqnarray}
Two independent Casimir operators can be defined in dimension 4: \begin{eqnarray}
C_1=-P_\m P^\m\quad{\textrm{and}}\quad C_2=-W_\m W^\m\quad,
\end{eqnarray} where $W_\m$ is the Pauli-Lubanski vector defined as: \begin{eqnarray}
W_\m=\frac{1}{2}\e_{\m\n\r\s}J^{\n\r}P^\s\quad.
\end{eqnarray}
The eigenvalues of the generators $P^\m$ are of course the four-momentum $p^\m$ of the particle, their spectrum is continuous. The rest of the quantum numbers is gathered in the notation $\s$: $P^\m|\psi_{p,\s}\rangle=p^\m|\psi_{p,\s}\rangle$. For each value of $p^2$ and, for $p^2<0$, each sign of $p^0$, one can then choose a standard four-momentum, say $q^\m$, and express any $p^\m$ of this class as $p^\m=L^\m_\n q^\n$, where $L(p)$ is a Lorentz transformation which is assumed not to act on the other quantum numbers: $|\psi_{p,\s}\rangle=N(p)U(L(p))|\psi_{q,\s}\rangle$, where $N(p)$ is an appropriate normalization factor. The precise expressions of $L(p)$, for the different values of $p^2$, can be found in \cite{Weinberg:1995mt}. Any Poincar\'e transformation is then decomposed into a rotation $L(p)$ and an element of the little group, the group of transformations $W$ that leaves $q^\m$ invariant: $q^\m=W^\m_\n q^\n$. The action of a general transformation on a state is given by the following formulas: \begin{eqnarray}U(\;\mathun,a)|\psi_{p,\s}\rangle &=&e^{-i p.a}|\psi_{p,\s}\rangle\\U(\L,0)|\psi_{p,\s}\rangle&=&\frac{N(p)}{N(\L p)}\sum_{\s'}D_{\s\s'}(W)|\psi_{\L p,\s'}\rangle\quad,\end{eqnarray} where $W=L^{-1}(\L p)\L L(p)$ belongs to the little group, and the coefficients $D_{\s\s'}$ define a representation of the little group: $U(W)|\psi_{q,\s}\rangle=\sum_{\s,\s'}D_{\s,\s'}(W)|\psi_{q,\s'}\rangle$. 

The little group depends of the case considered: In the massive case, where the reference momentum can be chosen $q^\m=(m,0,0,0)$, the little group is $SO(3)$, the Pauli-Lubanski vector takes the form $W_\m=m(0,S_1,S_2,S_3)$. The eigenvalues of $C_4$ are given by those of $\vec{S}^2$, which are of the form $s(s+1)$, $s$ being called the spin of the particle.

In the massless case, where the reference momentum can be chosen $q^\m=(E,0,0,E)$, the little group is the Euclidean group of $\mathbb{R}^2$: $ISO(2)$. The Pauli-Lubanski vector takes the form $W_\m = E(-J_{12},R_1,R_2,J_{12})$. The eigenvalues of $C_4=E^2(R_1^2+R_2^2)$ are arbitrary positive numbers $\m^2$. In the case $\m^2>0$, the representation of the little group is infinite-dimensional and corresponds to the ``infinite spin'' case. On the other hand, if $\m^2=0$, the representation of the little group is finite-dimensional and the last eigenvalue that can be considered is that of $J_{12}$: $J_{12}|\psi_{p,\s}\rangle=\s|\psi_{p,\s}\rangle$, where $\s$ is called the helicity of the particle, which must be (half-)integer in order for the representation to be single(double)-valued. 

Finally, if the fields are realized in the form of tensors covariant under the Lorentz group, it is possible to relate them to a value of the spin or of the helicity. We will only consider here bosonic fields, but a similar construction has been achieved for fermions. First, an arbitrary tensor with $d$ indices can be decomposed into different traceless components transforming under irreducible representations of the Lorentz group, and that correspond to distinct Young tableaux with $d$ boxes (see Appendix \ref{app:YD}). For tensors depending on the spacetime coordinates, the generators of the Poincar\'e algebra take the form $P_\m=-i\6_\m$ and $J_{\m\n}=-i(x_\m\6_\n-x_\n\6_\m)-iM_{\m\n}$ where $M_{\m\n}$ are generators of the Lorentz algebra. Let us consider a totally symmetric field $\phi_{\m_1...\m_s}$, which is visualized by a one-row Young diagram \begin{picture}(35,0)(0,0)\multiframe(0,0)(10.5,0){1}(30,6){$s$}\end{picture}. In the massive case, if it satisfies the relation $\6^\m\phi_{\m\m_2...\m_s}=0$ and the Klein-Gordon equation $\Box\phi_{\m_1...\m_s}=m^2$, then the eigenvalue of the quartic Casimir $W^2$ is $m^2 s(s+1)$, the field clearly corresponds to a spin-$s$ representation. The massless case is a bit more subtle. A tensor called the generalized curvature is selected among the components of the $s$'th derivatives of $\phi$ as corresponding to the Young diagram $\l$: \begin{picture}(37,0)(0,0)\multiframe(0,-4)(8.5,0){1}(8,8){}\multiframe(8.5,-4)(18.5,0){1}(18,8){\ldots}\multiframe(27,-4)(8.5,0){1}(8,8){}
\multiframe(0,4.5)(8.5,0){1}(8,8){}\multiframe(8.5,4.5)(18.5,0){1}(18,8){\ldots}\multiframe(27,4.5)(8.5,0){1}(8,8){}
\end{picture}. In the antisymmetric notation, the curvature $K$ is a tensor that bears $s$ pairs of antisymmetric indices \cite{Weinberg:1965rz,deWit:1979pe}. 
If a tensor with such symmetries satisfies the equations $T^\r_{\phantom{\r}\n_1|\r\n_2|...|\m_s\n_s}=0$ and $\6_{[\r}T_{\m_1\n_1]|\m_2\n_2|...|\m_s\n_s}=0$, it can be seen that it corresponds to an helicity $s$ representation. The second relation is identically satisfied by $K=Y_{\l}(\6^s\phi)$ and the first is the dynamical equation of the particle. Finally, it can be showed by some arguments of duality in the little group that the totally symmetric tensors are sufficient to describe every particle in dimension 4, in both the massive and helicity cases. Because of the similar form of the field in both cases, we will only use the word ``spin'' in the sequel, even when we consider massless fields. Finally, let us remark that, in the massless case, the observable is the curvature, also called ``field strength'', the field $\phi$ having a gauge freedom: any transformation of the type $\d_\xi \phi_{\m_1...\m_s}=s\6_{(\m_1}\xi_{\m_2...\m_s)}$ leaves $K$ invariant and thus does not modify the equations.

\subsection*{Extension to dimension $n$}

The definition of spin can be extended to any dimension where it becomes a set of Dynkin labels instead of the single Dynkin label of $so(2)$ or 
$so(3)$. In the massive case, the little group is $SO{(n-1)}$. In the massless case, it is $ISO(n-2)$. These groups provide several eigenvalues, which are not identified as a single ``spin''. Furthermore, the Poincar\'e group $ISO(n-1,1)$ has more than two Casimir operators. Finally, the symmetric tensors are not sufficient to describe every type of particle in arbitrary dimension, the other irreducible representations, corresponding to Young diagrams with several rows must be considered. The associated tensor components are referred to as ``mixed symmetry'' tensors.

In both massless and massive cases, the fields which will be called ``spin-$s$ fields'' are those whose Young diagram has $\l_1=s$ columns. This is mostly because the curvature $K$ built from such a massless field involves $s$ derivatives, in such a way that each antisymmetric group of indices of the tensor is antisymmetrised with one derivative. This amounts to adding a second line of length $\l_1$ in the diagram, corresponding to the indices of the derivatives. The antisymmetrization of an antisymmetric group of indices of $K$ with a $(s+1)$th derivative identically vanishes. Furthermore, the equation of motion for that kind of field is $Tr K=0$, where the trace consists in contracting one index of the first two columns. More details about the spin $s$ fields in any dimension can be found in \cite{Bekaert:2006py}, while the equation $Tr\ K=0$ has been discussed in \cite{Bekaert:2004dt,Bekaert:2003az,Bekaert:2003zq,Bekaert:2007ix}.

\subsection{Extension to $(A)dS$ spacetime}

The concept of spin-$s$ is directly extended to $dS$ or $AdS$ spacetime as being the number of columns of the Young diagram of the considered tensor fields. On the other hand, the tensors have to transform under representations of $SO(D,1)$ or $SO(D-1,2)$, which are semisimple groups. There are no more translations and the mass cannot be defined in the same way as in Minkowski spacetime. However, a ``massless'' case can be distinguished, because of its lower number of degrees of freedom. Some totally symmetric fields can be put in correspondence with the massless fields in Minkowski spacetime. On the other hand, what is call a ``massless'' tensor with mixed symmetry corresponds to several tensors with different symmetries in Minkowski spacetime \cite{Brink:2000ag}. Such a construction has first been addressed in \cite{Metsaev:1995re,Metsaev:1997nj} with the use of some gauge fixations. A more intrinsic formulation has been obtained in \cite{Angelopoulos:1997ij,Laoues:1998ik,Iazeolla:2008ix} for the case of totally symmetric fields and in \cite{Boulanger:2008up,Boulanger:2008kw} for the general case. Schematically, the argument in $AdS$ goes as follows: though there are no more translations in $S0(D-1,2)$, one can consider its maximal compact subalgebra $so(D-1)\oplus so(2)$. The generators of $so(D-1,2)$ can be decoupled into a set of rotations $M_{\m\n}$ and a set of transvections $P_\m=\l M_{n \n}$, where $\l^2=-\frac{2\L}{(n-1)(n-2)}$, they are such that $[P_a,P_b]=-i\l^2 M_{ab}$ and are thus rotations in $AdS$ but their flat limit are translations. Then the generator $M_{n0}$, proportional to $P_0$, is the energy operator. The operators $L^+_r$ and $L^-_r$, $r=1,2,3$, are then built in such a way that $P_r=\frac{i\l}{2}(L^+_r-L^-_r)$. They raise or lower the value of the energy. In the common case of a particle, it is assumed that there exists a lowest weight state $|E_0,\Theta\rangle$, such that $L^-_r|E_0,\Theta\rangle=0$. The notation $\Theta$ represents the tensorial behaviour of the field, that transforms under a irreducible representation of $SO(D-1)$, $\Theta$ being the associated Young diagram. The state $|E_0,\Theta\rangle$ can be viewed as the analogous of the state $|q,\s\rangle$ that has been written above in the Minkowski case. Arbitrary states of the particle can then be obtained by applying arbitrary Poincar\'e transformations on this state. In $AdS$, by applying the generators on the fundamental state, one obtains a Verma module of states: $(L^+_r)^t |E_0,\Theta\rangle=\bigoplus_{\Theta'}|E_0+t,\Theta'\rangle$. The Young diagrams $\Theta'$ are obtained by expanding the tensor product of $\Theta$ and the $t$ indices $r$ of the $L^+_r$ operators. By definition, the massless ``helicity'' case occurs when a null vector appears in $L^+_r|E_0,\Theta\rangle$: $\exists |E_0+1,\Theta'\rangle$ such that $\langle E_0+1,\Theta'|E_0+1,\Theta'\rangle=0$. It corresponds to the existence of pure gauge fields, that can be discarded. In $AdS$, one considers the coset space of the module of $|E_0,\Theta\rangle$ modulo that of $|E_0+1,\Theta'\rangle$. Though there is still an infinite number of states, there has been a shortening of the representation. The same kind of construction can be done in the $dS$ case, though the r\^ole of the energy is less clear than in the $AdS$ case. Many technical details and subtleties are provided in the references cited above. The totally symmetric case admits a local Lagrangian formulation, first obtained in \cite{Fronsdal:1979vb}, as well as its link with the Lagrangian in Minkowski spacetime.

\section{Fronsdal theory in Minkowski spacetime and antifield formulation}\label{FTM}

The totally symmetric massless spin-$s$ bosonic fields $\phi_{\m_1...\m_s}$ have been given a free Lagrangian in Minkowski spacetime by Fronsdal \cite{Fronsdal:1978rb}. The equations of motion are linear and involve two derivatives. Fronsdal's equations of motion, which need some constraints upon the fields and the gauge parameters, have been showed \cite{Bekaert:2004dt} to be equivalent to the unconstrained equations $Tr K=0$ that we have considered above. Let us note that an unconstrained action has been proposed in \cite{Francia:2002aa,Francia:2002pt}, which is nonlocal and whose equations are also equivalent to the equations $Tr\ K=0$ (see also \cite{Francia:2005bu}).

\subsection{Construction of the action}

First of all, some constraints are imposed on the fields and the gauge parameters, that will be justified below: the fields are assumed to be double traceless and the gauge parameters to be traceless. The trace of the fields is denoted $\phi'_{\mu_3...\mu_s}=\eta^{\mu_1\mu_2}\phi_{\mu_1...\mu_s}$, and their vanishing double trace $\phi''_{\mu_5...\mu_s}=0$. The gauge transformations that leave the curvature invariant are:
\begin{eqnarray}\d_{\xi}\phi_{\mu_1...\mu_s}=s\6_{(\mu_1}\xi_{\mu_2...\mu_s)}\quad,\label{FrGT}\end{eqnarray} where $\xi_{\mu_2...\mu_{s}}$ is an arbitrary tensor. The tracelessness constraint is: $\xi'_{\mu_4...\mu_s}=0$.  This constraint allows the following tensor to be gauge invariant:
 \begin{eqnarray}
F_{\mu_1...\mu_s}=\Box\phi_{\mu_1...\mu_s}-s\6_{(\mu_1}^{2\rho}\phi^{\phantom{\rho}}_{\mu_2...\mu_s)\rho}+\frac{s(s-1)}{2}\6^2_{(\mu_1\mu_2}\phi'_{\mu_3...\mu_s)}\quad.\end{eqnarray}It is called the Fronsdal tensor. Let us write down explicitly the curvature tensor:
\begin{eqnarray}K_{\mu_1\nu_1|...|\mu_s\nu_s}=2^s Y^{(s)}(\6_{\mu_1...\mu_s}\phi_{\nu_1...\nu_s})\quad,\end{eqnarray} where we have used the following permutation operator:
\begin{eqnarray}\nonumber Y^{(m)}=\frac{1}{2^m}[e-(\mu_1\nu_1)][e-(\mu_2\nu_2)]...[e-(\mu_m\nu_m)]\quad.\end{eqnarray} This operator is proportional to the Young symmetriser of $\l=(s,s)$ when acting on symmetric tensors, with a more natural weighting.
The following relation holds. It shows that only the traceless part of the curvature is independent of the Fronsdal tensor:
\begin{eqnarray}\eta^{\mu_{s-1}\mu_s}K_{\mu_1\nu_1|...|\mu_s\nu_s}=2^{s-2}Y^{(s-2)}(\6_{\mu_1...\mu_{s-2}}F_{\nu_1...\nu_s})\label{TrKurv}\quad.\end{eqnarray} Let us notice that this identity has first been established for spin-3 in \cite{Damour:1987vm}, then extended to spin-$s$ and mixed symmetry fields in \cite{Bekaert:2003az,Bekaert:2007ix}. The trace of the Fronsdal tensor is:
\begin{eqnarray}F'_{\mu_3...\mu_s}=2\Box\phi_{\mu_3...\mu_s}-2\6^{2\r\s}\phi_{\r\s\mu_3...\mu_s}+(s-2)\6^{2\r}_{(\mu_3}\phi^{\phantom{\r}}_{\mu_4...\mu_s)\r}\quad,\end{eqnarray}and the following identity holds:
\begin{eqnarray}\6^{\mu_s}F_{\mu_1...\mu_s}\equiv\frac{s-1}{2}\6^{\phantom{*}}_{(\mu_1}F'_{\mu_2...\mu_{s-1})}\label{Frrel}\quad.\end{eqnarray}
Then, we can define the generalized Einstein tensor, which is equivalent to the Fronsdal tensor:
\begin{eqnarray}G_{\mu_1...\mu_s}=F_{\mu_1...\mu_s}-\frac{s(s-1)}{4}\eta_{(\mu_1\mu_2}F'_{\mu_3...\mu_s)}\quad,\\ F_{\mu_1...\mu_s}=G_{\mu_1...\mu_s}-\frac{s(s-1)}{2(n+2s-6)}\eta_{(\mu_1\mu_2}G'_{\mu_3...\mu_s)}\quad.\end{eqnarray} Thanks to Eq.(\ref{Frrel}), it obeys the traceless identities: \begin{eqnarray}\6^{\mu_s}G_{\mu_1...\mu_s}-\frac{(s-1)(s-2)}{2(n+2s-6)}\eta_{(\mu_1\mu_2}\6^{\mu_s}G'_{\mu_3...\mu_{s-1})\mu_s}\equiv0\quad.\end{eqnarray}
These identities take the form $R^{\n_1...\n_s}_{\m_1...\m_s}G_{\n_1...\n_s}$ where $R$ is the generating set of the gauge transformations given in Eq.(\ref{FrGT}). Thus, $G_{\m_1...\m_s}$ is a natural choice for the equations of motion, since it would satisfy the Noether identities. Furthermore, the Einstein tensor is constituted by a symmetric operator acting on the fields. Hence, the Lagrangian \begin{eqnarray}\mathcal{\cl}_{Fs}=\frac{1}{2}\phi^{\mu_1...\mu_s}G_{\mu_1...\mu_s}\end{eqnarray} is gauge invariant modulo a divergence, and it yields the equations of motions:\begin{eqnarray}\frac{\d \mathcal{L}_{Fs}}{\d\phi^{\mu_1...\mu_s}}=G_{\mu_1...\mu_s}\approx 0\quad.\end{eqnarray}

These equations, together with the constraints on the fields and the gauge parameters, are equivalent to the unconstrained equation $Tr\ K\approx0$. First, since the tensors $F$ and $G$ are equivalent: $F_{\m_1...\m_s}\approx 0$. Then, Eq.(\ref{TrKurv}) tells that $Tr\ K\approx\ 0$. Let us remark that Eq.(\ref{TrKurv}) holds independently of the double-tracelessness condition on $\phi$. The equation $Y^{(s-2)}(\6_{\m_1...\m_{s-2}}F_{\n_1...\n_s})\approx 0$ admits the general solution $F_{\m_1...\m_s}\approx\6^3_{(\m_1\m_2\m_3}\ch_{\m_4...\m_s)}$. The tensor $\ch$ can be set to zero by a gauge fixation, since $\d_\xi F_{\m_1...\m_s}=\frac{s(s-1)(s-2)}{2}\6_{(\m_1\m_2\m_3}\xi'_{\m_4...\m_s)}$. This fixes the trace of the gauge parameter, therefore the residual gauge transformations can only involve a traceless parameter. After having done this, one may want to impose the de Donder gauge fixation $\6^\r\phi_{\r\m_2...\m_s}-\frac{s-1}{2}\6_{(\m_2}\phi'_{\m_3...\m_s)}=0$, in order to find the gauge-fixed equations $\Box\phi_{\m_1...\m_s}\approx 0$. However, this cannot be done if the gauge parameters are traceless, because the gauge variation of this condition is $\Box\xi_{\m_2...\m_s}$, which must be traceless by the first constraint. Only the traceless part of the de Donder gauge fixation can be achieved. Since the trace of the de Donder gauge fixation depends only on the double trace of the field, it is thus natural to impose the constraint $\phi''_{\m_5...\m_s}=0$. Let us notice that, since $\d_\xi\phi''_{\m_5...\m_s}=2\6^\r\xi'_{\r\m_5...\m_s}$, the two constraints hold together, since no further gauge transformations can modify the double trace of the fields. More details about this procedure can be found in \cite{Francia:2002aa,Bekaert:2004dt}. 

Finally, we can mention that those considerations have been extended to the mixed symmetry tensors. The action extending the Fronsdal action has been provided in \cite{Labastida:1989kw}. Furthermore, the equivalence between this formulation and the unconstrained equation $Tr\ K\approx 0$ has been generalized in \cite{Bekaert:2003az,Bekaert:2007ix}.

\subsection{Antifield formulation}

The Fronsdal Lagrangian is quadratic, hence the theory is linear, in the sense that we have defined in Section \ref{linth}. Thus, its antifield formulation is simple: The longitudinal derivative, which is a true differential in this case, is defined by: \begin{eqnarray}\g \phi_{\mu_1...\mu_s}=s\,\6_{(\mu_1}C_{\mu_2...\mu_{s})}\quad\label{actgamma},\end{eqnarray} where the ghosts $C_{\mu_2...\mu_{s-2}}$ are fermionic and traceless. The action of $\g$ on the ghosts and antifields gives 0.

\noindent The Koszul-Tate differential is related to the equations of motion and the Noether identities: \begin{eqnarray}\d \phi^{*\mu_1...\mu_s}=G^{\mu_1...\mu_s}\end{eqnarray} \begin{eqnarray}\d C^{*\mu_1...\mu_{s-1}}=-s\left[\6_{\mu_s}\phi^{*\mu_1...\mu_s}-\frac{(s-1)(s-2)}{2(n+2s-6)}\eta^{(\mu_1\mu_2}\6_{\mu_s}\phi^*{\phantom{}}'{\phantom{}}^{\mu_3...\mu_{s-1})\mu_s}\right]\quad.\end{eqnarray} Its action on the fields and the ghosts gives 0. Let us recall that the $antigh\ 1$ antifields $\phi^*$ are fermionic and the $antigh\ 2$ antifields $C^*$ are bosonic. The differentials $\d$ and $\g$ anticommute and both anticommute with $d$. Thus, the BRST-BV differential is: $\bs=\d+\g$ and $\forall F\ :\ \bs F=(W,F)$ where the generator is: \begin{eqnarray}W=\int_\mathcal{D}w^n =\int_\mathcal{D}\left[\mathcal{L}_{Fs}+s\phi^{*\mu_1...\mu_s}\6_{(\mu_1}C_{\mu_2...\mu_s)}\right]d^n x\quad.\label{wfrons}\end{eqnarray} We have denoted $w^n=w d^n x$ the $n$-form associated with $W$. In the jet space, the action of $\bs$ on a $n$-form $a^n=a d^n x$ is: $\bs\,a^n=(w,a) d^n x$.

\subsection{Cohomology of $\g$\label{cohgam}}

The computation of the cohomology of $\g$ is of great interest for various works using the antifield formalism. In particular, it plays an important r\^ole in the computation of consistent deformations of a linear theory. This is because the cohomology of $\g$ simultaneously selects gauge invariant expressions of the fields and discards expressions of the ghosts that are the gauge transformation of something. The determination of the cohomology of $\g$ for a particular spin-$s$ theory has been achieved in \cite{Bekaert:2005ka}. We recall this result and show that it extends naturally to the case of a theory involving fields with different spins.

The computation of $H^*(\g)$ consists in the study of Eq.(\ref{actgamma}) and its derivatives. These relations provide non-$\g$-closed combinations of the fields and $\g$-exact combinations of the ghosts. For example: in $\g \phi_{\m_1...\m_s}=s\6_{(\m_1}C_{\m_2...\m_s)}$, we see that the fields $\phi$ are not $\g$-closed, and the symmetrized first derivatives of the ghosts $C$ are $\g$-exact. Since the cohomology consists of cosets of $\g$-closed expressions modulo $\g$-exact ones, the undifferentiated fields clearly do not belong to it. On the other hand, a natural choice of representatives of the cosets is to identify with zero the manifestly $\g$-exact expression, for example: $\6_{(\m_1}C_{\m_2...\m_s)}\in [\,0\,]$. This double use of a coboundary is called ``cancellation by pairs''.

 The idea is, at any given degree of derivation $m$, to decompose the set of $m$th derivatives of the fields and the set of $(m+1)$th derivatives of the ghosts into different components, which can be done by using some Young tableaux (see Appendix \ref{as:gln}). Some of these components cancel by pairs, because of the equation: \begin{eqnarray}\g\6^m_{\r_1...\r_m}\phi^{\phantom{}}_{\m_1...\m_s}=s \6^{m+1}_{\r_1...\r_m(\m_1}C^{}_{\m_2...\m_s)}\quad,\end{eqnarray}the others are the gauge invariant functions and the non-$\g$-exact ghosts. As we already said before, there are two basic gauge-invariant functions, the Fronsdal ($F$) and curvature ($K$) tensors, that contain respectively $2$ and $s$ derivatives. It as been showed in \cite{Bekaert:2005ka} that every other gauge invariant functions are functions of $F$ and $K$. It can also be seen that the $s$th derivatives of the ghosts are all trivial (and thus all derivatives of higher orders). Let us introduce the de Wit--Freedman connections, presented in \cite{deWit:1979pe}:\begin{eqnarray}\forall\ m<s\ :\ \Gamma^{(m)}_{\r_1...\r_m;\m_1...\m_s}&=&\sum_{j=0}^{m} (-1)^j \left(\begin{array}{c}s\\j\end{array}\right)\6^m_{(\m_1...\m_j |(\r_{j+1}...\r_m}\phi^{}_{\r_1...\r_j)|\m_{j+1}...\m_s)}\quad,\label{dwfconn}\end{eqnarray} They are such that:\begin{eqnarray}\g \Gamma^{(m)}_{\r_1...\r_m;\m_1...\m_s}=(-1)^m(m+1)\left(\begin{array}{c}s\\m+1\end{array}\right)\6^{m+1}_{(\m_1...\m_{m+1}}C^{}_{\m_{m+2}...\m_s)\r_1...\r_m}\quad.\end{eqnarray}In particular, for $m=s-1$: $\g \Gamma^{(s-1)}_{\r_1...\r_{s-1};\m_1...\m_s}=-(-1)^s s \6^s_{\m_1...\m_s}C^{}_{\r_1...\r_{s-1}}$. Furthermore, if $s>2$, the Fronsdal tensor is the trace of the second connection: $F_{\m_1...\m_s}=\eta^{\r\s}\Gamma^{(2)}_{\r\s;\m_1...\m_s}$, and we see that it is $\g$-closed because the ghost is traceless: $\g F_{\m_1...\m_s}=\frac{s(s-1)(s-2)}{2}\eta^{\r\s}\6^m_{(\m_1\m_2\m_3}C_{\m_4...\m_s)\r\s}$, which is in agreement with the considerations made above. 
 
We now have to determine which parts of the derivatives of the ghosts are not $\g$-exact in degree $m<s-1$. First, the undifferentiated ghosts cannot be $\g$-exact and are thus in the cohomology. At $m=0$, no combination of the fields is $\g$-closed because the theory is irreducible. On the other hand, the derivatives of the ghosts decompose into the $\g$-exact totally symmetric part and a non exact traceless object, that contributes to the cohomology: \begin{eqnarray}U^{(1)}_{\mu_1\nu_1|\mu_2...\mu_{s-1}}=\6_{[\mu_1}C_{\nu_1]\mu_2...\mu_{s-1}}-\frac{(s-2)}{(n+s-4)}Y^{(1)}\left(\eta_{\mu_1(\mu_2}\6^{\r}C_{\mu_3...\mu_{s-1})\nu_1\r}\right)\quad.\end{eqnarray} It is antisymmetric in $\mu_1$ and $\nu_1$ and totally symmetric in the other $\mu_i$ indices. In terms of Young diagrams, the derivatives of the ghosts can be decomposed by making a product of two rows of boxes: \begin{picture}(17,0)(0,0)\multiframe(0,0)(20,0){1}(10,10){}\end{picture}$\otimes$\begin{picture}(40,0)(0,0)\multiframe(5,0)(20,0){1}(30,10){$s-1$}\end{picture}$=$\begin{picture}(40,0)(0,0)\multiframe(5,0)(20,0){1}(30,10){$s$}\end{picture}$\oplus$\begin{picture}(40,0)(0,0)\multiframe(5,5)(20,0){1}(30,10){$s-1$}\multiframe(5,-5.5)(5,0){1}(10,10){}\end{picture}. The first term cancels by pairs with the undifferentiated fields, and the traceless part of the second term is $U^{(1)}$. Let us remark that it is taken traceless, because the two terms in the sum have the same trace because of the vanishing of the trace of the ghosts. This trace is $\g$-exact, hence the traceless $U^{(1)}$ is the non-$\g$-exact tensor that has the lowest number of components.

At order $m$, the $m$th derivatives of the fields and the $(m+1)$th derivatives of the ghosts can both be visualized with Young diagrams:
\begin{eqnarray}\6^{m+1} C\ :&&\begin{picture}(37,0)(0,0)\multiframe(0,0)(20,0){1}(30,10){$m+1$}\end{picture}\otimes\begin{picture}(40,0)(0,0)\multiframe(5,0)(20,0){1}(30,10){$s-1$}\end{picture}=\begin{picture}(60,0)(0,0)\multiframe(5,0)(20,0){1}(50,10){$m+s$}\end{picture}\oplus\begin{picture}(60,0)(0,0)\multiframe(5,5)(20,0){1}(50,10){$m+s-1$}\multiframe(5,-5.5)(5,0){1}(10,10){}\end{picture}\oplus...\oplus\begin{picture}(50,0)(0,0)\multiframe(5,5)(20,0){1}(40,10){$s$}\multiframe(5,-5.5)(5,0){1}(30,10){$m$}\end{picture}\oplus\begin{picture}(50,0)(0,0)\multiframe(5,5)(20,0){1}(40,10){$s-1$}\multiframe(5,-5.5)(5,0){1}(30,10){$m+1$}\end{picture}\ ,\nonumber\\&&\nonumber \\ \6^m \phi\ :&&\begin{picture}(37,0)(0,0)\multiframe(0,0)(20,0){1}(30,10){$m$}\end{picture}\otimes\begin{picture}(40,0)(0,0)\multiframe(5,0)(20,0){1}(30,10){$s$}\end{picture}=\begin{picture}(60,0)(0,0)\multiframe(5,0)(20,0){1}(50,10){$m+s$}\end{picture}\oplus\begin{picture}(60,0)(0,0)\multiframe(5,5)(20,0){1}(50,10){$m+s-1$}\multiframe(5,-5.5)(5,0){1}(10,10){}\end{picture}\oplus...\oplus\begin{picture}(50,0)(0,0)\multiframe(5,5)(20,0){1}(40,10){$s$}\multiframe(5,-5.5)(5,0){1}(30,10){$m$}\end{picture}\quad.\label{expyt}\end{eqnarray} This decomposition does not keep track of the traces, but we already see that most of the components cancel by pairs. Anyway, the last term in the expansion of the derivatives of the ghosts has no correspondence. We call this object $\widetilde{U}^{(m+1)}$. This set of tensors is defined by:
\begin{eqnarray}\widetilde{U}^{(m)}_{\mu_1\nu_1|...|\mu_m\nu_m|\nu_{m+1}...\nu_{s-1}}=Y^{(m)}\left(\6^{m}_{\mu_1...\mu_m}C_{\nu_1...\nu_{s-1}}\right)\quad.\label{tildeu}\end{eqnarray} Their traces are $\g$-exact, thus we will consider their totally traceless parts $U^{(m)}$, that are the non-$\g$-exact ghost tensor involving $m$ derivatives that have the lowest number of components. We also call them the ``strictly'' non-$\g$-exact ghost tensors, because no linear combinations of them can yield a $\g$-exact expression. The other components are $\g$-exact. In fact, the only problem that happens with the other components in the expansion is that $\g$-closed combinations of the derivatives of the fields appear because the ghosts are traceless. The first one is the Fronsdal tensor, and traces of the other de Wit--Freedman connections share the same property. Furthermore, the connections are also defined recursively as: $\Gamma^{(m)}_{\r_1...\r_m;\m_1...\m_s}=\6_{\r_1}\Gamma^{(m-1)}_{\r_2...\r_m;\m_1...\m_s}-\frac{s}{m}\6_{(\m_1}\Gamma^{(m-1)}_{|\r_2...\r_m;\r_1|\m_2...\m_s)}$. Though the symmetry of the indices $\r$ is not manifest in the right-hand side, this relation is exact. Thus, we see that the trace $\eta^{\r_{m-1}\r_m}$ of $\G^{(m)}$ is a function of the trace of $\G^{(m-1)}$, and thus, the traces are all functions of the Fronsdal tensor. The extension of the definition of the connections to order $s$ is in fact the curvature tensor, written in symmetric convention. We know that it is gauge invariant independently of the tracelessness of the ghosts, and Eq.(\ref{TrKurv}) relates its traces to the Fronsdal tensor. By construction, the curvature tensor satisfies Bianchi identities: \begin{eqnarray}\6_{[\a} K_{\mu_1\nu_1]|...|\mu_s\nu_s}\equiv 0\quad.\end{eqnarray} By taking the trace of these identities, it is found that the divergence of the curvature is proportional to derivatives of its trace, and thus to antisymmetrised derivatives of the Fronsdal tensor:\begin{eqnarray}\6^{\m_s}K_{\m_1\n_1|...|\m_s\n_s}=2^{s-1}Y^{(s-1)}(\6^{s-1}_{\m_1...\m_{s-1}}F_{\n_1...\n_s})\quad.\end{eqnarray} We can choose between considering: the Fronsdal tensor, its symmetrized derivatives, the curvature tensor and all of its derivatives, or considering: the Fronsdal tensor, all of its derivatives and the traceless part of the curvature tensor and all of its derivatives. We have considered the first convention in the sequel.

Finally, we can establish that the cohomology of $\g$ is the set of functions of the following tensors: the antifields and their derivatives, the ghost tensors $U^{(k)}$, the Fronsdal tensor and its symmetrized derivatives $\6^m_{(\r_1...\r_m}\phi_{\m_1...\m_s)}$, and the curvature tensor and its derivatives. We denote a set of fields and their derivatives, considered together, by putting square brackets around the fields, we also denote the antifields $\phi^*$ and $C^*$ collectively as $\Phi^*_I$:\begin{eqnarray}\textrm{For any spin-$s$: }\ H^*(\g)=\left\{f\left([\Phi^*_I],[F]_{sym},[K],U^{(1)},...,U^{(s-1)}\right)\right\}\quad.\end{eqnarray} When local objects are considered, the total number of derivatives is bounded and the elements of the cohomology are polynomials in the ghosts and the antifields. Products of $U^{(k)}$ tensors provide a basis of the non-$\g$-exact polynomials in the ghosts, that we denote $\{\omega_J\}$. We can thus give the following presentation of the cohomology: \begin{eqnarray}H^i(\g)=\{\a^J\omega_J\ |\ \a^J\in H^0(\g)\ \textrm{and}\ pgh\ \omega_J=i\}\quad.\end{eqnarray} 
\subsubsection*{The $pgh$ 0 sector} 
A given $pgh$ 0 expression $f$ only depends on the fields and the antifields. The set of fields $\displaystyle\left\{[\phi]\right\}$ can be decomposed into a non-$\g$-closed sector $[\phi]^\Delta$ and the $\g$-closed tensors $[K]$ and $[F]_{sym}$. The action of $\g$ on $[\phi]^\Delta$ provides a basis of the $\g$-exact linear combinations of the derivatives of the ghosts, that we denote $\bar{\mathcal{C}}^\Delta$ (they can be built by constructing explicitly the tensors in the expansion of $\6^{m+1}\ C$ presented in Eq.(\ref{expyt})). Then, if we require $\g f=0$, since $\g f$ is linear in the $\bar{\mathcal{C}}^\Delta$, their coefficients must vanish. As a matter of fact, these are exactly the coefficients of the fields $[\phi]^\Delta$ in $f$, thus we have showed that $f$ does not depend on them:\begin{eqnarray}[f]\in H^0(\g)\Rightarrow f=f([\Phi^*_I], [K], [F]_{sym})\quad.\end{eqnarray} 
\subsubsection*{The $pgh\ k$ sector}
A given $pgh\ k$ object $f^k$ can be written: $f^k=f_\Delta \bar{\mathcal{C}}^\Delta+f_J\omega^J$ where $pgh f_\Delta=k-1$ and $pgh f_J=0$. Then, $\g f^k=0\Rightarrow \g(f_J\omega^J)=0\ \textrm{and}\ \g(f_\Delta\bar{\mathcal{C}}^\Delta)=0$. The two relations decouple because the first one is linear in the $\bar{\mathcal{C}}^\Delta$ while the second one is at least quadratic in them. Let us expand the first relation: $\g(f_J\o^J)=(\g f_J)\o^J=\r_{\D J}\bar{\mathcal{C}}^\D\o^J=0$. Since $\bar{\mathcal{C}}^\D$ and $\o^J$ are independent, the coefficients vanish and we obtain $\r_{\D J}\bar{\mathcal{C}}^\D=\g f_J=0\Rightarrow f_J=\a_J\in H^0(\g)$. In order to prove that the second part is trivial, let us first assume that $f_\Delta$ depends on the non-$\g$-closed fields at a given power $j$: $f_\Delta=f_{\Delta\Delta_1...\Delta_j}[\phi]^{\Delta_1}...[\phi]^{\Delta_{j}}$. The condition $\g(f_{\Delta}\bar{\mathcal{C}}^\Delta)=0$ provides that the coefficients must be totally symmetric: $f_{\Delta\Delta_1...\Delta_j}=f_{(\Delta\Delta_1...\Delta_j)}$. Then, we obtain that $$f_\Delta\bar{\mathcal{C}}^\Delta=\pm \g(f_{\Delta\Delta_1...\Delta_j}[\phi]^{\Delta_1}...[\phi]^{\Delta_{j}}[\phi]^\Delta)\mp\g(f_{\Delta\Delta_1...\Delta_j})[\phi]^{\Delta_1}...[\phi]^{\Delta_j}[\phi]^\Delta\quad.$$ However, $\g f_{\Delta\Delta_1...\Delta_j}=0$ because it does not depend on $[\phi]^\Delta$ by definition. Finally, the result extents automatically to a sum of terms at various powers in the $[\phi]^\Delta$, because of linear independence (the only problem that can arise is the limit as $j\rightarrow\infty$, but we will usually consider polynomials in the fields). We have thus established that $f_\D\bar{\mathcal{C}}^\D$ is $\g$-exact, hence the cohomology can be written in terms of the strictly non-$\g$-exact ghost tensors only, as announced.

\noindent{\bf Remark:} The cohomology class $H^0(\g)$ is usually related to the adjective ``invariant'', because it is composed of gauge invariant expressions. In the case where the expressions are polynomials in the fields as well as in the ghosts and antifields, which is the case for the theories under considerations and their deformations, the class is called the ``invariant polynomials''. Furthermore, a $\g$-closed expression is also called an invariant expression, and some (co)homologies restricted to $H^0(\g)$ are called invariant (co)homologies. 

\subsubsection*{Sum of several Fronsdal theories}

Let us consider a family of fields with various spins: $\{\phi^a_{\mu_1...\mu_{s_a}}\}$, such that the Lagrangian is the sum of their Fronsdal Lagrangians: \begin{eqnarray}\cl = \frac{1}{2}\sum_a\phi^{\mu_1...\mu_{s_a}}_a G^a_{\mu_1...\mu_{s_a}}\quad.\end{eqnarray} In the case where several fields are related to the same spin, the action could more generally contain an internal metric, but we always consider it positive definite, in such a way that the fields can be redefined in a diagonal way. By K\"unneth's formula, the cohomology of $\g$ is the direct product of the cohomologies of the individual theories. Let us provide some details about that fact. The gauge transformations, or the action of $\g$, are decoupled: \begin{eqnarray}\g \phi^a_{\mu_1...\mu_{s_a}}=s_a\6_{(\mu_1} C^a_{\mu_2...\mu_{s_a})}\quad,\end{eqnarray} so that $\g$ can always be seen as the sum of its restrictions for each field: $\g=\sum_a \g_a$. In order to define the cohomology of a given $\g_a$, let us denote $\{\phi\}^{\hat{a}}$ the set of all fields except $\phi^a$. Then it is obtained that: \begin{eqnarray}H^*(\g_a)=\left\{f(\{\phi\right\}^{\hat{a}})\}\otimes H^*(\g_a)\vert_{\{\phi\}^{\hat{a}}=0}\quad.\nonumber \end{eqnarray}
Given an arbitrary $\g$-closed function $f$ at $pgh$ 0, which does not depend on any ghost, we have $\g f = \sum_a \g_a f=0$. Since $\g_a f$ is linear in the ghosts $[C^a]$, the different terms are linearly independent and all of them vanish. This means that $f$ is in the intersection of the cohomologies of the different $\g_a$ so that: \begin{eqnarray}\nonumber[f]\in H^0(\g)\Rightarrow f=f([\phi^{*a}_I],[F^a]_{sym},[K^a])\quad.\end{eqnarray}The proof in $pgh>0$ remains valid if the definitions of $[\phi]^\Delta$, $\bar{\mathcal{C}}^\Delta$ and $\omega^J$ are properly extended. In particular, the $\omega^J$ are chosen to provide a basis of the products of the different ghost tensors $U^{a(j)}\ |j\leqslant s_a-1$. The $pgh\ i$ class is once again presented as: \begin{eqnarray}H^i(\g)=\{\a_J\omega^J\ |\ \a_J\in H^0(\g)\}\quad.\end{eqnarray}

\subsection{Further results about $\g$\label{furgam}}

In this section, we provide some cohomological results and definitions that will prove to be useful in the sequel. They are adapted from the corresponding results in \cite{Boulanger:2000rq}.

\begin{theorem}\label{hdinv} The cohomology of $d$ in the space of invariant functions $H^0(\g)$ has no positive $antigh$ sector in form degree less than $n$:\begin{eqnarray}\forall p<n\ :\ [a]\in H^p(d,H^0(\g))\Rightarrow antigh\ a=0\quad.\end{eqnarray}\end{theorem}

Let us first provide an example for spin-2: the set of curvature 2-forms $\Omega_{\a\b}=R_{\m\n|\a\b}dx^\m dx^\n$ is $d$-closed thanks to the Bianchi identities. The algebraic Poincar\'e lemma ensures that there exists 1-forms such that $\Omega_{\a\b}=d T_{\a\b}$, but these 1-forms cannot be invariant under $\g$ because $R_{\m\n|\a\b}$ is the invariant tensor with the lowest number of derivatives. Thus $d$ has some cohomology in $H^0(\g)$ in $antigh$ 0. However, it is not the case in $antigh\ k>0$. 

\noindent Proof: If $\g a_k=0$ and $d a_k=0$, then the Poincar\'e lemma ensures that, in form degree lower than $n$: $\exists b_k\ |\ a_k=db_k$. We can use an argument consisting in considering the antifields as ``foreground'' fields and the gauge invariant tensors as ``background fields'', which was described in \cite{Dubois-Violette:1991is}. The exterior differential can be split into two parts: $d_0$ acting only on the fields and the ghosts, and $d_1$ acting only on the antifields. These two derivatives are both differentials and have no cohomology in form degree $0<p<n$. Furthermore, $d_1$ has no cohomology in form degree 0 and strictly positive $antigh$, because a ``constant'' in the sense of $d_1$, i.e. a function of the fields and the ghosts only, is at $antigh\ 0$. Then, we can consider the grading corresponding to the number of derivatives acting on the antifields, say $j$, bounded by a maximal value $m$. We can expand $a_k$ according to that degree: $a_k=\sum_{j=0}^m a_k^{(j)}$. Clearly, $d_1$ raises that grading by one, while $d_0$ leaves it unchanged. We have to consider first the highest degree component of the cocycle: $d_1 a_k^{(m)}=0\Rightarrow a_k^{(m)}=d_1 b_k^{(m-1)}$, where $b_k^{(m-1)}$ is invariant, since $d_1$ does not act on the fields and thus cannot create tensors $[F]$ or $[K]$. We can now redefine $a'_k=a_k-db_k^{(m-1)}$: $a'_k$ is $d$-closed and its expansion in the number of derivatives of the antifields stops at degree $m-1$. The same argument can be repeated for each value of this degree, thus each component of $a_k$ at a given degree is removed by the addition of a $d$-exact expression whose object is invariant. The argument stop when reaching degree $0$. Thus $b_k$ can be chosen as being invariant $\Box$. Let us notice that this theorem holds for a sum of several Fronsdal theories.

\subsubsection*{The differential $D$}

We can now introduce the useful differential $D$, that is similar to $d$ but permits to isolate non-$\g$-exact ghosts while still being related to the form degree. First, the action of $D$ on the fields and the antifields is exactly the same as that of $d$. Its action on a ghost or a derivative of a ghost $\mathcal{C}^A$ differs from that of $d$ by a $\g$-exact object: $D\mathcal{C}^A=d\mathcal{C}^A+\g f^A$. Furthermore, $D$ is internal in the space of strictly non-$\g$-exact ghosts: $D\omega_J=A^I_J\omega^I$. More explicitly, its action is given by: \begin{eqnarray}DC_{\mu_2...\mu_{s}}&=&\frac{2(s-1)}{s}U^{(1)}_{\mu_1(\mu_2|\mu_3...\mu_{s})}dx^{\mu_1}\nonumber\\&...&\nonumber\\ DU^{(j)}_{\mu_1\nu_1|...|\mu_j\nu_j|\nu_{j+1}...\mu_{s-1}}&=&\frac{2(s-j-1)}{s-j}U^{(j+1)}_{\mu_1\nu_1|...|\mu_{j+1}(\nu_{j+1}|\nu_{j+2}...\mu_{s-1})}dx^{\mu_{j+1}}\nonumber\\&...&\nonumber\\ DU^{(s-1)}_{\mu_1\nu_1|...|\mu_{s-1}\nu_{s-1}}&=&0\quad. \label{defD}\end{eqnarray} The action of $D$ on the ghosts raises by one the number of derivatives acting on them. We define the $D$-degree as the total number of derivatives acting on the ghosts. It will often appear as a subindex of the ghost index, so that for example, the action of $D$ on the basis of the non exact ghosts can be written more precisely as $D\omega^{J_i}=A^{J_i}_{J_{i+1}}\omega^{J_{i+1}}$. It is a good grading that allows to distinguish linearly independent components in some equations. Of course, this definition extends automatically when considering several independent fields. Then, it is useful to prove the following consequence of Theorem \ref{hdinv}:
\newpage
\begin{theorem} If $\g a_k + d b_k = 0$ with $antigh\ a_k=k>0$, then $\exists\;e_k\ |\ \g a'_k=\g (a_k+de_k)=0$ .\label{gdg}\end{theorem}
\noindent Proof: First, a descent can be established. When acting with $\g$ on the given cocycle (let us assume that $a_k$ is a $p$-form), it is obtained that $d\g b_k=0\Rightarrow \exists\ c_k\ |\ \g b_k+d c_k=0$, thanks to the algebraic Poincar\'e lemma. The latter relation is a cocycle of the same structure than the first one, but in form degree $p-1$. By acting repeatedly with $\g$, the following equations are finally obtained: $\g m_k+ d n_k=0$ and $\g n_k=0$, either because $n_k$ is a 0-form or is naturally $\g$-closed. Let us notice that, once again, no constants appear in form degree 0 in strictly positive antifield number. Let us choose $n_k$ as a non-trivial representative of $H^*(\g)$: $n_k=\alpha_J\omega^J$. The second to last equation becomes: \begin{eqnarray}(d\alpha_J)\omega^J\pm\alpha_J d\omega^J+\g m_k=0\quad,\end{eqnarray} where the sign of the second term depends on the Grassmann parity of $\a_J$. Since $d\omega^J=D\omega^J+\g f^J$ and $\g\alpha_J=0$ , we find that $\a_Jd\o^J=\a_JD\o^J\pm \g(\a_Jf^J)$. Hence, $m_k$ can be redefined in such a way that: \begin{eqnarray}d\alpha_J\omega^J\pm\alpha_JD\omega^J=-\g m'=0\quad.\end{eqnarray}
Both expressions vanish because the left hand side has be written in a strictly non-$\g$-exact way. The $D$-degree 0 component is $d\alpha_{J_0}\omega^{J_0}=0$, it yields $d\alpha_{J_0}=0$ and thus $\exists\beta_{J_0}\in H^0(\g)\ |\  \alpha_{J_0}=d\beta_{J_0}$ thanks to theorem \ref{hdinv}. This implies that:\begin{eqnarray}\a_{J_0}\omega^{J_0}=d(\b_{J_0}\omega^{J_0})\mp\b_{J_0}D\omega^{J_0}+\g(...)\quad.\end{eqnarray} In other words, the object $n_k$ can be redefined by adding $\g$ and $d$-exact terms, in such a way that its $D$-degree expansion begins at 1. Then, the new bottom equation is $d\alpha'_{J_1}\omega^{J_1}=0$. By using the same arguments, it is now clear that all of the $D$-degree components of $n_k$ can be removed in the same way. Since the $D$-degree is bounded, because there is a finite number of generators $\omega^J$ in the basis of the non-$\g$-exact ghosts, it is found that $n_k$ is trivial. This implies that the second to last equation becomes $\g m_k=0$ and the same reasoning can be applied to remove one by one the equations of the descent. Finally, it is obtained that $\exists e_k\ | b_k = -\g e_k + d(...)\Rightarrow \g(a_k+ de_k)=0$ $\Box$.

\begin{theorem}The cohomology of $\d$ remains trivial in the space of invariant functions:\begin{eqnarray}\forall k>0\ : H_k(\d,H^0(\g))=0\quad.\end{eqnarray}\label{hdelinv}\end{theorem}
\noindent In other words, if $a\in H^0(\g)$ and $a=\d b$, then $b$ can be chosen invariant. Since $\d$ only acts on the antifields, it commutes with the operation that consists in setting to zero the non-$\g$-closed functions of the fields $[h]^\D$. Since $a$ is invariant, it does not depend on $[h]^\Delta$ and we obtain that: \begin{eqnarray}\nonumber a=a|_{[h]^\D=0}=(\d b)|_{[h]^\D=0}=\d(b|_{[h]^\D=0})\quad.\end{eqnarray}By definition, $b|_{[h]^\D=0}$ is invariant $\Box$.

\subsection{The homology class $H_2^n(\d|d)$}

It has been showed in Theorem \ref{hkndeltd} that, for an irreducible linear theory such as a Fronsdal theory, many classes of the homology of $\d$ modulo $d$ vanish: \begin{eqnarray}\forall k>2\ :\ H_k^n(\d|d)=0\quad.\end{eqnarray} Furthermore, it is possible to compute the class $H_2^n(\d|d)$ in the case of as sum of Fronsdal theories. We will review the proof given in \cite{Bekaert:2005ka} (see also \cite{Barnich:2005bn}). Let us consider a generic $antigh\ 2$ $n$-form: \begin{eqnarray} a_2^n=\sum_a f^a_{\mu_1...\mu_{s_a-1}}C^{*\mu_1...\mu_{s_a-1}}_a d^n x+\mu+ d(...)\quad,\end{eqnarray} where $\mu$ is quadratic in the antifields $\phi^{*(\mu)}_a$ or some of their derivatives. The functions $f^a$ are chosen traceless. This writing can be justified as follows: the terms linear in the derivatives of $C^{*(\m)}_a$ can always be written as the sum of an expression linear in the undifferentiated antifield and a $d$-exact term. It can also be seen that $\d\mu\approx 0$, because $\d$ acts on one of the two $antigh\ 1$ antifields. This brings in a set of equations of motion. It is then obtained that: 
\begin{eqnarray}\d a_2^n&\approx&-\sum_a s_a f^a_{\mu_1...\mu_{s_a-1}}\6_{\mu_{s_a}}\phi^{*\mu_1...\mu_{s_a}}_a + d(...)\\&\approx& \sum_a s_a \6^{\phantom{\mu_1}}_{(\mu_1}f^a_{\mu_2...\mu_{s_a})}\phi^{*\mu_1...\mu_{s_a}}_a + d(...)\quad.\end{eqnarray} 
On the other hand, as $a_2^n$ is in the homology of $\d$ modulo $d$: $\exists\ b_1^{k-1}\ |\ \d a_2^n+ d b_1^{k-1}=0$, and thus: 
\begin{eqnarray}\nonumber \exists v\ | \sum_a s_a \6^{\phantom{\mu_1}}_{(\mu_1}f^a_{\mu_2...\mu_{s_a})}\phi^{*\mu_1...\mu_{s_a}}_a\approx dv\quad.\end{eqnarray} 
Then, we can apply the variational derivative with respect to $\phi^*$, which gives 0 when acting on $dv$ and do not alter the on-shell equality. Since the different fields are independent, the following equations are obtained: 
\begin{eqnarray} \forall a\ : \6^{\phantom{\mu_1}}_{(\mu_1}f^a_{\mu_2...\mu_{s_a})}\approx 0\quad\label{weakill}.\end{eqnarray} 
The solution of this equation is weakly equal to a function of $x^\mu$ thanks to the fact that $H_0^0(d|\d)\cong\mathbb{R}$. To see this, let us apply $s-1$ derivatives to Eq.(\ref{weakill}), in a spin-$s$ case. Some appropriate combination yields the following equation:
\begin{eqnarray}\6_{\nu_1...\nu_s}f_{\mu_1...\mu_{s-1}}\approx 0\quad.\end{eqnarray} 
Thanks to Theorems \ref{desctheo}, \ref{ddeltheo} and \ref{hkndeltd}, we find that $H_n^n(\d|d)=0\Rightarrow H_0^0(d|\d)\cong \mathbb{R}$, hence it is found that: $\6_{\nu_1...\nu_{s-1}}f_{\mu_1...\mu_{s-1}}\approx \lambda_{\mu_1...\mu_{s-1}|\nu_1...\nu_{s-1}}\ |\ \lambda_{\mu_1...\mu_{s-1}|\nu_1...\nu_{s-1}}\in \mathbb{R}$. Furthermore, the coefficients $\l$ are chosen traceless in the $\mu$ indices. We can then proceed with the resolution of the equation, let us define: $\hat{f}_{\mu_1...\mu_{s-1}}=f_{\mu_1...\mu_{s-1}}-\l_{\mu_1...\mu_{s-1}|\nu_1...\nu_{s-1}}x^{\nu_1}...x^{\nu_{s-1}}$. It implies that $\6_{\nu_1...\nu_{s-1}}\hat{f}_{\mu_1...\mu_{s-1}}\approx 0$. Once again, since $H_0^0(d|\d)\cong\mathbb{R}$, the solution is $\6_1...\6_{\n_{s-2}}\hat{f}_{\mu_1...\mu_{s-1}}\approx \lambda_{\mu_1...\mu_{s-1}|\nu_1...\nu_{s-2}}\ |\ \lambda_{\mu_1...\mu_{s-1}|\nu_1...\nu_{s-2}}\in \mathbb{R}$. By going on like this repeatedly, it is finally obtained that \begin{eqnarray}f_{\mu_1...\mu_{s-1}}\approx\sum_{t=1}^{s-1}\l_{\mu_1...\mu_{s-1}|\nu_1...\nu_t}x^{\nu_1}...x^{\nu_t}\ \label{wksol},\end{eqnarray} where the sets of coefficients $\l$ are constants and are traceless in the $\mu$ indices. Finally, these coefficients must obey some further relations when the above expression for $f_{\mu_1...\mu_{s-1}}$ is brought back into Eq.(\ref{weakill}): \begin{eqnarray}\forall\ t\ : \l_{(\mu_1...\mu_{s-1}|\nu_1)\nu_2...\nu_t}=0\quad.\end{eqnarray} These relations are strongly vanishing because they concern only constants and not functions of the fields. They ensure that the coefficients $\l$ are represented by two-rows Young tableaux, with lengths $s-1$ and $t$. The right hand side of Eq.(\ref{wksol}) is in fact the solution of the strong ``Killing'' equation $\6_{(\mu_1}\xi_{\mu_2...\mu_s)}=0$ and we can finally write the result implicitly: \begin{eqnarray}\6_{(\mu_1}f_{\mu_2...\mu_s)}\approx0\Leftrightarrow f_{\mu_1...\mu_{s-1}}\approx \xi_{\mu_1...\mu_{s-1}}\ |\6_{(\mu_1}\xi_{\mu_2...\mu_{s-1})}=0\quad.\end{eqnarray} This can now be introduced into the expression of $a_2^n$:
\begin{eqnarray}a_2^n=\xi_{\mu_1...\mu_{s_a-1}}^a C^{*\mu_1...\mu_{s_a-1}}_a+\mu'+ \d(...)+ d(...)\quad,\end{eqnarray} 
where $\mu'$ is the sum of the old $\mu$ and of terms brought in by the weak equality $f^a_{\mu_1...\mu_{s_a-1}}\approx \xi^a_{\mu_1...\mu_{s_a-1}}$, which denotes the presence of a $\d\phi^*$. By construction, it is obvious that $\d (\xi_{\mu_1...\mu_{s_a-1}}^a C^{*\mu_1...\mu_{s_a-1}}_a)=d(...)$. Thus, we are left with the equation $\d\mu'=d(...)$. This cocycle is in fact trivial, as it has been showed in \cite{Barnich:1995db}. The proof goes along the same lines as the proof of Theorem \ref{hkndeltd}. We can summarize the result as follows:
\begin{theorem}In the case of a sum of Fronsdal theories, the homology class $H_2^n(\d|d)$ takes the form: \begin{eqnarray}H_2^n(\d|d)=\left\{[a_2^n]\ | a_2^n=\xi^a_{\mu_1...\mu_{s_a-1}}C^{*\mu_1...\mu_{s_a-1}}_a d^n x\ ,\ \forall\ a\ :\ \6^{\phantom{a}}_{(\mu_1}\xi^a_{\mu_2...\mu_{s_a-1})}=0\right\}\quad.\nonumber\end{eqnarray}\label{hdndeltd}\end{theorem}

\section{Fronsdal theory in a de Sitter or Anti de Sitter spacetime\label{FrAdS}}

The Fronsdal action can be extended to de Sitter/Anti de Sitter spacetimes, thanks to the expression of the Riemann tensor on these manifolds, as was showed in \cite{Fronsdal:1979vb}. The Riemann tensor satisfies the following relation: \begin{eqnarray}\car_{\a\m|\b\n}=\frac{2\L}{(n-1)(n-2)}\left(g_{\a\b}g_{\m\n}-g_{\a\m}g_{\b\n}\right)\quad,\label{riemds}\end{eqnarray} where $\L$ is the cosmological constant and $g_{\a\b}$ is the metric tensor of the considered $(A)dS$ spacetime. A $dS$ spacetime is characterized by a positive $\L$; an $AdS$ spacetime is characterized by a negative $\L$. Let us emphasize that the metric $g_{\m\n}$ is not a dynamical field. The gauge transformations become:\begin{eqnarray}\d_\xi \phi_{\mu_1...\mu_s}=s\N_{(\mu_1}\xi_{\mu_2...\mu_s)}\quad,\label{gtds}\end{eqnarray}where the covariant derivative is built with the Levi-Civita connexion of the metric $g_{\m\n}$. There are the same tracelessness constraints as in the Minkowski case: $\phi''_{\m_5...\m_s}=0$ and $\e'_{\m_4...\m_s}=0$. The covariant derivatives do not commute but, since the commutator brings in the Riemann tensor, we can always say that covariant derivatives commute up to terms involving a lower number of derivatives. For example, the Fronsdal tensor can be extended to $(A)dS$ by covariantizing the derivatives and adding some appropriate combination of the undifferentiated fields, in such a way that it is invariant under the above gauge transformations: \begin{eqnarray}F_{\mu_1...\mu_s}&=&\Box\phi_{\mu_1...\mu_s}-s\N_{(\mu_1}\N^\r\phi_{\mu_2...\mu_s)\r}+\frac{s(s-1)}{2}\N_{(\mu_1}\N_{\mu_2}\phi'_{\mu_3...\mu_s)}\nonumber\\&&-\frac{2\L}{(n-1)(n-2)}\Big[\left(s^2+(n-6)s-2n+6\right)\phi_{\mu_1...\mu_s}+s(s-1)\eta_{(\mu_1\mu_2}\phi'_{\mu_3...\mu_s)}\Big]\quad.\label{tfads}\end{eqnarray} The case of the curvature tensor is much more complicated. We can first consider the covariantization of $Y^{s}(\6_{\m_1...\m_s}\phi_{\n_1...\n_s})$. Its gauge transformation only involves terms with $s-1$ covariant derivatives. In order to correct them, an appropriate expression involving $(s-2)$th covariant derivatives of the fields must be added. The gauge transformation then involves terms with $s-3$ derivatives. The construction can be continued until one adds terms with $0$ or $1$ covariant derivatives. The determination of this expansion has been done in \cite{Bastianelli:2008nm}, we will only write it schematically: \begin{eqnarray}K_{\mu_1\nu_1|...|\mu_s\nu_s}=2^s Y^{(s)}\left(\N_{(\mu_1}...\N_{\mu_s)}\phi_{\nu_1...\nu_s}\right)+\textrm{lower order terms}\quad.\label{tkads}\end{eqnarray} Furthermore, the relation between the trace of $K$ and the derivatives of the Fronsdal tensor can be extended in the same way, it has been studied in \cite{Engquist:2007yk}: \begin{eqnarray}\eta^{\mu_{s-1}\mu_s}K_{\mu_1\nu_1|...|\mu_s\nu_s}=2^{s-2}Y^{(s-2)}(\N_{\mu_1}...\N_{\mu_{s-2}}F_{\nu_1...\nu_s})\ +\textrm{l. o. terms involving only $F$}\quad.\end{eqnarray} The other relations of section \ref{FTM} can be extended straightforwardly:
\begin{eqnarray}\N^{\mu_s}F_{\mu_1...\mu_s}=\frac{s-1}{2}\N^{\phantom{*}}_{(\mu_1}F'_{\mu_2...\mu_s)}\quad.\end{eqnarray} The generalized Einstein tensor is still
\begin{eqnarray}G_{\mu_1...\mu_s}=F_{\mu_1...\mu_s}-\frac{s(s-1)}{4}g_{(\mu_1\mu_2}F'_{\mu_3...\mu_s)}\quad,\label{tgads}\end{eqnarray}  and it satisfies the Noether identities of the above gauge transformations: \begin{eqnarray}\N^{\mu_s}G_{\mu_1...\mu_s}-\frac{(s-1)(s-2)}{2(n+2s-6)}g_{(\mu_1\mu_2}\N^{\mu_s}G'_{\mu_3...\mu_{s-1})\mu_s}\equiv 0\quad.\end{eqnarray} Finally, the Fronsdal Lagrangian describing the free massless spin-$s$ field is: \begin{eqnarray}\mathcal{L}_{Fs}=\frac{1}{2}\sqrt{-g}\phi^{\mu_1...\mu_s}G_{\mu_1...\mu_s}\quad,\label{flagads}\end{eqnarray} where $g=det\ g_{\mu\nu}$. The equations of motion are $G_{\m_1...\m_s}\approx 0$. Given this, the theory has the same number of physical degrees of freedom as in the Minkowski case. Let us remark that the limit $\L\rightarrow 0$ applies the $(A)dS$ Lagrangian on the Minkowski Lagrangian, while the generators of the $(A)dS$ algebra are applied on those of the Poincar\'e algebra. This process is referred to as an In\"on\"u-Wigner contraction \cite{Inonu:1953sp}.

%%%%%%%%%%%%%%%%%%%%%%%%%%%%%%%%%%%%%%%%%%%%%%%%
\chapter{The antifield consistent deformation scheme}\label{ch:antidef}
%%%%%%%%%%%%%%%%%%%%%%%%%%%%%%%%%%%%%%%%%%%%%%%%

The antifield formalism, that we have described in Chapter \ref{ch:anti}, can be used to reformulate the problem of the deformations of gauge theories as an expansion of the generator $W$, in such a way as to satisfy the master equation at every order in the deformation parameter. This point of view, developed in \cite{Barnich:1993vg}, allows one to completely solve some deformation problems under very few assumptions. For example, we have considered Fronsdal Lagrangians as a starting point and have obtained several results about the first order cubic vertices, as well as the first order of deformation of the gauge transformations and the gauge algebra. In the sequel, we describe this method, some general features and some results in the context of higher spin theories.

\section{Deformations of the master equation}
\subsection{Consistent deformations of a gauge theory}

First, let us define a nontrivial consistent deformation of a Lagrangian gauge theory: As a starting point, a zeroth order action $\stackrel{(0)}{S}$ is considered. At this stage, we do not consider the particular case of an $\stackrel{(0)}{S}$ quadratic in the fields. Some zeroth order gauge transformations are assumed to exist: $\stackrel{(0)}{\d}_\ve \phi^i=\stackrel{(0)}{R}\!\!\phantom{}^i_\a\ve^\a$, as well as reducibility relations and gauge algebra. The zeroth order operators will always be denoted with the index $(0)$. The deformations of this initial theory that are considered here consist in building expansions of the initial action, of the generating set of the gauge transformations, and of the different generations of reducibility relations, in powers of an arbitrary parameter $g$. A requirement for the consistency of these deformations is the preservation of the number of gauge transformations and the number of reducibility relations (for each generation of them). However, an Abelian theory, with $\stackrel{(0)}{C}\!\!{}^\a_{\b\g}=0$ and $\stackrel{(0)}{M}\!\!{}^{ij}_{\b\g}=0$, can become nonabelian. These expansions of the action, the generating set and the reducibility operators of the gauge transformations are denoted:
\begin{eqnarray}S&:=&\stackrel{(0)}{S}+g\stackrel{(1)}{S}+g^2\stackrel{(2)}{S}+...\\ R^i_\a&:=&\stackrel{(0)}{R}\!\!\phantom{}^i_\a+g\stackrel{(1)}{R}\!\!\phantom{}^i_\a+g^2\stackrel{(2)}{R}\!\!\phantom{}^i_\a+...\\ Z^\a_A&:=&\stackrel{(0)}{Z}\!\!\phantom{}^\a_A+g\stackrel{(1)}{Z}\!\!\phantom{}^\a_A+g^2\stackrel{(2)}{Z}\!\!\phantom{}^\a_A+...\\...&&\end{eqnarray} 
The complete action $S$ is required to be invariant under the complete gauge transformations, the commutator of which involves the complete structure operators. The gauge invariance of the action is equivalent to requiring that the Noether identities are satisfied. The Noether identities and the different reducibility relations are equalities of formal series in the deformation parameters. These series have to be decomposed into their components at any order in $g$. It is important to notice that there are some trivial ways of deforming a theory. First, redefinitions of the fields can always be done: \begin{eqnarray}\phi'{}^i=\phi^i+gF^i(\phi^j)\ .\end{eqnarray}The modification of the action that this induces does not alter the dynamics:
\begin{eqnarray}\stackrel{(0)}{S}[\phi^i+F^i]=\stackrel{(0)}{S}[\phi^i]+g\int_\mathcal{D}F^i\frac{\d\stackrel{(0)}{S}}{\d\phi^i}d^n x+...\quad.\end{eqnarray} Thus, any expansion of the action involving the equations of motion is trivial and can be discarded. On the same pattern, the gauge transformations can be modified trivially, either by adding trivial gauge transformations or by ``rotating'' the generating set: $R^i_\a=\stackrel{(0)}{R}\!\!\phantom{}^i_\a+g\e^\b_\a\stackrel{(0)}{R}\!\!\phantom{}^i_\b+g\mu^{ij}_\a\frac{\d\stackrel{(0)}{S}}{\d\phi^j}+...$. That kind of deformation can also be discarded.

\subsection{Deformations in the antifield formalism}

Let us now formulate these considerations in the antifield formalism. As we showed in Chapter \ref{ch:anti}, the generator $W$ carries all the information about the theory. The different operators of the theory can be read directly in the expression of $W$. First, the $antigh\ 0$ component of $W$ is the action: $W_0=S$. Then, the $antigh\ 1$ component is the only one linear in the ghosts and in the $antigh\ 1$ antifields, its coefficient is the generating set of the gauge transformations: $W_1=\phi^{*}_i R^i_\a C^\a$. In the next chapters, we will frequently consider the $antigh\ 1$ term of the first order deformations of $W$. Its determination is strictly equivalent to the determination of the non trivial deformations of the gauge transformations. Similarly, the $antigh\ 2$ terms linear in the $antigh\ 2$ antifields and quadratic in the ghosts provide the structure coefficients of the gauge algebra, while the terms quadratic in the $antigh\ 1$ antifields provide the operator $M^{ij}_{\a\b}$. Thus, the determination of the $antigh\ 2$ component of the first order deformations of $W$ is strictly equivalent to the determination of the non trivial deformations of the gauge algebra. The results obtained in the antifield formalism provide explicitly the expressions of the operators. Thus, these results can be translated in the gauge formalism with no difficulty.

The fact that the generator $W$ is the only object needed to describe the theory in the antifield formalism is particularly interesting for the problem of consistent deformations. The generator is the only object that has to be deformed, and the only relation that it must satisfy is the master equation. The initial theory is described by a zeroth order generator $\stackrel{(0)}{W}$ that satisfies the master equation: $\left(\stackrel{(0)}{W},\stackrel{(0)}{W}\right)=0$. The differential $\bs$ that is considered throughout this chapter is the one of the initial theory: \begin{eqnarray}\forall\ A\ : \bs A=\left(\stackrel{(0)}{W},A\right)\quad,\end{eqnarray} as well as its components: the Koszul-Tate differential $\d$ and the longitudinal derivative $\g$. The generator is expanded in powers of the same parameter $g$ as before:
\begin{eqnarray}
\stackrel{\phantom{(0)}}{W}:=\stackrel{(0)}{W}+g\stackrel{(1)}{W}+g^2\stackrel{(2)}{W}+...\quad.\end{eqnarray} The complete master equation reads: \begin{eqnarray}(W,W)=0\quad.\end{eqnarray} It has to be satisfied at all orders in the parameter $g$. In other words, each component of the expansion in $g$ of the master equation must vanish. The zeroth order of this expansion is the master equation of the initial theory and is satisfied by assumption. At first order in $g$, we get the equation: \begin{eqnarray}2\left(\stackrel{(0)}{W},\stackrel{(1)}{W}\right)=2\bs\stackrel{(1)}{W}=0\quad.\label{fofunc}\end{eqnarray} This equation is the one that provides the consistent first order deformations. Let us now consider the field redefinitions in the antifield formalism: \begin{eqnarray}\stackrel{\phantom{(0)}}{S}=\stackrel{(0)}{S}+g F^i\frac{\d\stackrel{(0)}{S}}{\d\phi^i}+...=\stackrel{(0)}{S}+\d(g F^i\phi^*_i)+...\quad,\end{eqnarray} by definition of the Koszul-Tate differential: $\d\phi^{*}_i=\frac{\d \stackrel{(0)}{S}}{\d\phi^i} $. In $antigh\ 0$, trivial deformations thus appear as $\d$-exact terms in the action. In fact, it can be showed (see \cite{Henneaux:1998i}) that the trivial deformations appear as $\bs$-exact terms in the generator $W$, whose $antigh\ 0$ terms precisely correspond to redefinitions of the fields in the action. This is another interesting feature of the antifield formalism, in which physically equivalent expressions are gathered in equivalence classes. In the case of first order consistent deformations, inequivalent solutions of Eq.(\ref{fofunc}) are spanned by the cohomology class $H^0(\bs)$, which is isomorphic to the set of gauge invariant functionals. 

The equation at second order in $g$ is: \begin{eqnarray}
2\left(\stackrel{(0)}{W},\stackrel{(2)}{W}\right)+\left(\stackrel{(1)}{W},\stackrel{(1)}{W}\right)=2\bs\stackrel{(2)}{W}+\left(\stackrel{(1)}{W},\stackrel{(1)}{W}\right)=0\quad.\label{sofunc}\end{eqnarray}
As a matter of fact, there always exists a functional $\stackrel{(2)}{W}$, but nothing ensures that it is local. In the same way, the construction of $W$ can be pursued without obstruction, but generally without preserving locality. The basic theorem underlying this construction can be found in \cite{Barnich:1993vg} and states that the map applying a representative of $H^0(\bs)$ on its antibracket with itself is $\bs$ trivial.

\subsection{Considerations in the local case}

As we have seen in Chapter \ref{ch:anti}, local functionals can be related to functions on the jet space or, equivalently, to $n$-forms on the jet space. Since the boundary terms are always thrown away when integrating a functional, these $n$-forms are defined up to $d$-exact terms. At first order of deformation, Eq.(\ref{fofunc}) becomes: \begin{eqnarray}
\exists\ f\ |\ 2\bs\stackrel{(1)}{w}=df\ ,\end{eqnarray} where $\stackrel{(1)}{W}=\int \stackrel{(1)}{w}$ is a local solution.
A local $\bs$-exact solution $\stackrel{(1)}{W}=\bs B$ is assumed to correspond to the $\bs$ modulo $d$ coboundary: $\stackrel{(1)}{w}=\bs b+dc$ where $B=\int b$. Thus, a local first order solution $\stackrel{(1)}{w}$ is a representative of the cohomology class $H^{0,n}(\bs|d)$. The computation of this cohomology class is the first step in the exhaustive determination of non-trivial local deformations. We provide some results about this for a Fronsdal initial theory in the next section. 

At second order in $g$, the local version of Eq.(\ref{sofunc}) is: \begin{eqnarray}\exists \stackrel{(2)}{w},\ e\ |\ (\stackrel{(1)}{w},\stackrel{(1)}{w})=-\frac{1}{2}\bs\stackrel{(2)}{w}+ d e\ .\label{secoeq}\end{eqnarray} This equation is not automatically satisfied by first order solutions. We have checked explicitly that obstructions can arise. More precisely, in the case of Fronsdal theories with spin up to 4, we show below that $\stackrel{(1)}{w}$ has only three components with antifield number 0,1 and 2. We have achieved the computation of the component of Eq.(\ref{secoeq}) with highest antifield number (which is 2), for the various first order solutions that we describe in Chapters \ref{ch:int23},\ref{ch:exo3} and \ref{ch:intmisc}. Those results are gathered in Chapter \ref{ch:socomp}.

\section{General results for the deformation of the Fronsdal theory}

\subsection{A bounded antifield number}

Let us consider the case of a sum of Fronsdal BRST-BV generators as a starting point. It reads (see Eq.(\ref{wfrons}) ): \begin{eqnarray}\stackrel{(0)}{W}=\sum_a\int_\cd\left(\frac{1}{2}\phi^{\mu_1...\mu_{s_a}}_a G^a_{\mu_1...\mu_{s_a}}+s_a\phi^{*\mu_1...\mu_{s_a}}_a\6_{(\mu_1}C^a_{\mu_2...\mu_{s_a})}\right)d^n x\ .\end{eqnarray} It provides the action of the differential $\bs=\d+\g$: \begin{eqnarray}\forall\ a\ :\ &\displaystyle\d C_a^{*\m_2...\m_{s_a}}=-s_a\left[\6_{\m_1}\phi_a^{*\m_1...\m_{s_a}}-\frac{(s_a-1)(s_a-2)}{2(n+2s_a-6)}\eta^{(\m_2\m_3}\6_{\m_1}\phi_a^*{}'{}^{\m_4...\m_{s_a})\m_1}\right]\quad,\nonumber&\\& \d \phi_a^{*\m_1...\m_{s_a}}=G_a^{\m_1...\m_{s_a}}\quad,\quad \g\phi^a_{\m_1...\m_{s_a}}=s\6_{(\m_1}C^a_{\m_2...\m_{s_a})}\quad.&\nonumber
\end{eqnarray}
Let us recall that this is an irreducible and Abelian theory. At first order in $g$, we must compute the cohomology class $H^{0,n}(\bs|d)$. A cocycle of this is as follows:\begin{eqnarray}\bs a+d b=0\label{smodd}\quad.\end{eqnarray} The $n$-form $a$ has a total ghost number 0, but is the sum of linearly independent terms of increasing antifield and pure ghost numbers (which are equal since $gh=pgh-antigh$). The $(n-1)$-form $b$ has a total ghost number 1 and can also be expanded according to the antifield number.\begin{eqnarray}a&=&a_0+a_1+a_2+...\\b&=&b_0+b_1+b_2+...\quad,\end{eqnarray} such that $antigh\ a_i=pgh\ a_i=i$ and $antigh\ b_i=pgh\ b_i-1=i$. Eq.(\ref{smodd}) can be split into independent components of definite antifield number: \begin{eqnarray}\forall\ i\geqslant 0\ :\ \d a_{i+1}+\g a_i+d b_i=0\quad.\end{eqnarray} Indeed, we are searching for local deformations, in the sense that any term added to the Lagrangian involves the fields and their derivatives up to a finite order. Even if these Lagrangian terms involve functions of derivatives of the fields, they can be considered as a series of polynomial terms. These terms have to satisfy the $\bs$ modulo $d$ cocycle independently because $\bs$ and $d$ do not alter the number of fields (in the sense of the extended jet space including the ghosts and the antifields). Let us notice that this is true only thanks to the linear nature of Fronsdal theory. Moreover, each of those terms involve a finite number of derivatives. 

We thus make the assumption that the first order Lagrangian deformation is a polynomial in the fields, involving a finite number of derivatives. The general case is simply a sum of those. We can now prove that the antifield number is bounded, using an argument similar to that of \cite{Barnich:1995db} for a Yang-Mills theory. It is possible to build an operator $K$ that is a counter of the number of derivatives minus the number of ghosts plus twice the number of antifields $\phi^*$ plus three times the number of antifields $C^*$: \begin{eqnarray}\nonumber K&=&\big(|\m|-1\big)\6_{(\m)}C^\a\frac{\6^L}{\6(\6_{(\m)}C^\a)}\,+|\m|\6_{(\m)}\phi^i\frac{\6^L}{\6(\6_{(\m)}\phi^i)}\\&&+\big(|\m|+2\big)\6_{(\m)}\phi^*_i\frac{\6^L}{\6(\6_{(\m)}\phi^*_i)}+\big(|\m|+3\big)\6_{(\m)}C^*_\a\frac{\6^L}{\6(\6_{(\m)}C^*_\a)}\quad,\end{eqnarray} where a summation over the multiindex $(\mu)$ is made (see below Eq.(\ref{multiind})), its length $|\m|$ being the current number of derivatives. It is obvious that $d$ raises the value of $K$ by 1. On the other hand, the weights of the antifields and the ghosts have been given in such a way that $\d$ and $\g$ do not modify the value of $K$. Thus: \begin{eqnarray} K a_0=\kappa a_0 \Longrightarrow\ \forall\ i\ : K a_i=\kappa a_i\ {\textrm{ and }}\ K b_i=(\kappa-1)b_i\quad.\end{eqnarray} Let us notice that a total ghost 0 couple $(\phi^*$,$C)$ as well as a total ghost 0 triplet $(C^*$,$C$,$C)$ both carry a $K$-number 1, hence the number of derivatives in the terms $a_i$ decreases with the antifield number $i$, since every $a_i$ carries the same $K$-number. On the other hand, we know that, at a given number of derivatives, the polynomial degree in the ghosts is bounded because of their fermionic behaviour. Since the number of derivatives in $a_0$ is bounded and decreases as the antifield and pure ghost numbers increase, we can conclude that the polynomial degree in the ghosts is bounded and thus that $\exists\ k\ |\ \forall k'>k\ :\ a_{k'}=0$. The same holds for $b$, and we can show that it can be chosen as finishing at $antigh\ (k-1)$: Thanks to the above boundary on the $a_i$, we get that $\forall j>k\ :\ db_j=0$. The algebraic Poincar\'e lemma for $(n-1)$-forms then implies that $\forall j>k\ ,\ \exists\ c_j\ |\ b_j=d c_j$. The modulo $d$ freedom in the equation allows to set those trivial components to 0. Then, let us consider the $antigh\ k$ component: $\g a_k+ d b_k=0$, because $a_{k+1}=0$. Theorem \ref{gdg} ensures that, if $k>0$, $\exists\ c_k\ |\g(a_k+dc_k)=0$. Once again, since $a$ is defined modulo $d$, the trivial term can be forgotten. Finally, the system of equations becomes: \begin{eqnarray}&\forall\ 0\leqslant i<k\ :\ \d a_{i+1}+ \g a_i+d b_i=0&\\&\g a_k=0\quad.\end{eqnarray} 

\subsection{Considerations about the top equations\label{topeq}}

The next step is to analyze the two top $antigh$ equations. For the top equation $\g a_k=0$, a representative of the cohomology of $\gamma$ can be chosen: \begin{eqnarray}a_k=\alpha_J\omega^J\quad,\end{eqnarray} where the $\{\omega^J\}$ are the basis of the products of $k$ non-$\g$-exact ghosts and $\a_J\in H^0(\g)$ are $n$-forms of $antigh\ k$ depending on the Fronsdal and curvature tensors of the different fields (see section \ref{cohgam}). We can then consider the second to last equation $\d a_k+\g a_{k-1}+ d b_{k-1}=0$. Since $\g$ and $\d$ anticommute, and both anticommute with $d$, the action of $\g$ on this equation yields $d\g b_{k-1}=0$. Thanks to the Poincar\'e lemma in form degree $n-1$, the solution of this equation is $\exists e_{k-1}\ |\g b_{k-1}+d e_{k-1}=0$ . Thanks to theorem \ref{gdg}, if $k\geqslant 2$, this equation can be rewritten $\exists c_{k-1}\ | \g(b_{k-1}+d c_{k-1})=0$, thus $b_{k-1}$ can be redefined as a $\g$-closed object. In the same way as for $a_k$, we choose it as being a representative of the cohomology of $\g$: $b_{k-1}=\b_J\omega^J$, where $\b_J$ are $(n-1)$-forms of $antigh\ (k-1)$ belonging to $H^0(\g)$. Let us insert the results about $a_k$ and $b_{k-1}$ in the second to last equation: \begin{eqnarray} (\d\alpha_J)\omega^J+\g a_{k-1}+ d(\b_J\omega^J)=0\quad.\end{eqnarray} We can now use the operator $D$ that we defined in section \ref{furgam}. It can be straightforwardly extended to the case of a sum of Fronsdal theories, as well as the associated $D$-degree, which counts the number of derivatives acting on the ghosts. The above equation becomes \begin{eqnarray}(\d\alpha_J+d\b_J)\omega^J+ \b_J D \omega^J=\g(...)=0\quad.\end{eqnarray} Both sides of this equation vanish, because the action of $D$ on the ghosts is defined as selecting the strictly non-$\g$-exact parts of their derivatives. The left-hand side can be expanded according to the $D$-degree. By definition of the $D$-degree, we know that $D\omega^{J_i}=A^{J_i}_{J_{i+1}}\omega^{J_{i+1}}$ where $A^{J_i}_{J_{i+1}}$ are constants multiplied by a basic 1-form. We get: \begin{eqnarray} &\d\a_{J_0}+ d \b_{J_0}=0&\label{Ddegzero}\\& \forall i>0\ :\ \d\a_{J_i}+ d\b_{J_i}+ \b_{J_{i-1}}A^{J_{i-1}}_{J_i}=0&\quad.\label{Ddegi}\end{eqnarray}
The bottom equation is a $\d$ modulo $d$ cocycle and we know form Theorem \ref{hkndeltd} that it is trivial if $k>2$. By making redefinitions of the other $\a_{J_i}$ and $\b_{J_i}$, the triviality can be propagated to the whole system. Unfortunately, this is not sufficient to ensure the triviality of $a_k$. Since $D$ is a differential, we obtain an identity on the coefficients $A$: $D^2\omega^{J_i}=A^{J_i}_{J_{i+1}}A^{J_{i+1}}_{J_{i+2}}\omega^{J_{i+2}}=0\Rightarrow A^{J_i}_{J_{i+1}}A^{J_{i+1}}_{J_{i+2}}=0$. The solution of Eq.(\ref{Ddegzero}) is $\alpha_{J_0}=\d f_{J_0}+d l_{J_0}$ and $\beta_{J_0}=\d l_{J_0}+d m_{J_0}$. Let us now redefine $\alpha'_{J_1}=\alpha_{J_1}+ l_{J_0}A^{J_0}_{J_1}$ and $\b'_{J_1}=\b_{J_1}+m_{J_0}A^{J_0}_{J_1}$, the equation in $D$-degree 1 becomes $\d \alpha'_{J_1}+d\beta'_{J_1}=0$ and is trivial. The same procedure can be done up to the maximum $D$-degree (that exists because the total number of derivatives is bounded). The problem is that, in general, the redefined objects are no longer invariant under $\g$. For example, let us reconstruct the $D$-degree 0 term: \begin{eqnarray}\alpha_{J_0}\omega^{J_0}&=&\d(f_{J_0}\omega^{J_0})+d(l_{J_0}\omega^{J_0})\mp l_{J_0}d\omega^{J_0}\nonumber\\&=&\d(...)+d(...)\mp l_{J_0}A^{J_0}_{J_1}\omega^{J_1}\mp l_{J_0}\g r^{J_0}\quad,\end{eqnarray}
where $d\omega^{J_0}=D\omega^{J_0}+\g r^{J_0}$. The first two terms are trivial, the third term is at $D$-degree one and corresponds to the redefinition $\alpha'_{J_1}$. However, the last term is not trivial unless $\g l_{J_0}=0$. The problem of finding invariant coboundaries of $\d$ modulo $d$ is a very technical one. It corresponds to the computation of the invariant cohomology of $\d$ modulo $d$ : $H_k^n(\d|d,H^0(\g))$ and is the object of the next section. We will prove that these classes always vanish for $k>n$. Quite generally, it can already be said that: \begin{eqnarray}{\textrm{If }}\ \forall k>j\geqslant 2\ :\ H_k^n(\d|d,H^0(\g))=0\ ,\quad {\textrm {then the expansion of $a$  stops at antigh $j$.}}\nonumber\end{eqnarray} The vanishing of those classes ensures that $l_{J_0}$ can be chosen in $H^0(\g)$ and thus that the $D$-degree 0 part of $a_k$ is trivial modulo invariant terms of $D$-degree 1. Then, at $D$-degree $1$, the redefined $\alpha'_{J_1}$ is invariant and can be written as an invariant coboundary of $\d$ modulo $d$: $\alpha'_{J_1}=\d f_{J_1}+d l_{J_1}$ and $\b'_{J_1}=\d l_{J_1}+d m_{J_1}$. The same type of redefinition can be done in $D$-degree 2 and thanks to the invariance of $l_{J_1}$, the term $\alpha'_{J_1}\omega^{J_1}$ is trivial modulo an invariant expression of $D$-degree 2. By repeating the same argument up to maximum $D$-degree, it is found that $a_k$ is trivial. Finally, a redefinition of $a_{k-1}$ using Theorem \ref{gdg} allows to obtain a new couple of last equations $\g a_{k-1}=0$ and $\d a_{k-1}+\g a_{k-2}+d b_{k-2}=0$ for which the same argument can be reproduced as long as the invariant classes of $\d$ modulo $d$ vanish.

\subsection{Invariant cohomology of $\d$ modulo $d$}

Throughout this section, we use Theorems $\ref{hdinv}$ and $\ref{hdelinv}$ that allow one to choose invariant objects in coboundaries of $d$ and $\delta$, in strictly positive $antigh$: \begin{eqnarray}& a\in H^0(\g)\textrm{ and }antigh\ a>0\ :\nonumber&\\ &\textrm{If }a=db\textrm{ then }\exists\ b'\in H^0(\g)\ |\ a=db'\quad,\nonumber&\\& \textrm{If }a=\d c\textrm{ then }\exists\ c'\in H^0(\g)\ |\ a=\d c'\quad.&\end{eqnarray}
For example, let us consider a cocycle of $\d$ modulo $d$ in $antigh\ k\geqslant 2$: \begin{eqnarray}\d a_k^n+d b_{k-1}^{n-1}=0\quad.\end{eqnarray} If $a_k^n$ is invariant, then $\d a_k^n$ is an invariant object of $antigh$ $(k-1)$, because the equations of motion and the antifields are invariant. Then, Theorem $\ref{hdinv}$ allows us to choose $b_{k-1}^{n-1}$ invariant as well. On the other hand, a coboundary of $\d$ modulo $d$ reads: \begin{eqnarray} a_k^n=\d \mu_{k+1}^n+d\mu_k^{n-1}\quad.\label{desctop}\end{eqnarray} If $a_k^n$ is invariant, nothing allows us to believe that the $\mu$ expressions can be chosen invariant. If it is not the case, some trivial objects in the whole algebra can become nontrivial in $H^0(\g)$ and $H_k^n(\d|d,H^0(\g))$ is larger than the restriction $H_k^n(\d|d)\cap H^0(\g)$. 

The first thing that can be done is to establish a descent of equations. For the moment, we have worked in form degree $n$. By acting on Eq.(\ref{desctop}) with $\d$, we obtain the relation $\d a_k^n=-d\d\mu_k^{n-1}$. Theorem \ref{hdinv} tells us that $\exists\ a_{k-1}^{n-1}\in H^0(\g)\ |\ \d a_k^n=-d a_{k-1}^{n-1}$. The combination of the two equations yields:\begin{eqnarray}d(a_{k-1}^{n-1}-\d \mu_k^{n-1})=0\Rightarrow \exists\ \mu_{k-1}^{n-2}\ | a_{k-1}^{n-1}=\d\mu_k^{n-1}+d\mu_{k-1}^{n-2}\quad,\end{eqnarray} thanks to the algebraic Poincar\'e lemma in form degree $(n-1)$. The last relation is similar to Eq.(\ref{desctop}) but in form degree $(n-1)$ and antifield number $(k-1)$. The same argument can be repeated until one reaches either form degree 0 or antifield number 1. The complete descent is thus: \begin{eqnarray}\nonumber a_k^n&=&\d \mu_{k+1}^n+d\mu_k^{n-1}\\a_{k-1}^{n-1}&=&\d \mu_{k}^{n-1}+d\mu_{k-1}^{n-2}\nonumber\\&\vdots&\nonumber\\ \textrm{either}\ a^0_{k-n}&=&\d\mu_{k-n+1}^0\quad\textrm{if}\ k>n\nonumber\\ \textrm{or}~~ a^{n-k+1}_1&=&\d\mu_2^{n-k+1}+d\mu_1^{n-k}\quad\textrm{if}\ k\leqslant n\quad.\end{eqnarray}We can now establish the following lemma: \begin{lemma}If one of the $\mu$'s is invariant then they can all be chosen invariant.\label{allthemus}\end{lemma}\noindent Let us assume that $\mu_b^{c-1}$ is invariant. It appears in two equations of the descent: \begin{eqnarray}a_b^c&=&\d\mu_{b+1}^c+d\mu_b^{c+1}\nonumber\\a_{b-1}^{c-1}&=&\d\mu_b^{c-1}+d\mu_{b-1}^{c-2}\quad.\end{eqnarray} The first equation tells us that $\d\mu_{b+1}^{c}$ is invariant. Thanks to Theorem \ref{hdelinv}, we can choose $\mu_{b+1}^c$ invariant. The other equation tells that $d\mu_{b-1}^{c-2}$ is invariant and Theorem \ref{hdinv} ensures that we can choose $\mu_{b-1}^{c-2}$ invariant. The same argument can then be applied to the next equations and the invariance propagate throughout the descent.$\Box$

This lemma also means that Theorem \ref{desctheo} can be restricted to $H^0(\g)$:\begin{eqnarray}\forall k\geqslant 2,\ \forall p>1\ :\ H_k^p(\d|d,H^0(\g))\cong H_{k-1}^{p-1}(\d|d,H^0(\g))\quad.\end{eqnarray} The following result can then be established: \begin{lemma} If $a_k^n$ is of antifield number $k>n$, then all the $\mu$'s in the descent can be taken to be invariant, and thus $H_k^n(\d|d,H^0(\g))=H_k^n(\d|d)\cap H^0(\g)$.\end{lemma} \noindent If $k>n$, the bottom equation of the descent is $a_{k-n}^0=\d\mu_{k-n+1}^0$ and Theorem \ref{hdelinv} allows us to choose $\mu_{k-n+1}^0$ invariant. Thanks to the previous lemma, the other $\mu$'s can also be taken to be invariant. Since any coboundary $a_k^n=\d \m_{k+1}^n+d \m_k^{n-1}$ remains valid in $H^0(\g)$, the invariant cohomology class $H^n_k(\d|d,H^0(\g))$ is the restriction to $H^0(\g)$ of the whole class.$\Box$

This automatically proves that \begin{eqnarray}\forall k>n\geqslant 2\ :\ H_k^n(\d|d,H^0(\g))=H_k^n(\d|d)\cap H^0(\g)=0\quad,\end{eqnarray} thanks to Theorem \ref{hkndeltd} for linear theories. Then, the considerations made in the previous section ensure that: \begin{prop}The first order deformations of the generator $W$ for a sum of Fronsdal theories have a finite expansion in the antifield number, that is bounded by the spacetime dimension $n$: \begin{eqnarray}\stackrel{(1)}{w}=a_0+a_1+...+a_n\ |\ antigh\ a_i=i\quad.\nonumber\end{eqnarray}\end{prop}

For antifield numbers from 2 to $n$, we have adopted a recursive strategy. As for Theorem \ref{hkndeltd}, the idea is to take variational derivatives of the highest $antigh$ component $a_k$. These no longer contain $d$-exact terms and the $\d$-exact terms can be replaced by invariant terms. Then $a_k$ can be reconstructed with an homotopy formula, but one term in the integral is not manifestly invariant and some technical computations are needed to finish the proof, for which a recurrence hypothesis on the antifield number is needed, and whose complexity seems to increase with the spin. Until now, this result has been showed up to spin-4. While it is rather obvious for spin-1 \cite{Barnich:1994}, the spin-2 proof was given in \cite{Boulanger:2000rq} and the spin-3 proof in \cite{Bekaert:2006jf}. We completed the spin-3 proof for the case of dimension 3 in \cite{Boulanger:2005br} and provided the spin-4 proof in \cite{Boulanger:2008tg}. In every case that has been solved, it is needed to assume that the result is true in $antigh\ (k+2)$. As an example, we present here the spin-4 proof after having set the general problem.

\subsubsection*{Variational derivatives and reconstruction of $a_k$}

Let us consider a $\d$ modulo $d$ coboundary in form degree $n$ and antifield number $k\geqslant2$. We can write the relation in dual notation: $a_k^n=a_k\,d^n x$ , $\m^n_{k+1}=b_{k+1}\,d^n x$ and $d\m^{n-1}_k=\6_\m J_k^\m\,d^n x$. The coefficients of the $n$-forms thus satisfy:\begin{eqnarray}a_k=\d b_{k+1}+\6_\mu J_k^\mu\quad.\label{cobkn}\end{eqnarray} The only assumption is the invariance of $a_k$, and we want to check if $b_{k+1}$ can be taken invariant, which is sufficient thanks to Lemma \ref{allthemus}. Let us denote the equations of motion with an operator $\cd_{ij}$ and the gauge transformations with an operator $\car^i_\a$ as in Section \ref{linth}: \begin{eqnarray}\frac{\d\cl}{\d\phi^{a\mu_1...\mu_{s_a}}}&=&G_{a\mu_1...\mu_{s_a}}=\cd^{\r\s}_{\mu_1...\mu_{s_a}\nu_1...\nu_{s_a}}\6^2_{\r\s}\phi^{\nu_1...\nu_{s_a}}_a\\ \g\phi^a_{\mu_1...\mu_{s_a}}&=&s\6_{(\mu_1} C^a_{\mu_2...\mu_{s_a})}=\car^{\r\nu_2...\nu_{s_a}}_{\mu_1...\mu_{s_a}}\6_\r C^a_{\nu_2...\nu_{s_a}}\quad.\end{eqnarray} The operator of the equations of motion is symmetric: $\cd^{\r\s}_{\mu_1...\mu_{s_a}\nu_1...\nu_{s_a}}=\cd^{\r\s}_{\nu_1...\nu_{s_a}\mu_1...\mu_{s_a}}$, and the Noether identities read: \begin{eqnarray}\car^{\r\nu_2...\nu_{s_a}}_{\mu_1...\mu_{s_a}}\cd^{\l\s|\mu_1...\mu_{s_a}}_{\phantom{\l\s|\mu_1...\mu_{s_a}}\tau_1...\tau_{s_a}}\equiv 0\quad.\end{eqnarray} The variational derivatives of Eq.(\ref{cobkn}) are taken as in Theorem \ref{hkndeltd}. The divergence term vanishes, and we have to take care of the (anti)commutation of $\d$ and the variational derivatives. The variational derivatives of $b_{k+1}$ are denoted $Z_{a|k-1}^{\mu_2...\mu_{s_a}}=\frac{\d^L b_{k+1}}{\d C^{*a}_{\mu_2...\mu_{s_a}}}$, $X_{a|k}^{\mu_1...\mu_{s_a}}=\frac{\d^L b_{k+1}}{\d \phi^{*a}_{\mu_1...\mu_{s_a}}}$ and $Y^a_{\mu_1...\mu_{s_a}|k+1}=\frac{\d^L b_{k+1}}{\d \phi_a^{\mu_1...\mu_{s_a}}}$. We get:
\begin{eqnarray}\frac{\d^L a_k}{\d C^{*a}_{\mu_2...\mu_{s_a}}}&=&\d Z_{a|k-1}^{\mu_2...\mu_{s_a}}\label{eqz}\\ \frac{\d^L a_k}{\d \phi^{*a}_{\mu_1...\mu_{s_a}}}&=&-\d X_{a|k}^{\mu_1...\mu_{s_a}}+\car^{\r\nu_2...\nu_{s_a}|\mu_1...\mu_{s_a}}\6_\r Z_{a\nu_2...\nu_{s_a}|k-1}\label{eqx}\\ \frac{\d^L a_k}{\d \phi_a^{\mu_1...\mu_{s_a}}}&=&\d Y^a_{\mu_1...\mu_{s_a}|k+1}+\cd^{\r\s}_{\mu_1...\mu_{s_a}\nu_1...\nu_{s_a}}\6^2_{\r\s}X^{a\nu_1...\nu_{s_a}}_k\label{eqy}\quad.
\end{eqnarray} Then, we can straightforwardly replace $X$, $Y$ and $Z$ by invariant objects, denoted $X\,{}'$, $Y\,{}'$ and $Z\,{}'$. First, thanks to Theorem \ref{hdelinv}, which ensures the triviality of the homology of $\d$ in the space of $\g$-invariant objects, we get that: \begin{eqnarray}\exists\ M^{\nu_2...\nu_{s_a}}_{a|k}\ |\ Z'\phantom{}^{\,\nu_2...\nu_{s_a}}_{a|k-1}=Z^{\nu_2...\nu_{s_a}}_{a|k-1}+\d M^{\nu_2...\nu_{s_a}}_{a|k}\in H^0(\g)\quad.\end{eqnarray} Then, by replacing this expression in Eq.(\ref{eqx}), we get that $\d(X+\car\6M)$ is invariant, hence Theorem \ref{hdelinv} yields that:\begin{eqnarray}\exists\ N^{\mu_1...\mu_{s_a}}_{a|k+1}\ |\ X'\phantom{}^{\,\mu_1...\mu_{s_a}}_{a|k}=X^{\mu_1...\mu_{s_a}}_{a|k}+\car^{\r\nu_2...\nu_{s_a}|\mu_1...\mu_{s_a}}\6_\r M_{a\nu_2..\nu_{s_a}|k}+\d N^{\mu_1...\mu_{s_a}}_{a|k+1}\in H^0(\g)\quad.\end{eqnarray} Finally, we can replace the last expression in Eq.(\ref{eqy}). The term depending on the tensor $M$ vanishes thanks to the symmetry of $\cd$ and the Noether identities. Thus, it is found that $\d(Y-\cd\6^2N)$ is invariant, and Theorem \ref{hdelinv} yields that: \begin{eqnarray}\exists\ P^a_{\mu_1...\mu_{s_a}|k+2}\ |\ Y'\phantom{}^a_{\mu_1...\mu_{s_a}|k+1}&=&Y^a_{\mu_1...\mu_{s_a}|k+1}-\cd^{\r\s}_{\mu_1...\mu_{s_a}\nu_1...\nu_{s_a}}\6^2_{\r\s}N^{a\nu_1...\nu_{s_a}}_{k+1}+\d P^a_{\mu_1...\mu_{s_a}|k+2}\nonumber\\ &\in& H^0(\g)\quad .\end{eqnarray} The system of equations becomes: 
\begin{eqnarray}\frac{\d^L a_k}{\d C^{*a}_{\mu_2...\mu_{s_a}}}&=&\d Z'\phantom{}_{a|k-1}^{\,\mu_2...\mu_{s_a}}\label{eqzp}\\ \frac{\d^L a_k}{\d \phi^{*a}_{\mu_1...\mu_{s_a}}}&=&-\d X'\phantom{}_{a|k}^{\,\mu_1...\mu_{s_a}}+\car^{\r\nu_2...\nu_{s_a}|\mu_1...\mu_{s_a}}\6_\r Z'\phantom{}_{a\nu_2...\nu_{s_a}|k-1}\label{eqxp}\\ \frac{\d^L a_k}{\d \phi_a^{\mu_1...\mu_{s_a}}}&=&\d Y'\phantom{}^{\,a}_{\mu_1...\mu_{s_a}|k+1}+\cd^{\r\s}_{\mu_1...\mu_{s_a}\nu_1...\nu_{s_a}}\6^2_{\r\s}X'\phantom{}^{\,a\nu_1...\nu_{s_a}}_k\label{eqyp}\quad.\end{eqnarray}
Finally, the reconstruction of $a_k$ can be done with an homotopy formula: \begin{eqnarray}a_k=\sum_a\int_0^1\left[C^{*a}_{\mu_2...\mu_{s_a}}\frac{\d^L a_k}{\d C^{*a}_{\mu_2...\mu_{s_a}}}+\phi^{*a}_{\mu_1...\mu_{s_a}}\frac{\d^L a_k}{\d \phi^{*a}_{\mu_1...\mu_{s_a}}}+\phi_a^{\mu_1...\mu_{s_a}}\frac{\d^L a_k}{\d \phi_a^{\mu_1...\mu_{s_a}}}\right](t)dt+div\quad.\end{eqnarray} The results of Eqs(\ref{eqzp})-(\ref{eqyp}) can be inserted in the formula. By working modulo $d$ and using the definition of $\d$ (let us remind that $\d C^{*a\nu_2...\nu_{s_a}}=-\6_\r[\car^{\r\nu_2...\nu_{s_a}|\mu_1...\mu_{s_a}}\phi^{*a}_{\mu_1...\mu_{s_a}}]$), we get: \begin{eqnarray}a_k=\d\left(\sum_a\int_0^1\left[C^{*a}_{\mu_2...\mu_{s_a}}Z'\phantom{}_{a|k-1}^{\,\mu_2...\mu_{s_a}}+\phi^{*a}_{\mu_1...\mu_{s_a}}X'\phantom{}_{a|k}^{\,\mu_1...\mu_{s_a}}+\phi_{a}^{\mu_1...\mu_{s_a}}Y'\phantom{}^{\,a}_{\,\mu_1...\mu_{s_a}|k+1}\right](t)dt\right)+div\ .\end{eqnarray} The first two terms in the expression under $\d$ are invariant, but the third one is not manifestly invariant. The remaining problem is to check if $\phi_{a}^{\mu_1...\mu_{s_a}}Y'\phantom{}^{\,a}_{\,\mu_1...\mu_{s_a}|k+1}$ is invariant modulo $d$ when $Y'\phantom{}^{\,a}_{\,\mu_1...\mu_{s_a}|k+1}$ obeys Eq.(\ref{eqyp}). We have the feeling that this conjecture is true for all spin, but it remains to be showed for $s>4$. 

\subsubsection*{Proof in the spin-4 case}

For the sake of argument, let us consider the case of a single spin-4 field $\phi_{\m\n\r\s}$. The ghosts are $C_{\n\r\s}$, the antifields are $\phi^{*\m\n\r\s}$ and $C^{*\m\n\r}$. Our goal is to prove that $Y'^{\m\n\r\s}_{k+1}\phi_{\m\n\r\s}$ is invariant if $Y'^{\m\n\r\s}_{k+1}$ satisfies the following equation:
\begin{eqnarray}
\frac{\d^L a_k}{\d \phi_{\m\n\r\s}} = \d Y'^{\m\n\r\s}_{k+1}
+{\cg}^{\m\n\r\s}{}_{\a\b\g\d}{X'}^{\a\b\g\d}_k
\label{2.47'}\quad,
\end{eqnarray} Where we have denoted ${\cg}^{\m\n\r\s}{}_{\a\b\g\d}={\cd}^{\a\b|\m\n\r\s}{}_{\a\b\g\d}\6^2_{\a\b}$.

Since $a_k$ is invariant, it depends on the fields only
through the curvature $K$, the Fronsdal tensor and their derivatives.
(We substitute $4\,\pa^{[\d}\pa_{[\g}F^{~\,\s]}_{\r]~~\m\n}$ for
$\eta^{\a\b}K^{\d\s}_{~~\,|\a\m|\b\n|\g\r}$ everywhere).
We then express the Fronsdal tensor in terms of the
Einstein tensor:
$F_{\m\n\r\s} = G_{\m\n\r\s} - \frac{6}{n+2}\,\eta_{(\m\n}G_{\r\s)}$,
so that we can write $a_k = a_k([\Phi^{*i}],[K],[G])$\,, where $[G]$
denotes the Einstein tensor and its derivatives.
We can thus write:
\begin{eqnarray}
\frac{\d^L a_k}{\d \phi_{\m\n\r\s}} = {\cg}^{\m\n\r\s}{}_{\a\b\g\d}
{A'}^{\a\b\g\d}_k + \pa_{\a}\pa_{\b}\pa_{\g}\pa_{\d}
{M'}^{\a\m\vert\b\n\vert\g\r\vert\d\s}_k
\label{2.49}
\end{eqnarray}
where $${A'}^{\a\b\g\d}_k\propto\frac{\d a_k}{\d G_{\a\b\g\d}}$$ and
$${M'}_k^{\a\m\vert\b\n\vert\g\r\vert\d\s}\propto{\frac{\d a_k}
{\d K_{\a\m\vert\b\n\vert\g\r\vert\d\s}}}$$
are both invariant and respectively have the same symmetry properties as
the Einstein and curvature tensors.

Combining Eq.(\ref{2.47'}) with Eq.(\ref{2.49}) gives:
\begin{eqnarray}
\d Y'^{\m\n\r\s}_{k+1} =
\pa_\a\pa_\b\pa_\g\pa_\d{M'}_k^{\a\m\vert\b\n\vert\g\r\vert\d\s}
                       + {\cg}^{\m\n\r\s}{}_{\a\b\g\d} {B'}^{\a\b\g\d}_k
\label{2.50}
\end{eqnarray}
with ${B'}^{\a\b\g\d}_k:={A'}^{\a\b\g\d}_k-{X'}^{\a\b\g\d}_k$.
Now, only the first term on the right-hand-side of Eq.(\ref{2.50}) is
divergence-free: $\pa_{\m}(\pa_{\a\b\g}{M'}_k^{\a\m\vert\b\n\vert\g\r})\equiv 0$ .
The second one instead obeys a relation analogous to the Noether identities:
\begin{eqnarray}
\partial^{\tau}G_{\m\n\r\tau}-\frac{3}{(n+2)}\,
\eta_{(\m\n}\partial^{\tau}G'_{\r)\tau}=0\,.
\nonumber
\end{eqnarray}
As a result, we have
$\displaystyle\d\left[\pa_{\m}({Y'}^{\m\n\r\s}_{k+1}-\frac{3}{n+2}\,\eta^{(\n\r}
{Y'}_{k+1}^{\s)\m})\right]=0\,$,
where ${Y'}_{k+1}^{\m\s}\equiv \eta_{\n\r}{Y'}_{k+1}^{\m\n\r\s}\,$.
By Theorem~\ref{hdelinv}, we deduce:
\begin{eqnarray}
        \pa_{\m}\Big({Y'}^{\m\n\r\s}_{k+1}-\frac{3}{n+2}\,\eta^{(\n\r}
{Y'}_{k+1}^{\s)\m}\Big)
        +\d {F'}_{k+2}^{\n\r\s}=0
        \,, \label{truc}
\end{eqnarray}
where ${F'}_{k+2}^{\n\r\s}$ is invariant and can be chosen symmetric and
traceless.
Eq.(\ref{truc}) determines a cocycle of $H^{n-1}_{k+1}(d\vert\d)$, for given
 $\n$, $\r$ and $\s\,$. Using Theorem \ref{ddeltheo}:
 $H^{n-1}_{k+1}(d\vert\d)\cong H^{n}_{k+2}(\d\vert d)\cong 0$ ($k\geqslant 1$) ,
  we deduce:
\begin{eqnarray}
        {Y'}^{\m\n\r\s}_{k+1}-\frac{3}{n+2}\,\eta^{(\n\r}{Y'}_{k+1}^{\s)\m}=
        \pa_{\a}T_{k+1}^{\a\m\vert\n\r\s} + \delta P^{\m\n\r\s}_{k+2}
        \,, \label{truc2}
\end{eqnarray}
where both $T^{\a\m\vert\n\r\s}_{k+1}$ and $P^{\m\n\r\s}_{k+2}$ are
invariant by the induction hypothesis. Moreover,
$T^{\a\m\vert\n\r\s}_{k+1}$ is antisymmetric in its first two
indices. The tensors $T^{\a\m\vert\n\r\s}_{k+1}$ and
$P^{\m\n\r\s}_{k+2}$ are both symmetric and traceless in $(\n,\r,\s)$.
This results easily from taking the trace of Eq.(\ref{truc2}) with
$\eta_{\n\r}$ and using Theorems \ref{desctheo} and \ref{ddeltheo}: 
$H^{n-2}_{k+1}(d\vert\d)\cong H^{n-1}_{k+2}(\d\vert d)\cong
H^{n}_{k+3}(\d\vert d)\cong 0$ which hold
since $k$ is positive. From Eq. (\ref{truc2}) we obtain
\begin{eqnarray}
        {Y'}^{\m\n\r\s}_{k+1} =
        \pa_{\a} [ T_{k+1}^{\a\m\vert\n\r\s}+
\frac{3}{n}\,T_{k+1}^{\a\vert\m(\n}\eta^{\r\s)} ]
        + \delta (...)
% + \delta [ P^{\m\n\r\s}_{k+2} + \frac{3}{n}\eta^{(\n\r}P^{\s)\m}_{k+2} ]\,,
         \label{truc3}
\end{eqnarray}
where $T_{k+1}^{\a\vert\m\n}\equiv \eta_{\t\r}T_{k+1}^{\a\t\vert\r\m\n}\,$.
We do not explicit the $\delta$-exact term since it plays no role in the
sequel.
Since $Y'^{\m\n\r\s}_{k+1}$ is symmetric in $\m$ and $\n$, we have also
$$\pa_{\a}\Big(T_{k+1~~\r\s}^{\a[\m\vert\n]}+
\frac{2}{n}\,T_{k+1~(\s}^{\a\vert[\m}\d^{\n]}_{\r)}\Big)
+\;\delta (...)=0\,.$$
The triviality of $H^{n-1}_{k+1}(d \vert \d)$ ($k>0$) implies again that
$T_{k+1~~\r\s}^{\a[\m\vert\n]}+\frac{2}
{n}\,T_{k+1~(\s}^{\a\vert[\m}\d^{\n]}_{\r)}$
is trivial, in particular,
\begin{eqnarray}
\partial_{\beta}S\,{}'{}^{\beta\a|\mu\nu|}_{\phantom{\b\a|\m\n}\,\r\s} + \d(...) =
T_{k+1~~\r\s}^{\a[\m\vert\n]}+\frac{2}{n}\,
T_{k+1~(\s}^{\a\vert[\m}\d^{\n]}_{\r)}
\label{derStoT}
\end{eqnarray}
where $S\,{}'{}^{\beta\a|\mu\nu|}_{\phantom{\b\a|\m\n}\,\r\s}$
is antisymmetric in the pairs of indices ($ \b, \a$) and ($\m,\n$),
while it is symmetric and traceless in ($ \r, \s$).
Actually, it is traceless in $\m, \n, \r\,\s$ as the right-hand side of
the above equation shows.
The induction assumption allows us to choose
$S\,{}'{}^{\beta\a|\mu\nu|}_{\phantom{\b\a|\m\n}\,\r\s}$ invariant, as well as the quantity under
the Koszul-Tate differential $\d\,$.
We now project both sides of Eq. (\ref{derStoT}) on
the following irreducible representation of the orthogonal group
\begin{picture}(35,16)(0,0)
\multiframe(1,4)(10.5,0){3}(10,10){$\a$}{$\m$}{$\s$}
\multiframe(1,-6.5)(10.5,0){2}(10,10){$\r$}{$\n$}
\end{picture} (see Appendix \ref{as:gln})
and obtain:
\begin{eqnarray}
        \partial_{\b}W^{'\b|\a\r|\m\n|\s}_{k+1}+ \d (\dots) = 0
\label{derW}
\end{eqnarray}
where $W^{'\b|\a\r|\m\n|\s}_{k+1}$ denotes the corresponding projection
of $S\,{}'{}^{\beta\a|\mu\nu|\r\s}\,$.
Eq.(\ref{derW}) determines, for given $(\m, \n, \a, \r,\s)\,$,
a cocycle of $H^{n-1}_{k+1}(d\vert\d,H^0(\gamma))$.
Using again the isomorphisms $H^{n-1}_{k+1}(d\vert\d)\cong H^{n}_{k+2}(\d\vert d)\cong 0$ ($k>0$)
and the induction hypothesis, we find:
\begin{eqnarray}
W^{'\b|\a\r|\m\n|\s}_{k+1}
 = \pa_{\l}\phi^{\l\b\vert\a\r|\m\n|\s}_{k+1} +
        \d (\dots)
        \label{Wintermsofphi}
\end{eqnarray}
where $\phi^{\l\b\vert\a\r|\m\n|\s}_{k+1}$ is invariant,
antisymmetric in $(\l, \b)$ and possesses the irreducible,
totally traceless symmetry
\begin{picture}(35,16)(0,0)
\multiframe(1,4)(10.5,0){3}(10,10){$\a$}{$\m$}{$\s$}
\multiframe(1,-6.5)(10.5,0){2}(10,10){$\r$}{$\n$}
\end{picture}
in its last five indices. The $\d$-exact term is invariant as well.
Then, projecting the equation (\ref{Wintermsofphi}) on the
totally traceless irreducible representation
\begin{picture}(35,16)(0,0)
\multiframe(1,4)(10.5,0){3}(10,10){$\a$}{$\m$}{$\s$}
\multiframe(1,-6.5)(10.5,0){3}(10,10){$\r$}{$\n$}{$\b$}
\end{picture}
and taking into account that $W^{'\b|\a\r|\m\n|\s}_{k+1}$
is built out from $S\,{}'{}^{\beta\a|\mu\nu|\r\s}\,$, we find
\begin{eqnarray}
\partial_{\l}\Psi^{'\l|\a\r|\m\n|\s\b}_{k+1} + \d (\dots)
 = 0
\end{eqnarray}
where $\Psi^{'\l|\a\r|\m\n|\s\b}_{k+1}$ denotes the
corresponding projection of $\phi^{\l\b\vert\a\r|\m\n|\s}_{k+1}\,$.
The same arguments used before imply
\begin{eqnarray}
\Psi^{'\l|\a\r|\m\n|\s\b}_{k+1} = \partial_{\t}
\Xi^{'\t\l|\a\r|\m\n|\s\b}_{k+1} + \delta(...)
\label{PsitoXi}
\end{eqnarray}
where the symmetries of $\Xi^{'\t\l|\a\r|\m\n|\s\b}_{k+1}$ on its
last 6 indices can be read off from the left-hand side and
where the first pair of indices is antisymmetric.
Again, $\Xi^{'\t\l|\a\r|\m\n|\s\b}_{k+1}$ can be taken to be invariant.

Then, we take the projection of $\Xi^{'\t\l|\a\r|\m\n|\s\b}_{k+1}$ on
the irreducible representation
\begin{picture}(45,16)(0,0)
\multiframe(1,4)(10.5,0){4}(10,10){$\t$}{$\a$}{$\m$}{$\s$}
\multiframe(1,-6.5)(10.5,0){4}(10,10){$\l$}{$\r$}{$\n$}{$\b$}
\end{picture}
of $GL(n)$ (here we do not impose tracelessness) and denote the
result by $\Theta^{'\t\l|\a\r|\m\n|\s\b}_{k+1}\,$. This invariant tensor possesses
the algebraic symmetries of the invariant spin-$4$ curvature tensor.
Finally, putting all the previous results together,
we obtain the following relation, using the symbolic
manipulation program \emph{Ricci} \cite{Lee}:
\be
        6\,{Y'}^{\m\n\r\s}_{k+1} =
\pa_{\a}\pa_{\b}\pa_{\c}\pa_{\d}\Theta^{'\a\m\vert\b\n\vert\c\r|\d\s}_{k+1}
               + {\cg}^{\m\n\r\s}{}_{\a\b\g\d}
  \widehat{X}_{k+1}^{'\a\b\g\d}{}+\d(\ldots)\,,
\label{result1}
\ee
with
\begin{eqnarray}
        \widehat{X}^{'}_{\a\b\g\d\vert k+1} &:=&
\frac{\cy^{\m\n\r\s}_{\a\b\c\d}}{n-2}\,
    \Big[-\frac{1}{3}\,\eta^{\t\l}
   S_{\t\m|\l\n|\r\s|k+1} + \frac{1}{3(n+1)}\,
\eta_{\m\n}\eta^{\t\l}\eta^{\kappa\z}
 (S_{\t\kappa|\l\z|\r\s|k+1} + 2\,S_{\t\kappa|\l\r|\z\s|k+1})
\nonumber \\
&&+\;\frac{2(n-2)}{n}\,\eta^{\kappa\t}\partial^{\l}
\phi_{\kappa\m|\l\n|\t\r|\s}
-\frac{4(n-2)}{n(n+2)}\,\eta_{\m\n}\eta^{\kappa\t}\eta^{\x\z}\partial^{\l}
\phi_{\kappa\x|\t\z|\l\m|\n}\Big]\;
        \label{result2}
\end{eqnarray}
being double-traceless and where $\cy^{\m\n\r\s}_{\a\b\c\d}$
projects on completely symmetric rank-4 tensors.\vspace{2mm}

Eq.(\ref{result1}) automatically implies that 
\begin{eqnarray}\phi_{\m\n\r\s}\,Y'^{\m\n\r\s}_{k+1}=\frac{1}{96}\,
\Theta^{'\a\m\vert\b\n\vert\c\r\vert\d\s}_{k+1}
K_{\a\m\vert\b\n\vert\c\r\vert\d\s}
+\frac{1}{6}G_{\a\b\g\d}\widehat{X}^{'\a\b\g\d}{}_{k+1}+\pa_\r \ell^\r+\d(\ldots)\,,\end{eqnarray} which is an invariant expression modulo trivial terms.$\Box$

\subsection{Cubic vertices \label{cbvrt}}

In the case when $\forall k>2\ :\ H_k^n(\d|d,H^0(\g))=H_k^n(\d|d)\cap H^0(\g)=0$, we have showed that the expansion of the first order deformation $\stackrel{(1)}{w}$ stops at antifield number 2 and that the $\bs$ modulo $d$ cocycle consists of the three following equations: \begin{eqnarray}\g a_2&=&0\label{eqagh2}\\ \d a_2+\g a_1+d b_1&=&0\label{eqagh1}\\ \d a_1+\g a_0+ d b_0&=&0\label{eqagh0}\quad.\end{eqnarray} Now, we know that this is the case for any combination of fields with spin up to 4, as well as the restriction of the antigh 2 class: \begin{eqnarray}\displaystyle H_2^n(\d|d,H^0(\g))=\left\{\left[\sum_a\xi^a_{\nu_2...\nu_{s_a}}C_a^{*\nu_2...\nu_{s_a}}d^n x\right]\ | \6_{(\nu_1}\xi^a_{\nu_2...\nu_{s_a})}=0\right\}\quad,\end{eqnarray}thanks to Theorems \ref{hkndeltd} and \ref{hdndeltd}, and since the representatives of $H_2^n(\d|d)$ were already chosen invariant. Until now, we have considered Lorentz-invariant objects. In fact, we make the usual assumption of the full Poincar\'e invariance, that also requires invariance under translations. That is why we will from now on forbid any explicit dependence on the coordinates. The only objects that will be allowed are the fields, antifields and ghosts, as well as the metric $\eta_{\mu\nu}$ (that is constant in the Cartesian coordinates that have been chosen), the Kronecker delta $\d_\mu^\rho$ and the Levi-Civita symbol $\e_{\mu_1...\mu_n}$. 
Most of the computations have been made in the parity-invariant case (without $\e$) but we have achieved the computation of the parity-breaking case for a set of spin-3 fields. The considerations made in Section \ref{topeq} hold in the particular case of $antigh\ 2$. Since $a_2$ and $b_1$ are $\g$-closed, we choose them as $a_2=\a_J\omega^J$ and $b_1=\b_J\omega^J$. Their coefficients must satisfy the system of equations (\ref{Ddegzero}) and (\ref{Ddegi}).
The solution of Eq.(\ref{Ddegzero}) is: \begin{eqnarray}\alpha_{J_0}&=&\sum_a \l^a_{\nu_2...\nu_{s_a}|J_0}C_a^{*\nu_2...\nu_{s_a}}d^n x\nonumber\\ \beta_{J_0}&=&-\frac{1}{(n-1)!}\left[\sum_a s_a \l^a_{\nu_2...\nu_{s_a}|J_0}\phi^{*\r\nu_2...\nu_{s_a}}\right]\e_{\r\mu_2...\mu_{n}}dx^{\mu_2}\wedge...\wedge dx^{\mu_{n}}\quad.\label{soljzero}\end{eqnarray}
As the parameters $\l$ are constants because of Poincar\'e invariance, it appears that $\b_{J_0}A^{J_0}_{J_1}$ cannot be $\d$ exact modulo $d$ unless it vanishes. This is because a $\d$-exact term is either proportional to the equations of motion or proportional to differentiated antifields $\phi^*$. Since $\b_{J_0}$ does not depend on the fields, the only possibility would be to remove the derivatives from the antifield with $d$-exact terms. However, this would require $\b_{J_1}$ to depend explicitly on $x^\m$, which is not possible since $\b_{J_1}$ is also assumed to be Poincar\'e invariant. Thus, the second equation decouples into $\d\alpha_{J_1}+d\beta_{J_1}=0$ and $\beta_{J_0}A^{J_0}_{J_1}=0$. The solution of the first one consists of expressions similar to Eq.(\ref{soljzero}) and the same argument can be repeated to decouple the third equation. This can then be done at any $D$-degree and we obtain the following system: \begin{eqnarray}\d\alpha_{J_i}+d\beta_{J_i}&=&0\label{alphabeta}\\ \beta_{J_i}A^{J_i}_{J_{i+1}}&=&0\quad.\label{betacond}\end{eqnarray} The first set of equations tells us that the coefficients of $a_2$ can all be written as the product of constants and undifferentiated antifields $C^*_a$. These coefficients are multiplied by the basis $\omega^J$, that consists of products of two strictly non-$\g$-exact ghost tensors $U$ that were defined in section \ref{cohgam}. This means that $a_2$ is cubic, it is a Lorentz-invariant combination of an undifferentiated antifield $C^*_a$ and two non-$\g$-exact ghost tensors. Let us recall that $a_2$ represents the first order of deformation of the gauge algebra in the antifield formalism. It is related to the deformation of the gauge transformations given by $a_1$ and to the vertex $a_0$. The cubic solutions $a_2$ will be related to cubic terms $a_1$ and $a_0$. However, we have to emphasize that homogeneous terms can appear, that are solution of the deformation problem stopping at antigh $1$ or $0$, and which are not necessarily cubic. We have thus established the following theorem: \begin{theorem}Under the assumptions of Poincar\'e invariance and locality, the first order of deformation of the sum of Fronsdal theories with spin up to 4 is bounded at antigh 2. Furthermore, the only possible nonabelian solutions are cubic.\label{thmcub}\end{theorem}

The second set of equations (\ref{betacond}) is a set of algebraic conditions on the coefficients $\b_{J_i}$. It tells us that every cubic $a_2$ is not automatically a solution of Eq.(\ref{eqagh1}). An equivalent way to find the consistent $a_2$'s is to list every $antigh\ 2$ cubic combinations and to test them in Eq.(\ref{eqagh1}). This provides a list of candidates for $a_1$, that have to be completed with cubic homogeneous terms involving the same field content and the same number of derivatives. Then, these expressions can be tested in Eq.(\ref{eqagh0}) to find the cubic vertices $a_0$, if they exist.

\noindent \textbf{Remark:}
Actually, even if the Theorem \ref{thmcub} can hardly be extended to 
$s>4$ for technical reasons, we can always assume that $a_2$ is
cubic, 
relax the limitation $s\leqslant 4$ 
and proceed with the determination of $a_1$ and $a_0$ according to
(\ref{eqagh1}) and (\ref{eqagh0}). 
In fact, it is impossible to build a ghost-zero cubic object with
$antigh>2$, so that a cubic deformation always stops at $antigh\ 2$. 
Moreover, a cubic element $a_2$ must be proportional to an $antigh\ 2$ antifield and quadratic in the ghosts. Then, modulo $d$ and $\g$, one can first remove the derivatives that the antifield could bear, and the $\g$-exact parts in the derivatives of the ghosts can be ignored. Thus, it is always possible to only consider the cubic form that we have described above.
Finally, combining the cohomological approach with other approaches
like the light-cone one \cite{Metsaev:2005ar,Metsaev:2007rn} 
may complete our results, as we actually show in the sequel. 
Such a combination of two different methods seems to us the most
powerful way to completely solve the first-order deformation problem. 

\subsection{A few words about deformations of (Anti)de Sitter Fronsdal theories}

The developments made above concern the case of a sum of Minkowski Fronsdal theories, in Cartesian coordinates. In that case, partial derivatives coincide with covariant derivatives. In the case of the $(A)dS$ spacetime, it is impossible to build a Cartesian system of coordinates, one thus have to deal with expressions involving covariant derivatives and depending explicitly on the coordinates. This is due to the absence of an Abelian subalgebra in $so(n-1,2)$ and $so(n,1)$, which denotes the absence of translations in the symmetry groups $SO(n-1,2)$ and $SO(n,1)$. Some of the results that we have established remain valid for $(A)dS$. The theory is still linear, its BRST-BV generator reads:\begin{eqnarray}\stackrel{(0)}{W}=\int_\cd \left(\frac{1}{2}\sqrt{-g}\phi^{\m_1...\m_{s_a}}_a G^a_{\m_1...\m_{s_a}}+s\phi^{*\m_1...\m_{s_a}}_a\N_{(\m_1}C^a_{\m_2...\m_{s_a})}\right)\ d^n x\quad.\end{eqnarray} It generates the differential $\bs$ through $\bs A=(\stackrel{(0)}{W},A)$, which is such that $\bs=\d+\g$. The linearity also implies that $\forall\ k>2\ : H_k^n(\d|d)=0$, thanks to Theorem \ref{hkndeltd}. Furthermore, \begin{eqnarray}H_2^n(\d|d)=\left\{\left[\xi^a_{\m_2...\m_{s_a}}C^{*\m_2...\m_{s_a}}_a d^n x\right]\ |\ \nabla_{(\mu_1}\xi^a_{\mu_2...\mu_{s_a})}=0\right\}\quad.\end{eqnarray} This result was proved in \cite{Barnich:2004ts,Memdea} for the spin-2 case. For the general case, the main result was obtained in \cite{Barnich:2005bn}: the weak equation $\N_{(\m_1}\L^a_{\m_2...\m_{s_a})}\approx 0$ admits the solution $\L^a_{\m_2...\m_{s_a}}\approx\xi^a_{\m_2...\m_{s_a}}$ where $\xi^a$ satisfies the strong ``Killing'' equation $\N_{(\m_1}\xi_{\m_2...\m_{s_a})}^a=0$. 

Let us now consider the problem of the cohomology of $\g$. Thanks to the linear nature of the theory, $\g$ is a true differential on $(A)dS$ as well. Its action on the fields, that reproduces the gauge transformations presented in Eq.(\ref{gtds}), read: \begin{eqnarray}\g \phi^a_{\m_1...\m_{s_a}}=s\N_{(\m_1}C^a_{\m_2...\m_{s_a})}\quad.\end{eqnarray} We have studied a similar problem in \cite{Memdea}, where we had to deal with covariant derivatives and to build the cohomology from several building blocks. What happens is that $\g$ and $\d$ have not a definite degree of derivation. However, $\g$ always raises the number of derivatives by at most 1. Thus, the process of cancellation by pairs that has been exposed in section \ref{cohgam} is no longer valid but can be adapted as follows: First, the undifferentiated ghosts $C_{\m_2...\m_s}$ are not exact, hence they belong to the cohomology. Then, let us consider the first cocycle: $\g\phi_{\m_1...\m_s}=s\N_{(\m_1}C_{\m_2...\m_s)}$. We can no longer say that $\phi_{\m_1...\m_s}$ and $\6_{(\m_1}C_{\m_2...\m_s)}$ cancel by pairs. However, we remark that the undifferentiated fields are not $\g$-closed, and the relation: $\6_{(\m_1}C_{\m_2...\m_s)}=\g(\frac{1}{s}\phi_{\m_1...\m_s})-(s-1)\stackrel{(0)}{\G}{}^{\!\!\r}_{\!\!\!\!(\m_1\m_2}C_{\m_3...\m_s)\r}$ tells us that the symmetrized first derivatives of the ghosts belong to the cosets of the undifferentiated ghosts, instead of being trivial. Anyway, the result is the same: this component of the first derivatives of the ghosts does not appear in the cohomology. On the other hand, the traceless part of the other component $\6_{[a}C_{\b]\m_3...\m_s}$ is strictly non-$\g$-exact. The covariant expression \begin{eqnarray}{U}^{(1)}_{\m_1\n_1|\m_2...\m_{s-1}}=\N_{[\m_1}C_{\n_1]\m_2...\m_{s-1}}-\frac{s-2}{n+2s-4}Y^{(1)}\left(\eta_{\m_1(\m_2}\N^\r C_{\m_3...\m_{s-1})\n_1}\right)\end{eqnarray} is equivalent because it differs of the expression with partial derivatives by terms linear in the undifferentiated ghosts, which are themselves strictly non-$\g$-exact. The rest of the construction can arguably be made along the same lines, by raising the maximal number of derivatives one by one. The Fronsdal and curvature tensors are the only gauge invariants in Minkowski spacetime and their equivalents on $(A)dS$ have the same terms involving the maximal number of derivatives. They are thus the only expressions that vanish under the terms in the gauge transformations involving the highest number of derivatives, which read as the Minkowski gauge transformations. Therefore, it is clear that $F_{\m_1...\m_s}$ is the only gauge invariant expression involving terms containing at most two derivatives. Then the covariant derivatives of $F$ are the only gauge invariant expression involving terms containing at most three derivatives, and so on. On the other hand, the components of the ghosts that are not corresponding to an $U^{(i)}$ tensor can be eliminated in favor of representatives of the cohomology involving fewer derivatives. The $U^{(i)}$ tensors can be built, for example as the traceless part of $Y^{(i)}\left(\N_{(\m_1}...\N_{\m_i)}C_{\n_1...\n_{s-1}}\right)$. Finally, the cohomology of $\g$ for a sum of Fronsdal theories on $(A)dS$ is the adaptation of that in Minkowski spacetime: \begin{eqnarray}H^*(\g)=\left\{\left[f([\Phi^*_I],[K^a],[F^a]_{\textrm{sym}},C^a,U^{a(i)})\right]\ |\ 1\leqslant i<s\right\}\ .\end{eqnarray}

Most of the other results, such as Theorem \ref{gdg} or the triviality of $H(d,H^0(\g))$ and $H(\d,H^0(\g))$ in strictly positive antifield number remain valid. However, since $\g$ has no definite degree of derivation, the differential $D$ can still be defined, but not the associated $D$-degree. The impossibility of splitting equations according to that degree prevents one to use the same arguments as above for the resolution of an $\bs$ modulo $d$ cocycle. Furthermore, the presence of covariant derivatives in every $SO(n-1,2)$ or $SO(n,1)$ covariant expressions prevents one to obtain $d$-exact terms. For example, a relation $\6_\m T^{\m\n}=0$ in Minkowski can be considered as a $d$-cocycle where the index $\n$ is fixed. That fixation of indices rely on the existence of a Cartesian system of coordinates. On the other hands, the covariant relation $\N_\m T^{\m\n}=0$ is not a $d$-cocycle, unless $T^{\m\n}$ is antisymmetric. Many steps in the proof of the theorem about the invariant cohomology of $\d$ modulo $d$ are based on the fixation of indices, and thus cannot be straightforwardly adapted to $(A)dS$ spacetime. The problem of directly computing consistent deformations in $(A)dS$ spacetime has not been solved yet. Fortunately, it is possible to make a limit when $\L\rightarrow0$, that allowed us to prove that some $(A)dS$ cubic vertices, described by Fradkin and Vasiliev, have a sense in Minkowski spacetime. Our main result about this is that the uniqueness of the Minkowski vertices can be extended to $(A)dS$ spacetime, thus providing some uniqueness results to Vasiliev theory for higher spins. The next chapter consists in a brief presentation of Fradkin--Vasiliev cubic vertices in $(A)dS$ spacetime, as well as the method to relate these vertices to Minkowski consistent first order solutions.

%%%%%%%%%%%%%%%%%%%%%%%%%%%%%%%%%%%%%%%%%%%%%%%%
\chapter{Fradkin--Vasiliev cubic vertices in $(A)dS$ spacetime and Minkowski limit\label{ch:FV}}
%%%%%%%%%%%%%%%%%%%%%%%%%%%%%%%%%%%%%%%%%%%%%%%%

As we explained in the end of the previous chapter, the cohomological method to exhaustively determine the first order nonabelian deformations of the Fronsdal action in a $dS$ or $AdS$ spacetime is not yet known. Fortunately, the difficulty can be circumvented. Some particular deformations can be straightforwardly defined. First, some cubic deformations with a $2-s-s$ spin configuration can be built, by completing an attempt of minimally deforming the Fronsdal action for a spin-$s$ field in an $(A)dS$ spacetime. We call this the Fradkin--Vasiliev procedure, it is presented in the first section of this chapter. Fradkin and Vasiliev did the computation for spin-3 in an unfortunately unpublished paper, the construction is explained in \cite{Vasiliev:2001ur}. Then, they built a more general cubic Lagrangian vertex in dimension 4, in an extended frame formalism containing auxiliary fields and constraints that allow one to eliminate them algebraically. The result is given as an action gauge invariant at first order, presented in \cite{Fradkin:1987ks,Fradkin:1987qy}. The same construction has more recently been made in dimension 5, in \cite{Vasiliev:2001wa}, but a formulation in arbitrary dimension has not yet been achieved. Furthermore, a Lagrangian formulation of a complete higher spin theory is not yet known. This is because the consistency of such a theory requires that every spin is present, which in turn implies the presence of an arbitrarily high number of derivatives and thus a non-local complete theory. For the moment, Vasiliev's theory for interacting higher spins is a set of equations of motions and constraints, presented in \cite{Vasiliev:1990en} for the 4-dimensional case, and in \cite{Vasiliev:2003ev} for the $n$-dimensional case. It is presented in the {\em unfolded formulation} of field theory, using free differential algebras, and that we do not review here (see \cite{D'Auria:1982my,D'Auria:1982nx,vanNieuwenhuizen:1982zf,Vasiliev:1988xc,Vasiliev:1988sa,Vasiliev:2001ur,Bekaert:2005vh,Skvortsov:2008vs} for some details). We briefly present the Fradkin--Vasiliev action in dimension 4 in the second section of this chapter. Finally, we establish one of our main arguments, which concerns the relation between the $(A)dS$ cubic vertices and the Minkowski cubic vertices. It allows us to prove the uniqueness of the Fradkin--Vasiliev procedure as well as that of any $(A)dS$ cubic vertex that would correspond to an unique deformation in Minkowski spacetime.

\section{Quasi-minimal deformation of the Fronsdal action in $(A)dS$}

The basic idea is to attempt to deform minimally a sum of Fronsdal Lagrangians for spin-2 and spin-$s$ fields in $(A)dS$, that we have defined in Section \ref{FrAdS} (mainly in Eqs (\ref{tfads}), (\ref{tgads}) and (\ref{flagads}) ). The free Lagrangian is thus: \begin{eqnarray}\stackrel{(2)}{\cl}=\cl_{F2}+\cl_{Fs}\quad.\end{eqnarray} What we call a minimal deformation consists in replacing the $(A)dS$ metric $g_{\m\n}$ by a dynamical full metric built with the spin-2 dynamical field: $\cg_{\m\n}=g_{\m\n}+\a h_{\m\n}$. This replacement must be made in the free action and in the gauge transformations. On one hand, the spin-2 Lagrangian naturally deforms into the full Einstein--Hilbert Lagrangian, this minimal deformation is consistent at all orders. On the other hand, the minimal deformation of the spin-$s$ Lagrangian is not gauge invariant at first order under the minimally deformed gauge transformations. The action and gauge transformations under consideration read:
\begin{eqnarray}&&\cl_{Fs,min}=\frac{1}{2}\sqrt{-\cg}\phi_{\a_1...\a_s}G^D_{\b_1...\b_s}\cg^{\a_1\b_1}...\cg^{\a_s\b_s}\ ,\\ &&\d_{\xi,min} \phi_{\m_1...\m_s}=s D_{(\m_1}\xi_{\m_2...\m_s)}\quad,\end{eqnarray} 
where $D$ is the covariant derivative built with the Lorentz connection of the full metric and $G^D$ is the generalized Einstein tensor defined in Eq.(\ref{tgads}), but where the $(A)dS$ covariant derivative $\N$ is replaced by $D$. The structure of $\cl_{Fs,min}$ and $\d_{\xi,min}$ is similar to that of their free versions. The only difference that arises in the gauge variation of the Lagrangian concerns the commutation of covariant derivatives. Let us recall that the Fronsdal action can be defined in $(A)dS$ spacetime thanks to the expression of its Riemann tensor:
\begin{eqnarray}\stackrel{(0)}{\car}\!\!{}^\m_{\phantom{\m}\n|\r\s}=\frac{2\L}{(n-1)(n-2)}(\d^\m_{\r}g_{\n\s}-\d^\m_{\s}g_{\n\r})\quad.\label{riemads}\end{eqnarray} 
Thanks to this, the commutator of two covariant derivatives of any tensor contains two less derivatives while being proportional to the cosmological constant. The deformed Riemann tensor takes the form:
\begin{eqnarray}\car^\m_{\phantom{\m}\n|\r\s}=\frac{2\L}{(n-1)(n-2)}(\d^\m_\r g_{\n\s}-\d^\m_{\s}g_{\n\r})+R^\m_{\phantom{\m}\n|\r\s}\ +\co(h^2)\quad,\end{eqnarray} where  \begin{eqnarray}&&R^\m_{\phantom{\m}\n|\r\s}=\a(\N_\r \l^\m_{\n\s}-\N_\s \l^\m_{\n\r})\ ,\\&& \l^\m_{\n\r}=-\frac{1}{2}\N^\m h_{\n\r}+ \N_{(\n}h^\m_{\phantom{\m}\r)}\quad.\end{eqnarray} 
The tensor $\l^\m_{\n\r}$ is the variation of the connection, and the tensor $R^\m_{\phantom{\m}\n\r\s}$, the first order of the Riemann tensor, appears in the commutation of covariant derivatives $D$. The variation of $\cl_{Fs,min}$ under the gauge transformations $\d_{\xi,min}$ does not vanish, because the gauge variation of the ``mass'' terms $\L\phi^{\m_1...\m_s}\phi_{\m_1...\m_s}$ and $\L\phi'{}^{\m_3...\m_s}\phi'_{\m_3...\m_s}$ does not coincide anymore with the terms proportional to the Riemann tensor coming from the commutation of covariant derivatives. The remaining terms are thus proportional to the difference between $R^\m_{\phantom{\m}\n\r\s}$ and the variation $\stackrel{(0)}{\car}\!\!{}^\m_{\phantom{\m}\n|\r\s}(\cg)-\stackrel{(0)}{\car}\!\!{}^\m_{\phantom{\m}\n|\r\s}(g)$. We call this difference the tensor $s_{\m\n|\r\s}$: 
\begin{eqnarray}s^\m_{\phantom{\m}\n|\r\s}=R^\m_{\phantom{\m}\n\r\s}-\a\frac{2\L}{(n-1)(n-2)}(\d^\m_\r h_{\n\s}-\d^\m_\s h_{\n\r})\quad.\end{eqnarray} 
As a matter of fact, this tensor is invariant under the zeroth order spin-2 gauge transformations: it is proportional to the curvature tensor $K_{\m\n|\r\s}$. This is no accident though. Since the Lagrangian $\cl_{Fs,min}$ is a scalar with respect to the full metric $\cg_{\m\n}$, multiplied by the density $\sqrt{-\cg}$, it is automatically invariant under full diffeomorphisms. Thus, the only problems that can arise come from the spin-$s$ gauge transformations. The non vanishing terms of the gauge variation thus depend of $h_{\m\n}$ through a gauge invariant tensor. Schematically: \begin{eqnarray}\d_{\xi,min}\cl_{Fs,min}\sim s_{\m\n|\r\s} \ca^{\m\n\r\s}(\phi\otimes\N\xi)\quad.\end{eqnarray} 

Then, the naive thing to do is to try to add by hand cubic terms, proportional to $s_{\m\n|\r\s}$ and quadratic in the undifferentiated fields. This quickly appears not to be sufficient to compensate the above anomaly, it is thus impossible to build a consistent cubic deformation that contains only two derivatives. However, it is possible to add terms containing more derivatives and proportional to an appropriate negative power of the cosmological constant. Only terms containing an even number of derivatives are considered. Some terms can also be added to the gauge transformations. Our claim is that there is a particular combination of such terms, containing at most $(2s-4)$ extra derivatives, that lead to a Lagrangian invariant at first order under en extension of the gauge transformations $\d_{\xi,min}$. The cubic part of the Lagrangian thus takes the form: \begin{eqnarray}\stackrel{(3)}{\cl}=\left[\sqrt{-\cg}(R-2\L)+\cl_{Fs,min}\right]\vert_{cubic}+\sum_{i=\frac{p+q}{2}=0}^{s-2}\L^{-i}s^{\m_1\m_2|\n_1\n_2}\ \beta_{\m_1\m_2\n_1\n_2}^{(\a)(\r)(\b)(\s)}\ \N^p_{(\a)}\phi_{(\r)}\N^q_{(\b)}\phi_{(\s)}\quad,\label{cubfv}\end{eqnarray} where $\b$ is a set of appropriate coefficients and where we have used some multiindices (whose lengths are $|\a|=p$, $|\b|=q$, $|\r|=|\s|=s$). The coefficients must take particular values such that the zeroth order gauge transformation of the extra terms only involves: terms with commutators of covariant derivatives, or terms proportional to the some zeroth order equations of motion. The latter can be compensated by adding the corresponding terms to the first order gauge transformations: If a term $-\frac{\d \stackrel{(2)}{\cl}}{\d\phi^i}F^i$ appears in $\stackrel{(0)}{\d} \stackrel{(3)}{\cl}$, then it can be compensated by: 
\begin{eqnarray} \stackrel{(1)}{\d}_F \stackrel{(2)}{\cl}=\frac{\d \stackrel{(2)}{\cl}}{\d\phi^i}F^i\Rightarrow \stackrel{(1)}{\d}_F\phi^i=F^i\quad.\end{eqnarray} 
The terms with commutators of covariant derivatives are then fitted to compensate the terms that were left at the previous degree of derivation. Since the number of possible Lorentz-invariant terms increases dramatically with the number of derivatives, it is clear to us that it becomes easier to fit the coefficients when increasing the number of derivatives and that there must exist a maximum number of derivatives at which the computation closes. This number is arguably $(2s-2)$ derivatives (i.e. $(2s-4)$ extra derivatives). We have some strong arguments in favor of that claim. The deformation of the gauge transformations appears to be nonabelian. 
Furthermore, we prove in Chapter $\ref{ch:intmisc}$ that there is only one possible nonabelian cubic deformation with the spin configuration $2-s-s$ in Minkowski spacetime, and we prove that it contains exactly $(2s-2)$ derivatives. In addition to that, the last section of this chapter is dedicated to building a special $\L\rightarrow0$ limit. We clearly see that the extra terms in the action involve a negative power of $\L$. However, by scaling the fields in a clever way, the quadratic terms in the action can be preserved, as well as the term in the action containing the highest number of derivatives. Thus, the quasi minimal construction presented here has to correspond to a consistent nonabelian deformation in Minkowski spacetime. Since the quasi minimal construction cannot possibly fail, it exists and must be related to the unique deformation in Minkowski spacetime, whose existence is thus established, and it contains at most $(2s-2)$ derivatives. 

The $2-3-3$ case had been computed a long time ago by Fradkin and Vasiliev. We have found back the result, with the help of the symbolic manipulation software {\em Ricci} \cite{Lee}. It is presented in Chapter $\ref{ch:int23}$, as well as its Minkowski limit. The computations soon become heavy when the number of covariant derivatives to commute and the number of spacetime indices on the fields increase. That is why this construction remains a bit theoretical.

\section{The Fradkin--Vasiliev action}

Let us briefly present the Fradkin--Vasiliev action in dimension 4 (this section is based on \cite{Fradkin:1987ks}). Though we have not made it explicitly, it allows one to find cubic vertices for any configuration $s-s'-s''$ in $(A)dS_4$ spacetime. This action is defined in the frame formalism, in which spacetime indices are replaced by indices of the tangent space. The latter are in turn replaced by Weyl spinor indices. The vielbein of $(A)dS_4$ is a tensor $e^a_{0\m}$ such that $g_{\m\n}=e^a_{0\m} e^b_{0\n} \eta_{ab}$. The Latin indices are flat indices of the tangent space of the manifold and are called frame indices. Frame tensors transform under the Poincar\'e transformations of the tangent space. The full vielbein $e^a_\m$ is related to the full metric instead of that of $(A)dS$. The spin-2 Lorentz connection is denoted $\omega^{a,b}_\m$ and is considered as an independent field. On the same pattern, the spin-$s$ field $\phi_{\m_1...\m_s}$ is replaced by a frame field $e_\m^{a_1...a_{s-1}}$ and a set of connections $\o_\m^{a_1...a_{s-1},b_1...b_t}\ ,\ t=1,...,s-1$. Those fields are all totally symmetric in their $a_i$ and $b_i$ indices, are traceless and obey the relation $\o_\m^{(a_1...a_{s-1},b_1)b_2...b_t}=0$. In terms of Young diagrams, the symmetry of the frame indices in $\o$ corresponds to a two-row diagram with lengths $(s-1)$ and $t$, written in symmetric convention (see Appendix \ref{app:YD}). The action that can be built for the free fields is such that the connections with $t>1$ are not dynamical (their variational derivatives identically vanish). The first connection is then found to be an auxiliary field. Upon elimination of the auxiliary fields and some gauge fixing, the theory can be reduced to the Fronsdal theory for the totally symmetric field: \begin{eqnarray}\phi_{\m_1...\m_s}=e_{(\m_1|\m_2...\m_s)}\quad,\end{eqnarray} where the frame indices have been replaced by spacetime indices by using the $(A)dS$ vielbein. 

Then, in dimension 4, the frame indices can be replaced by Weyl spinor indices, by using the Pauli matrices $(\s)_a^{\a\dot{\a}}$. The properties of the $\s$ matrices are such that: a single frame index is replaced by a pair $\a\dot{\a}$, an antisymmetric pair of frame indices is replaced by a pair $\a_1\a_2$ or a pair $\dot{\a}_1\dot{\a}_2$. The resulting fields have complex components, that have to be related by some hermiticity conditions in order for the frame tensors to be real. The frame fields $e_\m^{a_1...a_{s-1}}$ are replaced by tensors with the same number of dotted and undotted Weyl indices: $\o_\m^{\a_1...\a_{s-1},\dot{\b}_1...\dot{\b}_{s-1}}$. The connections $\o_\m^{a_1...a_{s-1},b_1...b_t}$ are replaced by tensors $\o_\m^{\a_1...\a_{s+t-1}\dot{\b}_1...\dot{\b}_{s-t-1}}$. The hermiticity conditions read: \begin{eqnarray}(\o_\m^{\a_1...\a_m,\dot{\b}_1...\dot{\b}_n})^\dagger=\o_\m^{\b_1...\b_n,\dot{\a}_1...\dot{\a}_m}\quad.\end{eqnarray}The Weyl indices are raised and lowered by using the antisymmetric matrices $\e_{\a\b}$, $\e^{\a\b}$, $\e_{\dot{\a}\dot{\b}}$ and $\e^{\dot{\a}\dot{\b}}$ such that: $\e_{12}=-1$ and $\e^{12}=1$. The convention used is: \begin{eqnarray}&&\psi^{\a}=\e^{\a\b}\psi_\b\quad \textrm{and}\quad \psi_\a=\e_{\a\b}\psi^\b\\&& \psi^{\dot{\a}}=\e^{\dot{\a}\dot{\b}}\psi_{\dot{\b}}\quad \textrm{and}\quad \psi_{\dot{\a}}=\e_{\dot{\a}\dot{\b}}\psi^{\dot{\b}}\quad.\end{eqnarray} Let us denote a set of $s$ indices: $\a(s)=\a_1...\a_s$ or $\dot{\a}(s)=\dot{\a}_1...\dot{\a}_s$.
Some curvatures are then defined: 
\begin{eqnarray}R_{\m\n|\a(n),\dot{\b}(m)}&=&\6^{}_\m\o_{\n|\a(n),\dot{\b}(m)}-\6^{}_\n\o_{\m|\a(n),\dot{\b}(m)}\nonumber\\&&+\sum_{p+q=n}\sum_{k+l=m}\sum_{s,t=0}^{+\infty}f(p,q,k,l,s,t)\omega_{\m|\a(p)\g(s),\dot{\b}(k)\dot{\d}(t)}\omega_{\n|\a(q)\phantom{\g(s)},\dot{\b}(l)}^{\phantom{\n|\a(q)}{\g}(s)\phantom{\dot{\b}(l)}\dot{\d}(t)}\quad.\label{curvframe}\end{eqnarray} The gauge transformations are defined as: \begin{eqnarray}\d_\e \o_{\n|\a(n),\dot{\b}(m)}=\6_\n\e_{\a(n),\dot{\b}(m)}+\sum_{p+q=n}\sum_{k+l=m}\sum_{s,t=0}^{+\infty}f(p,q,k,l,s,t)\omega_{\n|\a(p)\g(s),\dot{\b}(k)\dot{\d}(t)}\e_{\a(q)\phantom{\g(s)},\dot{\b}(l)}^{\phantom{\a(q)}{\g}(s)\phantom{\dot{\b}(l)}\dot{\d}(t)}\quad.\end{eqnarray}
In both definitions, the coefficients $f(p,q,k,l,s,t)$ take the form: \begin{eqnarray}&&f(p,q,k,l,s,t)=\frac{i^{s+t-1}}{s!t!}\left(\begin{array}{c}n\\p\end{array}\right)\left(\begin{array}{c}m\\k\end{array}\right)\l^{1+(|n-m|-|p+s-k-t|-|q+s-l-t|)/2}\d(g(p,q,k,l,s,t))\nonumber\\&& g(p,q,k,l,s,t)=[(p+k)(q+l)+(p+k)(s+t)+(q+l)(s+t)+1]mod\ 2\quad.\end{eqnarray} The curvatures appear to be depending quadratically on connections of arbitrarily high spin. The following action is then built:
\begin{eqnarray}S=\frac{1}{2}\sum_{n+m>0}\frac{i^{n+m+1}}{n!m!}\b(n+m)\epsilon(n-m)\l^{-|n-m|}\int\e^{\n\m\r\s}R_{\n\m|\a(n),\dot{\b}(m)}R_{\r\s}^{\phantom{\r\s}|\a(n),\dot{\b}(m)}d^4 x\quad,\label{actfv}\end{eqnarray} 
where $\b(n+m)$ are coefficients depending on the spin $2(s-1)=n+m$, $\l^2=-\L$ and $\epsilon(x)=\theta(x)-\theta(-x)$. Because of the presence of negative powers of $\L$ in the action, it is defined for a strictly non vanishing cosmological constant. This action is a generalization of the MacDowell--Mansouri action for the spin-2 field, presented in \cite{MacDowell:1977jt}. Let us emphasize that the tensors $\o$ have a vanishing zeroth order component and are thus purely at first order, except those of spin-2: \begin{eqnarray}\o_{\n|\a,\dot{\b}}=e_{0\n|\a,\dot{\b}}+\o'_{\n|\a,\b}\ ,\ \o_{\n,\a(2)}=\o_{0\n,\a(2)}+\o'_{0\n,\a(2)}\textrm{ and }\o_{\n,\dot{\b}(2)}=\bar{\o}_{0\n,\dot{\b}(2)}+\o'_{0\n,\dot{\b}(2)}\quad.\nonumber\end{eqnarray} The ``0'' components are directly related to the vielbein and the Lorentz connection of $(A)dS$. The components bearing a prime are all of first order. In the same way, the curvatures contain a first order part and a higher order part: $R=R'+\co((\o')^2)$. It can be showed that the quadratic part of the action given in Eq.(\ref{actfv}) is equivalent to a sum of Fronsdal actions. The correspondence between the cubic part of this action and cubic vertices in the spacetime formalism requires the introduction of some constraints: \begin{eqnarray}&\e^{\n\m\r\s}R'_{\n\m|\a(n),\dot{\b}(m-1)\dot{\d}}e_{0\r\a_{n+1}}^{\dot{\d}}=0\ ,\ if n\geqslant m&\\& \e^{\n\m\r\s}R'_{\n\m|\a(n-1)\g,\dot{\b}(m)}e_{0\r\dot{\b}_{m+1}}^{\g}=0\ ,\ if m\geqslant n\quad.&\end{eqnarray} Under these assumptions, the extra fields (those with $|n-m|>1$) can be eliminated at first order in favor of the frame and first connection fields. The action is also invariant under the gauge transformations at first order. We will not enter into the details, but it can already be seen that cubic vertices involving various combinations of tensors $\o$ of different spins appear in the action, which yield consistent cubic Lagrangian vertices in $(A)dS$ if one eliminates the extra fields as Fradkin and Vasiliev suggested. To finish this section, let us sketch how the elimination of the auxiliary fields is achieved: the equations of motion and the extra constraints impose the vanishing of some components of the curvatures. The linear part of the curvatures presented in Eq.(\ref{curvframe}) take the form $R'{}^{(t)}=d\o^{(t)}+e_0\wedge \o^{(t+1)}+e_0^a\wedge\o^{(t-1)}$ where $e^a_0=e_{0\m}^a dx^\m$ is the vielbein 1-form and where the index $(t)$ is the length of the second line of the Young diagrams associated to the frame indices. The exterior product of $\o^{(t+1)}$ with $e_0^a$ (the index $a$ being contracted with one of the $t+1$ indices) can also be seen as the action of a nilpotent operator $\s^-$. The last term can be seen as the action of another operator $\s^+$. Thus, non-$\s^-$-closed components of $\o^{(t+1)}$ can be eliminated in favor of the ones that bears less indices: $\o^{(t)}$ et $\o^{(t-1)}$. On the other hand, the gauge transformations read: $\d \o^{(t)}=d\e^{(t)}+\s^-\e^{(t+1)}+\s^+\e^{(t-1)}$. Thus, we see that every $\s^-$-exact term in the connections is pure gauge. The fixation of $\e^{(t+1)}$ allows one to remove the $\s^-$-exact terms. Finally, after using the equations of motion, the extra conditions and the gauge fixation of every gauge parameter but the last, all that is left is representatives of the $\s^-$ cohomology, which appear to be the totally symmetric fields, the Fronsdal and curvature tensors and the totally symmetric gauge parameter.

\section{Minkowski limit \label{minlim}}

We can now present how the cubic construction for a $2-s-s$ configuration can be related to cubic vertices in the Minkowski spacetime. We introduce some dimension factors, that are powers of the Planck length $l_p$. The parameter $\a$ inserted in the expression of the metric, as well as in the cubic terms, can be chosen, for example, as being $\a=\sqrt{2}(l_p)^{\frac{n-2}{2}}$. The quadratic and cubic terms of the Lagrangian can be gathered in the following limited action: \begin{eqnarray}S=\frac{1}{l_p^{n-2}}\int\sqrt{-g}\left(\stackrel{(2)}{\cl}+\stackrel{(3)}{\cl}\right)d^n x\quad.\end{eqnarray} We know that it is invariant under some extended first order gauge transformations. Our goal is to establish a limit that relates the $(A)dS$ spacetime and the Minkowski spacetime, i.e. a limit $\L\rightarrow0$. However, a simple $\L\rightarrow0$ limit cannot be done, since the non minimal cubic terms in the Lagrangian involve negative powers of $\L$. What can be done is to make the fields and the Planck length tend to 0 as well, with definite weights: \begin{eqnarray}\L=\e^2 \widetilde{l}_p^{-2}\ ,\ l_p=\e^{\D_p}\widetilde{l}_p\ ,\ h_{\m\n}=\e^{\D_h}\widetilde{h}_{\m\n}\ ,\ \phi_{\m\n\r}=\e^{\D_\phi}\widetilde{\phi}_{\m\n\r}\quad.\end{eqnarray} Once this is made, the limit $\e\rightarrow 0$ is taken. The quadratic parts are quadratic in $h$ or $\phi$ and proportional to $l_p^{2-n}$. Thus, if $\D_h=\D_\phi=\frac{n-2}{2}\D_p$, the limit preserves the quadratic Fronsdal action for $\widetilde{h}$ and $\widetilde{\phi}$ in Minkowski spacetime. We must now consider the limit of the cubic extra terms, let us recall their expression: \begin{eqnarray}V^{(p,q)}=\nonumber l_p^{2-n}\L^{-i}s^{\m_1\m_2|\n_1\n_2}\ \beta_{\m_1\m_2\n_1\n_2}^{(\a)(\r)(\b)(\s)}\ \N^p_{(\a)}\phi_{(\r)}\N^q_{(\b)}\phi_{(\s)}\ ,\ 0\leqslant i=\frac{p+q}{2}\leqslant s-2\quad.\end{eqnarray} The powers of $\e$ brought in by the Planck length and the spin-$s$ fields compensate, which leaves us with the powers brought in by the spin-2 fields, contained in $s_{\m\n|\r\s}$, and by the power of $\L$: $V^{(p,q)}=\e^{-2i+\D_h}\widetilde{V}^{(p,q)}$. In order for the limit to exist, we have to impose the condition $\D_h\geqslant max_i(2i)=2s-4$. If $\D_h>2s-4$, the extra terms all vanish. The interesting case is thus: \begin{eqnarray}\D_h=\D_\phi=\frac{n-2}{2}\D_p=2s-4\quad,\end{eqnarray} 
which allows one to preserve the terms containing the highest number of derivatives $p+q+2=2s-2$. This choice also preserves the terms in the gauge transformations that contain the highest number of derivatives. By construction, the deformation terms with two derivatives arising in $\cl_{Fs,min}$ are similar to $V^{(0,0)}$ and thus vanish when $\e\rightarrow 0$ as well. Finally, the only terms that remains in Minkowski spacetime are the free Fronsdal Lagrangians and the cubic terms with $2s-2$ derivatives. An interesting fact is that there is no trace of the minimal coupling attempt in the corresponding Minkowski cubic vertex and linear gauge transformations. 

Furthermore, let us emphasize that this limit only works for a particular value of the spin-$s$. In a sum of $2-s-s$ $(A)dS$ deformations, the only cubic terms that can be preserved are those containing the highest number of derivatives in the highest spin deformation. In the infinite sum of all these deformations, no Minkowski limit can be defined. 

Finally, we can claim that the same kind of limit can be defined for any $(A)dS$ deformation, for example obtained in dimension 4 or 5 from the Fradkin--Vasiliev action. It will always be possible to allocate weights to the different fields, to the Planck length and to the cosmological constant, in such a way as to preserve the free Lagrangians and the cubic terms containing the highest number of derivatives. This type of limit establishes that any $(A)dS$ first order consistent deformation must correspond to a Minkowski first order consistent deformation. The antifield formalism allows one to determine every nonabelian Minkowski consistent cubic deformations. Furthermore, we provide in Chapter \ref{ch:intmisc} some arguments that indicate that there are few distinct nonabelian cubic deformations for a given spin configuration $s-s'-s''$. We can conjecture that there is a bijection between the $(A)dS$ and the Minkowski nonabelian cubic deformations, the latter consisting of the terms containing the highest number of derivatives in the expansion of the former.

More precisely, we can already prove that the Minkowski limit is injective. Suppose that there are two different nonabelian cubic deformations in $(A)dS$, whose vertices are $S^\L(\phi)$ and $S\,{}'{}^\L(\phi)$, that admit the same nonabelian Minkowski limit $S^{\L=0}(\widetilde{\phi})$. The first order equations of the consistent deformation problem are linear (in the antifield formalism, they read $s\stackrel{(1)}{w}+db=0$), hence any linear combination of consistent deformations is a consistent deformation. Thus, the difference between $S^\L$ and $S'{}^\L$ is a vertex $S\,{}''{}^\L$. Since both deformations admit the same limit, they share the same terms involving the highest number of derivatives and their difference thus contains less derivatives. We can now choose the appropriate weights such that the limit of $S\,{}''{}^\L$ corresponds to a Minkowski vertex involving its highest number of derivatives. Hence, instead of considering $S$ and $S\,{}'$, one may choose $S$ and $S\,{}''$ that correspond to different Minkowski vertices. The argument repeats at any degree of derivation, thus it is always possible to choose an appropriate basis of deformations corresponding to distinct Minkowski vertices. Once every of them are related to an $(A)dS$ vertex, any new deformation in $(A)dS$ can be decomposed according to the basis. We can also prove that the possibility of scaling away the nonabelianess while at the same time retaining the vertex is ruled out. Consider a generator in $(A)dS$: $W_\L=\stackrel{(0)}{W_\L}+g\stackrel{(1)}{W_\L}+\cdots$ with $\stackrel{(1)}{W_\L}=\int (a^\L_2+a^\L_1+a^\L_0)$ where $a^\L_2$, $a^\L_1$ and $a^\L_0$ are cubic and contain, respectively, the nonabelian deformation of the gauge algebra, the corresponding gauge transformations and vertices. The master equation amounts to $\c^\L a^\L_2=0$, $\c^\L a_1^\L+\d^\L a_2^\L=d c_1^\L$ and $\c^\L a_0^\L+\d^\L a^\L_1=d c_0^\L$ where $\c^\L$ and $\d^\L$ have $\L$-expansions starting at order $\L^0$. Since the system is linear and determines $a_1^\L$ and $a_0^\L$ for a given $a_2^\L$, it follows that all $a_i^\L$ scale with $\L$ in the same way in the limit $\L\rightarrow 0$.

The question of showing that this correspondence is surjective, i.e. that every cubic nonabelian Minkowski deformation can be covariantized in such a way as to obtain a consistent $(A)dS$ deformation, remains to be addressed. As a matter of fact, in the case of the $1-s-s$ and $2-s-s$ configurations, we prove in Chapter \ref{ch:intmisc} that there is only one possible deformation in Minkowski spacetime, involving respectively $(2s-1)$ and $(2s-2)$ derivatives. We are thus already sure that the Fradkin-Vasiliev construction is the unique consistent deformation for a $2-s-s$ configuration in $(A)dS$ spacetime. In the same way, we know that there is at most one unique cubic vertex in a $1-s-s$ configuration in $(A)dS$. Let us notice that, in general, the deformations include some coefficients depending on the internal indices of the fields (that we denoted $a$ on $\phi^a_{\m_1...\m_{s_a}}$). In the case of a $1-s-s$ deformation, these coefficients are antisymmetric in the spin-$s$ internal indices. The brief presentation of the Fradkin-Vasiliev action that we made in the previous section does not include such indices, but the more general presentation of \cite{Fradkin:1987qy} does. The uniqueness of the deformations must of course be understood modulo those internal coefficients. Another remark that can be formulated is that the map defined by the Minkowski limit only requires first order consistency. For example, there are two spin-3 consistent deformations in Minkowski spacetime, involving three and five derivatives, as was showed in \cite{Bekaert:2006jf}. We prove in Chapter \ref{ch:socomp} that the deformation with three derivatives is inconsistent at second order. Anyway, the two vertices might correspond to distinct $(A)dS$ deformations, that could be part of a complete $(A)dS$ theory. As we already stated for the $2-s-s$ deformation, the limit only works at a given number of derivatives. Several vertices with various spin content can survive to the same limit, but it is impossible to establish a Minkowski limit for a complete $(A)dS$ theory, that involves every spin and thus an arbitrary number of derivatives. Conversely, one may wonder if a full Minkowski theory really exists. If it is the case, we are already sure that it does not correspond to the flat limit of an $(A)dS$ theory. 

As a conclusion of this chapter, we can now discuss the question of the equivalence principle and the compatibility with Einstein--Hilbert gravitation. As we recalled in the introduction, it is known for some time that there is an incompatibility between higher-spin cubic vertices in Minkowski spacetime and gravitation. We prove in Chapter \ref{ch:socomp} that it is actually the case for the $2-3-3$ and $1-2-2$ deformations that we have found. This means that the possible nonabelian Minkowski higher spin theory is exotic, in the sense that it cannot exist in the current universe, in which gravitation is dominant. This is in fact in relation with the fact that the minimal coupling terms are absent from the Minkowski couplings between spin-2 and spin-$s$ fields. Another proof of the incompatibility has been provided in \cite{Porrati:2008rm}, which is based on $S$-matrix arguments, involving a limit when the energy goes to 0. This work shows that a Minkowski theory violates the equivalence principle. On the other hand, everything remains possible in $(A)dS$ spacetime: the fact that $2-s-s$ deformations begin like a minimal coupling is a sign that the $(A)dS$ theory satisfies the equivalence principle. Furthermore, the presence of a minimal energy related to the $(A)dS$ mass of the gauge fields, which depends on $\L$, prevents one to consider the low energy limit by setting an infrared cutoff, in addition to the absence of an $S$-matrix formalism in $(A)dS$. These considerations about the $(A)dS$ consistency should be published in a near future \cite{IP}.

%%%%%%%%%%%%%%%%%%%%%%%%%%%%%%%%%%%%%%%%%%%%%%%%
\chapter{Consistent cubic vertices involving spin-2 and spin-3 fields}\label{ch:int23}
%%%%%%%%%%%%%%%%%%%%%%%%%%%%%%%%%%%%%%%%%%%%%%%%
In this chapter, we study the first order deformation problem for a collection of spin-2 and spin-3 fields. First, using the results of chapter \ref{ch:antidef}, we find the two unique Poincar\'e invariant nonabelian cubic Lagrangian vertices. One of them involves two spin-2 fields, the other involves two spin-3 fields. In the second part, we recover the quasi-minimal deformation of the Fronsdal action in (Anti)de Sitter spacetime described in Chapter \ref{ch:FV}. Finally, we conclude that it is the unique $2-3-3$ consistent nonabelian deformation in $(A)dS$, thanks to the Minkowski limit. This Chapter is mostly based on results that we obtained in \cite{Boulanger:2006gr,Boulanger:2008tg}. Let us note that the $2-3-3$ $(A)dS$ cubic deformation has been presented, at the same time and independently, in \cite{Zinoviev:2008ck}.

\section{Complete study at first order in Minkowski spacetime}

We shall consider in this chapter an arbitrary number of spin-2 fields and an arbitrary number of spin-3 fields. The spin-2 fields are denoted $h^a_{\mu\nu}$, which is a more usual notation than $\phi^a_{\m\n}$ (in this particular study, the index $a$ is the internal index of the spin-2 family, while the spin-3 fields bear capital Latin letters). The first order Riemann tensor reads: \begin{eqnarray}
R^a_{\a\m|\b\n}=-2K^a_{\a\m|\b\n}=-\frac{1}{2}(\6^2_{\a\b}h^a_{\m\n}-\6^2_{\a\n}h^a_{\m\b}-\6^2_{\m\b}h^a_{\a\n}+\6^2_{\m\n}h^a_{\a\b})\quad,
\end{eqnarray} and is of course related to the curvature tensor for a generic spin-$s$ field. Similarly, the Ricci tensor $R^a_{\m\n}=R^{a\a}_{\phantom{a\a}\m|\a\n}$ is proportional to the spin-$2$ Fronsdal tensor, so that the Riemann tensor is the only gauge invariant tensor in the particular case of spin-$2$.

The spin-3 fields are denoted by $\phi^A_{\mu\nu\rho}$, their Fronsdal and curvature tensors read: \begin{eqnarray}F^A_{\m\n\r}=\Box\phi^A_{\m\n\r}-3\6^{2\s}_{(\m}\phi^A_{\n\r)\s}+3\6^2_{(\m\n}\phi'{}^A_{\r)}\end{eqnarray}
\begin{eqnarray}
K^A_{\m_1\n_1|\m_2\n_2|\m_3\n_3}=6Y^{(3)}(\6_{\m_1\m_2\m_3}\phi^A_{\n_1\n_2\n_3})\quad,
\end{eqnarray} and satisfy the identity $K^{A\a}_{\phantom{A\a}\m|\a\n|\r\s}\equiv 2\6_{[\r}F^A_{\s]\m\n}$. 

The initial Fronsdal Lagrangian for this collection of fields is \begin{eqnarray}
\stackrel{(0)}{\cl}= -h^a_{\m\n}H^{\m\n}_a+\frac{1}{2}\phi^A_{\m\n\r}G^{\m\n\r}_A\quad,
\end{eqnarray} where $H^a_{\m\n}=R^a_{\m\n}-\frac{1}{2}\eta_{\m\n}R^a$ , $R^a=\eta^{\a\b}R^a_{\a\b}$ and $G^A_{\m\n\r}=F^A_{\m\n\r}-\frac{3}{2}\eta_{(\m\n}F'{}^A_{\r)}$. In the antifield formalism, the BRST-BV generator $\stackrel{(0)}{W}=\int \stackrel{(0)}{w}$ is local and is given by (see Eq.(\ref{wfrons})):
\begin{eqnarray}
\stackrel{(0)}{w}=\left[\stackrel{(0)}{\cl}+2h^{*\m\n}_a\6_{(\m}C^a_{\n)}+3\phi^{*\m\n\r}\6_{(\m}C^A_{\n\r)}\right]d^n x\quad.
\end{eqnarray}Let us remind that the differential $\bs=\d+\g$ is defined by $\bs A=(\stackrel{(0)}{W},A)$. More explicitly, the actions of $\g$ and $\d$ are: \begin{eqnarray}
\g h^a_{\m\n}=2\6_{(\m} C_{\n)}^a\quad,\quad\g\phi^A_{\m\n\r}=3\6_{(\m} C_{\n\r)}^A
\end{eqnarray}
\begin{eqnarray}
&\d h^{*\m\n}_a=-2H^{\m\n}_a \quad,\quad \d \phi^{*\m\n\r}_A=G^{\m\n\r}_A&\\& \d C^{*\m}_a=-2\6_\n h^{*\m\n}_a \quad,\quad \d C^{*\m\n}_A=-3\left[\6_\r\phi^{*\m\n\r}_A-\frac{1}{n}\eta^{\m\n}\6_\r\phi'{}^{*\r}_A\right]&\quad.
\end{eqnarray}
The undifferentiated spin-2 ghosts $C^a_\m$ and their antisymmetrized first derivatives $\6_{[\m}C^a_{\n]}$ are the only non-$\g$-exact spin-2 ghosts (the latter could be called $U^{(1)a}_{\m\n}$ as in Section \ref{cohgam}). Furthermore, the undifferentiated spin-3 ghosts $C^A_{\m\n}$, the traceless part of their antisymmetrized first derivatives $T^A_{\a\b|\m}:=U^{(1)A}_{\a\b|\m}=\6_{[a}C^A_{\b]\m}-\frac{1}{n-1}\eta_{\m[\a}\6^\r C^A_{\b]\r}$, and the traceless part of their twice antisymmetrised second derivatives $U^A_{\a\m|\b\n}:=U^{(2)A}_{\a\m|\b\n}$ provide us with a minimal set of non-$\g$-exact spin-3 ghost tensors (see Section \ref{cohgam}).

\subsection{Deformation of the gauge algebra}

Theorem \ref{thmcub} holds for a collection of spin-2 and spin-3 fields. We thus know that the first order of deformation of the generator $w=\stackrel{(0)}{w}+g\stackrel{(1)}{w}+...$ admits an expansion in the antifield number that stops at most at 2, and the top term $a_2$ must be cubic. Furthermore, the general study in section \ref{cbvrt} shows that $a_2$ can be taken proportional to an undifferentiated antifield $C^*$ and quadratic in the non-$\g$-exact ghost tensors. This is because, if $a_2$ is cubic, it is linear in the $antigh\ 2$ antifields $C^*$ and quadratic in the ghosts. The first order deformation $\stackrel{(1)}{w}$ is a representative of a coset of $H^{0,n}(\bs|d)$ and is thus defined up to a $d$-exact and a $\bs$-exact term. The modulo $d$ freedom allows one to remove the derivatives that $C^*$ could bear, while the modulo $\bs$ freedom allows to add (or remove) any $\g$-exact term to $a_2$. A convenient way to build a complete list of independent deformations is to work with the tensors $U^{(i)}$. However, most of the times, the rest of the computations is much easier if one adds some $\g$-exact terms. Let us recall the equations needed to compute the first order deformations: if $\stackrel{(1)}{w}=a_0+a_1+a_2$, where $antigh\ a_i=i$, then $\exists b=b_0+b_1$ such that:
\begin{eqnarray} \g a_2&=&0\label{eqag2}\\ \d a_2+\g a_1&=&db_1 \label{eqag1}\\ \d a_1+\g a_0&=&db_0\label{eqag0}\quad.\end{eqnarray} Let us notice that the addition of a $\g$-exact term to $a_2$ is related to the addition of a $\d$-exact term to $a_1$ (those two terms forming an $\bs$-exact term in $\stackrel{(1)}{w}$). This $\d$-exact term does not alter Eq.(\ref{eqag0}), so that the different choices for $a_2$ lead to the same $a_0$ (which is, in turn, defined modulo $\g$ and $d$).

Poincar\'e invariant solutions only involve the Minkowski metric and the Kronecker delta (we only determine here the parity-invariant deformations), in such a way that $a_2$ is Lorentz-invariant. We can now establish a list of possible terms for $a_2$. These terms will be characterized by their field content and the number of derivatives involved. Terms with similar field and derivative contents may be linearly dependent modulo $d$, and in fact have to mix with precise weights in order to provide a consistent solution of Eq.(\ref{eqag0}). Each of those terms will be given a set of internal ``structure constants'', that bear a Latin index for each of the three fields. These are not constrained at this stage, but relations are likely to appear later. Finally, let us remind that $a_2$ contains the information about the deformation of the gauge algebra, as it contains the first order of the structure operator. Conversely, ``homogeneous'' solutions, stopping at antigh 1, correspond to Abelian deformations of the gauge transformations. We have not addressed the computation of these solutions, focusing only on the nonabelian deformations.

\subsubsection{$2-2-3$ candidates}

First, let us establish the complete list of $a_2$ terms quadratic in the spin-2 fields and linear in the spin-3 fields (``field'' has to be understood here in the extended sense of field, antifield or ghost). Spin-2 undifferentiated ghosts $C^a_\m$ and $antigh\ 2$ antifields $C^{*\m}_a$ both bear one spacetime index. Those of spin-3, $C^A_{\m\n}$ and $C^{*\m\n}_A$, bear 2. In order for the contraction of indices to be Lorentz-invariant (i.e. there are no more free spacetime indices), the total number of indices must be even. In the $2-2-3$ configuration, an antifield and two ghosts bring in four indices, thus the $a_2$ terms must contain an even number of derivatives. The non-$\g$-exact spin-2 ghosts tensors are $C^a_\mu$ and $\6_{[\mu}C^a_{\nu]}$, while the non-$\g$-exact spin-3 ghosts tensors are $C^A_{\mu\nu}$, $T^A_{\mu\nu|\a}$ and $U^A_{\mu\a|\nu\b}$. A product of two ghost tensors contains at most 3 derivatives, thus the highest even number is 2. Let us enumerate the candidates with no derivatives: \begin{eqnarray}\stackrel{(1)}{a}_2&=&\stackrel{(1)}{f}_{abC}C^{a*\mu}C^{b\nu}C^C_{\mu\nu}\,d^n x\,,\nonumber\\ \stackrel{(2)}{a}_2 &=&\stackrel{(2)}{f}_{A[bc]}C^{*A\mu\nu}C^b_{\mu}C^c_{\nu}\,d^n x\quad.\end{eqnarray} Then, let us enumerate the candidates with two derivatives :
\begin{eqnarray} \stackrel{(3)}{a}_2 &=& \stackrel{(3)}{f}_{abC}\,C^{*a\m}\,\6^{[\n}C^{\r] b}\,T^C_{\n\r|\m}\,d^n x+\g\!\!\stackrel{(3)}{c}=\stackrel{(3)}{f}_{abC}\,C^{*a\m}\,\6^{[\n}C^{\r] b}\,\6_{[\n}C^C_{\r]\m}\,d^n x\quad,\nonumber\\ \stackrel{(4)}{a}_2 &=& \stackrel{(4)}{f}_{A[bc]}\,C^{*A\m\n}\,\6_{[\m}C_{\a]}^b \,\6_{[\n}C_{\b]}^c\,\eta^{\a\b} \,d^n x +\g\!\!\stackrel{(4)}{c}\quad.\end{eqnarray} At first, we have kept the modulo $\g$ freedom, then we have chosen particular values for later convenience. We have not written $\stackrel{(3)}{c}$ explicitly, but it exists because the divergence of the spin-3 ghost is $\g$-exact. For the other candidate, we take $\stackrel{(4)}{c}=0$.

\subsubsection{$2-3-3$ candidates}

The other candidates are quadratic in the spin-3 fields and linear in the spin-2 fields. In order to build Lorentz-invariant expressions, we have to consider this time an odd number of derivatives, because the product of two ghosts and one $antigh\ 2$ antifield bring in five indices. A product of two spin-3 ghosts contains at most 4 derivatives, thus the highest odd number is 3. Let us enumerate the candidates with one derivative: \begin{eqnarray}\stackrel{(5)}{a}_2 &=& \stackrel{(5)}{f}_{ABc}\,C^{*A\m\n}\,\6_{[\n}C^B_{\s]\m}\, C^{c\s}\,d^n x\,+\g\!\!\stackrel{(5)}{c}\quad,\nonumber \\ \stackrel{(6)}{a}_2 &=& \stackrel{(6)}{f}_{ABc}\,C^{*A\m\n}\,C^{B\,\a}_{~\m}\,\6_{[\n}C^c_{\a]}\,d^n x\,+\g\!\!\stackrel{(6)}{c}\quad,\nonumber \\ \stackrel{(7)}{a}_2 &=& \stackrel{(7)}{f}_{aBC}\,C^{*a\a}\,C^{B\m\n}\,\6_{[\n}C^C_{\a]\m}\,d^n x\,+\g\!\!\stackrel{(7)}{c}\quad,\end{eqnarray} Finally, let us provide the only candidate with three derivatives : \begin{eqnarray}\stackrel{(8)}{a}_2 &=& \stackrel{(8)}{f}_{aBC}\,C^{*a\m}\,T^{B\a\b|\n}U^C_{\a\b|\m\n}\,d^n x\, +\g\!\!\stackrel{(8)}{c}\nonumber\\&=&\stackrel{(8)}{f}_{aBC}\,C^{*a\m}\,\6^{[\a}C^{|B|\b]\n}\,\6_{\m}\6_{[\a}C_{\b]\n}^C\,d^n x\quad.\end{eqnarray}We have already written the candidates in the most convenient way, $\stackrel{(8)}{c}$ is not written explicitly while the other $\g\stackrel{(i)}{c}$ are considered as vanishing.

\subsection{Deformation of the gauge transformations}

The next step is to compute Eq.(\ref{eqag1}). We thus have to compute $\d \stackrel{(i)}{a}$ for the different $a_2$ terms that we have enumerated above. The candidates have different field and derivative contents, except numbers 5 and 6. We will compute candidates number 5 and 6 together, and the others individually. If it appears that any of the $\d\!\stackrel{(i)}{a}_2$ , ($i=1,2,3,4,7,8$) cannot be $\g$-exact modulo $d$, then we can already conclude that it is obstructed. On the other hand, if they are $\g$-exact modulo $d$, then the object of $\g$ can be identified with the inhomogeneous part of a candidate $a_1$. These inhomogeneous parts will be denoted $\widetilde{a}_1$. Some homogeneous terms with the same field and derivative contents as $\widetilde{a}_1$ can also be needed for the last equation (Eq.(\ref{eqag0})). They are denoted $\bar{a}_1$ and are such that $\g \bar{a}_1=0$. More precisely, the homogeneous equation is $\g \bar{a}_1+d \bar{b}_1=0$, however, Theorem \ref{gdg} tells us that, in $antigh\ 1$, the $d$-exact term is also $\g$-exact and can thus be reabsorbed in $\bar{a}_1$.

\subsubsection{$2-2-3$ candidates}
Let us proceed with the computation for the $2-2-3$ candidates:
\begin{itemize}
\item $\begin{array}{rcl}
\displaystyle\d\!\stackrel{(1)}{a}_2&=&-2\stackrel{(1)}{f}_{abC}\6_\rho h^{a*\mu\rho}C^{b\nu}C^C_{\mu\nu}\,d^n x\\&=&d(...) + 2 \stackrel{(1)}{f}_{abC}h^{a*\mu\rho}(\6_\rho C^b_{\nu}C^{C\phantom{\mu}\nu}_{\phantom{C}\mu}+C^{b\nu}\6_\rho C^C_{\mu\nu})\,d^n x \end{array}$\\ The latter term cannot be $\g$-exact unless $\stackrel{(1)}{f}$ vanishes, because nothing causes the symmetrization of the derivatives of the ghosts. 
\item $\begin{array}{rcl}\displaystyle\d\!\stackrel{(2)}{a}_2 &=&-3\stackrel{(2)}{f}_{A[bc]}\6_\rho (\phi^{*A\mu\nu\rho}-\frac{1}{n}\eta^{\mu\nu}\phi^{*}{}'{}^{A\rho})C^b_{\mu} C^c_{\nu}\,d^n x\\&=&d(...)+6\stackrel{(2)}{f}_{A[bc]}(\phi^{*A\m\n\r}\6_{(\r} C^b_{\m)}C^c_{\n}-\frac{1}{n}\phi^{*}{}'{}^{A\r}\6_{\r}C^b_{\s}C^{c\s})\, d^n x\end{array}$\\ The first term in the brackets is $\g$-exact, but the second (the trace part) is not, thus this candidate also vanishes: $\stackrel{(2)}{f}=0$.
\item $\begin{array}{rcl}\d\!\stackrel{(3)}{a}_2 &=&-2\stackrel{(3)}{f}_{abC}\,\6^{\phantom{1}}_\s h^{*a\m\s}\,\6^{[\n}C^{\r] b}\,\6^{\phantom{1}}_{[\n}C^C_{\r]\m}\,d^n x\\ &=&d(...)+2\stackrel{(3)}{f}_{abC}h^{*a\m\s}\Big[\6_\s^{2[\n}C^{\r]b}_{\phantom{\r}}\,\6_\n C^C_{\r\m} +\6^{\phantom{1}}_{[\n}C^b_{\r]}\6^{2\phantom{\s}[\n}_{(\s}C_{\phantom{\r C}\m)}^{\r]C}\Big]\, d^n x\\ &=& d(...) - \g\left( \stackrel{(3)}{f}_{abC}h^{*a\m\s}\Big[2\6_{\phantom{1}}^{[\n}h^{\r]b}_{\phantom{\r b}\s}\,\6^{\phantom{1}}_\n C^C_{\r\m} - \6^{\phantom{1}}_{[\n}C^b_{\r]}\6_{\phantom{1}}^{[\n}\phi_{\phantom{\r]C}\m\s}^{\r]C}\Big]\, d^n x \right)
\end{array}$ \\ $\Rightarrow \stackrel{(3)}{a}_1=\stackrel{(3)}{f}_{abC}h^{*a\m\s}\Big[2\6_{\phantom{1}}^{[\n}h^{\r]b}_{\phantom{\r b}\s}\,\6^{\phantom{1}}_\n C^C_{\r\m} - \6^{\phantom{1}}_{[\n}C^b_{\r]}\6_{\phantom{1}}^{[\n}\phi_{\phantom{\r]C}\m\s}^{\r]C}\Big]\,d^n x+ \stackrel{(3)}{\bar{a}}_1\quad|\quad \g\stackrel{(3)}{\bar{a}}_1=0$
\item $\begin{array}{rcl}\d\!\stackrel{(4)}{a}_2 &=& -3\stackrel{(4)}{f}_{A[bc]}\,\6_\s(\phi^{*A\m\n\s}-\frac{1}{n}\eta^{\m\n}\phi^{*'A\s})\,\6^{\phantom{1}}_{[\m}C_{\a]}^b \,\6^{\phantom{1}}_{[\n}C_{\b]}^c\,\eta^{\a\b} \,d^n x\\ &=& d(...) + 6 \stackrel{(4)}{f}_{A[bc]}\,(\phi^{*A\m\n\s}-\frac{1}{n}\eta^{\m\n}\phi^{*'A\s})\,\6^2_{\s[\m}C_{\a]}^b \,\6^{\phantom{1}}_{[\n}C_{\b]}^c\,\eta^{\a\b} \,d^n x\\ &=& d(...) - \g \left( 6 \stackrel{(4)}{f}_{A[bc]}\,[\phi^{*A\m\n\s}-\frac{1}{n}\eta^{\m\n}\phi^{*'A\s}]\,\6^{\phantom{1}}_{[\m}h_{\a]\s}^b \,\6^{\phantom{1}}_{[\n}C_{\b]}^c\,\eta^{\a\b} \,d^n x \right)
\end{array}\\ \Rightarrow \stackrel{(4)}{a}_1=6 \stackrel{(4)}{f}_{A[bc]}\,[\phi^{*A\m\n\s}-\frac{1}{n}\eta^{\m\n}\phi^{*'A\s}]\,\6^{\phantom{1}}_{[\m}h_{\a]\s}^b \,\6^{\phantom{1}}_{[\n}C_{\b]}^c\,\eta^{\a\b} \,d^n x+\stackrel{(4)}{\bar{a}}_1\quad|\quad \g\stackrel{(4)}{\bar{a}}_1=0$
\end{itemize}
Two candidates appear at this stage. In fact, they will have to be considered together when solving Eq.(\ref{eqag0}). This is because their images under $\d$ are not independent modulo $d$. Since they constitute the only candidate involving two derivatives, let us denote their sum $a_{1,2}=\stackrel{(3)}{a}_1+\stackrel{(4)}{a}_1$. Similarly, the sum of the two homogeneous terms, that are still unconstrained at this stage, will be denoted $\bar{a}_{1,2}=\stackrel{(3)}{\bar{a}}_1+\stackrel{(4)}{\bar{a}}_1$. 

\subsubsection{$2-3-3$ candidates}

Let us first separately compute $\d\!\stackrel{(5)}{a}_2$ and $\d\!\stackrel{(6)}{a}_2$:\\
\noindent $\begin{array}{rcl}
\d \stackrel{(5)}{a}_2 &=&\displaystyle -3 \stackrel{(5)}{f}_{\!\!\!ABc}\6_\r\Big[\phi^{*A\m\n\r}-\frac{1}{n}\eta^{\m\n}\phi^{*}{}'{}^{A\r}\Big]\6^{\phantom{1}}_{[\n}C^B_{\s]\m} C^{c\s} d^n x\\&=&\displaystyle d(...)+\!\frac{3}{2n}\!\stackrel{(5)}{f}_{\!\!\!ABc}\6_\r \phi^{*}{}'{}^{A\r}\6^{\phantom{1}}_\n C_\s^{B\n} C^{c\s} d^n x + 3\! \stackrel{(3)}{f}_{\!\!\!ABc}\phi^{*A\m\n\r}\Big[\6^2_{\r[\n}C^B_{\s]\m}C^{c\s}+\6_{[\n}^{\phantom{1}}C^B_{\s]\m}\6^{\phantom{1}}_\r C^{c\s}\Big]d^n x\\ &=& d(...) + 3 \stackrel{(5)}{f}_{\!\!ABc}\phi^{*A\m\n\r}\6^{\phantom{1}}_{[\n}C^B_{\s]\m}\6_{[\r}^{\phantom{1}}C^c_{\t]}\eta^{\s\t}d^n x\\&& \displaystyle -\g\stackrel{(5)}{f}_{\!\!ABc}\left\{\frac{3}{4n}\6_\r\phi^{*'A\r}\phi^{'B}_\s C^{c\s}+\frac{3}{2}\phi^{*A\m\n\r}\6_{[\n}^{\phantom{1}}\phi^B_{\s]\m\r}C^{c\s}-\frac{3}{2}\phi^{*A\m\n\r}\6^{\phantom{1}}_{[\n}C^B_{\s]\m}h^{c\s}_\r\right\}d^n x
\end{array}$\\
\noindent $\begin{array}{rcl}
\d \stackrel{(6)}{a}_2 &=& \displaystyle -3 \stackrel{(6)}{f}_{\!\!ABc}\6_\r\Big[\phi^{*A\m\n\r}-\frac{1}{n}\eta^{\m\n}\phi^{*'A\r}\Big]C_\m^{B\a}\6^{\phantom{1}}_{[\n}C^c_{\a]} d^n x\\ &=& 3\stackrel{(6)}{f}_{\!\!ABc}\phi^{*A\m\n\r}\Big[\6^{\phantom{1}}_\r C_\m^{B\a}\6^{\phantom{1}}_{[\n}C^c_{\a]}+C_\m^{B\a}\6^2_{\r[\n}C^c_{\a]}\Big]d^n x\\ &=&-\g \stackrel{(6)}{f}_{\!\!ABc}\phi^{*A\m\n\r}\Big[\phi^{B\phantom{\m}\a}_{\r\m}\6^{\phantom{1}}_{[\n}C^c_{\a]}-3C^{B\a}_\m\6^{\phantom{1}}_{[\n}h^c_{\a]\r}\Big]d^n x +2\stackrel{(6)}{f}_{\!\!ABc}\phi^{*A\m\n\r}\6^{\phantom{1}}_{[\r}C^B_{\b]\m}\6^{\phantom{1}}_{[\n}C^c_{\a]}\eta^{\a\b}d^n x
\end{array}$\vspace{2mm}\\
The non-$\gamma$-exact modulo $d$ terms are the same in the two expressions. Thus, we must set the relation $\stackrel{(6)}{f}_{\!\!ABc}=-\frac{3}{2}\stackrel{(5)}{f}_{\!\!ABc}$ in order to obtain a consistent $a_1$. This is the only candidate involving one derivative, let us denote it $a_{2,1}=\stackrel{(5)}{a}_2+\stackrel{(6)}{a}_2$. We finally obtain: \begin{eqnarray}a_{1,1}&=&\stackrel{(5)}{f}_{\!\!ABc}\frac{3}{2}\phi^{*A\m\n\r}\left[\6_{[\n}^{\phantom{1}}\phi^B_{\s]\m\r}C^{c\s}-\6^{\phantom{1}}_{[\n}C^B_{\s]\m}h^{c\s}_\r-\phi^{B\phantom{\m}\a}_{\r\m}\6^{\phantom{1}}_{[\n}C^c_{\a]}+3C^{B\a}_\m\6^{\phantom{1}}_{[\n}h^c_{\a]\r}\right]d^n x\nonumber\\&&+\stackrel{(5)}{f}_{\!\!ABc}\frac{3}{4n}\6_\r\phi^{*'A\r}\phi^{'B}_\s C^{c\s}d^n x\label{a1d1}\quad.\end{eqnarray} There is no homogeneous part $\bar{a}_{1,1}$ because a $\g$-closed term linear in the fields contains at least two derivatives (in a Riemann or Fronsdal tensor).\\ Let us consider the seventh candidate: \\ $\begin{array}{rcl}
\d \stackrel{(7)}{a}_2&=&-2\stackrel{(7)}{f}_{\!\!aBC}\6_\r h^{*a\a\r}C^{B\m\n}\6^{\phantom{1}}_{[\n}C^C_{\a]\m}d^n x\\&=&d(...)+\gamma(...)+2\stackrel{(7)}{f}_{\!\!aBC}h^{*a\a\r}\Big[-\frac{2}{3}T^{B\phantom{\m}\n}_{\m\r|}T^{C\phantom{\n}\m}_{\n\a|}-\frac{2}{3}T^{B\m|\n}_{\r}T^C_{\a\m|\n}+C^{B\m\n}U^C_{\n\a|\r\m}\Big]d^n x
\end{array}$\\ We have not written explicitly the $\gamma$-exact part, because the non $\gamma$-exact modulo $d$ terms cannot be eliminated. The first two terms into brackets could vanish by imposing $\stackrel{(7)}{f}_{\!\!aBC}=\stackrel{(7)}{f}_{\!\!a(BC)}$, but the last one is definitely an obstruction, as one readily convinces oneself by taking the Euler-Lagrange derivative with respect to the antifields.\\
Finally, let us study the only candidate involving three derivatives:\\
$\begin{array}{rcl}
\d\stackrel{(8)}{a}_2&=&-2\stackrel{(8)}{f}_{\!\!aBC}\,\6_\r h^{*a\m\r}\,\6^{[\a}C^{|B|\b]\n}\,\6_{\m}\6_{[\a}C_{\b]\n}^C\,d^n x\\ &=&d(...)+2\stackrel{(8)}{f}_{\!\!aBC}h^{*a\m\r}\Big[\6_{\r}^{2[\a}C^{|B|\b]\n}\,\6_{\m}\6_{[\a}C_{\b]\n}^C+ \6_{\phantom{1}}^{[\a}C^{|B|\b]\n}\,\6^3_{\m\r[\a}C_{\b]\n}^C\Big]d^n x
\end{array}$\\
The first term into brackets is clearly antisymmetric in $B$ and $C$, while the second is $\g$-exact because it involves third derivatives of $C^B_{\m\n}$. Thus, by imposing the relation $\stackrel{(8)}{f}_{aBC}=\stackrel{(8)}{f}_{a(BC)}$, we obtain a consistent $a_1$, that can be denoted $a_{1,3}$ as it is the only one involving three derivatives: \begin{eqnarray}
a_{1,3}=\stackrel{(8)}{f}_{aBC}h^{*a\m\r}\6_{\phantom{1}}^{\a}C^{B\b\n}\Big[\6^2_{\n[\a}\phi^C_{\b]\m\r}-2\6^2_{\m[\a}\phi^C_{\b]\n\r}\Big]d^n x+\bar{a}_{1,3}\quad|\quad \g\bar{a}_{1,3}=0\quad.
\end{eqnarray}

\subsection{Deformation of the Lagrangian\label{deflag233}}

Let us proceed with the computation of the cubic vertices. Because of the length of the expressions appearing, for example when introducing the equations of motions, it would be hard to make a straight computation. Instead of that, we have listed every possible Lorentz-invariant cubic terms for a Lagrangian vertex $a_{0,test}$, with the appropriate field and derivative contents. We have also considered the most general homogeneous part of $a_1$, when it is possible to have one, let us denote it $\bar{a}_{1,test}$. We have explicitly computed the following quantity: \begin{eqnarray}\d(\widetilde{a}_1+\bar{a}_{1,test})+\g a_{0,test}\quad,\end{eqnarray} where $\widetilde{a}_1$ is the inhomogeneous part of $a_1$, that has been determined in the previous section. This expression is linear in the ghosts or their derivatives. The modulo $d$ freedom is given away by taking the Euler-Lagrange derivatives with respect to the ghosts, which is equivalent to ``integrating by parts'' in order to make appear the undifferentiated ghosts. The result must vanish in order to obtain a consistent vertex. This yields a system of equations for each candidate $a_{1,i}$. We do not give all the details of the resolution of these systems nor the explicit lists of terms, but we expose the procedure in the next paragraphs.

\subsubsection{Candidate with one derivative}

The $a_0$ terms corresponding to $a_{1,1}$ are linear in the spin-2 fields, quadratic in the spin-3 fields and contain two derivatives. 
As the Lagrangian is defined modulo $d$, it is sufficient to make a list where, for example, the spin-2 fields bear no derivatives. The terms can then be classified into two categories: 25 terms with the two derivatives on the same spin-3 field ($h^a_{.\,.}\6_{.\,.}\phi^B_{.\,.\,.}\phi^C_{.\,.\,.}$) and 24 with one derivative on each spin-3 field ($h^a_{.\,.}\6_{.}\phi^B_{.\,.\,.}\6_{.}\phi^C_{.\,.\,.}$). The test vertex $a_{0,test}$ is the sum of these 49 terms contracted with arbitrary coefficients $\stackrel{(j)}{g}_{aBC}$, $j=1,...,49$. 

We have used the symbolic manipulation software FORM \cite{FORM} to deal with the considered expressions. After having encoded $a_{0,test}$ and $\widetilde{a}_{1,1}$ in a file, we have computed their images under $\g$ and $\d$. Then, it is rather easy to take the Euler-Lagrange derivative and to obtain the system of equations. Because of the complicated nature of the coefficients $\stackrel{(j)}{g}_{aBC}$, we have solved the system by making substitutions by hand. This is lengthy but systematic. Finally, we have found that the solution is trivial: the coefficients all vanish and the candidate with one derivative is thus obstructed.

\subsubsection{$2-2-3$ vertex with three derivatives}

The other candidates go along the same lines, except for the non-vanishing $\bar{a}_1$. The $a_0$ terms corresponding to $a_{1,2}$ are quadratic in the spin-2 fields, linear in the spin-3 fields and contain three derivatives. This vertex is defined modulo $d$, thus we have chosen terms involving an undifferentiated spin-3 field. They can be classified into two categories: 25 terms with three derivatives on the same spin-2 field ($\phi^A_{.\,.\,.} \6_{.\,.\,.}h^b_{.\,.}h^c_{.\,.}$), and 50 terms with two derivatives on a spin-2 field and one on the other ($\phi^A_{.\,.\,.}\6_{.\,.}h_{.\,.}\6_.h_{.\,.}$). The test vertex $a_{0,test}$ is the sum of these terms contracted with arbitrary coefficients $\stackrel{(j)}{k}_{Abc}$, $j=1,...,75$.
For this candidate, it is important not to forget $\bar{a}_{1,test}$. With two derivatives, we can build terms that are linear in the undifferentiated Riemann or Fronsdal tensor (the spin-3 curvature contains three derivatives and cannot appear here). Let us recall that the Fronsdal and Ricci tensors are $\d$-exact, so that any term linear in the Fronsdal tensor can be eliminated in favor of a term linear in the Ricci tensor by adding a $\d$-exact expression to $\bar{a}_1$. For example : $C^{a\a} G^B_{\a\b\c} h^{*c\b\c}=-\d [ C^{a\a} \phi^{*B}_{\a\b\c} h^{*c\b\c} ] + C^{a\a} \phi^{*B}_{\a\b\c} G^{c\b\c}$. For the same reason, some terms linear in the Ricci tensor are not independent modulo $\d$ : $h^{*a\a\b}G'{}^{b}C^C_{\a\b}=\d(...) + G^{a\a\b}h^{*}{}'{}^{b}C^C_{\a\b}$. This allows one to consider a short list of six terms : $h^{*a\a\c}R^b_{\a\b|\c\d}C^{C\b\d}$ , $h^{*a\a\b}R^{b\c}_\a C^C_{\b\c}$ , $h^{*a\a\b}R^b C^C_{\a\b}$ , $\phi^{*A\a\b\c}R^b_{\a\b}C^c_\c$ , $\phi^{*}{}'{}^{A\a}R^b_{\a\b}C^{c\b}$ and $\phi^{*}{}'{}^{A\a}R^b C^c_{\a}$, that are contracted with some other arbitrary coefficients $\stackrel{(j)}{l}_{Abc}\ ,\ j=1,...,6$.

We have gathered these ingredients in a file, and have achieved the computation in the same way as for the first candidate. We obtained a system of hundreds of equations depending on the 82 variables bearing family indices. We solved it by substitution and found that it actually admits a nontrivial solution. First, the relation $\stackrel{(4)}{f}_{Abc}=-\frac{1}{3}\stackrel{(3)}{f}_{bcA}$ is obtained. It means that $\stackrel{(3)}{f}_{bcA}$ must be antisymmetric. We also found that the only $\bar{a}_1$ term needed is one of those linear in the Riemann tensor, while it is possible to keep only terms of the second category in $a_0$. This leaves us with 48 terms with coefficients proportional to $\stackrel{(3)}{f}_{bcA}$. Let us denote the only coefficient $g_{Abc}=\stackrel{(3)}{f}_{bcA}$. Finally, the complete solution can be written: 
\begin{eqnarray}a_{2,2}= g_{Abc}\left[\,C^{*b\m}\,\6^{\n}C^{c\r}\,\6_{[\n}C^A_{\r]\m}\,
 -\frac{1}{3}\,C^{*A\m\n}\,\6_{[\m}C_{\a]}^b \,\6_{[\n}C_{\b]}^c\,\eta^{\a\b}\right]
  \,d^n x\quad,
\label{a72} 
\end{eqnarray}
\begin{eqnarray}a_{1,2} &=& g_{Abc}h^{*b\m\s}\left[2\6^\n h^{c\r}_\s 
\6_{[\n}C^A_{\r]\m}-\6^\n C^{c\r}\6_{[\n}h^A_{\r]\m\s}\right]d^n x
\nonumber \\ 
&& 
 - 3 g_{Abc}h^{*b\a\b}R^c_{\a\m|\b\n}C^{A\m\n}d^n x 
\nonumber\\ 
&& 
-2 g_{Abc}\left[\phi^{*A\m\n\r}-\frac{1}{n}\eta^{\m\n}\phi^{*'A\r}\right]
\6_{[\m}h^b_{\a]\r}\6_{[\n}C^c_{\b]}\eta^{\a\b} d^n x\quad,
\end{eqnarray}

\begin{eqnarray}
  a_{0,2} = \stackrel{(3)}{\cal{L}}d^n x = g^A_{\ bc}\,U^{bc}_{A}\, d^n x \quad,
\label{vert223}  
\nonumber
\end{eqnarray}
where, denoting $h^b=\eta^{\a\b}h^b_{\a\b}$ :
\begin{eqnarray}
	U^{bc}_{A} &=& -  \frac{1}{2}\, \phi_A^{'\a} \Box  h^b \6_{\a} h^c 
	        +  \frac{1}{2}\, \phi_A^{\a\b\c} \6_{\b}\6_{\c}h^b \6_{\a}h^c
	        +  \frac{1}{2}\,\phi_A^{\a\b\c} \Box  h^b_{\b\c} \6_{\a}h^c
\nonumber \\       
	   &&+\; \frac{1}{2}\,\phi_A^{'\a} \6^{\b}\6^{\c} h^b_{\b\c} \6_{\a}h^c
         +              \phi_A^{'\a} \6^{\b}\6^{\c} h^b  \6_{\a}h_{\b\c}^c
	       + \frac{1}{2}\,\phi_A^{'\a} \Box  h^{b\b\c} \6_{\a}h_{\b\c}^c
\nonumber \\       
	   && -\; \phi_A^{\a\b\c} \6_{\b\d}h_{\c}^{b\,\d}\6_{\a}h^c
	        - \phi_A^{\a\b\c} \6_{\b\d}h^b \6_{\a}h_{\c}^{c\,\d}
          - \frac{1}{2}\,\phi_A^{\a\b\c} \Box  h^b_{\b\d}\6_{\a}h_{\c}^{c\,\d}
\nonumber \\       
	   &&-\; \frac{3}{2}\,\phi_A^{'\a} \6^{\b}\6^{\c} h^b_{\b\d} \6_{\a}h_{\c}^{c\,\d}
	        - \frac{1}{2}\,\phi_A^{\a\b\c} \6_{\b}\6_{\c} h^{b\m\n} \6_{\a}h_{\m\n}^c
          - \phi_A^{\a\b\c} \6^{\m}\6^{\n} h^b_{\b\c} \6_{\a}h_{\m\n}^c
\nonumber \\       
	   &&+\; 
	         \frac{1}{2}\,\phi_A^{\a\b\c} \6_{\c}\6_{\d} h^{b\d\e} \6_{\a}h^c_{\b\e}
         + \frac{3}{2}\,\phi_A^{\a\b\c} \6_{\c}\6_{\d} h^b_{\b\e} \6_{\a}h^{c\d\e}
	       + \phi_A^{\a\b\c} \6^{\d}\6^{\e} h^b_{\b\d} \6_{\a}h^c_{\c\e}
\nonumber \\       
	   &&-\; \frac{1}{4}\,\phi_A^{'\a} \6_{\a}\6_{\c}h^b  \6^{\c}h^c
	        - \frac{1}{2}\,\phi_A^{\a\b\c} \6_{\a}\6_{\e}h^b_{\b\c}\6^{\e}h^c
          + \phi_A^{\a\b\c} \6_{\a}\6_{\e}h^b \6^{\e}h_{\b\c}^c
\nonumber \\       
	   &&+\; \frac{1}{4}\,\phi_A^{'\a} \6_{\a}\6_{\e}h^{b\m\n} \6^{\e}h_{\m\n}^c
	        - \frac{1}{2}\,\phi_A^{\a\b\c} \6_{\a}\6_{\e}h^b_{\b\d} \6^{\e}h_{\c}^{c\,\d}
	        + \phi'_{A\m} \6_{\a}\6_{\e}h^{b\a\m} \6^{\e}h^c 
\nonumber \\       
	   &&-\; \phi_{A\m\b\c} \6_{\a}\6_{\e}h^{b\a\m} \6^{\e}h^{c\b\c} 
	        + \frac{1}{2}\,\phi_A^{'\t} \6_{\a}\6_{\e}h^{b\a\m} \6^{\e}h^c_{\m\t} 
	        - \phi'_{A\m} \6_{\a}\6_{\e}h^b \6^{\e}h^{c\a\m} 
\nonumber \\       
	   &&+\; \phi_{A\m\b\c} \6_{\a}\6_{\e}h^{b\b\c} \6^{\e}h^{c\a\m} 
          - \frac{1}{2}\,\phi_A^{'\t} \6_{\a}\6_{\e}h^b_{\m\t} \6^{\e}h^{c\a\m} 
	        - \frac{1}{2}\,\phi_A^{'\a} \Box  h^b_{\a\c} \6^{\c}h^c  
\nonumber \\       
	   &&+\; \frac{1}{2}\,\phi_A^{\a\b\c} \6_{\b}\6_{\c} h^b_{\a\r} \6^{\r}h^c  
	        +\frac{1}{2}\,\phi_A^{\a\b\c} \Box  h^b_{\a\r} \6^{\r}h_{\b\c}^c  
          +\frac{1}{2}\,\phi_A^{'\a} \6_{\b}\6_{\c} h^b_{\a\r} \6^{\r}h^{c\b\c}
\nonumber \\       
	   &&-\; \phi_A^{\a\b\c} \6_{\b}\6_{\d} h^b_{\a\r} \6^{\r}h_{\c}^{c\,\d}  
	        -\frac{1}{4}\,\phi'_{A\m} \6^{\b}\6^{\m} h^b_{\b\c} \6^{\c}h^c  
          -\frac{1}{2}\,\phi_{A\m\n\r} \6^{\b}\6^{\m} h^b_{\b\c} \6^{\c}h^{c\n\r}
\nonumber \\       
	   &&+\;   \phi_A^{'\n} \6^{\b}\6^{\m} h^b_{\b\c} \6^{\c}h_{\m\n}^c  
	        - \frac{1}{2}\,\phi'_{A\m} \Box  h^b_{\b\c} \6^{\c}h^{c\b\m}  
          + \frac{1}{2}\,\phi_{A\m\n\r} \6^{\n}\6^{\r} h^b_{\b\c} \6^{\c}h^{c\b\m} 
\nonumber \\       
	   &&-\;\frac{1}{4}\,\phi_A^{'\t} \6_{\m}\6_{\t} h^b_{\b\c} \6^{\c}h^{c\b\m}  
	        + \frac{1}{2}\,\phi_A^{'\a} \Box  h^b \6^{\c}h_{\a\c}^c  
          - \frac{1}{2}\,\phi_A^{\a\m\n} \6_{\m}\6_{\n}h^b \6^{\c}h_{\a\c}^c
\nonumber \\       
	   &&-\; \frac{1}{2}\,\phi_A^{\a\m\n} \Box  h^b_{\m\n} \6^{\c}h_{\a\c}^c  
	        - \frac{1}{2}\,\phi_A^{'\a} \6^{\m}\6^{\n} h^b_{\m\n} \6^{\c}h^c_{\a\c}  
          + \phi_A^{\a\m\n} \6_{\n}\6_{\r} h_{\m}^{b\,\rho} \6^{\c}h^c_{\a\c}
\nonumber \\       
	   &&+\; \frac{1}{4}\,\phi'_{A\b} \6^{\a}\6^{\b} h^b \6^{\c}h_{\a\c}^c  
	        + \frac{1}{2}\,\phi_{A\b\m\n} \6^{\a}\6^{\b} h^{b\m\n} \6^{\c}h^c_{\a\c}  
          - \phi_A^{'\t} \6^{\a}\6^{\m} h^b_{\m\t} \6^{\c}h_{\a\c}^c
\nonumber \\       
	   &&+\; \frac{1}{2}\,\phi'_{A\m} \Box  h^{b\a\m}\6^{\c}h_{\a\c}^c
	        - \frac{1}{2}\,\phi_{A\m\n\r} \6^{\n}\6^{\r} h^{b\a\m}\6^{\c}h^c_{\a\c}
          + \frac{1}{4}\,\phi_A^{'\t} \6_{\m}\6_{\t} h^{b\a\m}\6^{\c}h^c_{\a\c}  \quad. 
\nonumber
\end{eqnarray}
This is the only first order nonabelian cubic deformation involving two spin-2 fields and one spin-3 field. In fact, we show that it is obstructed at second order in deformation in Chapter \ref{ch:socomp}.

\subsubsection{$2-3-3$ vertex with four derivatives}

We must now consider the most promising candidate. It is promising because it is the only consistent cubic deformation involving two spin-3 fields and one spin-2 field and it is not yet known if it is obstructed or not at higher orders in deformation. It is also interesting because it is related it to the Fradkin--Vasiliev deformation in $(A)dS$. First, in \cite{Boulanger:2006gr}, we obtained the vertex in a inconvenient form: a long list of terms involving an undifferentiated spin-2 field. Then, in \cite{Boulanger:2008tg}, we could rewrite it as a short list of terms proportional to the undifferentiated Riemann tensor, modulo $d$ and $\d$ (let us notice that it is always possible to add a $\g$-exact term to $a_1$ along with a $\d$-exact term in $a_0$). We exhibit the shorter version below.

We used, once again, the same method to find the vertex. We listed all possible terms involving an undifferentiated spin-2 field, two spin-3 fields and four derivatives acting on them. They can be divided into three categories: 41 terms with all derivatives acting on the same spin-3 field, 93 terms with three derivatives acting one one field and 77 terms with two derivatives acting on both fields. These 211 terms are contracted with arbitrary coefficients $\stackrel{(j)}{k}_{aBC}$, $j=1,...,211$. 
As for the $\bar{a}_1$: it contains different kinds of terms. There are three derivatives so it could be possible to use the spin-3 curvature, but in fact, only its trace can appear, which is equal to antisymmetrized derivatives of the Fronsdal tensor. The terms involving the spin-3 ghost and the Ricci tensor can be eliminated in favor of terms involving the Fronsdal tensor by adding an appropriate $\d$-exact expression. Furthermore, most of the terms involving the Fronsdal tensor and the spin-2 ghosts are either antisymmetric modulo $d$ in the spin-3 indices or related to the following term : $\phi^{*}{}'{}^{B\a}\6_\a^{\phantom{\a}}G^C_{\b}C^{a\b}$. Since we already know that the structure coefficient of $a_{1,3}$ is symmetric in the spin-3 indices, it is obvious that antisymmetric $\bar{a}_1$ terms cannot be related to $a_{1,3}$. They could rather correspond to an independent deformation not deforming the gauge algebra and we have not studied them here. On the other hand, there are eight independent terms proportional to the spin-3 ghosts: two proportional to the Riemann tensor et six proportional to the Fronsdal tensor or its trace. Once again, most of them appeared not to be used.

This time, the computation is really heavy: more than a thousand equations in which we had to seek interesting relations among the coefficients. After that lengthy calculus, we obtained the vertex and the associated $\bar{a}_1$. The latter finally does not involve the spin-2 ghost, the only terms that remain are those proportional to the Riemann tensor and a third one proportional to the trace of the Fronsdal tensor. In $a_0$, we obtained a sum of 136 non vanishing terms. All the coefficients are proportional to $\stackrel{(8)}{f}_{aBC}$, that we will now denote simply $f_{aBC}$. We will not exhibit that list here, especially because we have found a better expression in the light of the work in $(A)dS$ described in Chapter \ref{ch:FV} and applied in the next section. In fact, the choice of undifferentiated spin-2 fields is poor because everything can be written in terms of the Riemann tensor (modulo some redefinitions of the fields which appear as $\d$-exact terms in $a_0$). The solution in $(A)dS$ that we provide in the next section contains a sum of terms linear in the undifferentiated Riemann tensor and quadratic in the spin-3 fields, bearing two covariant derivatives. That sum, in which the covariant derivatives are replaced by partial derivatives, is equivalent to the Minkowski vertex. To show that, we introduced an arbitrary sum of $\d$-exact terms in another file and solved the system of equations that arises. The coefficients have definite values in terms of $f_{aBC}$. We provide this expression of the vertex, but without writing explicitly the $\d$-exact part, which is long and not instructive. The complete solution $\stackrel{(1)}{w_3}$ is:

\begin{eqnarray} 
a_{2,3} = f_{aBC}\,C^{*a\m}\,\6^{\a}C^{B\b\n}\,\,
\6_{\m}\6_{[\a}C_{\b]\n}^C\,d^n x\quad,
\label{a82}
\end{eqnarray}
\begin{eqnarray} a_{1,3} &=& 
f_{aBC}\left[\frac{3}{8}h^{*a\m\n}\6^\r F^{'B}_\r C^C_{\m\n}+\frac{3}{2}\phi^{*B\a\b\g}\6_\a 
R^a_{\b\m|\g\n}C^{C\m\n}+\frac{2}{n}\phi^{*'B\a}R_{\a\m|\n\r}^a \6^\n C^{C\r\m}\right]d^n x
\nonumber\\ &&- 
\;f_{aBC}\,h^{*a\m\r}\,\6^\a 
C^{B\b\n}\left[2\6_{\m[\a}\phi^C_{\b]\n\r}-\6_{\n[\a}\phi^C_{\b]\m\r}\right]d^n x\quad,
\label{a81}
\end{eqnarray}

\begin{eqnarray}&a_{0,3}=\frac{1}{2}f^a_{\phantom{a}BC}R_a^{\a\b|\c\d}&\left[\6^\r\phi^{B}_{\r\a\c}\6^\s\phi^C_{\s\b\d}-2\6^\r\phi^{B\s}_{\phantom{B\s}\a\c}\6_{(\r}\phi^C_{\s)\b\d}+2\phi'{}^{B\r}\6^2_{\a\c}\phi^C_{\r\b\d}+\phi^{B\r\a\c}\6^{2\d}_\r\phi'{}^{C\b}\right.\nonumber\\&&-3\phi'{}^{B\a}\6_\r^{2\d}\phi^{C\r\b\c}+2\phi^{B\phantom{\s}\a}_{\r\s}\6^{2\r\d}\phi^{C\s\b\c}-2\phi^{B\s\a\c}\6^{2\r\d}\phi^{C\phantom{\s}\b}_{\r\s}+\phi'{}^{B\a}\6^{2\b\d}\phi'{}^{C\c}\nonumber\\&&\left.-\phi^{B\a\r\s}\6^{2\b\d}\phi^{C\c}_{\phantom{C\c}\r\s}-\phi^{B\a\c\r}\6^2_{\r\s}\phi^{C\b\d\s}\right]+\d c_1\quad.\label{minvert}
\end{eqnarray}
This solution is the only local, Poincar\'e invariant and nonabelian first order deformation that is linear in the spin-2 field and quadratic in the spin-3 field.

\section{Quasi-minimal deformation in $(A)dS$ for a $2-3-3$ configuration}

In the light of chapter \ref{ch:FV}, we know that a possible cubic deformation involving one spin-2 field and two spin-3 fields is the sum of the minimal deformation and some correcting terms linear in the Weyl tensor $w_{\m\n|\r\s}$ (which is the traceless part of the tensor $s_{\m\n|\r\s}$). We also know that a consistent Minkowski limit of this vertex can be taken in such a way as to keep the Fronsdal Lagrangian and the terms involving the highest number of derivatives while the other terms vanish. 
The computation that we have achieved, with the help of the Mathematica package Ricci \cite{Lee}, consists of the following steps: First, let us consider the sum of Fronsdal Lagrangians in $(A)dS$ for a spin-2 and a spin-3 fields. The spin-2 part deforms naturally into an Einstein--Hilbert Lagrangian, thus the part that interests us is the minimal deformation terms of the spin-3. We use the alternative definition of the Fronsdal Lagrangian that is quadratic in the first derivatives of the fields:
\begin{eqnarray}-\frac{\stackrel{(2)}{\cl}_{s}}{\sqrt{-g}} &=&\frac{1}{2}\,
\N_\m \phi_{\a_1...\a_s}\N^\m \phi^{\a_1...\a_s} -
\frac{1}{2}\,s\N^\m \phi_{\m\a_1...\a_{s-1}}\N_\n
\phi^{\n\a_1...\a_{s-1}}\nonumber\\&& + \frac{1}{2}\,s(s-1)\N_\a
\phi'_{\b_1...\b_{s-2}}\N_\m \phi^{\m\a\b_1...\b_{s-2}}
 - \frac{1}{4}\,s(s-1)\N_\m \phi'_{\a_1...\a_{s-2}}\N^\m
\phi'^{\a_1...\a_{s-2}}\nonumber\\&& -\frac{1}{8}s(s-1)(s-2)\N^\m
\phi'_{\m\a_1...\a_{s-3}}\N_\n \phi'^{\n\a_1...\a_{s-3}}
\nonumber\\&&
+\frac12 \,\l^2\left[s^2+(n-6)s-2n+6\right]\phi_{\a_1...\a_s}
\phi^{\a_1...\a_s}\nonumber\\&&-\frac14 \l^2 s(s-1)
\left[s^2+(n-4)s-n+1\right]\phi'_{\a_1...\a_{s-2}}\phi'^{\a_1...\a_{s-2}}\quad.
\label{Fs}\end{eqnarray}

In this expression, indices are contracted with the $(A)dS$ metric $g_{\m\n}$. The minimal deformation consists in replacing this metric by $\cg_{\m\n}=g_{\m\n}+\a h_{\m\n}$ where $h_{\m\n}$ is the spin-2 field. Three types of cubic terms appear: First, some terms related to the determinant: $\sqrt{-\cg}=\sqrt{-g}(1+\frac{1}{2}\a h'+...)$. The second type arises because contractions of indices must be made with $\cg_{\m\n}$, and the third type arises because of the modification of the connection and thus of the covariant derivatives (let us denote $D_\mu$ the covariant derivative involving the Levi-Civita connection of $\cg_{\m\n}$). These terms form the minimal part of the cubic vertex that we can denote $\stackrel{(3)}{\cl}_{min}$. The minimal gauge transformations are $\d_{\ve,min}h_{\m\n}=2D_{(\m}\ve_{\nu)}$ and $\d_{\xi,min} \phi_{\m\n\r}=3D_{(\m}\xi_{\n\r)}$. We can check their action on the Lagrangian at first order: $\stackrel{(0)}{\d_\xi}\stackrel{(3)}{\cl}_{min}+\stackrel{(1)}{\d_{\xi,min}}\stackrel{(2)}{\cl}$. As we know, this does not vanish, we thus have to keep the expression and to compare it with the zeroth order gauge transformation of the Weyl terms. The only possible Weyl term with no derivatives on the spin-3 fields is $w^{\a\b|\c\d}\phi_{\a\c\r}\phi_{\b\d}^{\phantom{\b\d}\r}$, but it is not sufficient to remove the obstruction, a general expression involving two covariant derivatives has to be introduced, with arbitrary coefficients. This expression is invariant under the spin-2 gauge transformations by construction, we thus have to check the spin-3 gauge transformations. This computation involves covariant derivatives of tensors bearing several indices and is thus heavy, which is why we used a software. The result does not vanish but is proportional to the equations of motion and can thus be understood as a complement to the gauge transformations, that thus do not remain minimal either. The cubic vertex, that has to be understood as lying on $(A)dS$ (thus the contractions are made here with $g_{\a\b}$) takes the following form:
\begin{eqnarray}\frac{\stackrel{(3)}{\cl}_{FV}}{
\sqrt{- g}}&\approx&\frac{11}{2}~
w_{\a\b|\c\d}~\phi^{\a\c}_{\phantom{\a\c}\m}\phi^{\b\d\m}
+ \frac{n-2}{2\Lambda} ~w_{\a\b|\c\d}~\Big[2\ \phi'_\m
 \N^{(\b} \N^{\d)}\phi^{\a\c\m}
+ \phi^{\a\c}_{\phantom{\a\c}\m} \N^{(\d}
\N^{\m)}\phi'^{\b}\label{FVvertex}\\[5pt]
&&- 3\
\phi'^{\a} \N^{(\d}
\N^{\m)}\phi^{\b\c\phantom{\m}}_{\phantom{\b\c}\m}
+ 2\ \phi^{\a}_{\phantom{\a}\m\n} \N^{(\d}
\N^{\n)}\phi^{\b\g\m}
+  \N_\m\phi^{\a\c\m}_{\phantom{\a\c\m}}
\N_\n\phi^{\b\d\n}_{\phantom{\b\d\n}}
- \phi^{\a\c\m} \N_{(\m}
\N_{\n)}\phi^{\b\d\n}_{\phantom{\b\d\n}}\nonumber\\[5pt]&&
- 2\  \N^{(\m}\phi^{\n)\a\c}
\N^{\phantom{\m}}_\m\phi^{\b\d}_{\phantom{\b\d}\n}
- 2\
\phi^{\a\c}_{\phantom{\a\c}\m} \N^{(\d} \N^{\n)}
\phi^{\b\m\phantom{\n}}_{\phantom{\b\m}\n}
+ \phi'^{\a} \N^{(\b} \N^{\d)}\phi'^{\g}
- \phi^{\a}_{\phantom{\a}\m\n} \N^{(\b}
\N^{\d)}\phi^{\c\m\n}\Big]\quad.\nonumber
\end{eqnarray}
The computation does not yield the deformation of the gauge transformations in a convenient way, its determination would require some complementary computations.

\section{Uniqueness of the $(A)dS$ deformation}

The relation between the Minkowski vertex presented in Eq. (\ref{minvert}) and the $(A)dS$ vertex presented in Eq. (\ref{FVvertex}) is obvious: we have obtained explicitly that the terms with two covariant derivatives of the latter are similar to those of the former, with coefficients that are proportional. As we showed in section \ref{minlim}, a scaling limit can be taken, that precisely selects the terms with the highest number of derivatives. Let us recall the argument that we provided about the correspondence between the cubic vertices, in the particular case of the $2-3-3$ construction, for which we make the proof a bit more precise.

Let us denote the at most cubic parts of the Minkowski action $S^{\L=0}[\widetilde{h},\widetilde{\phi}]$ and of the $(A)dS$ action: $S^{\L}[h,\phi]$. The uniqueness of $S^{\L=0}[\widetilde{h},\widetilde{\phi}]$ is instrumental in showing the
uniqueness of its $(A)dS$ completion $S^{\L}[h,\phi]\,$,
due to the linearity of the perturbative deformation scheme and the smoothness
of the flat limit at the level of cubic actions. The proof goes as follows.
First suppose that there exists another action
$S^{\prime\L}[h,\phi]=S^{\Lambda}_{free} + g\,S^{\prime\Lambda}_{cubic}$
that admits a nonabelian gauge algebra and whose top vertex, denoted
$V_{\Lambda}^{\,\prime {\rm top}}(2,3,3)$, involves $n_{\rm top}$ derivatives
with $n_{\rm top}\neq 4$. Then, this action would scale to a nonabelian
flat-space action whose cubic vertex would involve $n_{\rm top}$ derivatives.
This is impossible, however, because the \emph{only} nonabelian cubic vertex
in flat space is the one presented in Eq.(\ref{minvert}) and that contains four derivatives.
Secondly, suppose there exists a nonabelian action
$S^{\prime\prime\L}[h,\phi]=S^{\Lambda}_{free}
+g\,S^{\prime\prime\Lambda}_{cubic}$
whose top vertex contains $4$ derivatives but is otherwise different from the terms in $\stackrel{(3)}{\cl}_{FV}$ with four derivatives.
Then its flat limit would yield a theory with a cubic vertex, involving
4 derivatives, but different from $a_{0,3}$ , which is
impossible due to the uniqueness of the latter deformation.
Thirdly, and finally, suppose there exists a cubic Lagrangian deformation whose top vertex is the same as that of $\stackrel{(3)}{\cl}_{FV}$
 but differing from it in the terms with lesser numbers of derivatives.
By the linearity of the BRST-BV deformation scheme, the difference between
this coupling and $\stackrel{(3)}{\cl}_{FV}$ would lead to a nonabelian theory in $(A)dS$
with top vertex involving less than 4 derivatives.
Its flat-space limit would therefore yield a nonabelian action
whose top vertex would possess less than 4 derivatives, which is impossible
due to the uniqueness of $S^{\L=0}[\widetilde{h},\widetilde{\phi}]\,$ $\Box$.

This proof in fact holds in any case where the Minkowski deformation is unique and an $(A)dS$ deformation in known. Though the computations have not been achieved explicitly for every $2-s-s$ case, the results obtained in Chapter \ref{ch:intmisc} ensure that the $(A)dS$ uniqueness extends to the $2-s-s$ Fradkin-Vasiliev construction as well.

%%%%%%%%%%%%%%%%%%%%%%%%%%%%%%%%%%%%%%%%%%%%%%%%
\chapter{Spin-3 self interactions}\label{ch:exo3}
%%%%%%%%%%%%%%%%%%%%%%%%%%%%%%%%%%%%%%%%%%%%%%%%

In this chapter, we review the problem of the self-interacting spin three field (and more generally, interactions between different spin-3 fields), at first order in the deformation parameter. The parity-invariant case has been addressed in the antifield approach in \cite{Bekaert:2006jf}. One of the nonabelian solutions given in this paper had been found earlier by Berends, Burgers and van Dam in \cite{Berends:1984wp}, this solution is thus called the BBvD deformation. We briefly recall their results, for completeness and because some important considerations about the BBvD deformation at second order are discussed in Chapter \ref{ch:socomp}. Then, we achieve the computation of the nonabelian first order spin-3 deformations in the parity-breaking case, which leads to two new consistent vertices, in dimensions 3 and 5.

In both cases, the initial Lagrangian is the sum of spin-3 Fronsdal Lagrangians involving the fields $\phi^A_{\m\n\r}$ (the notation is the same as in Chapter \ref{ch:int23}). Just as in the former chapter, Theorem \ref{thmcub} ensures that the nonabelian deformations must be cubic, in both parity-invariant and parity-breaking cases. In terms of the antifield formalism, the deformation of the BRST-BV generator $W$ is, once again, the integral of a sum of three terms $a_0$, $a_1$ and $a_2$, indexed by the antifield number, and corresponding to the deformations of the Lagrangian, of the gauge transformations and of the gauge algebra. The considerations made in section \ref{cbvrt} ensure that the non equivalent solutions for $a_2$ are linear in the undifferentiated antifields $C^{*\m\n}_A$ and quadratic in the ghosts $C^A_{\m\n}$ or the non-$\g$-exact components of their derivatives, which are minimally described by the tensors $T_{\a\b|\m}^A$ and $U^A_{\a\m|\b\n}$ that have be defined in Chapter \ref{ch:int23}. In the parity-invariant case, the Lorentz-invariant contractions are made with the sole metric $\eta_{\m\n}$. In the parity-breaking case, the expressions are linear in the Levi-Civita symbol $\e_{\m_1...\mu_n}$ and thus depend on the spacetime dimension.

\section{Parity-invariant deformations}

Let us recall the results obtained in \cite{Bekaert:2006jf}. As usual, the first order deformations are $n$-forms $\stackrel{(1)}{w}=a_2+a_1+a_0$ such that $s\stackrel{(1)}{w}+db=0$. Two cubic solutions for $a_2$ have been obtained. One with three derivatives in the vertex (which corresponds to two in $a_2$), and one with five derivatives. The first one is equivalent to the solution given in \cite{Berends:1984wp}, that is why it is denoted $\stackrel{(1)}{w}_{BBvD}$. The solutions for $a_2$ are as follows: \begin{eqnarray}a_{2,BBvD}=f^A_{\phantom{A}BC}C^{*\m\n}_A\left[T^B_{\mu\a|\b}T^{C\,\a|\b}_\n-2T^B_{\mu\a|\b}T^{C\,\b|\a}_\n+\frac{3}{2}C^{B\a\b}U^C_{\m\a|\n\b}\right]d^n x+\g c_2\ ,\end{eqnarray} \begin{eqnarray}a_{2,5}=g^A_{\phantom{A}BC}C^{*\m\n}_A U^B_{\m\a|\b\c}U^{C\,\a|\b\c}_\n d^n x\ +\g e_2\ , \end{eqnarray} where $f_{ABC}$ and $g_{ABC}$ are arbitrary coefficients totally antisymmetric in their indices. These are related to $a_1$ solutions, that in both cases are the sum of inhomogeneous expressions $\widetilde{a}_1$ such that $\d a_2+\g\widetilde{a}_1+db_1=0$ and  homogeneous, $\g$-closed expressions $\bar{a}_1$, involving the same number of derivatives, and required in order for $a_0$ to exist. If the $\g$-exact term $\g c_2$ is chosen in such a way as to restore the traces of the derivatives of the ghosts, the following expression is found: \begin{eqnarray}&\tilde{a}_{1,BBvD}=\displaystyle-\frac{3}{2}f^A_{\phantom{A}BC}&\left[(\phi_A^{*\m\n\r}-\frac{1}{n}\eta^{\m\n}\phi^*_A{}'{}^\r)\left(2\6_{[\m}\phi^B_{\a]\b\r}(\tilde{T}^{C\a|\b}_\n-2\tilde{T}^{C\b|\a}_\n)+\phi^{B\a\b}_{\r}\6_{[\n}C^C_{\b][\a,\,\m]}\right.\right.\nonumber\\ &&\left.\left.-3C^{B\a\b}\6_{[\n}\phi^C_{\b]\r[\a,\,\m]}\right)+\frac{1}{n}\phi^*_A{}'{}^\r\6_{[\r}C^B_{\a]\b}(\6_\r\phi^{C\s\a\b}-2\6^{(\a}\phi'{}^{|C|\b)})\right]d^n x\ ,
\end{eqnarray} where $\tilde{T}^A_{\m\n|\a}=\6_{[\m}C^A_{\n]\a}$ and where indices after a comma denote partial derivatives. The required homogeneous expression is \begin{eqnarray}\bar{a}_{1,BBvD}=-\frac{9}{8}f_{ABC}\phi^*{}'{}^{A\m}G^B_{\m\n\r}C^{C\n\r}d^n x\ .\end{eqnarray} In the same way, by choosing an appropriate $\g e_2$, the following result is obtained for the 5-derivative case: \begin{eqnarray}a_{1,5}=-6g^A_{\phantom{A}BC}\left[\phi_A^{*\m\n\r}-\frac{1}{n}\eta^{\m\n}\phi_A^*{}'{}^\r\right]\6_{[\b|[\m}\phi^B_{\a]\r|\l]}\6^{2\b}_{\phantom{2\b}[\n}C_{\t]}^{C\l}\eta^{\a\t} d^n x\ .\end{eqnarray} Finally, the vertices can be computed from Eq.(\ref{eqagh0}). We will only recall the BBvD vertex here. The vertex with five derivatives consists of a sum of hundreds of terms, its expression can be found in the appendix of \cite{Bekaert:2006jf}. The BBvD vertex can be written as $a_{0,BBvD}=-\frac{3}{8}f_{ABC}L^{ABC}$ where: \begin{eqnarray}L^{ABC}&=&-\frac{3}{2}\phi'{}^{A\a}\phi'{}^{B\b,\,\c}\phi'{}^C_{\b,\,\a\c}+3\phi'{}^{A\a,\,\b}\phi^{B\c}\phi'{}^C_{\g,\,\a\b}+6\phi^{A\a\b\c,\,\d}\phi'{}^B_{\a}\phi'{}^C_{\b,\,\c\d}+\frac{1}{2}\phi'{}^{A\a}\phi^{B\b\c\d,\,\eta}\phi^C_{\b\c\d,\,\a\eta}\nonumber\\&&
+\phi^{A\a}_{\phantom{A\a},\,\a\b}\phi^B_{\c\d\eta}\phi^{C\c\d\eta,\,\b}+\phi'{}^{A\a,\,\b}\phi^{B\c\d\eta}\phi^C_{\c\d\eta,\,\a\b}-3\phi^A_{\a\b\c}\phi^{B\a\b}_{\phantom{B\a\b}\d,\,\eta}\phi'{}^{C\d,\,\g\eta}-3\phi^A_{\a\b\c}\phi^{B\a\b\d,\,\c\eta}\phi'{}^C_{\d,\,\eta}\nonumber\\&&
+3\phi^A_{\a\b\c,\,\d}\phi^{B\a\b\eta}\phi'{}^{C,\,\c\d}_\eta+3\phi^{A\phantom{\b\c},\,\c\d}_{\a\b\c}\phi^{B\a\b\eta}\phi'{}^C_{\eta,\,\d}-\frac{9}{4}\phi'{}^A_{\a,\,\b\c}\phi'{}^{B\b}\phi'{}^{C\c,\,\a}-\frac{1}{4}\phi'{}^A_{\a,\,\b}\phi'{}^{B\b,\,\c}\phi'{}^{C\,,\,\a}_{\,\c}\nonumber\\&&
-3\phi^A_{\a\b\c}\phi'{}^{B\d,\,\a}\phi'{}_{\,\d}^{C\,,\,\b\c} -\frac{3}{2}\phi'{}^{A\,,\,\a}_{\,\a}\phi^{B\b,\,\c}\phi^{C\phantom{\c\d},\,\d}_{\b\c\d}+3\phi'{}^A_{\a}\phi'{}^B_{\b,\,\c}\phi_\d^{C\b\c,\,\a\d}+\frac{3}{2}\phi'{}^{A,\,\a\b}_\a\phi'{}^{B\c,\,\d}\phi^C_{\b\c\d}\nonumber\\&&
+3\phi'{}^A_{\a,\,\b}\phi'^B_{\c,\,\d}\phi^{C\b\c\d,\,\a}-\frac{3}{2}\phi'{}^A_\a\phi^{B\phantom{\c\d},\,\b}_{\b\c\d}\phi_\eta^{C\c\d,\,\a\eta}-6\phi^{A\phantom{\b\c},\,\a\d}_{\a\b\c}\phi'{}^{B\b,\,\eta}\phi^{C\ \c}_{\d\eta}+6\phi^{A\phantom{\b\c},\,\a\d}_{\a\b\c}\phi'{}^{B\b}\phi^{C\ \c,\,\eta}_{\d\eta}\nonumber\\&&
-2\phi^A_{\a\b\c,\,\d}\phi_\l^{B\a\d,\,\eta}\phi_\eta^{C\l\b,\,\c} +\phi^A_{\a\b\c}\phi^{B\phantom{\eta\l},\,\a}_{\d\eta\l}\phi^{C\d\eta\l,\,\b\c}-3\phi^{A\phantom{\b\c},\,\a}_{\a\b\c}\phi_\d^{B\b\c,\eta}\phi_{\eta\l}^{C\ \d,\,\c}\nonumber\\&&
+3\phi^{A\phantom{\b\c},\,\a\d}_{\a\b\c}\phi^{B\b\c\eta,\,\l}\phi^C_{\eta\d\l}+6\phi^A_{\a\b\c,\,\d}\phi^{B\a\b\eta,\,\l}\phi^{C\ \d,\,\c}_{\eta\l}\end{eqnarray}

\section{Parity-breaking deformations}

The parity-breaking case can now be studied. We prove in the sequel that there are two consistent nonabelian solutions, in dimension 3 with 2 derivatives, and in dimension 5 with 4 derivatives. An interesting result is that there are neither nonabelian parity-breaking deformations in dimension 4 nor in dimension greater than 5. Some Schouten identities have to be taken into account and play an important r\^ole in our computations. Let us recall that these consist in antisymmetrizing $n+1$ indices in dimension $n$. In the present case of parity-breaking expressions involving a Levi-Civita symbol, the indices that are considered are invariably the $n$ indices contracted with the Levi-Civita symbol and another one.

\subsection{Deformations of the gauge algebra}
%------------------------------------------------
%
Let us determine the possible $a_2$, dimension by dimension. The deformation obeys Eqs(\ref{eqagh2})-(\ref{eqagh0}). This time, we will first check the expressions explicitly along the $D$-degree, hence we use Eqs(\ref{alphabeta})-(\ref{betacond}).

\subsubsection*{Dimension 3}

In dimension 3, any tensor with the symmetry of the Weyl tensor, such as $U^A_{\m\a|\n\b}$, identically vanishes. This implies that the second derivatives of the ghosts are all $\g$-exact. Thus, the highest $D$-degree of a quadratic $pgh\ 2$ $\omega^{J_i}$ is 2. Furthermore, there is no possible Lorentz-invariant contractions of the indices in even $D$-degree: the $antigh\ 2$ antifields and the ghosts bear two indices, the Levi-Civita symbol bears three, hence the total number of indices is the $D$-degree plus 9. Thus, we just have to determine the $D$-degree one candidates. We can check at this stage if Eq.(\ref{eqagh1}) is satisfied. In order to do that, we first compute $\beta_{I_1}$ such that $\d \a_{I_1}+ d \b_{I_1}=0$ with $a_2=\a_{I_1}\omega^{I_1}$. Then, we check if it satisfies Eq.(\ref{betacond}): $\beta_{I_1}A^{I_1}_{I_2}=0$ where $D\omega^{I_1}=A^{I_1}_{I_2}\omega^{I_2}$. There are four ways to contract the indices in $D$-degree one, but they are all proportional. Let us enumerate them: \begin{eqnarray}&&f^A_{\phantom{A}BC}\ve^{\m\n\r}C^{*\a\b}_A C_{\m\a}^B T^C_{\n\r|\b}d^3 x\ ,\quad f_{ABC}\ve^{\m\n\r}C^{*A}_{\m\a}C^{B\a\b} T^C_{\n\r|\b}d^3 x\ ,\quad f_{ABC}\ve^{\m\n\r}C^{*A}_{\m\a}C_{\n\b}^B T^{C\a\b}_{\phantom{C\a\b}\,|\r}d^3 x\ ,\quad \nonumber\\&&f_{ABC}\ve^{\m\n\r}C^{*A}_{\m\a}C_{\n\b}^B T^{C\a|\b}_{\r}d^3 x\nonumber\ .\end{eqnarray} To see that they are proportional, some Schouten identities have to be used: First, in the last expression, let us antisymmetrize the index $\m$ of the antifield, the index $\n$ of the ghost, and the indices $\r$ and $\a$ of the tensor $T^C$: \begin{eqnarray}0&\equiv&f_{ABC}\e^{\m\n\r}\left[C^{*A}_{\m\a}C_{\n\b}^B T^{C\a|\b}_{\r}-C^{*A\a}_{\phantom{*A\a}\a}C_{\m\b}^B T^{C\phantom{\r}|\b}_{\n\r}+C^{*A}_{\r\a}C_{\phantom{B\a}\b}^{B\a} T^{C\phantom{\r}|\b}_{\m\n}-C^{*A}_{\n\a}C_{\r\b}^B T^{C\a\phantom{\r}|\b}_{\phantom{C\a}\m}\right]d^3 x\nonumber\\&\equiv& 2f_{ABC}\e^{\m\n\r}C^{*A}_{\m\a}C_{\n\b}^B T^{C\a|\b}_{\r}d^3 x+f_{ABC}\e^{\m\n\r}C^{*A}_{\m\a}C^{B\a\b} T^C_{\n\r|\b}d^3 x\ .\nonumber\end{eqnarray} Then, let us make the same operation, but this time considering $\b$ instead of $\a$ on $T^C$: \begin{eqnarray}0&\equiv&f_{ABC}\e^{\m\n\r}\left[C^{*A}_{\m\a}C_{\n\b}^B T^{C\a|\b}_{\r}-C^{*A\b}_{\phantom{*A\b}\a}C_{\m\b}^B T^{C\a}_{\n\phantom{C}\,|\r}+C^{*A}_{\r\a}C_{\phantom{B\b}\b}^{B\b} T^{C\a}_{\m\phantom{C}\,|\n}-C^{*A}_{\n\a}C_{\r\b}^B T^{C\b\a}_{\phantom{C\b\a}|\m}\right]d^3 x\nonumber\\&\equiv& f_{ABC}\e^{\m\n\r}C^{*A}_{\m\a}C_{\n\b}^B T^{C\a|\b}_{\r}d^3 x-\frac{1}{2}f_{ABC}\e^{\m\n\r}C^{*A\a\b}C_{\m\a}^B T^C_{\n\r|\b}d^3 x+f_{ABC}\e^{\m\n\r}C^{*A}_{\m\a}C_{\n\b}^B T^{C\a\b}_{\phantom{C\a\b}|\r}d^3 x\ .\nonumber\end{eqnarray} Finally, in the second expression, let us antisymmetrize the indices $\m$ of the antifield, $\b$ of the ghost and $\n$,$\r$ of the tensor $T^C$: \begin{eqnarray}0&\equiv&f_{ABC}\e^{\m\n\r}\left[C^{*A}_{\m\a}C^{B\a\b} T^C_{\n\r|\b}-C^{*A\b}_{\phantom{*A\b}\a}C^{B\a}_{\phantom{B\a}\n} T^{C}_{\r\m|\b}+C^{*A}_{\n\a}C^{B\a}_{\phantom{B\a}\r} T^{C\b}_{\m\phantom{C}\,|\b}-C^{*A}_{\r\a}C^{B\a}_{\phantom{B\a}\m} T^{C\b}_{\phantom{C\b}\n|\b}\right]d^3 x\nonumber\\&\equiv&f_{ABC}\e^{\m\n\r}C^{*A}_{\m\a}C^{B\a\b} T^C_{\n\r|\b}d^3 x-f_{ABC}\e^{\m\n\r}C^{*A\a\b}C^B_{\m\a}T^C_{\n\r|\b}d^3 x\nonumber\ .\end{eqnarray} Thus, since we can choose any of the four expressions for $a_2$, we have considered the first one: \begin{eqnarray} a_{2,1}=f^A_{\phantom{A}BC}\ve^{\m\n\r}C^{*\a\b}_A C_{\m\a}^B T^C_{\n\r|\b}d^3 x\ .\label{exo3a21}\end{eqnarray} The associated $b_1$ is: \begin{eqnarray}b_{1,1}=\beta_{I_1}\omega^{I_1}=- 3\,f^A_{\phantom{A}BC}\ve^{\m\n\r} \left(\phi_A^{*\a\b\l}-\frac{1}{3}\eta^{\a\b}
  \phi^*_A{}'{}^{\l}\right)C^{B}_{\m\a}T^{C}_{\n\r|\b}\frac{1}{2}\e_{\l\s\t}dx^\s  dx^\t\, .\end{eqnarray} We know that $a_1$ exists if Eq.(\ref{betacond}) is satisfied. In this case, we get:  
\begin{eqnarray} \b_{I_{1}}A^{I_{1}}_{I_{2}}\o^{I_2}&=&-3\,f^A_{\phantom{A}BC}\e^{\m\n\r}\left[\phi_A^{*\a\b\l}-\frac{1}{3}\eta^{\a\b}\phi^*_A{}'{}^\l\right] \left(\frac{4}{3} T^{B}_{\l (\a|\m)} T^{C}_{ \n\r|\b}\right) \frac{1}{2} \e_{\l\s\t} dx^\h dx^\s  dx^\t\nonumber \\ &=&-2\,f^A_{\phantom{A}(BC)}\e^{\m\n\r} \left[\phi_A^{*\a\b\l}-\frac{2}{3}\eta^{\a\b}\phi^*_A{}'{}^\l\right] T^{B}_{\l \m\vert\a} T^{C}_{ \n\r\vert \b}\ d^3 x\,.\label{betAd3d1}\end{eqnarray}
This holds thanks to some other Schouten identities: First, there is no term proportional to $C U$ in $D(C T)$ because $U$ vanishes. Secondly, the following relation holds: $\ve^{\m\n\r}T^B_{\l\a|\m}T^{C\phantom{\r}|\a}_{\n\r}=\ve^{\m\n\r}T^B_{\l\m|\a}T^{C\phantom{\r}|\a}_{\n\r}$. It is obtained by antisymmetrizing over the lower $\m$, $\n$, $\r$ and $\a$ indices. Then, the symmetry on the coefficient is due to the fermionic behaviour of the ghost tensors and to the relation: \begin{eqnarray}\ve^{\m\n\r}\left[\phi_A^{*\a\b\l}-\frac{2}{3}\eta^{\a\b}
  \phi^*_A{}'{}^\l\right](T^B_{\l\m|\a}T^C_{\n\r|\b}+T^B_{\n\r|\a}T^C_{\l\m|\b})d^3 x=0\ ,\end{eqnarray} which is obtained by antisymmetrizing over the lower $\m$, $\n$, $\r$ and $\l$ indices. Finally, we find that $a_{1,1}$ exists if the coefficients are antisymmetric over the last two indices : $f^A_{\phantom{A}BC}=f^A_{\phantom{A}[BC]}$.

\subsubsection*{Dimension 4}

There is no nontrivial deformation of the gauge algebra in dimension 4. The number of indices in a parity-breaking $a_2$ is equal to the $D$-degree plus 10, so that Lorentz-invariant contractions can only be obtained in even $D$-degree. Furthermore, in $D$-degree 0, the product $C^*CC$ involves three symmetric pairs, that cannot be contracted with the four antisymmetric indices of $\ve^{\m\n\r\s}$. Thus, we have to check what happens in $D$-degrees 2 and 4 (let us recall that $D$-degree 4 is the maximum $D$-degree of an $\omega^{J_i}$ for spin-3 in $pgh\ 2$).

\vspace{3mm}
\noindent{\bfseries{$D$-degree two:}}

Four terms can be obtained in $D$-degree 2: three of the form $\ve C^* T T$ and one of the form $\ve C^* C U$. There is no way to antisymmetrize $C^* C U$ over five indices, thus there does not exist a Schouten identity for the last one. Let us study the Schouten identities for the first three terms. The different contractions read:
\begin{eqnarray}
&T_1^{A[BC]}\!=\!\ve^{\m\n\r\s}\,C^{*A\a\b}\,T^{B}_{\m\n|\a}\,T_{\r\s|\b}^C\,,\,
T_2^{ABC}\!=\!\ve^{\m\n\r\s}\,C^{*A\a}_{\m}\,T^{B\ \b}_{\n\r|}\,T_{\s\a|\b}^C\,,\,
T_3^{ABC}\!=\!\ve^{\m\n\r\s}\,C^{*A\a}_{\m}\,T^{B\ \b}_{\n\r|}\,T_{\s\b\vert \a}^C\,.&\nonumber\end{eqnarray}
There are two Schouten identities relating these terms. To find them, let us contract a product $\ve C^* T T$ with antisymmetrized products of Kronecker deltas: $\d^{[\a\b\g\d\e]}_{[\m\n\r\s\t]}=\d^{[\a}_{[\m}\d^\b_\n\d^\g_\r\d^\d_\s\d^{\e]}_{\t]}\,$. The two different ways of doing that are:
\begin{eqnarray}
\d^{[\a\b\g\d\e]}_{[\m\n\r\s\t]}\ve^{\m\n\r\s} C^{*\t}_{A~\a} T^B_{\b\g|\l}T^{C~|\l}_{\d\e}=0\;,\;
\d^{[\a\b\g\d\e]}_{[\m\n\r\s\t]}\ve^{\m\n\r\s} C^{*\l}_{A~\a} T^{B~\,|\t}_{\b\g}T^C_{\d\e|\l}=0\,.
\nonumber \end{eqnarray}
The first identity implies that $T_2^{ABC}$ is symmetric in its last two indices: $T_2^{ABC}=T_2^{A(BC)}\,$,
while the second one relates $T_1^{A[BC]}$ and $T_3^{ABC}\,$: $T_3^{ABC}=T_1^{A[BC]}\,$. We can now write the general form of $a_{2,2}$: 
\begin{eqnarray}
a_{2,2}&=&\stackrel{(1)}{k^A}_{[BC]}\ve^{\m\n\r\s}\,C_A^{*\a\b}\,T^{B}_{\m\n|\a}\,T_{\r\s|\b}^C d^4 x 
+\stackrel{(2)}{k^A}_{(BC)}\ve^{\m\n\r\s}\,C^{*\a}_{A~\,\m}\,T^{B\,~|\b}_{\n\r}\,T_{\s\a|\b}^C d^4 x\nonumber\\&&
+\stackrel{(3)}{k^A}_{BC}\ve^{\m\n\r\s}\,C^{*}_{A\m\a}\,C^B_{\n\b}\,U^{C\ \ |\a\b}_{\r\s}d^4 x\,.\nonumber\end{eqnarray}
The associated $b_{1,2}$ reads:
\begin{eqnarray}
b_{1,2}&=&-3\ \ve^{\m\n\r\s}\Big[ ( \phi_A^{*\l\a\b}-\frac{1}{4} 
\h^{\a\b}\phi^*_A{}'{}^\l)\stackrel{(1)}{k^A}_{[BC]}T^B_{\m\n|\a}T^C_{\r\s|\b}\nonumber \\
&&+(\phi^{*\l\a}_{A\phantom{\l\a}\m}-\frac{1}{4}\d^\a_{\m}\phi^*_A{}'{}^\l)(\stackrel{(2)}{k^A}_{(BC)}
T^{B \,~|\b}_{\n\r}\,T^C_{\s\a\vert \b}+\stackrel{(3)}{k^A}_{BC}\ C^B_{\n\b}\,U^{C\phantom{\r\s|\a}\b}_{\phantom{C}\r\s|\a})\Big]
\frac{1}{3!}\ve_{\l \r\s\t}dx^\r dx^\s dx^\t\,.\nonumber\end{eqnarray} Let us then check the consistency condition:
\begin{eqnarray} 
\b_{I_2}A_{I_3}^{I_2}\o^{I_3}
&=&-\displaystyle\frac{3}{2}\stackrel{(1)}{k^A}_{[BC]}\ve^{\m\n\r\s}\phi^*_A{}'{}^\l T^{B\phantom{\n}|\a}_{\,\m\n}U^C_{\l\a|\r\s}d^4 x\nonumber\\&&
-3\stackrel{(2)}{k^A}_{(BC)}\ve^{\m\n\r\s}\phi_{a\phantom{\a\l}\m}^{*\a\l}\Big(T^{B\phantom{\n}|\b}_{\,\n\a}U^C_{\l\b|\r\s}-T^{B\phantom{\n}|\b}_{\,\n\r}U^C_{\l\b|\s\a}\Big)d^4 x\nonumber\\&&
+4\stackrel{(3)}{k^A}_{BC}\ve^{\m\n\r\s}\phi_{A\phantom{\a\l}\m}^{*\a\l}T^B_{\l(\b|\n)}U^{C\phantom{\r\s|\a}\b}_{\phantom{C}\r\s|\a}d^4 x\nonumber\\
&=&\Big[ -\frac{3}{2}(\stackrel{(1)}{k^A}_{[BC]}+\stackrel{(2)}{k^A}_{(BC)})
\ve^{\b\g\r\s}\phi^*_A{}'{}^\m\h^{\a\n}-(6\stackrel{(2)}{k^A}_{(BC)}+4\stackrel{(3)}{k^A}_{BC})\ve^{\m\n\l\b}
\phi^{*\c\r}_{A\phantom{\c\r}\l} \h^{\a\s}\Big]\nonumber\\
&&\times T^B_{\b\c|\a}U^C_{\m\n|\r\s}d^4x\label{betAd4d2}\end{eqnarray} 
The latter equality is obtained by using the following Schouten identities: Let us consider the various terms of the form $\ve \phi^* T U$ (we omit the internal indices in the next expressions, since they are not affected by Schouten identities): 
\begin{eqnarray}
&T_1=\ve^{\m\n\r\s}\phi_{\m}^{*\a\b}T_{\n\g|\b}U_{\r\s|\a}^{\phantom{\r\s|\a}\g} \;,\;
T_2=\ve^{\m\n\r\s}\phi_{\m}^{*\a\b}T_{\n\b|\g}U_{\r\s|\a}^{\phantom{\r\s|\a}\g}\;,\;
T_3=\ve^{\m\n\r\s}\phi^*{}'{}^{\a}T^{\phantom{\m\n}|\b}_{\m\n}U_{\r\s|\a\b}\;,& \nonumber \\
&T_4=\ve^{\m\n\r\s}\phi^*{}'{}_{\m}T^{\a\b}_{\phantom{\a\b}|\n}U_{\r\s|\a\b}\;,\;
T_5=\ve^{\m\n\r\s}\phi_{\m}^{*\a\b}T_{\n\r\vert \g}U_{\s\a|\b}^{\phantom{\s\a|\b}\g}\;.&\nonumber 
\end{eqnarray} There are three Schouten identities:
\begin{eqnarray}
&\d^{[\a\b\g\d\e]}_{[\m\n\r\s\t]}\ve^{\m\n\r\s} \phi^{*~\,\t}_{\a\l}  T_{\b\g\vert \h}U_{\d\e\vert}^{~~\l\h}=0\;,\;
\d^{[\a\b\g\d\e]}_{[\m\n\r\s\t]}\ve^{\m\n\r\s} \phi^{*}_{\a}  T_{\b\g\vert \l}U_{\d\e\vert}^{~~\t\l}=0\;,&\nonumber \\
&\d^{[\a\b\g\d\e]}_{[\m\n\r\s\t]}\ve^{\m\n\r\s} \phi^{*~\,\l}_{\a\h}  T_{\b\g\vert \l}U_{\d\e\vert}^{~~\t\h}=0
\;.&\nonumber 
\end{eqnarray}
An explicit expansion of these identities yields the relations
$$T_3+2T_2+2T_5=0\;,\ T_3-T_4=0\;,\ T_1=0\;,$$ that we just have to apply to Eq.(\ref{betAd4d2}). The expression of $\beta_{I_2}A^{I_2}_{I_3}$ cannot vanish unless $$\stackrel{(1)}{k^A}_{[BC]}=\stackrel{(2)}{k^A}_{(BC)}=\stackrel{(3)}{k^A}_{BC}=0\ .$$
Thus $a_{2,2}$ is trivial and can be set to zero, as well as $b_{1,2}$.

\vspace{3mm}
\noindent {\bfseries{$D$-degree 4:}} In $D$-degree 4, the candidates are of the form $\ve C^* U U$. There are two different ways to contract the indices: $T_1^{A[BC]}=\ve^{\m\n\r\s}C_A^{*\a\b}U^B_{\m\n|\a\c}U_{\phantom{C}\r\s|\b}^{C\phantom{\r\s|\b}\c}$ and $T_2^{ABC}=\ve^{\m\n\r\s}C^{*}_{A\m\a}U^B_{\n\r|\b\g}U_\s^{C\a|\b\g}$, but both expressions vanish because of the following Schouten identities:
\begin{eqnarray}
\d^{[\a\b\g\d\e]}_{[\m\n\r\s\t]}\ve^{\m\n\r\s}C^{*A\l}_{\a}U^{B\phantom{\b\c}|\t\h}_{\phantom{B}\b\c}U^{C}_{\d\ve|\l\h}=0\;,\;
\d^{[\a\b\g\d\e]}_{[\m\n\r\s\t]}\ve^{\m\n\r\s}C^{*A\t}_{\a}U^{B\phantom{\b\c}|\l\h}_{\phantom{B}\b\c}U^{C}_{\d\ve|\l\h}=0\;.
\nonumber \end{eqnarray}
They imply that
$T_1^{A[BC]}+T_2^{ABC}=0$ and $T_2^{ABC}=T_2^{A(BC)}$, which can be satisfied only if $T_1^{A[BC]}=T_2^{A(BC)}=0\,$. Thus $a_{2,4}$ is trivial and can be set to zero, as well as $b_{2,4}$.

\subsubsection*{Dimension 5}

In dimension 5, the number of indices in a possible $a_2$ term is equal to the $D$-degree plus 11. Therefore, it is only possible to write a Lorentz-invariant expression in odd $D$-degree. Furthermore, in $D$-degree 1, an expression of the form $C^* C T$ cannot be antisymmetrized over five indices (because $T^C_{[\a\b|\c]}=0$). Thus, we only have to study the $D$-degree 3 case. In fact, there is only one nontrivial term in $D$-degree 3:
  \begin{eqnarray}a_{2,3}=g^A_{\phantom{A}BC}\ve^{\m\n\r\s\t}C^{*}_{A\m\a}T^B_{\n\r|\b}U^{C\a\b}_{\phantom{C\a\b}|\s\t}d^5 x\ .\label{exo3a23}\end{eqnarray} 
The associated $b_1$ reads:
\begin{eqnarray}
b_{1,3}&=&\b_{I_3}\o^{I_3}=-3g^a_{~bc}\ve^{\m\n\r\s\t}\phi^{*\phantom{\m\a}\,\l}_{A\m\a}\,T^B_{\n\r\vert\b}\,U^{C\a\b}_{\phantom{C\a\b}|\s\t}\frac{1}{4!}\ve_{\l \g\d\h \xi}dx^\g dx^\d dx^\h 
dx^{\xi}\,.\nonumber \end{eqnarray} This candidate leads to a consistent $a_1$ if
\begin{eqnarray}\b_{I_3}A^{I_3}_{I_4}\o^{I_4}=-3g^A_{\phantom{A}[BC]}\ve^{\m\n\r\s\t}\phi^{*\a\l}_{A\phantom{\a\l}\m}U^B_{\l\b|\n\r}U^{C\phantom{A}\b}_{\phantom{C}\a\phantom{\b}|\s\t}d^5 x=0\ ,\end{eqnarray} which is true only if $g_{ABC}=g_{A(BC)}$.

\subsubsection*{Dimension $\mathbf{n>5}$}

It is impossible to build a non-vanishing parity-breaking $a_2$ term in dimension $n>5$, because an expression $C^*\omega^{J_i}$ involves at most 5 indices that can be antisymmetrized all together (remember that $T^A_{[\a\b|\c]}\equiv0$ and $U^A_{[\a\b|\c]\d}\equiv0$).

\vspace{4mm}
The two candidates that we have written in Eq.(\ref{exo3a21}) and Eq.(\ref{exo3a23}) are strictly non-$\g$-exact. The usual freedom to add $\g$-exact expressions of course holds in this case. We have established the following theorem, in which we have chosen $\g$-exact terms in order to get expressions more simple to treat in the following sections.
\begin{theorem}\label{defa2}
If the last term $a_2$ of the first order deformation of a sum of spin-3 theories is parity-breaking and Poincar\'e-invariant, then it is trivial except
in three and five dimensions. In those cases,  modulo trivial terms, it can be written respectively
\begin{eqnarray} \label{deftrois} a_{2}=f^A_{~[BC]} \e^{\m\n\r}C^{* \a\b}_A C^B_{\m\a}
\pa_{[\n}C^C_{\r]\vert \b}d^3x\,\end{eqnarray}
and
\begin{eqnarray} a_{2} = g^A_{~(BC)}\ve^{\m\n\r\s\t}C^{*\a}_{A~\;\m}
\pa_{[\n}C_{\r]}~^{\!\!\!\!\!\!\!B\,\b} \pa^2_{\a[\s}C^C_{\t]\b}d^5 x\,.
\label{a2n5}
\end{eqnarray}
The structure constants $f^A_{~[BC]}$ define an internal, 
anticommutative algebra $\ca$ while the structure constants  $g^A_{~(BC)}$ define an 
internal, commutative algebra $\cb\,$.  

\end{theorem}

\subsection{Deformation in 3 dimensions}
\label{dim3peu}

In the previous section, we determined that the only nontrivial first-order deformation of the free theory in three dimensions deforms the gauge algebra by the term (\ref{deftrois}). 
We now check that this deformation can be lifted and leads to a consistent first-order deformation of the Lagrangian. The check of Eq.(\ref{betacond}) ensures that $a_1$ exists, we just have to compute it now by using Eq.(\ref{eqagh1}). The same type of Schouten identities as those used to prove Eq.(\ref{betAd3d1}) must be used here: \begin{eqnarray}\ve^{\m\n\r}\6^{\phantom{C}}_{[\l}C_{\m]\a}^B\6^{\phantom{C}}_{[\n}C^{C\,\a}_{\r]}-\ve^{\m\n\r}\6^{\phantom{C}}_{[\l}C_{\a]\m}^B\6^{\phantom{C}}_{[\n}C^{C\,\a}_{\r]}-\ve^{\m\n\r}\6^{\phantom{C}}_{[\l}C_{\m]\n}^B\6^{\phantom{C}}_{\a}C^{C\,\a}_{\r}\equiv 0\ ,\end{eqnarray} \begin{eqnarray}\ve^{\m\n\r}\6_{[\l}^{\phantom{(B}} C_{\m]\a}^{(B} \6^{\phantom{C)}}_{[\n}C^{C)}_{\r]\b}\equiv 0\end{eqnarray} and \begin{eqnarray}\ve^{\m\n\r}\phi^*_A{}'{}^\l C^{B\a}_\m \6^{\phantom{C}}_{[\n}C^C_{\r][\a,\l]}-\ve^{\m\n\r}\phi^*_A{}'{}_\r C^{B\l\a} \6^{\phantom{C}}_{[\m}C^C_{\n][\a,\l]}+2\eta^{\l\kappa}\ve^{\m\n\r}\phi^*_A{}'{}_\n C^{B\a}_\r \6^{\phantom{C}}_{[\kappa}C^C_{\m][\a,\l]}\equiv 0\ .\end{eqnarray} The computation of $\d a_2$ leads to: \begin{eqnarray}\d a_2= 3 f^A_{\phantom{A}(BC)}\ve^{\m\n\r}\left[\phi^{*\a\b\l}_A-\frac{1}{3}\eta^{\a\b}\phi^*_A{}'{}^\l\right]\left(\6_\l C_{\m\a}^B \6_{[\n}C^C_{\r]\b}+ C_{\m\a}^B \6^2_{\l[\n}C^C_{\r]\b}\right)d^n x+d(...)\ .\label{d3pbda2}\end{eqnarray}Thanks to the above Schouten identities and to the relations $\6_\l C^B_{\m\a}=\frac{1}{3}\g\phi^B_{\l\m\a}+\frac{2}{3}\left[\6_{[\l}C_{\m]\a}+\6_{[\l}C^B_{\a]\m}\right]$, $\g\phi'{}^A_\r=2\6^\s C^A_{\r\s}$ , $\g \6_{[\r}\phi_{\s]\a\b}^A=2\6_{[\r}C_{\s](\a,\b)}$ and $\g(\6^\r\phi^A_{\a\b\r}-2\6_{(\a}\phi'{}^A_{\b)})=4\eta^{\kappa\l}\6^{\phantom{C}}_{[\kappa}C^C_{\a][\b,\l]}$, it is found that the ghost part of Eq.(\ref{d3pbda2}) is $\g$-exact and thus we get the expression of $a_1$: 
\begin{eqnarray}
a_1=f^A_{~[BC]}\e^{\m\n\r} \Big[ 3\, (\phi_A^{*\a\b\l}-\frac{1}{3} \h^{\a\b} \phi^*_A{}'{}^\l)\,(\frac{1}{3}\phi^{B}_{\a\m\l}\6_{[\n}C^C_{\r]\b}
+\frac{1}{2} C^{B}_{\a\m} \6_{[\r}\phi^{C}_{ \n] \b\l})\nonumber&& \\
+\frac{1}{3}\phi^*_a{}'{}^\l \6_{[\l}C^B_{\n]\m}\phi'_\r{}^C+\phi^*_A{}'{}_\m C^{B\,\a}_{\,\n}(-\frac{1}{2} \6^\l \phi^C_{\l\a\r}+\6_{(\a}\phi'{}^C_{\r)})\Big] d^3 x\,.\nonumber&& 
\end{eqnarray}
There is no possible homogeneous part $\bar{a}_1$ in $D$-degree 1 because $G^A_{\m\n\r}$ involves two derivatives. 

This expression of $a_1$ can then be put in Eq.(\ref{eqagh0}). In the same way as for the spin-2--spin-3 deformations, we have achieved this computation with the help of the symbolic manipulation software FORM \cite{FORM}. We have first established a list of possible terms for $a_0$. They have the structure $\ve \6^2 \phi\,\phi\,\phi$ or $\ve\6 \phi\,\6 \phi\, \phi$. Since the Lagrangian is defined modulo $d$, the first ones are not independent of the second ones. Furthermore, Schouten identities can be written for each term. They allow, for example, to choose only terms of the structure $\ve^{\m\n\r}\6_\m\phi^A_{\n..}\6_.\phi^B_{...}\phi^C_{...}$. We have written a sum of the remaining terms, which we denote $a_{0,test}$, and have computed $\d a_1+\g a_{0,test}$. The modulo $d$ freedom can be fixed, for example, by removing the derivatives that the ghosts bear (this can be done by taking the Euler-Lagrange derivative $\frac{\d^L}{\d C^A_{\a\b}}$ then by (left)multiplying the result by the undifferentiated $C^A_{\a\b}$). The expression that is obtained involves terms of the structure $\e C \pa^3 \phi \phi$ or $\e C \pa^2 \phi \pa \phi$. At this stage, it is once again crucial to take Schouten identities into account. There would seem to be respectively 45 and 130 expressions in the two sets of terms (which would mean 175 equations in the system), but there are 108 identities between them, which leaves us with a much smaller system that admits a solution. We have obtained the following expression for $a_0$:
\begin{eqnarray}
a_0=f_{[ABC]}\e^{\m\n\r}\Big[
\frac{1}{4} \6_\m \phi^A_{\n\a\b}\6^\a \phi'{}^{B\b}\phi'{}^{C}_\r
+\frac{1}{4} \6_\m \phi^A_{\n\a\b}\6^\a \phi^{B\b\g\d}\phi^{C}_{\r\g\d}
-\frac{5}{4} \6_\m \phi^A_{\n\a\b}\6^\a \phi'{}^{B\g}\phi^{C\b}_{\r\g}\nonumber \\
-\frac{3}{8} \6_\m \phi'{}^A_{\n}\6^\a \phi'{}^{B}_{\a}\phi'{}^{C}_{\r}
+\frac{1}{4}\6_\m \phi^{A\a\b}_{\n}\6^\g \phi'{}^{B}_{\g}\phi^{C}_{\r\a\b}
-\6_\m \phi'{}^{A}_{\n}\6^\g \phi^{B}_{\a\b\g}\phi_{\r}^{C\a\b}\nonumber \\
+\frac{1}{2} \6_\m \phi^{A}_{\n\a\b}\6^\g \phi^{B}_{\a\g\d}\phi_{\r}^{C\b\d}
+2 \6_\m \phi'{}^{A}_{\n}\6^\b \phi'{}^{B\g}\phi_{\r}^{C\b\g}
-\frac{1}{4}\6_\m \phi^{A}_{\n\a\b}\6^\g \phi^{B\a\b\d}\phi_{\r\g\d}^{C}\nonumber \\
-\frac{1}{4}\6_\m \phi^{A}_{\n\a\b}\6^\g \phi^{B\b}\phi_{\r\g}^{C~\a}
-\frac{5}{8}\6_\m \phi'{}^{A}_{\n}\6_\r \phi'{}^{B\b}\phi'{}_{\b}^{C}
+\frac{7}{8}\6_\m \phi^{A}_{\n\a\b}\6_\r \phi^{B\a\b\g}\phi'{}_{\g}^{C}\nonumber \\
+\frac{1}{4}\6_\m \phi^{A}_{\n\a\b}\6_\g \phi_\r^{B\a\g}\phi'{}^{C\b}
+\frac{1}{4}\6_\m \phi'{}^{A}_{\n}\6^\a \phi_{\r\a\b}^{B}\phi'{}^{C\b}
-\frac{1}{4}\6_\m \phi^{A}_{\n\a\b}\6^\g \phi_{\r\g\d}^{B}\phi^{C\a\b\d}\nonumber \\
-\frac{1}{8}\6_\m \phi'{}^{A}_{\n}\6^\a \phi'{}_{\r}^{B}\phi'{}^{C}_{\a}
-\frac{1}{8}\6_\m \phi^{A}_{\n\a\b}\6^\g \phi_{\r}^{B\a\b}\phi'{}^{C}_{\g}
\Big]d^3x\,.\nonumber 
\end{eqnarray}
Furthermore, we have obtained the condition that the structure coefficients must be totally antisymmetric in order for that solution to exist: $f_{ABC}=f_{[ABC]}$. This can be seen as defining an invariant norm on the algebra $\ca$ defined by $f^A_{\phantom{A}[BC]}$.

We have achieved some second-order computations about this vertex, that will be considered in chapter \ref{ch:socomp}. We can already announce that it is inconsistent by itself.

\subsection{Deformation in 5 dimensions}

Let us perform the same analysis for the candidate in five dimensions. First, let us compute $a_1$ from $a_2$ (given by Eq.(\ref{a2n5})), using Eq.(\ref{eqagh1}): 
\begin{eqnarray}
\d a_2&=&-3g^A_{~(BC)}\e^{\m\n\r\s\t}\pa_\l
\phi^{*\a\l}_{A~~\m}\6_{[\n}C^B_{\r]\b}\6_{\a[\s}C^{C\b}_{\t]}d^5 x\nonumber\\
&=&-db_1+3g^A_{~(BC)}\e^{\m\n\r\s\t}\phi^{*\a\l}_{A~~\m}\Big[\pa_{\l[\n}C^B_{\r]\b}\pa_{\a[\s}C^{C\b}_{\t]}
+\pa_{[\n}C^B_{\r]\b}\pa_{\l\a[\s}C^{C\b}_{\t]}\Big]d^5 x\,.
\nonumber \end{eqnarray}
The first term between square bracket vanishes because of the symmetries of 
the structure constants $g^A_{~BC}$ and we obtain:  
\begin{eqnarray}a_1=\frac{3}{2}g^A_{~(BC)}\e^{\m\n\r\s\t}\phi^{*\a\l}_{A~~\,\m}\pa_{[\n}^{\ }C^{B\ \b}_{\r]}\left[\pa_{\b[\s}\phi^C_{\t]\l\a}-2\pa_{\l[\s}\phi^C_{\t]\a\b}\right]d^5x\,.\nonumber \end{eqnarray}
The element $a_1$ gives the first-order deformation of the gauge transformations. 
By using the definition of the generalized de Wit--Freedman connections 
\cite{deWit:1979pe} that we have recalled in Eq.(\ref{dwfconn}), we get the following simple expression for $a_1$: 
\begin{eqnarray}
        a_1=g^A_{~(BC)}\e^{\m\n\r\s\t}\phi^{*\a\b}_{A~~\,\m}\pa_{[\n}^{\ }C^{B\ \l}_{\r]} 
        {\Gamma}^C_{\l[\s;\t]\a\b}  d^5x\,,
        \label{a1dWF}
\end{eqnarray}
where ${\Gamma}^{(2)C}_{\l\s;\t\a\b}$ is the second de Wit--Freedman spin-3 connection 
\begin{eqnarray}
        {\Gamma}^{(2)C}_{\l\s;\t\a\b} =3\, \pa_{(\t}\pa_{\a}\phi^c_{\b)\l\s} 
        + \pa_{\l}\pa_{\s}\phi^C_{\t\a\b}   
- \frac{3}{2}\,\big(\pa_{\l}\pa_{(\t}\phi^C_{\a\b)\s}
        +\pa_{\s}\pa_{(\t}\phi^C_{\a\b)\l}\big)\nonumber
\end{eqnarray}
transforming under a gauge transformation $\d_{\l}\phi^A_{\m\n\r}=3\,\pa_{(\m}\l^A_{\n\r)}$ 
according to 
\begin{eqnarray}
        \d_{\l} {\Gamma}^{(2)C}_{\r\s;\t\a\b} = 3\, \pa_{\t} \pa_{\a} \pa_{\b}\l^C_{\r\s}\,. \nonumber
\end{eqnarray}
The expression (\ref{a1dWF}) for $a_1$ implies that the deformed gauge transformations are 
\begin{eqnarray}
        \stackrel{(1)}{{\d_{\l}}} \phi^A_{\m\a\b} = 3\, \pa_{\m}\l^A_{\a\b} + g^A_{~(BC)} \,
        \e_{\m}^{~\,\n\r\s\t}\,{\Gamma}^B_{\g\n;\r\a\b}\,\pa_{\s}^{\ }\l^{C\ \g}_{\;\t}\,,
\label{defogt}
\end{eqnarray}
where the right-hand side must be completely symmetrized over the indices $(\m\a\b)\,$.       

Then, the cubic deformation of the free Lagrangian $a_0$ is obtained from $a_1$
by solving Eq.(\ref{eqagh0}). Once again, we consider the most general cubic
expression involving four derivatives. The modulo $d$ freedom allows one to consider terms of the structure $\ve \6^3\phi\,\6\phi\,\phi$ or $\ve \6^2\phi\,\6^2\phi\,\phi$, which are not constrained by Schouten identities (terms of the structure $\ve \6^2\phi\6\phi\6\phi$ would). There are 17 terms of the first structure and 29 terms of the second. We can build an $a_{0,test}$ with those 46 terms. Then, we compute $\d a_1+\g a_{0,test}$, take the (left)Euler-Lagrange derivative with respect to $C_{\a\b}$, and (left)multiply  by $C_{\a\b}$. The result must be 0. We get a sum of terms of the structure $\ve C\pa^4 \phi \pa \phi$ or $\ve C\pa^3 \phi\pa^2 \phi$. These are not related by Schouten identities and are therefore independent, all coefficients of the obtained equation thus have to vanish. When solving this system of equations with the software,
we found that a consistent solution exists if the internal coefficients are completely symmetric: $g_{ABC}=g_{(ABC)}$.
In other words, the corresponding internal commutative algebra $\cb$ 
possesses an invariant norm. As for the algebra $\ca$ of the $n=3$ case, 
the positivity of energy requirement
imposes a positive-definite internal metric with respect to which the norm is defined. Finally, the solution for the cubic vertex $a_0$ reads:
\begin{eqnarray}\nonumber
a_0&=&\frac{3}{2}g_{(ABC)}\e^{\m\n\r\s\t} \Big\{-\frac{1}{8}\6_\m\Box
\phi'{}^A_\n\6_\r \phi'{}^B_\s \phi'{}^C_\t 
+\frac{1}{2}\6^3_{\m\a\b} \phi'{}^A_\n\6_\r
\phi^{B\a\b}_\s \phi'{}^C_\t
+\frac{1}{4}\6_\m\Box \phi^{A\a\b}_\n\6_\r
\phi^B_{\s\a\b}\phi'{}^C_\t\nonumber \\
&&\qquad+\frac{3}{8}\6_\m\Box \phi'{}^A_\n\pa_\r \phi^{B\a\b}_\s \phi^C_{\t\a\b}
-\frac{1}{2}\6_\m\Box \phi^{A\a\b}_\n\6_\r \phi^B_{\s\a\g} \phi^{C\ \g}_{\t\b}
-\frac{1}{2}\6^{3\a\b}_\m \phi'{}^A_\n\6_\r \phi^B_{\s\a\g} \phi^{C\ \g}_{\t\b}
\nonumber \\
&&\qquad -\frac{1}{2}\6^{3\a\b}_\m \phi^A_{\n\a\g}\6_\r \phi'{}^B_\s \phi^{C\ \g}_{\t\b}
-\frac{1}{4}\6^{3\a\b}_\m \phi^A_{\n\a\b}\6_\r \phi^{B\g\d}_\s
\phi^C_{\t\g\d}
-\frac{1}{2}\6^{3\a\b}_\m \phi^A_{\n\g\d}\6_\r \phi^B_{\s\a\b}
\phi^{C\g\d}_\t
\nonumber \\
&&\qquad +\6^{3\a\b}_\m \phi^A_{\n\b\g}\6_\r \phi^{B\g\d}_\s \phi^C_{\t\a\d}
+\frac{1}{2}\6^2_{\m\a}\phi_\n^{A\a\b}\6_\r^{2\g}\phi'{}^B_\s
\phi^C_{\t\b\g}
-\6^2_{\m\a}\phi_{\n\b\g}^A\pa_{\r\d}^2\phi^{B\a\b}_\s
\phi^{C\g\d}_\t \Big\}d^5 x\,.\nonumber \end{eqnarray}

We have achieved second-order computations about this vertex, that will also be considered in Chapter \ref{ch:socomp}. We can announce that, as far as we have investigated, this is consistent if the internal coefficients $g_{ABC}$ can be taken diagonal.

\subsection{Discussion}\label{sec:concl}

In this chapter, we obtained the only two consistent nonabelian parity-breaking first-order deformations of a sum of spin-3 Fronsdal theories. 
The first one is defined in $n=3$ and involves a multiplet of gauge fields  
$\phi^A_{\m\n\r}$ taking values in an internal, anticommutative, invariant-normed 
algebra $\ca\,$. 
The second one lives in a space-time of dimension $n=5$, the fields taking value in an 
internal, commutative, invariant-normed algebra $\cb\,$. As will be showed in Chapter 8, taking the metrics which define the inner product in $\ca$ and $\cb$ positive-definite 
(which is required for the positivity of energy), the $n=3$ candidate gives rise to inconsistencies when continued to perturbation order two, whereas the $n=5$ one passes the 
 test and can be assumed to involve only \emph{one} kind of 
self-interacting spin-3 gauge field $\phi_{\m\n\r}$, bearing no internal ``color'' index.

Remarkably, the cubic vertex of the $n=5$ deformation is rather simple. 
Furthermore, the Abelian gauge transformations are deformed by the addition of a term involving the second de Wit--Freedman connection in a straightforward way, cf. 
Eq.(\ref{defogt}). 
The relevance of this second generalized Christoffel symbol in relation to a hypothetical spin-3 covariant derivative was already stressed in \cite{Bengtsson:1983bp}.     

It is interesting to compare the results of the present spin-3 analysis with those found 
in the spin-2 case first studied in \cite{Boulanger:2000ni}. 
There, two parity-breaking first-order consistent nonabelian deformations of 
Fierz-Pauli theory were obtained, also living in dimensions $n=3$ and $n=5$.
The massless spin-2 fields in the first case bear a color index, the internal algebra 
$\widetilde{\ca}$ 
being commutative and further endowed with an invariant scalar product. 
In the second, $n=5$ case, the fields take value in an anticommutative, invariant-normed 
internal algebra $\widetilde{\cb}$. 
It was further shown in \cite{Boulanger:2000ni} that the $n=3$ first-order consistent 
deformation could be continued to \emph{all} orders in powers of the coupling constant, the 
resulting full interacting theory being explicitly written down \footnote{Since the 
deformation is consistent, starting from $n=3$ Fierz-Pauli, the complete $n=3$ 
interacting theory of \cite{Boulanger:2000ni} describes no propagating physical 
degree of freedom. 
On the contrary, the topologically massive theory in \cite{Deser:1981wh,Deser:1982vy} 
describes a massive graviton with \emph{one} propagating degree of freedom (and not 
\emph{two}, as 
was erroneously typed in \cite{Boulanger:2000ni}).}. 
However, it was not determined in \cite{Boulanger:2000ni} whether the $n=5$ candidate could 
be continued to all orders in the coupling constant. Very interestingly, 
this problem was later solved in \cite{Anco:2003pf}, where a consistency condition was 
obtained at second order in the deformation parameter, \emph{viz} the algebra 
$\widetilde{\cb}$ must be nilpotent of order three. 
Demanding positivity of energy and using  
the results of \cite{Boulanger:2000ni}, the latter nilpotency condition implies      
that there is actually no $n=5$ deformation at all: the structure constant of the 
internal algebra $\cb$ must vanish \cite{Anco:2003pf}. 
Stated differently, the $n=5$ first-order deformation 
candidate of \cite{Boulanger:2000ni} was shown to be inconsistent \cite{Anco:2003pf} when continued at second order in powers of the coupling constant, in analogy with the spin-3 
first-order deformation written in \cite{Berends:1984wp}.   

In the present spin-3 case, the situation is somehow the opposite.
Namely, it is the $n=3$ deformation which shows inconsistencies when going to second order, 
whereas the $n=5$ deformation passes the first test. Also, in the $n=3$ case the fields take 
values in an anticommutative, invariant-normed internal algebra $\ca$ whereas the fields 
in the $n=5$ case take value in a commutative, invariant-normed algebra $\cb\,$. 
However, the associativity condition deduced from a second-order consistency condition is 
obtained for the latter case, which implies that the algebra $\cb$ is a direct sum of 
one-dimensional ideals. 
We summarize the previous discussion in Table \ref{table1}. 
\begin{table}[!ht]
\centering
\begin{tabular}{|c||c|c|}
\hline 
 & $s=2$  & $s=3$ \\ \hline \hline
$n=3$& $\widetilde{\ca}$ commutative, invariant-normed 
&  $\ca$ anticommutative, invariant-normed and \\
 &  & nilpotent of order $3$ \\ \hline
$n=5$ & $\widetilde{\cb}$ anticommutative, invariant-normed and & $\cb$ commutative, invariant-normed and\\ 
&  nilpotent of order $3$ & associative \\
\hline
\end{tabular}
\caption{\it Internal algebras for the parity-breaking first-order 
deformations of spin-$2$ and spin-$3$ free gauge theories.\label{table1}}
\end{table}

It would be of course very interesting to investigate further the $n=5$ deformation 
exhibited here. 

%%%%%%%%%%%%%%%%%%%%%%%%%%%%%%%%%%%%%%%%%%%%%%%
\chapter{Results about $1-s-s$, $2-s-s$ and $s-s'-s''$ consistent deformations}\label{ch:intmisc}
%%%%%%%%%%%%%%%%%%%%%%%%%%%%%%%%%%%%%%%%%%%%%%%%

In this chapter, we study some more general first order deformation problems. First, we explicitly compute cubic vertices involving one spin-1 field and two spin-$s$ fields. Then we compute the unique vertex involving a spin-2 field and two spin-4 fields. Furthermore, we determine the only possible cubic first order deformation of the gauge algebra involving a spin-2 field and two spin-$s$ fields. These are arguably related to a vertex, that is the Minkowski limit of the vertex obtained in $(A)dS$ using the Fradkin--Vasiliev procedure, though we have not computed it explicitly for any spin-$s$. The uniqueness of the Minkowski deformation implies the uniqueness of the Fradkin--Vasiliev deformation, by taking the appropriate limit that has been discussed in Section \ref{minlim}. Then, we make some considerations about the possibility of constructing a cubic deformation of the gauge algebra involving any three fields of spins $s\leqslant s'\leqslant s''$. An important result is that it is not possible if $s''\geqslant s+s'$, which implies, for example, that there are no nonabelian cubic deformations for a $1-1-s$ configuration with $s\geqslant 2$ or a $2-2-s$ configuration with $s\geqslant 4$. The unique vertex $2-2-3$ that we have presented in Chapter \ref{ch:int23} is thus the only one of its kind.

\section{Consistent cubic deformations of the type $1-s-s$\label{sec:1ss}}

Let us first say a few words about the $1-2-2$ case. It is the first one that we computed, we realized later that it could be straightforwardly extended to the $1-s-s$ configuration. The strange thing is that it is in fact a ``lower spin'' interaction, that had not been considered before. One may wonder if it is compatible with usual lower spin interactions. We have established that it is not compatible with Einstein--Hilbert theory at second order (the details are given in Chapter \ref{ch:socomp}), which suggests that there would be different types of spin-$2$ fields. We provide below the general spin-$s$ result. In the spin-$2$ case, the notation $h^a_{\m\n}$ and $K^a_{\m\n|\r\s}=-2 R^a_{\m\n|\r\s}$ is once again used. We also use the usual convention $A_\m$ for a spin 1 field. Since we consider an interaction linear in the spin-1 field, it will not be denoted with a family index as no relation could arise. The spin-$s$ expressions bear the usual capital letters. The Faraday tensor is proportional to the spin-$1$ curvature, that is the basic gauge-invariant tensor: $F_{\m\n}=\6_\m A_\n -\6_\n A_\m = 2K_{\m\n}$. Let us precise the expression of $\g$ and $\d$ in the spin-$1$ case: \begin{eqnarray}\g A_\m=\6_\m C\quad,\quad\d A^{*\n}=\6_\m F^{\m\n}=\frac{\d L}{\d A_\n}\quad{\textrm{and}}\quad\d C^*=-\6_\n A^{*\n}\quad.\end{eqnarray}

\subsection{Determination of $a_2$}

We can now determine the possible first-order deformations of the gauge algebra. As we already explained in the previous chapters, a $gh\ 0$ cubic expression only contains terms of $antigh\leqslant2$. For example, an expression of $antigh\ 3$ is at least a polynomial of degree 5 ($C^* h^* C C C$ or $h^*h^*h^* C C C$). Furthermore, a cubic $antigh\ 2$ expression has to be linear in the $antigh\ 2$ antifields and quadratic in the ghosts, or their derivatives. Then, since the $n$-form $w^n$ corresponding to the BRST-BV generator $W$ is defined modulo $d$, one can always add a $d$-exact expression to the $antigh\ 2$ term of $\stackrel{(1)}{w}{}^{\!\!\!n}$ in such a way as to remove the derivatives that $C^*$ could bear. Finally, the first order deformation $\stackrel{(1)}{w}{}^{\!\!\!n}=a_0+a_1+a_2$ is a $\bs$ modulo $d$ cocycle. As usual, $a_2$ is defined modulo $\g$, thus the only relevant derivatives of the ghosts are the tensors $U^{(i)A}$ defined in section \ref{cohgam}. For $s\leqslant 4$, we know that those cubic expressions are the only possible deformations. For $s>4$, though we are not sure if the first order could be a polynomial of degree greater than 3, at least we establish the unique cubic solutions.

The possible terms for $a_2$ are either proportional to the spin-1 antifields $C^*$ or to the spin-$s$ antifields $C^{*A}_{\mu_1...\mu_{s-1}}$. In the latter case, the expression must be proportional to the undifferentiated ghost $C$, because all its derivatives are $\g$-exact. In order to contract all indices, the only possibility is: \begin{eqnarray}a_{2,0}=g_{AB}C^{*A}_{\mu_1...\mu_{s-1}}C C^{B\mu_1...\mu_{s-1}}d^n x\quad.\end{eqnarray} In the case of an expression proportional to $C^*$, the two ghost tensors have to bear the same number of indices, that have to be contracted among themselves. This leaves us with a set of $s-1$ candidates: \begin{eqnarray}a_{2,2i}=\stackrel{(i)}{f}_{[AB]}C^*U^{(i)A}_{\mu_1\nu_1|...|\mu_i\nu_i|\nu_{i+1}...\nu_{s-1}}U^{(i)B\mu_1\nu_1|...|\mu_i\nu_i|\nu_{i+1}...\nu_{s-1}}d^n x\quad,\quad i\leqslant s-1\quad.\end{eqnarray}

\subsection{Determination of $a_1$}

We must now check if Eq.(\ref{eqagh1}) admits solutions for the above candidates. We get easily that $a_{1,0}$ exists:
\begin{eqnarray}\d a_{2,0}&=&-s g_{AB}\6_{\mu_s}\phi^{*A\m_1...\mu_s}C C^B_{\m_1...\mu_{s-1}} d^n x
\nonumber \\
&=& s g_{AB}\phi^{*A\m_1...\mu_s}\left[\6_{\mu_s} C\ C^B_{\m_1...\mu_{s-1}} + C\ \6_{(\mu_1} C^B_{\m_2...\mu_s)} \right]d^n x + d(...)\nonumber \\
&=& -\g\left(g_{AB}\phi^{*A\m_1...\mu_s}\left[s A_{\mu_s} C^B_{\mu_1...\mu_{s-1}}-C \phi^B_{\m_1...\m_s}\right]d^n x\right) +d(...)\quad.\end{eqnarray}
Since there are no homogeneous solutions with no derivatives,
we can conclude that:
\begin{eqnarray} a_{1,0}=g_{AB}\phi^{*A\m_1...\mu_s}\left[s A_{\mu_s} C^B_{\mu_1...\mu_{s-1}}-C \phi^B_{\m_1...\m_s}\right]d^n x+\g(...)\quad.\end{eqnarray}
For the other candidates, we get: \begin{eqnarray}\d a_{2,2i}&=&-\stackrel{(i)}{f}_{[AB]}\6_\r A^{*\r}U^{(i)A}_{\mu_1\nu_1|...|\mu_i\nu_i|\nu_{i+1}...\nu_{s-1}}U^{(i)B\mu_1\nu_1|...|\mu_i\nu_i|\nu_{i+1}...\nu_{s-1}}d^n x\nonumber\\&=&2\stackrel{(i)}{f}_{[AB]}A^{*\r}\ D{U}^{(i)A}_{\mu_1\nu_1|...|\mu_i\nu_i|\nu_{i+1}...\nu_{s-1}}{U}^{(i)B\mu_1\nu_1|...|\mu_i\nu_i|\nu_{i+1}...\nu_{s-1}}\frac{1}{(n-1)!}\ve_{\r\s_1...\s_{n-1}}dx^{\s_1}...dx^{\s_{n-1}}\nonumber\\&&+\,d(...) + \g(...)\nonumber\quad,\end{eqnarray} where we have isolated the strictly non-$\g$-exact part by using the differential $D$ that has been defined in section \ref{furgam}. Then, Eq.(\ref{defD}) tells us that $D U^{(i)} \div U^{(i+1)}$ and we see that there is an obstruction, unless $i$ is maximal. We have thus established that only the candidate with the highest number of derivatives admits a solution $a_1$. Let us choose a more simple expression for $a_2$ by adding an appropriate $\g$-exact term: \begin{eqnarray}a_{2,2s-2}=f_{[AB]}C^* Y^{(s-1)}(\6^{(s-1)}_{\mu_1...\mu_{s-1}}C^A_{\nu_1...\nu_{s-1}})\6^{(s-1)\mu_1...\mu_{s-1}}C^{B\nu_1...\nu_{s-1}}d^n x\quad,\nonumber\end{eqnarray} where we recall that the operator $Y^{(s-1)}$ antisymmetrizes the pairs $\m_i\n_i$: $\displaystyle Y^{(i)}=\frac{1}{2^i}\prod_{j=1}^i[e-(\m_j\n_j)]$. Then, let us compute $a_{1,2s-2}$:
\begin{eqnarray}
\d a_{2,2s-2}&=&-f_{[AB]}\6_\r
A^{*\r}Y^{(s-1)}(\6^{(s-1)}_{\mu_1...\mu_{s-1}}C^A_{\nu_1...\nu_{s-1}})\6^{(s-1)\mu_1...\mu_{s-1}}C^{B\nu_1...\nu_{s-1}}d^n x\nonumber \\
&=& 2 f_{[AB]}A^{*\r}Y^{(s-1)}(\6_\r\6^{(s-1)}_{\mu_1...\mu_{s-1}}C^A_{\nu_1...\nu_{s-1}})\6^{(s-1)\mu_1...\mu_{s-1}}C^{B\nu_1...\nu_{s-1}} d^n x+d(...)\nonumber \\
&=& -\g\left[2f_{[AB]}A^{*\r}Y^{(s-1)}(\6^{(s-1)}_{\mu_1...\mu_{s-1}}\phi^A_{\nu_1...\nu_{s-1}\r})\6^{(s-1)\mu_1...\mu_{s-1}}C^{B\nu_1...\nu_{s-1}}d^n x\right]+d(...)\ ,\end{eqnarray} and finally: \begin{eqnarray}&a_{1,2s-2}=\tilde{a}_{1,2s-2}+\bar{a}_{1,2s-2}\quad \textrm{where}\quad \g \bar{a}_{1,2s-2}=0\quad \textrm{and}&\nonumber\\&\tilde{a}_{1,2s-2}=2f_{[AB]}A^{*\r}Y^{(s-1)}(\6^{(s-1)}_{\mu_1...\mu_{s-1}}\phi^A_{\nu_1...\nu_{s-1}\r})\6^{(s-1)\mu_1...\mu_{s-1}}C^{B\nu_1...\nu_{s-1}}d^n x\quad.&\end{eqnarray}

\subsection{Determination of $a_0$}

Finally, we can compute the possible vertices $a_0$, that have to be solutions of Eq.(\ref{eqagh0}). For the candidate $a_{1,0}$, we get: \begin{eqnarray}\displaystyle\d a_{1,0}=g_{AB}G^{A\m_1...\mu_s}\left[s A_{\m_s} C^B_{\mu_1...\mu_{s-1}} - C \phi^B_{\m_1...\m_s}\right]d^n x\ .\label{dela10}\end{eqnarray} Let us then consider the most general candidate for $a_0$. Since $a_1$ contains no derivatives, the $a_0$ candidate contains only one. Then, the modulo $d$ freedom allows us to consider only terms proportional to the undifferentiated $A_\mu$. There are only eight terms of the structure $A \6\phi^A\phi^B$ (and only six in the particular case of spin-2). The action of $\g$ on a linear combination of those terms can be split into a part proportional to the spin-1 ghosts and a part proportional to the spin-$s$ ghosts, that would have to correspond to the two terms of Eq.(\ref{dela10}). Modulo $d$, the first part further splits into something quadratic in the first derivatives of the spin-$s$ fields and something linear in the second derivatives of a field and the undifferentiated other. The first expression has to vanish, which imposes that most of the internal coefficients are antisymmetric over $A$ and $B$. The second expression must coincide with the second term in Eq.(\ref{dela10}). This establishes that the coefficients are all proportional to $g_{AB}$, modulo a term linear in the Faraday tensor in $a_0$. Then, we can proceed with the computation of the part proportional to the spin-$s$ ghost. Modulo $d$, this part decomposes into terms of the structure $C^A A \6^2\phi^B$, $C^A\6 A\6\phi^B$ and $C^A\6^2 A \phi^B$. The first set coincides with the term in $a_1$ but unfortunately, the other two sets bring in obstructions, though they turn out to be proportional to the Faraday tensor.

Let us now check Eq.(\ref{eqagh0}) for $a_{1,2s-2}$. We first compute $\d \tilde{a}_{1,2s-2}$:
\begin{eqnarray}
\d\tilde{a}_{1,2s-2}&=&2f_{[AB]}\6_{\s}F^{\s\r}Y^{(s-1)}(\6^{s-1}_{\mu_1...\mu_{s-1}}\phi^A_{\nu_1...\nu_{s-1}\r})\6^{(s-1)\mu_1...\mu_{s-1}}C^{B\nu_1...\nu_{s-1}}d^n x\nonumber\\
&=&-2^{2-s}f_{[AB]}\6^{\s}A^{\r}K^A_{\m_1\n_1|...|\m_{s-1}\n_{s-1}|\s\r}\6^{(s-1)\m_1...\m_{s-1}}C^{B\n_1...\n_{s-1}}d^n x\nonumber\\&&-\g\left[f_{[AB]}F^{\s\r}Y^{(s-1)}(\6^{s-1}_{\m_1...\m_{s-1}}\phi^A_{\n_1...\n_{s-1}\r})\6^{(s-1)\m_1...\m_{s-1}}\phi^{B\n_1...\n_{s-1}}_{\phantom{B\n_1...\n_{s-1}}\s}d^n x\right]+d(...)\nonumber\\
&=&2 f_{[AB]}A^{\r}Y^{(s-1)}(\6^{s-1}_{\m_1...\m_{s-1}}F^A_{\n_1...\n_{s-1}\r})\6^{(s-1)\m_1...\m_{s-1}}C^{B\n_1...\n_{s-1}}d^n x\nonumber\\&&-2f_{[AB]} C Y^{(s-1)}(\6^{s-1}_{\m_1...\m_{s-1}}F^A_{\n_1...\n_{s-1}\r})\6^{(s-1)\m_1...\m_{s-1}}\phi^{B\n_1...\n_{s-1}\r}d^n x\nonumber\\
&&-\g a_{0,2s-2} + d(...)\,.\nonumber
\end{eqnarray} where the vertex is 
\begin{eqnarray}
a_{0,2s-2} &=& -f_{[AB]}F^{\r\s}Y^{(s-1)}(\6^{s-1}_{\mu_1...\mu_{s-1}}\phi^A_{\nu_1...\nu_{s-1}\r})\6^{(s-1)\mu_1...\mu_{s-1}}
\phi^{B\nu_1...\nu_{s-1}}_{\phantom{B\nu_1...\nu_{s-1}}\s}\ d^n x\nonumber\\
&&+f_{[AB]}\frac{1}{2^{s-2}}A^\r K^A_{\mu_1\nu_1|...|\mu_{s-1}\nu_{s-1}|\r\s}\6^{(s-1)\mu_1...\mu_{s-1}}\phi^{B\nu_1...\nu_{s-1}\s}d^n x\,.
\end{eqnarray}
The first two terms of $\d \tilde{a}_{1,2s-2}$ are $\d$-exact, because $F^A_{\m_1...\m_s}=\d(\phi^{*A}_{\m_1...\mu_s}-\frac{s(s-1)}{2(n+2s-6)}\eta^{}_{(\m_1\m_2}\phi^*{}'{}^A_{\m_3...\m_s)})$. They correspond to the following homogeneous terms:
\begin{eqnarray}
\bar{a}_{1,2s-2} = 2f_{[AB]}\6^{(s-2)\mu_2...\mu_{s-1}}
\phi^{*A\rho_1\rho_2\nu_3...\nu_{s-1}\t}D^{\nu_1\nu_2\s}_{\rho_1\rho_2\t} F^{\mu_1}_{\phantom{\mu_1}\s} Y^{(s-1)}
(\6^{s-1}_{\mu_1...\mu_{s-1}}C^B_{\nu_1...\nu_{s-1}})d^n x\ ,
\end{eqnarray}where
\begin{eqnarray}
D^{\nu_1\nu_2\s}_{\rho_1\rho_2\t}&=&\d^{(\nu_1}_{(\rho_1}\d^{\nu_2)}_{\rho_2\phantom{)}}
\d^{\s\phantom)}_{\t)}-\frac{2s-3}{2(n+2s-6)}\,\eta_{(\rho_1\rho_2}^{\phantom{)}}
\eta_{\phantom{)}}^{\s(\nu_1}\d_{\t)}^{\nu_2)}\ .
\nonumber
\end{eqnarray}
Let us notice that some terms of the structure $f_{[AB]}C\6^{s-2}\phi^{*A}\6^{s-2}G^{B}$ appear at first, but they are simultaneously $\d$-exact and $\g$-closed and have thus been omitted in the previous expression since they correspond to trivial gauge transformations. Let us write down explicitly the gauge transformations given by $a_1$ (the inhomogeneous part affects only the spin-1 while the homogeneous part affects the spin-$s$):
\begin{eqnarray}&&
\bullet\stackrel{(1)}{\d}_\xi A_\rho=f_{[AB]}Y^{(s-1)}(\6^{s-1}_{\mu_1...\mu_{s-1}}
\phi^A_{\nu_1...\nu_{s-1}\r})Y^{(s-1)}(\6^{(s-1)\mu_1...\mu_{s-1}}
\xi^{B\nu_1...\nu_{s-1}})\\&&
\bullet\stackrel{(1)}{\d}_{\xi}
\phi_{A\rho_1\rho_2\nu_3...\nu_{s-1}\t}=2(-1)^{s-1}
f_{[AB]}D^{\nu_1\nu_2\s}_{\rho_1\rho_2\t}\times\nonumber 
\6^{(s-2)\mu_2...\mu_{s-1}}
\Big[ F^{\mu_1}_{\phantom{\mu_1}\s} Y^{(s-1)}(\6^{s-1}_{\mu_1..\mu_{s-1}}
\xi^B_{\nu_1...\nu_{s-1}})\Big]\quad,
\end{eqnarray}
where the right-hand side of the latter equation must be symmetrized over all the free indices.

\subsubsection{Alternative form of the vertex}

The expression obtained for $a_{0,2s-2}$ is very compact, but it depends on the undifferentiated $A_\mu$. This could seem a bit strange, because the deformation of the gauge transformations depends only on the spin-$s$ gauge parameter. And indeed, as it should, it is in fact possible to add a $\d$-exact expression to the previously displayed vertex in such a way as to obtain an equivalent vertex proportional to the Faraday tensor. Let us consider the second term of $a_{0,2s-2}$:
\begin{eqnarray}&&\frac{1}{2^{s-2}}f_{[AB]}A^\r K^A_{\mu_1\nu_1|...|\mu_{s-1}\nu_{s-1}|\r\nu_s}\6^{(s-1)\mu_1...\mu_{s-1}}\phi^{B\nu_1...\nu_s}d^n x\\&=&4 f_{[AB]}A^\r \6^s_{\mu_1...\mu_{s-1}[\r}\phi^A_{\nu_s]\nu_1...\nu_{s-1}}Y^{(s-1)}\left(\6^{(s-1)\mu_1...\mu_{s-1}}\phi^{B\nu_1...\nu_s}\right)d^n x\nonumber\quad.\end{eqnarray} 
The result can be obtained by making three integrations by parts and by using the following relation to make appear $\d$-exact expressions:
\begin{eqnarray}
K^{A\phantom{\mu_1\nu_1|...|\mu_{s-1}\nu_{s-1}|}\mu_{s-1}}_{\phantom{A}\mu_1\nu_1|...|\mu_{s-1}\nu_{s-1}|\phantom{\mu_{s-1}}\nu_s}&=& 2^{s-1}Y^{(s-2)}(\6^{s\s}_{\phantom{s\s}\mu_1...\mu_{s-2}[\s}\phi^A_{\nu_{s-1}]\nu_1...\nu_{s-2}\nu_s}-\6^{s}_{\nu_s\mu_1...\mu_{s-2}[\s}\phi^{A\s}_{\phantom{A\s}\nu_{s-1}]\nu_1...\nu_{s-2}})\quad.\nonumber\\&&\label{sprelk}\end{eqnarray} First, we can integrate by parts the derivative $\6_{\mu_{s-1}}$ of $\phi^A$. After applying the above relation on the term where $\6_{\mu_{s-1}}$ acts on $\phi^B$, the rest involves a derivative $\6^{\nu_s}$ on $\phi^B$. The second step is to integrate by parts that derivative. Once again, Eq.(\ref{sprelk}) can be applied on the term where $\phi^A$ bears the derivative $\6^{\nu_s}$ in such a way as to exchange it with the index $\nu_{s-1}$. The rest thus involves the derivative $\6_{\nu_{s-1}}$, which can finally be integrated by parts, which make appear a Faraday term and a $\d$-exact term. The three terms involving derivatives of $A_\mu$ are antisymmetrized thanks to the structure of the spin-$s$ factors, and are thus the wanted Faraday writing. Let us exhibit the result:
\begin{eqnarray}a_{0,2s-2}=-f_{[AB]}\ct^{AB}d^n x+\d(...)+d(...)\end{eqnarray}where
\begin{eqnarray}\ct^{AB}&=&F^{\r\s}Y^{(s-1)}\left(\6^{s-1}_{\mu_1...\mu_{s-1}}\phi^A_{\nu_1...\nu_{s-1}\r}\right)\6^{(s-1)\mu_1...\mu_{s-1}}
\phi^{B\nu_1...\nu_{s-1}}_{\phantom{B\nu_1...\nu_{s-1}}\s}\nonumber\\&&+2F_{\mu_{s-1}}^{\phantom{\mu_{s-1}}\r}\6^{s-1}_{\mu_1...\mu_{s-2}[\r}\phi^A_{\nu_s]\nu_1...\nu_{s-1}}Y^{(s-1)}\left(\6^{(s-1)\mu_1...\mu_{s-1}}\phi^{|B|\nu_1...\nu_{s-1}\nu_s}\right)\nonumber\\&& + 2 F^{\r\nu_s} Y^{(s-2)}\left(\6^{s-1}_{\mu_1...\mu_{s-2}[\r}\phi^A_{\nu_s]\nu_1...\nu_{s-1}}\right)\6^{(s-1)\mu_1...\mu_{s-2}[\mu_{s-1}}\phi^{|B|\nu_{s-1}]\nu_1...\nu_{s-2}}_{\phantom{|B|\nu_{s-1}]\nu_1...\nu_{s-2}}\mu_{s-1}}\nonumber\\&&+2F_{\nu_{s-1}}^{\phantom{\nu_{s_1}}\r} \6^{s-1}_{\mu_1...\mu_{s-2}[\r}\phi^{A\phantom{A\nu_1...\nu_{s-2}}\nu_s}_{\nu_s]\nu_1...\nu_{s-2}}Y^{(s-2)}\left(\6^{(s-1)\mu_1...\mu_{s-2}[\mu_{s-1}}\phi^{|B|\nu_{s-1}]\nu_1...\nu_{s-2}}_{\phantom{|B|\nu_{s-1}]\nu_1...\nu_{s-2}}\mu_{s-1}}\right)\nonumber\quad.\end{eqnarray}

\subsection{Exhaustive list of interactions $1-s-s$}

We have determined the only nonabelian cubic interaction. We can then consider the results obtained in~\cite{Metsaev:2005ar,Metsaev:2007rn}
using a powerful light-cone method. 
We learn from the work \cite{Metsaev:2005ar}
that there exist only \textit{two} possible
cubic couplings between one spin-$1$ and two spin-$s$ fields. 
The first coupling involves $2s-1$ derivatives in the cubic vertex 
whereas the other involves $2s+1$ derivatives. 
Therefore we conclude that the first coupling corresponds to 
the nonabelian deformation obtained in the previous subsection. 
The other one simply is the Born-Infeld-like coupling
\begin{eqnarray}
 \stackrel{(3)}{\cl}&=&g_{[AB]}\,F^{\r\s}\,
\eta^{\l\t}\,K^A_{\mu_1\nu_1|...|\mu_{s-1}\nu_{s-1}|\r\l}
\;K_{\s\t|}^{B~\;\,\mu_1\nu_1|...|\mu_{s-1}\nu_{s-1}} \quad,
\end{eqnarray}
which is strictly invariant under the Abelian gauge transformations. 

\section{Consistent cubic deformations of the type $2-s-s$}\label{sec:2ss}

Let us now consider the similar problem of a $2-s-s$ cubic coupling. As for the $1-s-s$ study, we are not sure if a consistent deformation could begin at polynomial degree greater than 3, but we can determine every possible cubic couplings, which are described in the antifield formalism by the first order term of the generator $\stackrel{(1)}{w}=a_0+a_1+a_2$, that has to satisfy Eqs(\ref{eqagh2})-(\ref{eqagh0}). Once again, a short list of candidates $a_2$ can be established and only two expressions are related to an $a_1$, one with only one derivative in $a_2$ and $a_1$, the other with $2s-3$ derivatives. Furthermore, we have established a general proof that the candidate with one derivative is obstructed in any dimension and for any spin-$s$. The cubic vertex related to the second candidate is much more complicated to build explicitly, although our results of Chapter \ref{ch:FV} show that this must be the flat limit of the Fradkin--Vasiliev cubic coupling in $(A)dS$. This follows by the uniqueness of the 
nonabelian deformation that we have obtained. We have achieved the computation of the $2-s-s$ deformation in the particular cases of spin-3 and 4 only, the former is the unique $2-3-3$ vertex exposed in Chapter \ref{ch:int23}, the latter is presented below. However, we have obtained the deformation of the gauge algebra and gauge transformations in the general case with arbitrary $s$.

\subsection{Determination of $a_2$}

We do not consider a spin-2 family index in this computation, since no relation could arise about it. The spin-$s$ fields bear the usual capital letters.
The building blocks are thus: \begin{itemize}\item The $antigh\ 2$ antifields $C^{*\m}$ and $C^{*A\mu_1...\mu_{s-1}}$ 
\item The spin-2 ghosts $C_\m$ and the ghost tensor $\6_{[\m}C_{\n]}$
\item The spin-$s$ ghosts $C^A_{\mu_1...\mu_{s-1}}$ and the ghost tensors $U^{(j)A}_{\m_1\n_1|...|\m_j\n_j|\n_{j+1}...\n_{s-1}}$ , $j\leqslant s-1$\end{itemize} The possible $a_2$ terms split into two categories: those 
proportional to $C^{*A\mu_1...\mu_{s-1}}$ and those proportional to 
$C^{*\mu}$.

Let us study the first category: $C^{*A\mu_1...\mu_{s-1}}$ carries $(s-1)$ symmetric indices. The spin-2 ghost tensor bears at most 2.
Since no traces can be made, the spin-4 ghost tensor can bear at most $(s+1)$ indices. However,
$U^{(2)A}_{\mu_1\nu1|\mu_2\nu_2|\nu_3...\nu_{s-1}}$ bears two antisymmetric 
pairs. One could be contracted with $\6_{[\a}C_{\b]}$ but the other cannot be contracted with the symmetric antifields. The only possible combination involving $\6_{[\a}C_{\b]}$ is thus 
\begin{eqnarray}f_{AB}C^{*A\mu_1...\mu_{s-1}}
\6_{[\mu_1}C_{\a]}C^{B\a}_{\phantom{B\a}\mu_2...\mu_{s-1}}d^n x\quad .\label{2ssa21}\end{eqnarray}
If we consider the undifferentiated $C_\a$, the only possibility is:
\begin{eqnarray}g_{AB}C^{*A\mu_1...\mu_{s-1}}C^\a U^{(1)A}_{\a\mu_1|\mu_2...\mu_{s-1}}
d^n x\quad.\label{2ssa22}\end{eqnarray}
For the second category, the structure has to be $C^* U^{(i)A} U^{(j)B}$. One can assume, for example, that $i\leqslant j$. Since $C^{*\m}$ bears one index, $U^{(i)A}$ bears $i+s-1$ and no traces can be taken, we find that $U^{(j)B}$ has to bear $i+s$ indices and thus $i=j-1$. This leaves us with a set of 
candidates: \begin{eqnarray}a_{2,2j-1}=l_{AB}C^{*\a}U^{(j-1)A\mu_1\nu_1|...|
\mu_{j-1}\nu_{j-1}|\nu_j...\nu_{s-1}}U^{(j)B}_{\mu_1\nu_1|...|
\mu_{j-1}\nu_{j-1}|\a\nu_j|\nu_{j+1}...\nu_{s-1}}d^n x\quad,\quad j\leqslant s-1\ \label{2ssa23} .\end{eqnarray}

\subsection{Determination of $a_1$}

The two possible $a_2$ terms given in Eqs(\ref{2ssa21})-(\ref{2ssa22}) are not independent modulo $d$ and thus have to be considered together, we call their sum $a_{2,1}$. It has to satisfy Eq.(\ref{eqagh1}). The variation $\d a_{2,1}$ reads:
\begin{eqnarray}\d a_{2,1}&=&d(...) + s\phi^{*A\mu_1...\mu_s}\6_{\mu_s}\Big[g_{AB}C^\a U^{(1)B}_{\a\mu_1|\mu_2...\mu_{s-1}}+f_{AB}
\6_{[\mu_1}C_{\a]}C^{B\a}_{\phantom{B\a}\mu_2...\mu_{s-1}}\Big]d^n x\quad.\end{eqnarray} The derivative of the expression between square brackets, symmetrized over all the $\mu$ indices, has to be $\g$-exact. The first term of this expression contains the term $g_{AB}\6_{[\mu_s}C_{\a]}U^{(1)B\a}_{\phantom{(1)B\a}\mu_1|\mu_2...\mu_{s-1}}$, that constitutes an obstruction to the existence of $a_1$. The non $\g$-exact part of the second expression can be easily computed by using the definition of the differential $D$ in Eq.(\ref{defD}): $\6_{(\mu_s}C^B_{\mu_2...\mu_{s-1})\a}=\g(...)-\frac{2}{s}U^{(1)B}_{\a(\mu_s|\mu_2...\mu_{s-1})}$. The unwanted term thus have the same structure as the first one, they cancel if $f_{AB}=\frac{s}{2}g_{AB}$. Let us provide the final expression of $a_{2,1}$, to which we have added a $\g$-exact term, and the result for $a_{1,1}$:
\begin{eqnarray}a_{2,1}=g_{AB}C^{*A\mu_1...\mu_{s-1}}\left[C^\a \6_{[\a}C^B_{\mu_1]\mu_2...\mu_{s-1}} +\frac{s}{2}\6_{[\mu_1}C_{\a]}C^{B\a}_{\phantom{B\a}\mu_2...\mu_{s-1}}\right]d^n x\ \end{eqnarray} and
\begin{eqnarray}a_{1,1}&=&g_{AB}\phi^{*A\m_1...\mu_s}\left[\frac{s}{2}h_{\m_s}^{\phantom{\m_s}\a}\6_{[\a} C^B_{\m_1]\m_2...\m_{s-1}}-\frac{s}{s-1}C^\a\6_{[\a}\phi^B_{\m_1]\m_2...\m_s}\right.\nonumber\\&&\left.\qquad\qquad\quad\;\;\;+\frac{s^2}{2}\6_{[\m_1}h_{\a]\m_s}C^{B\a}_{\phantom{B\a}\m_2...\m_s}-\frac{s}{2}\6_{[\m_1} C_{\a]}\phi^{B\a}_{\phantom{B\a}\m_2...\m_s}\right]\nonumber\\&&+\frac{s(s-2)}{4(n+2s-6)}g_{AB}\6_\s\phi'^{*A\m_3...\m_s}C^\a\phi'^B_{\a\m_3...\m_{s-1}}\quad.\end{eqnarray} 
It remains now to check if the candidates $a_{2,2j-1}$, presented in Eq.(\ref{2ssa23}), satisfy Eq.(\ref{eqagh1}). Using once again the definition of $D$ (Eq.(\ref{defD})), one gets for $\d a_{2,2j-1}$ an expression of the structure:
\begin{eqnarray}\d a_{2,2j-1}=d(...)+\g(...)+l_{AB}h^* U^{(j)A}U^{(j)B}+l_{AB}h^* 
U^{(j-1)A}U^{(j+1)B}\quad.\end{eqnarray} The first nontrivial term vanishes upon imposing 
$l_{AB}=l_{(AB)}$, but the second cannot be removed unless $j=s-1$. The only candidate that survives is thus the one with the highest number of derivatives:
\begin{eqnarray}
a_{2,2s-3}&=&l_{(AB)}C^{*\a}U^{(s-2)A\mu_1\nu_1|...|\mu_{s-2}\nu_{s-2}|\nu_{s-1}}U^{(s-1)B}_{\mu_1\nu_1|...|\mu_{s-2}\nu_{s-2}|\a\nu_{s-1}}d^n x +\g(...)\nonumber\\
&=&l_{AB}C^{*\a}\6^{(s-2)\mu_1...\mu_{s-2}}C^{A\nu_1...\nu_{s-1}}Y^{(s-2)}(\6^{s-1}_{\mu_1...\mu_{s-2}[\a} C^B_{\nu_{s-1}]\nu_1...\nu_{s-2}})d^n x\quad. \label{a22ss}\end{eqnarray} The corresponding $a_1$ is
\begin{eqnarray}
a_{1,2s-3}&=&l_{(AB)}h^{*\a}_{\phantom{*\a}\b}\Big[\6^{(s-2)\mu_1...\mu_{s-2}}\phi^{A\nu_1...\nu_{s-1}\b}Y^{(s-2)}(\6^{s-1}_{\mu_1...\mu_{s-2}[\a}C^B_{\nu_{s-1}]\nu_1...\nu_{s-2}})\nonumber\\&&-2 Y^{(s-2)}(\6^{(s-2)\mu_1...\mu_{s-2}}C^{A\nu_1...\nu_{s-1}})\6^{s-1}_{\mu_1...\mu_{s-2}[\a}\phi^B_{\nu_{s-1}]\nu_1...\nu_{s-2}\b}\Big]d^n x+\bar{a}_{1,2s-3}\ .
\end{eqnarray}

\subsection{Inconsistency of the candidate with one derivative}

We have checked that the candidate $a_{1,1}$ is not related to any vertex. To do so, we have used the same method as for the other computations of $a_0$. First, there is no homogeneous part $\bar{a}_{1,1}$ that could be added to $a_{1,1}$, because a nontrivial $\g$-closed tensor contains at least two derivatives. Then, we have to consider the most general candidate for $a_0$ and to test it. Just as for the $1-s-s$ case with zero derivative, it is found that there are exactly 55 possible different monomials involving a spin-2 field, two derivatives and two spin-$s$ fields. This is due to the double tracelessness of the spin-$s$ field: at least $s-4$ indices of both fields are contracted, which leaves us with two sets of terms of the structure $h^{}_{..}\6^2_{..}\phi^A_{....(\mu)}\phi^{B\phantom{..}(\mu)}_{....}$ and $h^{}_{..}\6_{.}\phi^A_{....(\mu)}\6_{.}\phi^{B\phantom{..}(\mu)}_{....}$, where $(\mu)\equiv\mu_1...\mu_{s-4}$. Both sets have the same number of elements $\forall s\geqslant 4$: the first contains 28 terms, the second contains 27.
These 55 terms are then combined in an expression $a_{0,test}$. Finally, the computation of $\d a_{1,1}+\g a_{0,test}$ can be made. This is obviously tedious by hand, so that we again resorted to FORM. This computation is similar to the one that we had made before in the $2-3-3$ case, and that we have presented in section \ref{deflag233}. That one involved only 49 terms, but the proof can be easily extended to the $2-4-4$ case. Then, the difficult point was the arbitrary number of indices of an arbitrary spin-$s$ field. We had to consider a multiindex for the $s-4$ indices that are always contracted (just like the $(\mu)$ that we wrote above). This is not as easy as it seems, because the actions of $\d$ and $\g$ involve all the indices, so that we had to compute the different expressions and to find ``effective'' coefficients depending on the spin. This was quite technical, but did not modify the result obtained in the $2-4-4$ case: it is impossible to match the coefficients, the obstruction we meet is impossible to circumvent and these candidates are definitely rejected, for all value of the spin $s$. 

\subsection{Exhaustive list of cubic $2-s-s$ couplings}

Using the results of \cite{Metsaev:2005ar}, we learn that there exist
only \textit{three} cubic couplings with the $2-s-s$ configuration. They involve
a total number of derivatives in the vertex being respectively 
$2s+2$, $2s$ and $2s-2\,$. Moreover, it is indicated 
\cite{Metsaev:2005ar} that the $\,2s\,$-derivative coupling only 
exists in dimension $n>4\,$. 
{}From our results of the last subsection, 
we conclude that the last coupling is the nonabelian coupling
with $2s-2$ derivatives. The coupling with $2s+2$ derivatives is 
simply the strictly-invariant Born-Infeld-like vertex
\begin{eqnarray}
\stackrel{(3)}{\cl}_{BI}&=& t_{(AB)}\,K^{\a\b|\c\d}\,
\;K_{\a\b|}^{A~\;\,\mu_1\nu_1|...|\mu_{s-1}\nu_{s-1}} 
\;K^B_{\c\d|\mu_1\nu_1|...|\mu_{s-1}\nu_{s-1}} \quad,
\end{eqnarray}
whereas the vertex with $2s$ derivatives is given by 
\begin{eqnarray}
\stackrel{(3)}{\cl}_{2s}&=& u_{(AB)}\;
\delta^{[\mu\n\r\s\l]}_{[\a\b\c\d\e]}\;
h^{\a}_{\phantom{\a}\m}
\;K^{A\,\b\c|~~~\!|\mu_1\nu_1|...|\mu_{s-2}\nu_{s-2}}_{\phantom{A\,\b\c|}\n\r}
\;K^{B \,\d\e|}_{\phantom{B \,\d\e|}\s\l|\mu_1\nu_1|...|\mu_{s-2}\nu_{s-2}}\quad.
\end{eqnarray}
It is easy to see that this vertex is not identically zero
and is gauge-invariant under the Abelian transformations, up
to a total derivative. 

\subsection{Computation of the unique $2-4-4$ vertex}
\label{sec:244}

The general computation of the unique nonabelian $2-s-s$ vertex is still to be done, but we have achieved the particular case of spin $4$. Let us recall the expressions of $a_2$ and $a_1$ with $s=4$ (we have renamed the internal coefficient $f_{(AB)}$):
\begin{eqnarray}
a_{2,5}&=&f_{AB}C^{*\m_3}\6^{2\mu_1\mu_2}C^{A\nu_1\nu_2\nu_{3}}Y^{(2)}(\6^{3}_{\mu_1\mu_2\mu_3} C^B_{\nu_1\nu_2\nu_3})d^n x\quad. \label{a2244}
\end{eqnarray} 
\begin{eqnarray}
a_{1,5}&=f_{(AB)}h^{*\a}_{\phantom{*\a}\b}\Big[&\6^{2\mu_1\mu_2}\phi^{A\nu_1\nu_2\nu_3\b}Y^{(2)}(\6^{3}_{\mu_1\mu_2[\a}C^B_{\nu_3]\nu_1\nu_2})\nonumber\\&&-2 Y^{(2)}(\6^{2\mu_1\mu_{2}}C^{A\nu_1\nu_2\nu_3})\6^{3}_{\mu_1\mu_2[\a}\phi^{B\phantom{\nu_3\nu_1\nu}\b}_{\nu_3]\nu_1\nu_2}\Big]d^n x+\bar{a}_{1,5}\quad.
\end{eqnarray}
This time, we have made an hypothesis on the possible structure of the vertex, which is of course based on the Fradkin--Vasiliev approach. We know that the ``quasi-minimal'' perturbation of the $(A)dS$ Fronsdal Lagrangian for spin-2 and spin-4 involves only terms proportional to the Riemann (or Weyl) tensor, and that the Minkowski limit of that expression is its part that involves the maximal number of derivatives. That is why we have written the most general expression involving the undifferentiated Weyl tensor and four derivatives acting on the two spin-2 fields, for a total of 6 derivatives that match with the 5 contained in $a_{2,5}$ and $a_{1,5}$. Modulo $d$ and $\d$, we have established a list of 64 such terms. At that stage, it is crucial to keep in mind that the vertex $a_{0,5}$ is defined modulo $\d$ (or modulo redefinitions of the fields, which is the same). We had to consider the most general expression $c_1$ linear in an undifferentiated antifield $h^*$ or $\phi^{*A}$ and quadratic in the fields. There are 203 terms of the structure $h^*\6^4\{\phi^A\phi^B\}$. For the other set, we have made the assumption that $h$ appears only through the Weyl tensor, and there are 21 terms of the structure $\phi^{*A}\6^2\{w \phi^B\}$. Finally, the most general candidate for $\bar{a}_{1,test}$ has to be considered, which could be reduced, modulo $d$ and $\d$, to a set of 26 terms. The system of equations, obtained by computing $\d (\tilde{a}_{1,5}+\bar{a}_{1,test})+\g(a_{0,test}+\d c_1)$, and then by applying variational derivatives with respect to $C_\mu$ and $C^A_{\mu_1...\mu_{s-1}}$, contains thousands of equations, most of them redundant, for over 300 coefficients bearing two internal indices. We could solve this system of equations with the software and found a solution in which all the coefficients are symmetric in the internal indices.
An interesting fact is that $\bar{a}_1$ consists only of terms linear in the Riemann tensor:
\begin{eqnarray}\bar{a}_{1,5}=\frac{4}{D+2}\,
f_{AB}\phi^{*A\a}_{\phantom{*A\a}\b}\6^\t R^{\m\n|\a\s} 
U^B_{\m\n|\b\s|\t}d^Dx-2f_{AB}\phi^{*A\m\r}_{\phantom{*A\m\r}\a\b}
\6^\t R^{\a\n|\b\s} U^B_{\m\n|\r\s|\t}d^n x\nonumber\quad.
\end{eqnarray}
Here is the vertex that we obtained:
\begin{eqnarray}
a_{0,5}&=&f_{AB}w_{\m\n|\r\s}\Big[\,
 \frac{1}{2}\,\6^{\m\r\a}\phi'^{A\n\b}\6_\a\phi'^{B\s}_{\phantom{,B\s}\b}
-\frac{1}{3}\,\6^{\m\r\a}\phi^{A\n\b\c\d}\6_\a\phi^{B\s}_{\phantom{B\s}\b\c\d}
+\frac{1}{4}\,\6^{\m\r\a}\phi^{A\n\b\c\d}\6_\b\phi^{B\s}_{\phantom{B\n}\a\c\d}
\nonumber \\ &&
+\frac{3}{4}\,\6^{\m\a\b}\phi'^{A\n\r}\6_\a\phi'^{B\s}_{\phantom{,B\s}\b}
+\frac{3}{4}\,\6^{\m\a\b}\phi^{A\n\r\c\d}\6_\a\phi^{B\s}_{\phantom{B\s}\b\c\d}
-\frac{3}{2}\,\6^{\m\a\b}\phi^{A\n\r}_{\phantom{A\n\r}\b\c}\6_\a\phi'^{B\s\g}
\nonumber \\&&
-\frac{1}{2}\,
\6^{\m}_{\phantom{\m}\b\c}\phi^{A\n\r\a}_{\phantom{A\n\r\a}\d}\6_\a
\phi^{B\s\b\c\d}-\frac{3}{4}\,
\6^{\m\a\b}\phi^{A\s}_{\phantom{A\s}\b\c\d}\6_\a\phi^{B\n\r\c\d}
+\frac{3}{2}\,\6^{\m\a\b}\phi'^{A\s\g}\6_\a\phi^{B\n\r}_{\phantom{B\n\r}\b\c}
\nonumber \\&&
-\,\6^{\m}_{\phantom{\m}\b\c}\phi'^{A\s\a}\6_\a\phi^{B\n\r\b\c}
+\frac{1}{2}\,
\6^{\m}_{\phantom{\m}\b\c}\phi^{A\s\a\c}_{\phantom{A\s\a\c}\d}\6_\a
\phi^{B\n\r\b\d}-\frac{1}{2}\,\6_{\a\b\c}\phi^{A\m\r\a\d}\6_\d\phi^{B\n\s\b\c}
\nonumber \\&&
+\frac{1}{2}\,
\6_{\a\b\c}\phi^{A\m\r\a\t}\6^\b\phi^{B\n\s\c}_{\phantom{B\n\s\c}\t}
+\frac{1}{8}\,\6^{\a\b}\phi'^{A\m\r}\6_{\a\b}\phi'^{B\n\s}
+\frac{3}{8}\,
\6^{\a\b}\phi^{A\m\r\c\d}\6_{\a\b}\phi^{B\n\s}_{\phantom{B\n\s}\c\d}
\nonumber \\&&
-\frac{1}{2}\,\6^{\a}_{\phantom{\a}\b}\phi^{A\m\r\b\c}\6_{\a\c}\phi'^{B\n\s}
+\frac{1}{2}\,
\6_{\a\b}\phi^{A\m\r\b\t}\6^{\a\c}\phi^{B\n\s}_{\phantom{B\n\s}\c\t}
-\frac{3}{4}\,
\6^{\a\b}\phi^{A\m\r\c\d}\6_{\a\c}\phi^{B\n\s}_{\phantom{B\n\s}\b\d}
\nonumber\\&&
+\frac{1}{4}\,\6_{\a\b}\phi^{A\m\r\c\d}\6_{\c\d}\phi^{B\n\s\a\b}\;
\Big]+\d(...)\quad.
\label{v244}
\end{eqnarray}
We have not provided the expression of the redefinitions of the fields needed for $a_1$ and $a_0$ to match, since it is rather long and not very useful (but the computation provides it). 
This $2-4-4$ vertex should correspond to the flat limit of the corresponding Fradkin--Vasiliev 
vertex. The uniqueness of the former can be used to prove the uniqueness of 
the latter, as we did explicitly in the $2-3-3$ case, and as we already mentioned
previously.

\section{General considerations about $a_2$}

To conclude this chapter, let us provide general arguments that should simplify the classification of the nonabelian deformations in any case that we have not considered. These results will be published elsewhere \cite{IP2}.For any cubic configuration $s-s'-s''$, with $s\leqslant s'\leqslant s''$, there are not many possibilities of building consistent $a_2$, and only some of them are related to a consistent $a_1$. As usual, a cubic $a_2$ can always be written in the form $$a_2=C^* U^{(i)} U^{(j)}d^n x+\g(...)\quad,$$ where the $U^{(i)}$ are non-$\g$-exact ghost tensors.

The first thing that can be said is that there are no nonabelian deformations if $s''\geqslant s+s'$. For example, there is no way of building a $1-1-s$ deformation if $s\geqslant 2$, or a $2-2-s$ deformation with $s\geqslant 4$. This is due to the number of symmetric indices that are present in those expressions. More precisely, let us consider a product of two ghost tensors $U^{(i)}$ and $U^{(j)}$, respectively of spin $s_1$ and $s_2$ (with $s_1<s_2$). We are interested in the minimum number of free indices of that product (in other words, the maximum number of contracted indices). \begin{itemize}\item If $i\leqslant j$, all of the indices of $U^{(i)}$ can be contracted with $s_1+i-1$ indices of $U^{(j)}$. Let us visualize this in terms of Young diagrams:
\begin{eqnarray}U^{(i)}: \begin{picture}(60,0)(0,0)\multiframe(0,0)(10,0){1}(50,10){$s_1-1$}\multiframe(0,-10.5)(10,0){1}(20,10){$i$}\end{picture},\ U^{(j)}:\begin{picture}(70,0)(0,0)\multiframe(0,0)(10,0){1}(60,10){$s_2-1$}\multiframe(0,-10.5)(10,0){1}(40,10){$j$}\end{picture}\nonumber\end{eqnarray}\begin{eqnarray} &\Rightarrow  {\textrm{Maximal contraction}}:&\bullet\ \begin{picture}(36,0)(0,0)\multiframe(0,-2)(10,0){1}(35,10){$s_2-s_1$}\end{picture}\otimes\begin{picture}(30,0)(0,0)\multiframe(0,-2)(10,0){1}(30,10){$j-i$}\end{picture}\textrm{ if $j<s_1$}\nonumber\\&&\bullet\ \bigoplus_a\begin{picture}(95,0)(0,0)\multiframe(0,1)(10,0){1}(90,10){$s_2-s_1+j-i-a$}\multiframe(0,-9.5)(10,0){1}(20,10){$a$}\end{picture}\textrm{ if $j-i\geqslant a\geqslant j-s_1-1>0$ }\nonumber\quad.\end{eqnarray}
Since $U^{(j)}$ bears $s_2+j-1$ indices, the minimum number of free indices is $s_2-s_1+j-i$. The indices can be symmetrized if $j<s_1$ since there is a component \begin{picture}(85,0)(0,0)\multiframe(0,-2)(10,0){1}(80,10){$s_2-s_1+j-i$}\end{picture} in the tensor product. If $j\geqslant s_1$, no contraction of the two tensors $U$ can be symmetrized and thus no Lorentz invariant $a_2$ can be built.
\item If $j<i<s_1<s_2$, let us visualize the ghost tensors: 
\begin{eqnarray}U^{(i)}: \begin{picture}(80,0)(0,0)\multiframe(0,0)(10,0){1}(75,10){$s_1-1$}\multiframe(0,-10.5)(10,0){1}(20,10){$j$}\multiframe(20.5,-10.5)(10,0){1}(40,10){$i-j$}\end{picture},\ U^{(j)}:\begin{picture}(120,0)(0,0)\multiframe(0,0)(10,0){1}(75,10){$s_1-1$}\multiframe(75.5,0)(10,0){1}(40,10){$s_2-s_1$}\multiframe(0,-10.5)(10,0){1}(20,10){$j$}\end{picture}\nonumber\quad.\end{eqnarray}
The maximal contraction is obtained by contracting the $s_1-1$ boxes and the $j$ boxes, which leaves one with a product: \begin{picture}(36,0)(0,0)\multiframe(0,-2)(10,0){1}(35,10){$s_2-s_1$}\end{picture} $\otimes$ \begin{picture}(30,0)(0,0)\multiframe(0,-2)(10,0){1}(30,10){$i-j$}\end{picture} , which always involves a totally symmetric component. Explicitly, this reads:
\begin{eqnarray}
U^{(i)\mu_1\nu_1|...|\mu_j\nu_j|\mu_{j+1}\b_1|...|\mu_i\b_{i-j}|\mu_{i+1}...\mu_{s_1-1}}U^{(j)}_{\mu_1\nu_1|...|\mu_j\nu_j|\mu_{j+1}...\mu_{s_2-1}}\quad.
\end{eqnarray} 
The $\b$ indices are free and there are $s_2-s_1$ free $\mu$ indices. The minimum number in this case is thus $s_2-s_1+i-j$.\end{itemize} The two cases can be gathered as $N_{min}=s_2-s_1+|i-j|$. 

We are also interested in the maximum number of free indices that can be symmetrized in a product $U^{(i)}U^{(j)}$. \begin{itemize} \item If $i\leqslant j<s_1$, $j$ pairs have to be contracted:\begin{eqnarray}U^{(i)}:\begin{picture}(95,0)(0,0)\multiframe(0,0)(10,0){1}(30,10){$j$}\multiframe(30.5,0)(10,0){1}(60,10){$s_1-j-1$}\multiframe(0,-10.5)(10,0){1}(20,10){$i$}\end{picture},\ U^{(j)}:\begin{picture}(115,0)(0,0)\multiframe(0,0)(10,0){1}(110,10){$s_2-1$}\multiframe(0,-10.5)(10,0){1}(30,10){$j$}\end{picture}\quad.\end{eqnarray}If one contracts less than $j$ pairs, some indices remain in the second line of $U^{(j)}$ and the result does not contain a totally symmetric component. This leaves us with $s_1+s_2+i-j-2$ free indices. \item If $i\geqslant j$, on the same way, $i$ pairs have to be contracted, leaving $s_1+s_2+j-i-2$ free indices.\end{itemize} Thus, the maximum number is $N_{max}=s_1+s_2-|i-j|-2$.

Let us now consider a candidate for $a_2$, for a configuration $s-s'-s''$ with $s\leqslant s'\leqslant s''$. Three cases have to be studied, related to the spin of the antifield. In the case of a spin-$s$ antifield $C^{*\mu_1...\mu_{s-1}}$, the minimum number of free indices in the product $U^{(i)}U^{(j)}$ is $s''-s'+|i-j|$. In order for $a_2$ to be Lorentz-invariant, every index must be contracted, hence the former number must be lower or equal than the number of indices of the antifield. We thus obtain the relation: $s''-s'+|i-j|\leqslant s-1$. In the case of a spin $s'$ antifield, the same argument can be applied, it leads to the relation $s''-s+|i-j|\leqslant s'-1$, which is the same as the first one. Finally, in the case of a spin $s''$ antifield, the minimal condition $s'-s+|i-j|\leqslant s''-1$ is always satisfied, since $i<s$ and $j<s$, which imply $|i-j|-s<0$, in order for the contraction to be symmetrizable. On the other hand, in this case, we have to consider the fact that the maximum number of free symmetrizable indices must be greater or equal than the number of indices of the antifield: $s+s'-|i-j|-2\geqslant s''-1$, and we obtain once again the same condition. Thus for any combination of the fields, the spins have to satisfy the inequality: \begin{eqnarray}s+s'-s''>|i-j|\geqslant0\quad.\end{eqnarray} This shows the announced property. Furthermore, it provides an upper bound on the difference between the numbers of derivatives in the two ghost tensors.

Then, if we want to build Lorentz-invariant expressions, the total number of indices has to be even. For an antifield of spin $s_3$ and ghost tensors $U^{(i)}$ of spin $s_1$ and $U^{(j)}$ of spin $s_2$, the numbers of indices are $s_3-1$, $s_1+i-1$ and $s_2+j-1$, for a total of $s_1+s_2+s_3+i+j-3$. Thus, we find that, for a configuration $s\leqslant s'\leqslant s''$: \begin{eqnarray}s+s'+s''+i+j\equiv 1(mod\ 2)\quad.\end{eqnarray}

Finally, let us emphasize that the total number of derivatives $i+j$ is bounded. As we already mentioned before, $i$ and $j$ must be strictly lower than the spins of the two ghost tensors in order for the free indices to be symmetrizable. If $U^{(i)}$ is of spin $s_1$ and $U^{(j)}$ is of spin $s_2$ with $s_1\leqslant s_2$, then $i\leqslant s_1-1$ and $j\leqslant s_1-1$. Thus, we obtain the condition $i+j\leqslant 2s_1-2$. If we consider now a candidate for $a_2$, this condition immediately tells that \begin{eqnarray}i+j\leqslant 2s'-2\quad.\end{eqnarray} 
More, precisely, if the spin-$s$ antifield is considered, then the boundary is $2s'-2$. If either the spin $s'$ or $s''$ antifield is considered, the boundary is lower: $i+j\leqslant 2s-2$. 

Let us summarize these considerations in the following theorem:
\begin{theorem}\label{thma2}Given a cubic setup of fields with spins $s\leqslant s'\leqslant s''$, the possible Poincar\'e invariant $a_2$ terms are contractions of an undifferentiated $antigh\ 2$ antifield and of two ghost tensors, involving $i$ and $j$ derivatives. The spins and the numbers of derivatives have to satisfy the following properties:\begin{itemize}
\item $|i-j|<s+s'-s''$
\item $s+s'+s''+i+j$ is odd
\item In the case of a spin-$s$ antifield: $i+j\leqslant 2s'-2$\\ In the case of a spin $s'$ or $s''$ antifield: $i+j\leqslant 2s-2$
\end{itemize}
\end{theorem}

To end this chapter, let us show that the candidate with the highest number of derivatives $2s'-1$ always satisfies Eq.(\ref{eqagh1}). We do not provide the corresponding $a_1$ explicitly but the equation ensures that it exists. In the case of an even number of derivatives (in other words when the sum $s+s'+s''$ is odd), the candidate with $2s'-2$ derivatives reads: 
\begin{eqnarray}a_2&=&C^{\mu_1...\mu_{s-1}}U^{(s'-1)}_{\a_1\r_1|...|\a_\l\r_\l|\mu_1\r_{\l+1}|...|\m_{s'-\l-1}\r_{s'-1}}\nonumber\\&&\times U^{(s'-1)\a_1\r_1|...|\a_\l\r_\l|\phantom{\m_{s'-\l}}\r_{\l+1}|...|\phantom{\mu_{2s'-2\l-2}}\r_{s'-1}|}_{\,\phantom{(s'-1)\a_1\r_1|...|\a_\l\r_\l|}\m_{s'-\l}\phantom{\r_{\l+1}|...|}\m_{2s'-2\l-2}\phantom{\r_{s'-1}|}\m_{2s'-2\l-1}...\m_{s-1}}\,d^n x\quad,\end{eqnarray}
 where $\l=\frac{s'+s''-s-1}{2}$. In terms of Young diagrams, this contraction can be seen as follows: \begin{eqnarray}
C^*:\begin{picture}(125,0)(0,0)\multiframe(0,0)(10,0){1}(50,10){$s'-\l-1$}\multiframe(50.5,0)(10,0){1}(70,10){$s''-\l-1$}\end{picture} , U^{(s'-1)}_{s'}:\begin{picture}(75,0)(0,0)\multiframe(0,0)(10,0){1}(20,10){$\l$}\multiframe(20.5,0)(10,0){1}(49.5,10){$s'-\l-1$}\multiframe(0,-10.5)(10,0){1}(70,10){$s'-1$}\end{picture} , U^{(s'-1)}_{s''}:\begin{picture}(85,0)(0,0)\multiframe(0,0)(10,0){1}(20,10){$\l$}\multiframe(20.5,0)(10,0){1}(70,10){$s''-\l-1$}\multiframe(0,-10.5)(10,0){1}(70,10){$s'-1$}\end{picture}\quad.\end{eqnarray} 
The variation of this expression under delta takes the form: \begin{eqnarray}\nonumber\d a_2=d(...)+s\Big[\phi^{*\mu_1...\mu_s}-\frac{(s-1)(s-2)}{2(n+2s-6)}\eta^{(\m_1\m_2}\phi^*{}'{}^{\m_3...\m_{s-1})\m_s}\Big]\6_{\m_s}\Big[U^{(s'-1)}U^{(s'-1)}\Big]\quad.\end{eqnarray} The action of $\6_{\m_s}$ on the spin $s'$ tensor $U^{(s'-1)}$ is automatically $\g$-exact. The only possible traces that can be taken between the two tensors just create one more contracted antisymmetric pair, thanks to the relation $U^{(i)}_{[\m_1\n_1|\m_2]\n_2|...}=0$. The symmetric indices of the spin $s''$ tensor are thus all $\m$ indices contracted with the trace of $\phi^*$. Then, since the index $\m_s$ of the derivative is also contracted with the antifield, the relation $\g[ Y^{(i)}(\6^i_{\m_1...\m_i}\phi_{\n_1...\n_s})]\div Y^{(i)}(\6^{i+1}\6_{\m_1...\m_i(\n_{i+1}}C_{\n_{i+2}...\n_s)\n_1...\n_i})$, together with the definition of the $U^{(i)}$'s, ensures that $\d a_2$ is actually $\g$-exact modulo $d$. 
 
The case of an even sum $s+s'+s''$ is a bit more complicated. There are two possible terms, that have to be proportional in order for $a_1$ to exist: \begin{eqnarray}a_2&=&f C^{\mu_1...\mu_s}U^{(s'-1)}_{\a_1\r_1|...|\a_\l\r_\l|\mu_1\r_{\l+1}|...|\m_{s'-\l-1}\r_{s'-1}}\nonumber\\&&\times U^{(s'-2)\a_1\r_1|...|\a_\l\r_\l|\phantom{\m_{s'-\l}}\r_{\l+1}|...|\phantom{\mu_{2s'-2\l-3}}\r_{s'-2}|\r_{s'-1}}_{\,\phantom{(s'-2)\a_1\r_1|...|\a_\l\r_\l|}\m_{s'-\l}\phantom{\r_{\l+1}|...|}\m_{2s'-2\l-3}\phantom{\r_{s'-2}|\r_{s'-1}}\m_{2s'-2\l-2}...\m_{s-1}}\,d^n x\ \nonumber\\&&
+g C^{\mu_1...\mu_s}U^{(s'-2)}_{\a_1\r_1|...|\a_\l\r_\l|\mu_1\r_{\l+1}|...|\m_{s'-\l-2}\r_{s'-2}|\r_{s'-1}}\nonumber\\&&\ \,\times U^{(s'-1)\a_1\r_1|...|\a_\l\r_\l|\phantom{\m_{s'-\l}-1}\r_{\l+1}|...|\phantom{\mu_{2s'-2\l-2}}\r_{s'-1}|}_{\,\phantom{(s'-1)\a_1\r_1|...|\a_\l\r_\l|}\m_{s'-\l-1}\phantom{\r_{\l+1}|...|}\m_{2s'-2\l-3}\phantom{\r_{s'-1}|}\m_{2s'-2\l-2}...\m_{s-1}}\,d^n x\quad,\end{eqnarray} where $\l=(s'+s''-s-2)/2$. In terms of Young diagrams, these contractions read:\begin{eqnarray}&&
C^*:\begin{picture}(125,0)(0,0)\multiframe(0,0)(10,0){1}(50,10){$s'-\l-1$}\multiframe(50.5,0)(10,0){1}(70,10){$s''-\l-2$}\end{picture} , U^{(s'-1)}_{s'}:\begin{picture}(75,0)(0,0)\multiframe(0,0)(10,0){1}(20,10){$\l$}\multiframe(20.5,0)(10,0){1}(50,10){$s'-\l-1$}\multiframe(0,-10.5)(10,0){1}(60,10){$s'-1$}\multiframe(60.5,-10.5)(10,0){1}(10,10){1}\end{picture} , U^{(s'-2)}_{s''}:\begin{picture}(100,0)(0,0)\multiframe(0,0)(10,0){1}(20,10){$\l$}\multiframe(20.5,0)(10,0){1}(70,10){$s''-\l-2$}\multiframe(91,0)(10,0){1}(10,10){1}\multiframe(0,-10.5)(10,0){1}(60,10){$s'-1$}\end{picture}\nonumber\quad,\\ \textrm{and}&&\nonumber\\&& C^*: \begin{picture}(125,0)(0,0)\multiframe(0,0)(10,0){1}(50,10){$s'-\l-2$}\multiframe(50.5,0)(10,0){1}(70,10){$s''-\l-1$}\end{picture} , U^{(s'-2)}_{s'}:\begin{picture}(85,0)(0,0)\multiframe(0,-10.5)(10,0){1}(20,10){$\l$}\multiframe(20.5,-10.5)(10,0){1}(49.5,10){$s'-\l-2$}\multiframe(0,0)(10,0){1}(80,10){$s'-1$}\end{picture} , U^{(s'-1)}_{s''}:\begin{picture}(90,0)(0,0)\multiframe(0,0)(10,0){1}(20,10){$\l$}\multiframe(20.5,0)(10,0){1}(70,10){$s''-\l-1$}\multiframe(0,-10.5)(10,0){1}(80,10){$s'-1$}\end{picture}\quad.\nonumber
\end{eqnarray}
This time, the computation of $\d a_2$ consists of four terms. The term involving the derivative of the spin-$s'$ tensor $U^{(s'-1)}$ is automatically $\g$-exact, and the term where the spin-$s''$ tensor $U^{(s'-1)}$ is differentiated is $\g$-exact, thanks to the same arguments as for the odd case. On the other hand, the terms where the $U^{(s'-2)}$ tensors are differentiated are problematic. Fortunately, the non-$\g$-exact terms that appear are the same in the two expressions and the coefficients $f$ and $g$ can be fitted to obtain a $\g$-exact result. 

%%%%%%%%%%%%%%%%%%%%%%%%%%%%%%%%%%%%%%%%%%%%%%%%%%%%
\chapter{Second order computations}\label{ch:socomp}
%%%%%%%%%%%%%%%%%%%%%%%%%%%%%%%%%%%%%%%%%%%%%%%%%%%%

The computations made so far consisted in the determination of the solutions of the master equation at first order in perturbation. As we emphasized in Chapter \ref{ch:antidef}, the existence of a local first order solution does not imply the existence of a full solution, local at any order. Indeed, in the cases that we considered, for a given cubic configuration of spins and a given number of derivatives, there is either zero or one solution at first order in perturbation, which is satisfactory. However, some further restrictions can appear at higher order in perturbation. These restrictions appear as some new relations among the internal coefficients of the vertices. In the case where the internal coefficients vanish, the vertex can be obstructed. The purpose of this chapter is to compute the component of highest antifield number of the second order part of the master equation, for the different cubic deformations that have been considered previously. The motivation is to check one of the main features of higher spin theories: the necessity of considering every value of the spin and thus an arbitrary number of derivatives in order for a full consistent theory to exist. However, as far as we investigated, this condition might not be sufficient in Minkowski spacetime. For example, the Berends--Burgers--van Dam deformation \cite{Berends:1984wp} has been showed to be inconsistent when considered alone. We prove that it is still obstructed in arbitrary dimension when it is considered together with deformations involving spin-4 and spin-5 fields, thus invalidating the hopes expressed by Berends, Burgers and van Dam in \cite{Berends:1984rq}. Once again, this is due to the fact that the number of derivatives is a good grading in Minkowski spacetime: in accordance with the considerations that we made in Chapter \ref{ch:intmisc}, the number of derivatives in the possible deformations increases linearly with the spins involved. Thus, cubic terms involving higher values of the spin involve more derivatives and remain independent of the terms involving lower values when making second order computations. On the other hand, $(A)dS$ deformations are more likely to be part of a full consistent theory, since they contain terms with various numbers of derivatives, that could remove some obstructions. This is in agreement with the discussion of Section \ref{minlim}: a full consistent $(A)dS$ theory does not admit a flat limit. We can also emphasize that it is still possible that a full theory in Minkowski spacetime exists. For example, no obstructions appear on the $1-s-s$ and $2-s-s$ deformations in the computations that we achieve in the sequel. We can conjecture that cubic $s-s'-s'$ deformations, involving the maximum number of derivatives $2s'-1$, could exist and might be the first order of an expansion local at all orders. However, we show that the $2-3-3$ and $1-2-2$ deformations are not compatible with Einstein--Hilbert's theory. This was predictable, since no ``quasi-minimal'' procedure can be built in Minkowski spacetime. This indicates that there could be two types of spin-2 fields. Some of them being of the Einstein-Hilbert type, and strictly self-interacting, as was addressed in \cite{Boulanger:2000rq}. The possible Minkowski theory could correspond to the tensionless limit of string field theory, the non-EH spin-2 fields being part of the spectrum of the open string, while the EH fields are usually considered in string theory as being part of the spectrum of the closed string.

\section{The second order equation}

We have presented in Chapter \ref{ch:antidef} the general deformation scheme. A local deformation of a free theory described by the initial local generator $\stackrel{(0)}{w}$, satisfying the initial master equation $(\stackrel{(0)}{w},\stackrel{(0)}{w})d^n x=d(...)$, is an expansion $w=\stackrel{(0)}{w}+g\stackrel{(1)}{w}+g^2\stackrel{(2)}{w}+...$, that has to satisfy the master equation to all orders in the parameter $g$. The first order local solutions that we have considered are cubic, and their antifield expansion stops at 2: $\stackrel{(1)}{w}=a_0+a_1+a_2$. We know that the first order of the master equation decomposes into Eqs(\ref{eqagh2})-(\ref{eqagh0}).

Let us now consider the second order of the master equation for a local generator: \begin{eqnarray}2\left(\stackrel{(0)}{w},\stackrel{(2)}{w}\right)d^n x+\left(\stackrel{(1)}{w},\stackrel{(1)}{w}\right)d^n x=d(...)\quad\Longleftrightarrow\quad \exists\ e\ |\ \left(\stackrel{(1)}{w},\stackrel{(1)}{w}\right)d^n x=-\frac{1}{2}\bs\stackrel{(2)}{w}+de\quad.\label{sogen}\end{eqnarray} 
{\bf Remark:} We consider antibrackets of $n$-forms, which are just the antibrackets of their components and are thus functions. That is why we have introduced $d^n x$ factors in Eq.(\ref{sogen}). 

This equation can then be decomposed according to the antifield number. The second order deformation $\stackrel{(2)}{w}$ can contain an homogeneous part $\stackrel{(2)}{\bar{w}}$, satisfying $\bs\stackrel{(2)}{\bar{w}}=2 d\bar{e}$. The non trivial solutions belong to $H^{0,n}(\bs|d)$, just as $\stackrel{(1)}{w}$, and are thus the same as for the first order. 

Let us now consider the inhomogeneous equation. The components of the cubic $\stackrel{(1)}{w}$ have a definite field content: $a_2$ is linear in the $antigh\ 2$ antifields $C^*$ and quadratic in the ghosts $C$ or their derivatives; $a_1$ is linear in the $antigh\ 1$ antifields $\phi^*$, in the fields $\phi$ and in the ghosts $C$; and $a_0$ is cubic in the fields. Then, we know that the antibracket takes variational derivatives of its two arguments with respect to conjugate fields. For example, in the antibracket of an $a_2$ with another $a_2$, a $C^*$ and a $C$ are differentiated, leaving an $antigh\ 2$ object linear in $C^*$ and cubic in the ghosts. In the antibracket of an $a_2$ with an $a_1$, the only possibility is to differentiate the $C^*$ in $a_2$ and the ghost in $a_1$, leaving an $antigh\ 1$ result linear in $\phi^*$ and $\phi$ and quadratic in the ghosts. The antibracket of two $a_1$'s is also an $antigh\ 1$ expression, since one has to differentiate one of the antifields $\phi^*$ and one of the fields. Finally, the antibracket of an $a_1$ and an $a_0$ is an $antigh\ 0$ expression, cubic in the fields and linear in the ghosts. Thus, the top component of the inhomogeneous equation is at $antigh\ 2$, which requires that the expansion of the inhomogeneous part of the second order deformation, $\stackrel{(2)}{\tilde{w}}=-2(c_0+c_1+...)$, stops at most at $antigh\ 3$. Furthermore, it is quite obvious that an antibracket $(a_2,a_2)$ cannot contain $\d$-exact terms, because it is neither proportional to the antifields $\phi^*$ nor to the fields. This leads us to discard $\d c_3$ terms and to consider a second order deformation stopping at $antigh\ 2$. Finally, we obtain the following system of equations:\begin{eqnarray}&(a_2,a_2)=\g c_2+d e_2&\label{a2a2}\\&2(a_2,a_1)+(a_1,a_1)=\d c_2+\g c_1+ d e_1&\\& 2(a_1,a_0)=\d c_1+\g c_0+ d e_0&\quad.\end{eqnarray}
The resolution of the whole system of equations would of course provide the expression of the second order solution of the deformation, whose inhomogeneous part is quartic when the first order is cubic. We have not addressed the resolution of the $antigh$ 0 and 1 equations. On the other hand, checking if $(a_2,a_2)$ is $\g$-exact modulo $d$ can be done rather quickly and can bring in interesting restrictions on the first order solutions. Before we proceed with those computations, let us remark that the first order cubic solutions are only independent at first order. They are expected to ``interact'' at higher orders. In other words, obstructions can appear when taking antibrackets of first order solutions with themselves, which can sometimes be cured by considering a sum of several vertices with different field contents. In that case, crossed antibrackets have to be considered. They do not automatically vanish and can bring in further obstructions. In fact, this is one of the most expected feature of higher spin theories: considering individual fields is only possible at first order. It was advocated to be the solution of the higher spin problem by Berends, Burgers and van Dam \cite{Berends:1984rq}. When building a second order (or higher) deformation, it was suggested to introduce an infinite number of fields, taking any value of the spin. In that case, the Lagrangian would be a formal sum of an infinity of terms containing an arbitrary number of derivatives. The latter fact has already been mentioned when discussing the first order terms: the vertices that seem to behave well are those with the maximum number of derivatives, and this number increases with the spin. The impossibility of obtaining a full local Lagrangian also appears in the Vasiliev formalism: the cubic Lagrangian nonabelian vertices can be built in dimension 4 and 5, it is the Fradkin--Vasiliev action that we described in Chapter \ref{ch:FV}. However, the full Vasiliev theory is not Lagrangian, it consists in a set of equations of motions involving an infinite set of independent fields and connections, corresponding to the different symmetric spin-$s$ fields. We were interested in showing that this requirement is actually needed in a cohomological study in Minkowski, but it does not appear in the $antigh\ 2$ component. Let us finally notice that, because of the relation between $a_2$ and the first order deformation of the gauge algebra, Eq.(\ref{a2a2}) is equivalent to computing the lowest order of the Jacobi identity. In the next sections, we consider antibrackets of the $a_2$ solutions that we have obtained in the previous chapters, both between themselves or with some of the other known solutions, such as the Einstein--Hilbert and Berends--Burgers--van Dam solutions. It results in inconsistencies or in the impossibility of the coexistence of some of the solutions.

\noindent {\textbf{Remark:}} Eq.(\ref{sogen}) has a meaning only for first order solutions. If $\stackrel{(1)}{w}$ were not, the modulo $s$ freedom would introduce non-$\bs$-exact terms. Let us consider the antibracket of an $\bs$-exact term $\bs c$ with an arbitrary $\stackrel{(1)}{w}$:
\begin{eqnarray}(\stackrel{(1)}{w},\bs c)=(\stackrel{(1)}{w},(\stackrel{(0)}{w},c))&=&(\stackrel{(0)}{w},(c,\stackrel{(1)}{w}))-(c,(\stackrel{(0)}{w},\stackrel{(1)}{w}))\\&=&\bs(c,\stackrel{(1)}{w})-(c,\bs\stackrel{(1)}{w})\quad.\end{eqnarray} 
The second term is not manifestly $\bs$-exact. If $\stackrel{(1)}{w}$ is a first order solution, then $\bs\stackrel{(1)}{w}=db$ and $(c,db)=0$ because the variational derivatives of a $d$-exact expression always vanish. This is important to notice. For example, if $\stackrel{(1)}{w}$ is not a consistent first order solution, then, in the computation of $(a_2,a_2)$, the addition of a $\g$-exact term to $a_2$ can introduce non-$\g$-exact terms. So, this has to be taken into account in the case where we would consider the antibracket of a confirmed $a_2$ with a candidate, essentially in order to prove that some obstructions could be removed or not.
Fortunately, we can prove that the addition of cubic $\g c_2$ terms to $a_2$'s that admits an $a_1$ does not create such terms.
To see this, we must consider the commutation of $\g$ with variational derivatives with respect to the antifields and the ghosts: \begin{eqnarray}\g\frac{\d f}{\d C^*_{\mu_1...\mu_{s-1}}}=\frac{\d}{\d C^*_{\mu_1...\mu_{s-1}}}\g f\quad,\quad \g\frac{\d^R f}{\d \phi^*_{\mu_1...\mu_s}}=-\frac{\d^R}{\d \phi^*_{\mu_1...\mu_s}}\g f\quad,\end{eqnarray}
\begin{eqnarray}
\frac{\d^L}{\d C_{\mu_1...\mu_{s-1}}}\g f=-\g\frac{\d^L f}{\d C_{\mu_1...\mu_{s-1}}}-s\6_{\r}\frac{\d f}{\d \phi_{\r\mu_1...\mu_{s-1}}}\quad.\end{eqnarray}
Thus, if we consider a modulo $\g$ modification of an $a_2$: \begin{eqnarray}(a_2+\g c_2,a_2+\g c_2)&=&(a_2,a_2)+\g(...)\mp 2s \frac{\d^R a_2}{\d C^{*\mu_1...\mu_{s-1}}}\6_\r\frac{\d^L c_2}{\d\phi_{\r\mu_1...\mu_{s-1}}}\\&=&\ (a_2,a_2)+\g(...)+div.\pm 2s \6_\r\frac{\d^R a_2}{\d C^{*\mu_1...\mu_{s-1}}}\frac{\d^L c_2}{\d\phi_{\r\mu_1...\mu_{s-1}}}\quad.\end{eqnarray} The last term is not $\g$-exact in general. It is in fact related to the commutation of $\d$ with the variational derivatives with respect to the antifields $\phi^*$ (see for example Eq.(\ref{eqx})). Since $a_2$ does not depend on these antifields, we get: \begin{eqnarray} -s\6_{(\r}\frac{\d a_2}{\d C^{*\mu_1...\mu_{s-1})}}=\frac{\d^L}{\d\phi^{*\r\mu_1...\mu_{s-1}}}\d a_2\quad.\end{eqnarray} Therefore, if $\d a_2=-\g a_1-d b_1$, the commutation of $\g$ and the variational derivatives with respect to $\phi^*$ ensures that \begin{eqnarray}(a_2+\g c_2,a_2+\g c_2)=(a_2,a_2)+\g(...)+div.\pm\g\Big(\frac{\d^L a_1}{\d\phi^{*\mu_1...\mu_s}}\Big)\frac{\d c_2}{\d\phi_{\mu_1...\mu_s}}\quad.\end{eqnarray} Finally, thanks to the cubic nature of the considered $a_2$, we are sure that the variational derivatives of $c_2$ with respect to the fields only depend on the ghosts and antifields and are thus $\g$-closed. Thus, $(a_2+\g c_2,a_2+\g c_2)$ and $(a_2,a_2)$ differ by $\g$-exact and $d$-exact terms, which do not modify the structure of Eq.(\ref{a2a2}).

\section{Computation of $(a_2,a_2)$ expressions}

\subsection{General considerations}

Let us provide some general rules about those computations. First, as we already emphasized in the previous section, the antibracket consists in performing various variational differentiations with respect to conjugate fields. We can first consider the antibracket of an $a_2$ with itself, it reads: \begin{eqnarray}(a_2,a_2)=\sum_s 2\frac{\d^R a_2}{\d C_{\m_1...\m_s}}\frac{\d^L a_2}{\d C^{*\m_1...\m_s}}\quad.\end{eqnarray} This vanishes in any case where an $a_2$ does not contain a ghost and its conjugate $antigh\ 2$ antifield. For example, let us consider an $s-s'-s''$ configuration, with $s<s'\leqslant s''$. Theorem \ref{thma2} ensures that the only possible nontrivial $a_2$ containing $2s'-1$ or $2s'-2$ derivatives is linear in the spin-$s$ antifield, and linear in the spin $s'$ and $s''$ ghosts. Since we have considered a spin-$s$ strictly lower than the others, this candidate satisfies automatically $(a_2,a_2)=0$. Furthermore, we have proved in Chapter \ref{ch:intmisc} that those deformations are related to an $a_1$. Thus, the relation still holds when adding a $\g$-exact term to $a_2$. This quite general rule applies for the $1-s-s$ and $2-s-s$ deformations that have been considered in Chapter \ref{ch:intmisc}. The fact is, that the only vertices that have been obtained so far are either cubic or quadratic in the same spin-$s$ field, and most of them contain the highest possible number of derivatives for the given setup. The only exception so far is the BBvD spin-3 vertex, that involves three derivatives, and coexists with another vertex with five derivatives. It is quite obvious that the antibracket does not alter the number of derivatives: if $a_{2,1}$ contains $k_1$ derivatives and $a_{2,2}$ contains $k_2$ derivatives, then $(a_{2,1},a_{2,2})$ contains $k_1+k_2$ derivatives. Then, if we consider two configurations $s-s_1-s_2$ and $s-s_3-s_4$, the terms of $(a_2,a_2)$ where the spin-$s$ ghosts and antifields are differentiated is quartic in the fields, with a configuration $s_1-s_2-s_3-s_4$. These two very simple rules can be used to seek which $a_2$ candidates can be considered to try to remove an obstruction. 

\subsection{Computation for the spin-3 parity-breaking vertices}

Before we compute the second order $(a_2,a_2)$ test for the various parity-invariant deformations, let us complete the argument of Chapter \ref{ch:exo3} about the parity-breaking deformations in dimension 3 and 5. Let us recall the expressions for $a_2$ (see Eqs(\ref{deftrois})-(\ref{a2n5})). The dimension 3 candidate is:

\begin{eqnarray} a_{2,1}=f^A_{\phantom{A}[BC]}\ve^{\m\n\r}C^{*\a\b}_A C_{\m\a}^B \6_{[\n}C^C_{\r]\b}d^3 x\quad,\end{eqnarray} and the dimension 5 candidate is:
\begin{eqnarray}a_{2,3}=g^A_{\phantom{A}(BC)}\ve^{\m\n\r\s\t}C^{*}_{A\m\a}\6^{}_{[\n}C_{\r]}^{B\,\b} \6^2_{\a[\s}C^C_{\t]\b}d^5 x\quad.\end{eqnarray} We have not chosen the strictly non-$\g$-exact expressions, because the $(a_2,a_2)$ computations are a bit easier with these ones. Since these are consistent first order deformations, the modulo $\g$ freedom of $a_2$ only alters $(a_2,a_2)$ by $\g$- and $d$-exact terms. Let us recall that, in the particular case of dimension 3, the spin-3 ghost tensor $U$, which has the same symmetry as a Weyl tensor, identically vanishes. More generally, for a spin-$s$ ghost, traceless tensors $U^{(i)}$ with $i\geqslant 2$ vanish, which leaves us with a maximum of two derivatives in $a_2$, and thus three derivatives in $a_0$. So, in fact $a_{2,1}$ is the candidate with the highest number of derivatives in dimension 3 (in order to build Lorentz-invariant objects, the total number of indices must be even. This imposes an odd number of derivatives in odd dimension). The same considerations will be made later about the BBvD vertex. Of course, the two parity-breaking deformations do not exist in the same Minkowski spacetimes, thus only their antibrackets with themselves have to be considered. The computation for the dimension 3 candidate yields:
\begin{eqnarray}
        (a_{2,1},a_{2,1})&=& 2 \frac{\d^R a_2}{\d C^{*\a\b}_A}\frac{\d^L a_2}{\d C_{\a\b}^A}
  \nonumber \\
  &=&\g\mu + d\n + 2f^A_{~BC}f_{EAD}\ve^{\m\n\r}\ve_{\a\l\t}
  \Big[\,
  \frac{1}{2}\,C^{*E\s\x}C_{\m}^{B\,\a}{T}^C_{\n\r|\s}
  {T}^{D\l\t|}_{~~~~~\x}+
  \frac{1}{2}\,C^{*E\s\x}C_{\m\s}^{B}{T}_{\n\r|}^{C~~\a}
  {T}^{D\l\t|}_{~~~~~\x}
  \nonumber \\
  &&-\frac{1}{3}\,C^{*E\a\x}C_{\m}^{B\,\s}{T}^C_{\n\r|\s}
  {T}^{D\l\t|}_{~~~~~\x}-
  \frac{2}{3}\,C^{*E\s\x}{T}^{B\l}_{~~(\m|\a)}
  {T}^C_{\n\r|\s}C_{\x}^{D\,\t}-
  \frac{2}{3}\,C^{*E\s\x}{T}^{B\l}_{~~(\m|\s)}
  {T}^{C~~\a}_{\n\r|}C_{\x}^{D\,\t}
  \nonumber \\
  &&+\frac{4}{9}\,C^{*E\a\x}{T}^{B\l}_{~~(\m|\s)}
  {T}^{C~~\s}_{\n\r|}C_{\x}^{D\,\t}\,
  \Big]\quad.
\end{eqnarray}

The use of the variable $\widetilde{T}_{\a\b}:=\ve^{\m\n}_{~~\a}T_{\m\n|\b}$
in place of $T_{\m\n|\r}(=-\frac{1}{2}\ve^{\a}_{~\m\n}\widetilde{T}_{\a\r})$ 
simplifies the calculations. We find, after expanding the products of $\ve$-densities,  
\begin{eqnarray}
        (a_{2,1},a_{2,1})&=&\g\mu + d\n + f^A_{~BC}f_{EAD}C^{*E\s\t}
  \Big[\,
    C^{B\m\a}{\widetilde{T}}^C_{\m\s}{\widetilde{T}}^D_{\a\t}    
  + C^{B\m}_{~~\s}{\widetilde{T}}^C_{\m\a}{\widetilde{T}}^{D\,\a}_{\;\t}
  \nonumber \\  
&&
 - \frac{2}{3}\,C^{B\m\a}{\widetilde{T}}^C_{\m\a}{\widetilde{T}}^D_{\s\t}   
  + C^{D\m}_{~~\s}{\widetilde{T}}^B_{\m\a}{\widetilde{T}}^{C\,\a}_{\;\t}  
  - \frac{1}{3}\,C^D_{\s\t}{\widetilde{T}}^{B\a\m}{\widetilde{T}}^C_{\a\m} 
  \Big]\quad.
  \label{inter1}
\end{eqnarray}
We then use the only possible Schouten identity 
\begin{eqnarray}
        0 &\equiv& C^{*E\,\t}_{~[\s}C^{B\,\m}_{\;\a}{\widetilde{T}}^{C\,\s}_{\;\m}
        {\widetilde{T}}^{D\,\a}_{\;\t]}\nonumber \\
        &=& \frac{1}{24}\Big[
         - C^{*E\s\t}C^{B\m\a}{\widetilde{T}}^C_{\s\t}{\widetilde{T}}^D_{\m\a}
         + 2 C^{*E\s\t}C^{B\m\a}{\widetilde{T}}^C_{\s\m}{\widetilde{T}}^D_{\a\t}
   + 2 C^{*E\s\t}C^B_{\s\m}{\widetilde{T}}^C_{\t\a}{\widetilde{T}}^{D\,\a\m}
  \nonumber \\  
  && \qquad  - C^{*E\s\t}C^B_{\s\t}{\widetilde{T}}^C_{\m\n}{\widetilde{T}}^{D\,\m\n}
      - C^{*E\s\t}C^{B\m\n}{\widetilde{T}}^C_{\m\n}{\widetilde{T}}^D_{\s\t}
   + 2 C^{*E\s\t}C_{\s}^{B\,\m}{\widetilde{T}}^C_{\m\a}{\widetilde{T}}^{D\,\a}_{~\t}
        \Big]\quad, \label{Sca2a2}
\end{eqnarray}
in order to substitute in Eq.(\ref{inter1}) the expression of $C^{*e\s\t}C^{b\m\a}{\widetilde{T}}^c_{\m\s}{\widetilde{T}}^d_{\a\t}$ in terms of 
the other summands appearing in Eq.(\ref{Sca2a2}).  
Consequently, the following expression contains only linearly independent terms: 
\begin{eqnarray}
        &(a_{2,1},a_{2,1}) = \g\mu + d\n + C^{*E\s\t}&\Big[
         {\textstyle\frac{1}{2}}f^A_{~BC}f_{DEA}C^{B\m\a}{\widetilde{T}}^C_{\s\t}
   {\widetilde{T}}^D_{\m\a}+{\textstyle \frac{1}{6}}f^A_{~BC}f_{DEA}C^{B\m\a}{\widetilde{T}}^D_{\s\t}
                {\widetilde{T}}^C_{\m\a}
         \nonumber \\
          &&+ {\textstyle{2}} f^A_{~C(B}f_{D)EA}C^{B\,\m}_{\s}{\widetilde{T}}^C_{\t\a}
        {\widetilde{T}}^{D\,\a}_{~\m}
        +{\textstyle \frac{1}{2}}f^A_{~B[C}f_{D]EA}C^{B}_{\s\t}{\widetilde{T}}^C_{\m\a}
        {\widetilde{T}}^{D\m\a}\nonumber\\
        &&+{\textstyle \frac{1}{3}}f^A_{~BC}f_{DEA}C^{D}_{\s\t}{\widetilde{T}}^C_{\m\a}
        {\widetilde{T}}^{B\m\a}
        \Big]\nonumber\quad,
\end{eqnarray}
where we used that the structure constants of $\ca$ obey 
$f_{ABC}\equiv \d_{AD}f^D_{~BC}=f_{[ABC]}$. 

Therefore, the above expression is a $\g\,$-coboundary modulo $d$ if and only if 
 $f^A_{~BC}f_{DEA}=0$, meaning that the internal algebra $\ca$ is nilpotent of order three. 
In turn, this implies \footnote{The internal metric $\d_{AB}$ being Euclidean, the 
condition $ f^A_{~BC}f_{AEF}\equiv \d_{AD} f^A_{~BC}f^D_{~EF} = 0$ can be seen as 
expressing the vanishing of the norm of a vector in Euclidean space (fix $E=B$ and $F=C$), 
leading to $f^A_{~BC}=0$. } that $f^A_{~BC}=0$ and the deformation is trivial. 

Let us now compute the antibracket for the dimension 5 candidate:
\begin{eqnarray}  (a_{2,3},a_{2,3}) &=&-g^A_{BC}g_{DEA}^{}\ve^{\bar{\m}\bar{\n}\bar{\r}\bar{\s}\bar{\t}}\ve_\m^{\ \,\n\r\s\t}\d^{(\m}_{\bar{\t}} \d^{\a)}_\d 
(4\pa_{\bar{\m}}C_{\bar{\n}}^{*D\g}\pa_{\g\bar{\r}}C_{\bar{\s}}^{E\d}+2\pa_{\g\bar{\m}}C_{\bar{\n}}^{*D\g}\pa_{\bar{\r}}C_{\bar{\s}}^{E\d} )
 \pa_\n^{\ }C_\r^{B\b}\pa_{\a\s}C_{\t\b}^{C}\nonumber \\
&=&  -12 g^A_{B[C}g_{D]EA}^{} C^{*B\a\b}U_{\ \a}^{C\ \g|\m\n}U_{\b\g}^{D\ \,|\r\s}U^E_{\m\n|\r\s}
+ \g c_2 + \pa_{\m}j^{\m}_2\quad. \nonumber\end{eqnarray}
The first term appearing in the right-hand side of the
above equation is a nontrivial element of $H(\g\vert d)$ . 
Its vanishing implies that the structure constants $g_{(ABC)}$ of the 
commutative invariant-normed algebra $\cb$ 
must obey the associativity relation $g_{~\;B[C}^A g_{D]EA}^{\ }=0$. As for the spin-2 deformation problem (see \cite{Boulanger:2000rq}, Sections 5.4 and 6), this means that, modulo 
redefinitions of the fields, there is no cross-interaction between different kinds of spin-3 gauge fields provided the internal metric in $\cb$ is positive-definite, which is required by the positivity of energy. 
The cubic vertex $a_0$ can thus be written as a sum of independent self-interacting vertices, one for each field $\phi_{\m\n\r}^A\,$, $A=1,\ldots,N\,$. 
Without loss of generality, we may drop the internal index $a$ and consider only one \emph{single} self interacting spin-3 gauge field $\phi_{\m\n\r}\,$. 

\section{Inconsistency of the Berends-Burgers-van Dam spin-3 vertex}

We have recalled the BBvD first order deformation in chapter \ref{ch:exo3}. The $antigh\ 2$ term of this deformation is
\begin{eqnarray}a_{2,BBvD}=f^A_{\phantom{A}BC}C^{*\m\n}_A\left[T^B_{\mu\a|\b}T^{C\,\a|\b}_\n-2T^B_{\mu\a|\b}T^{C\,\b|\a}_\n+\frac{3}{2}C^{B\a\b}U^C_{\m\a|\n\b}\right]d^n x+\g c_2\quad.\end{eqnarray} It has been showed \cite{Bekaert:2006jf} that $(a_{2,BBvD},a_{2,BBvD})$ presents an obstruction containing terms of the structure $C^*TTU$ and $C^*CUU$ that cannot be eliminated. The coefficient of the obstruction is $f_{ABC} f^A_{DE}$, whose vanishing implies the vanishing of the deformation itself. Let us notice that, in dimension 3, since $U^A_{\a\b|\c\d}\equiv0$, the BBvD candidate passes the test. In dimension 4, some Schouten identities could imply the weaker associativity condition $f_{AB[C}f^A_{D]E}=0$, however, this still implies the vanishing of $f_{ABC}$ in the end. Furthermore, in dimension 3, the deformation with 5 derivatives vanishes \cite{Bekaert:2006jf}, in that case the BBvD candidate involves the maximum possible number of derivatives. In dimension 4, it has been showed that Schouten identities imply the vanishing of $a_2$, thus the 5-derivative deformation is Abelian. It is important to notice that we thus prove that there is no pure spin-3 cubic deformation in Minkowski spacetime in dimension 4. Let us remark that the case of dimension 3 is a bit particular, since traceless tensors associated to a Young diagram whose first two columns have length 2 identically vanish\footnote{More generally, in dimension $n$, any tensor associated to an irreducible representation of $O(n)$ (and thus traceless), whose Young diagram is such that the sum of the heights of the first two columns is greater than $n$, identically vanish (see \cite{Hamermesch}, page 394). Let us remark that, for $n\geqslant 4$, two-row tensors, such as the traceless part of the curvature or the strictly non-$\g$-exact ghost tensors, are never constrained by this condition.} which implies that every gauge invariant tensor vanishes on-shell, which only allows topological theories.

The idea of this section is to prove that no other $a_2$'s can provide the same kind of terms. First, $(a_{2,BBvD},a_{2,BBvD})$ is of course quartic in the spin-3 fields (in the extended meaning of fields, ghosts or antifields), and contains four derivatives. The only possibility of getting terms quartic in the spin-3 fields is to take the $(a_2,a_2)$ of two expressions with the same spin configuration $s-3-3$. Then, the first rule of theorem \ref{thma2} ensures that $1\leqslant s \leqslant 5$. We know everything about the $1-3-3$ and $2-3-3$ cases: in both cases, there is only one solution, whose $a_2$ is linear in the spin 1/spin-2 $antigh\ 2$ antifield. These candidates satisfy trivially $(a_2,a_2)=0$, hence they cannot help for the BBvD obstruction. 

To complete the argument, we have to investigate the $3-3-4$ and $3-3-5$ cases. It is rather simple: we will prove that the only $a_2$ candidates that are related to an $a_1$ contain more than two derivatives, which is sufficient to be sure that no obstruction containing four derivatives will arise. The results about those two cases are presented in the next two subsections. The results are sufficient to establish the inconsistency of the BBvD deformation in dimension greater than three, in the parity-invariant case, thereby invalidating the hopes expressed by the authors BBvD concerning a possible solution of their problem by the addition of higher spin contributions. Therefore, in flat spacetime, their spin-3 self coupling is definitely inconsistent and no higher spin can cure this problem contrary to the general belief. It is only in $(A)dS$ that this candidate can play a role, as suggested by the FV solution.

\subsection{Study of $a_2$ in the $3-3-4$ case}

Let us use theorem $\ref{thma2}$, with $s=s'=3$ and $s''=4$. The sum of the spins is even, thus the number of derivatives in $a_2$ has to be odd. The maximum is $2s'-3=3$. Furthermore, the difference between the numbers of derivatives acting on the two ghosts obeys $|i-j|<s+s'-s''=2$, and is thus equal to 1. The possible strictly non-$\g$-exact Lorentz-invariant expressions with one derivative read: \begin{eqnarray}\stackrel{(1)}{t}{}^{\!\!AB}=C^{*\m\n\r}C^{A\a}_{\m}T^B_{\a\n|\r}\ ,\ \stackrel{(2)}{t}{}^{\!\!AB}=C^{*A\m\n} T^B_{\m\a|\b} C_\n^{\ \a\b}\ ,\ \stackrel{(3)}{t}{}^{\!\!AB}=C^{*A\m\n} C^{B\r\s} U^{(1)}_{\m\r|\n\s}\quad.\end{eqnarray} Those with three derivatives are: \begin{eqnarray}\stackrel{(4)}{t}{}^{\!\!AB}=C^{*\m\n\r}T_{\m}^{A\a|\b}U_{\n\a|\r\b}^B\ ,\ \stackrel{(5)}{t}{}^{\!\!AB}=C^{*A\m\n}T^{B\a\b|\c}U^{(2)}_{\a\b|\c\m|\n}\ ,\ \stackrel{(6)}{t}{}^{\!\!AB}=C^{*A\m\n}U^B_{\a\b|\c\m}U^{(1)\a\b|\c}_{\phantom{(1)\a\b|\c}\n}\quad.\end{eqnarray}
Let us check that the candidates with three derivatives are related to an $a_1$: \begin{eqnarray}\d\stackrel{(4)}{t}{}^{\!\!AB}=div+\g(...)+4\phi^{*\m\n\r\s}U^{A\a|\phantom{\s}\b}_{\m\phantom{\a|}\s}U^B_{\n\a|\r\b}-\frac{6}{n+2}\phi^*{}'{}^{\r\s}U^{A\n\a|\phantom{\s}\b}_{\phantom{A\n\a|}\s}U^B_{\n\a|\r\b}\quad.\end{eqnarray} This term is antisymmetric in $AB$, thus a symmetric set of coefficients ensures the vanishing of the non-$\g$-exact terms. The variation under $\d$ of the two other terms provides the same non-$\g$-exact term $\phi^{*A\m\n\r}U^{B\a\b|\phantom{\r}\c}_{\phantom{B\a\b|}\r}U^{(2)}_{\a\b|\c\m|\n}$, they vanish if $\stackrel{(5)}{t}{}^{\!\!AB}$ and $\stackrel{(6)}{t}{}^{\!\!AB}$ have opposite coefficients. Finally, we get: \begin{eqnarray}a_{2,3}=k_{(AB)}C^{*\m\n\r}T_{\m}^{A\a|\b}U_{\n\a|\r\b}^B d^n x+l_{AB}C^{*A\m\n}\Big[T^{B\a\b|\c}U^{(2)}_{\a\b|\c\m|\n}-U^B_{\a\b|\c\m}U^{(1)\a\b|\c}_{\phantom{(1)\a\b|\c}\n}\Big]d^n x\quad.\end{eqnarray} On the other hand, the candidates involving one derivative are obstructed: \begin{eqnarray}\d \stackrel{(1)}{t}{}^{\!\!AB}=div.+\g(...)+4\Big(\phi^{*\m\n\r\s}-\frac{3}{n+2}\eta^{(\m\n}\phi^*{}'{}^{\r)\s}\Big)\Big[-T^{A\a}_{\phantom{A\a}\s|\m}T^B_{\a\n|\r}+C^{A\a}_{\m}U^B_{\a\n|\s\r}\Big]\quad.\end{eqnarray} All the terms vanish if $\stackrel{(1)}{t}{}^{\!\!AB}$ is multiplied by a symmetric coefficient, except one proportional to the trace of $\phi^*$: $\frac{-4}{n+2}\phi^*{}'{}^{\n\s}C^{A\a\r}U^{B}_{\a\n|\s\r}$. This obstruction cannot be removed. The variation under $\d$ of $\stackrel{(2)}{t}{}^{\!\!AB}$ and $\stackrel{(3)}{t}{}^{\!\!AB}$ contains the obstructions $\phi^{*A\m\n\r}U^B_{\m\a|\r\b}C_\n^{\phantom{\n}\a\b}$ and $\phi^{*A\m\n\r}C^{\a\b}U^{(2)}_{\m\a|\n\b|\r}$. Finally, the only possible $3-3-4$ deformation contains three derivatives in $a_2$. Even if the vertex exists, which is not sure, the only terms in $(a_{2,3},a_{2,3})$ contain six derivatives. This cannot remove the obstruction of the BBvD deformation.

\subsection{Study of $a_2$ in the $3-3-5$ case}

Theorem $\ref{thma2}$ ensures that the number of derivatives in $a_2$ is even, and is maximum 4. Furthermore the two ghosts bear the same number of derivatives, since $|i-j|<3+3-5=1$. There are candidates with four, two and zero derivatives. Once again, only the candidates with four derivatives satisfy Eq.(\ref{eqagh1}). The possible terms with no derivatives are: \begin{eqnarray} \stackrel{(1)}{t}{}^{\!\!AB}=C^{*\m\n\r\s}C^A_{\m\n}C^B_{\r\s}\quad{\textrm{and}}\quad \stackrel{(2)}{t}{}^{\!\!AB}=C^{*A\m\n}C^{B\r\s}C_{\m\n\r\s}\quad.\end{eqnarray} 
Those with two derivatives are: \begin{eqnarray}\stackrel{(3)}{t}{}^{\!\!AB}=C^{*\m\n\r\s}T^{A\a}_{\phantom{A\a}\m|\n}T^B_{\a\r|\s}\quad{\textrm{and}}\quad \stackrel{(4)}{t}{}^{\!\!AB}=C^{*A\m\n}T^{B\a\b|\c}U^{(1)}_{\a\b|\c\m\n}\quad.\end{eqnarray} Those with four derivatives are: \begin{eqnarray} \stackrel{(5)}{t}{}^{\!\!AB}=C^{*\m\n\r\s}U^{A\a|\phantom{\n}\b}_{\m\phantom{\a|}\n}U^B_{\r\a|\s\b}\quad{\textrm{and}}\quad\stackrel{(6)}{t}{}^{\!\!AB}=C^{*\m\n}U^{\a\b|\c\d}U^{(2)}_{\a\b|\c\d|\m\n}\quad.\end{eqnarray}Let us notice that $\stackrel{(1)}{t}{}^{\!\!AB}$, $\stackrel{(3)}{t}{}^{\!\!AB}$ and $\stackrel{(5)}{t}{}^{\!\!AB}$ are naturally antisymmetric over $AB$. It is quite obvious that $\d\stackrel{(5)}{t}{}^{\!\!AB}$ is $\g$-exact modulo $d$ because the third derivative of the spin-3 ghost are $\g$-exact. Then, we can consider $\d\stackrel{(6)}{t}{}^{\!\!AB}$, which is $\g$-exact modulo $d$, for the same reason than the previous one and because $\6_{(\r}U^{(2)\a\b|\c\d|}_{\phantom{(2)\a\b|\c\d|}\m\n)}$ is $\g$-exact. On the other hand, obstructions arise for any of the other candidates. For $\stackrel{(3)}{t}{}^{\!\!AB}$, one of the trace term remains, which is proportional to $\phi^*{}'{}^{\m\n\t}U^A_{\a\m|\t\s}T^{B\a\ |\s}_{\phantom{B\a}\n}$. For $\stackrel{(4)}{t}{}^{\!\!AB}$, the obstruction consists of two terms, proportional to $\phi^{*A\m\n\r}T^{B\a\b|\c}U^{(2)}_{\a\b|\c\m|\n\r}$ and $C^{*A\m\n\r}U^{B\a\b|\phantom{\r}\c}_{\phantom{B\a\b|}\r}U^{(1)}_{\a\b|\c\m\n}$. Finally, with no derivatives, the obstruction of $\stackrel{(1)}{t}{}^{\!\!AB}$ arises once again in the trace terms, it is proportional to $\phi^*{}'{}^{\m\n\r}T^A_{\a\m|\n}C^{B\a}_{\phantom{B\a}\r}$. The obstruction of $\stackrel{(2)}{t}{}^{\!\!AB}$ consists of two terms proportional to $\phi^{*A\m\n\r}C^{B\a\b}U^{(1)}_{\r\a|\b\m\n}$ and $\phi^{*A\m\n\r}T^{B\a|\b}_{\r}C_{\m\n\a\b}$. None of those obstructions can be removed, the only possible cubic $a_2$ thus involves four derivatives. Thus, any $(a_2,a_2)$ term involves eight derivatives, this can of course not remove the BBvD obstruction. Finally, since we have considered a spin greater than four, we are not sure if the cubic deformations are the only possible ones, but any solution of degree higher than three will provide terms of power higher than four in $(a_2,a_2)$, which can not compensate the BBvD obstruction either.

\section{Inconsistency of the $2-2-3$ deformation and incompatibility of the $2-3-3$ deformation with Einstein--Hilbert theory}

Let us now consider the second order condition for the $2-2-3$ deformation presented in Eqs(\ref{a72})-(\ref{vert223}). Let us recall the expression of $a_2$:
\begin{eqnarray}a_{2,2}= g_{Abc}\left[\,C^{*b\m}\,\6^{\n}C^{c\r}\,\6_{[\n}C^A_{\r]\m}\,
 -\frac{1}{3}\,C^{*A\m\n}\,\6_{[\m}C_{\a]}^b \,\6_{[\n}C_{\b]}^c\,\eta^{\a\b}\right]
  \,d^n x\quad.
\end{eqnarray} The computation of its antibracket with itself yields:
\begin{eqnarray}(a_{2,2},a_{2,2})&=&
2g_{Cae}g^{\phantom{D}e}_{D\phantom{e}b}\left[C^{*a\m}\6_{[\b} C^b_{\g]} T^{C}_{\t\a|\m}U^{D\t\a|\b\g}
-\frac{2}{3}C^{*c\m\t}\6_{[\m}C^a_{\a]}\6_{[\b} C^b_{\g]}U_\t^{C\a|\b\g}\right]
\nonumber\\&&
+\g(...)+div.\end{eqnarray}
The interesting fact is that the obstruction does not involve terms quartic in the spin-2 (there are four derivatives for three spin-2 ghosts, thus some second derivatives of them have to be present, which leads to $\g$-exact expressions). The terms of the obstruction cannot be eliminated upon imposing some appropriate symmetry property on the coefficients $g_{Cae}$. Let us remark that, once again, the condition is fulfilled in dimension 3 thanks to the vanishing of $U^A$.

We may now determine which other deformations could compensate this obstruction. In order to obtain a $2-2-3-3$ expression, we must choose either a $s-2-2$ and a $s-3-3$ deformations or two $s-2-3$ deformations. In the first case, the spin-$s$ is lower than 4 because there is no $2-2-s$ nonabelian deformations for $s\geqslant4$. If $s=1$, we know that the antibracket of the only possible $1-2-2$ and $1-3-3$ deformations is vanishing. If $s=2$, we have to check the antibracket of the Einstein-Hilbert and of the $2-3-3$ deformation $a_2$'s (that we denote here $a_2^{EH}$ (\cite{Boulanger:2000rq}) and $a_{2,3}$). If $s=3$, we can consider $a_{2,2}$ with the spin-3 deformations. The BBvD deformation is the only one with the good number of derivatives, and we have already showed that it is obstructed, thus we will not consider this. In the second case, the spin is comprised between 2 and 4, but the spin-2 and spin-3 cases are the antibrackets of $a_{2,2}$ and $a_{2,3}$ with themselves. Thus, we only have to check if there are any $2-3-4$ deformation candidates.
Let us first check the antibracket of $a_{2,3}$ and $a_2^{EH}$: \begin{eqnarray}a_{2}^{EH}=a_{abc}C^{*a\mu}C^{b\nu}\6_{[\mu}C^c_{\nu]}d^n x\quad,\end{eqnarray}
\begin{eqnarray} 
a_{2,3} = f_{aBC}\,C^{*a\m}\,\6^{\a}C^{B\b\n}\,\,
\6_{\m}\6_{[\a}C_{\b]\n}^C\,d^n x\quad,\end{eqnarray}
\begin{eqnarray}(a_{2}^{EH},a_{2,3})&=&-a^a_{\ ef}f_{aBC}C^{*e\n}
\6_{[\n}C^f_{\m]}T^B_{\r\s|\t}U^{C\r\s|\m\t}
+a^a_{\ ef}f_{aBC}C^{*e\t}C^f_\m U^B_{\r\s|\t\n}U^{C\r\s|\m\n}\nonumber\\&&
+\g(...)+div.\end{eqnarray} The obstructions are not the same as those of $(a_{2,2},a_{2,2})$, though they have the right field content and number of derivatives. Thus, this seems to prevent the coexistence of the consistent $2-3-3$ deformation and Einstein-Hilbert's theory. Both obstructions can only be removed by adding $2-3-4$ deformations. We will now prove that no terms proportional to the spin-2 $antigh\ 2$ antifields and involving four derivatives can appear in the antibrackets of two $2-3-4$ candidates related to an $a_1$. In order to obtain $2-2-3-3$ terms, we have to differentiate the spin-4 antifield $ C^{*\m\n\r}$ and the spin-4 ghost $C_{\m\n\r}$. In the $2-3-4$ deformation problem, the number of derivatives in $a_2$ must be even, and both ghosts have to bear the same number of derivatives. The maximum number of derivatives is $4$ if the terms are proportional to the spin-2 antifield, and it is 2 in the other cases. Thus, the only nontrivial $a_2$ terms proportional to $C^{*\m\n\r}$ are $C^{*\m\n\r}C_{\m\n}C_\r$ and $C^{*\m\n\r}T^\a_{\phantom{\a}\m|\n}\6_{[\a}C_{\r]}$. Only the second one is related to an $a_1$, the first one presents an obstruction proportional to the trace of the spin-4 antifield: $\phi^*{}'{}^{\m\n}C^\r_{\phantom{\r}\m}\6_{[\r}C_{\n]}$. The only possible expression contains two derivatives, thus we have to combine it with terms proportional to another $antigh\ 2$ antifield also containing two derivatives. These are $C^{*\m}T^{\a\b|\c}U^{(1)}_{\a\b|\c\m}$ and $C^{*\m\n}\6^{[\a}C^{\b]}U^{(1)}_{\a\b|\m\n}$. The one proportional to $C^{*\m\n}$ is related to an $a_1$, but the one proportional to $C^{*\m}$ presents the obstructions $\phi^{*\m\n}T^{\a\b|\c}U^{(2)}_{\a\b|\c\n|\m}$ and $\phi^{*\m\n}U^{\a\b|\phantom{\n}\c}_{\phantom{\a\b|}\n}U^{(1)}_{\a\b|\c\m}$. The only candidate proportional to $C^{*\m}$ that is related to an $a_1$ contains four derivatives, it is $C^{*\m}U^{\a\b|\c\d}U^{(2)}_{\a\b|\c\d|\m}$. Thus, while it is possible to compensate some terms proportional to $C^{*\m\n}$ in $(a_{2,2},a_{2,2})$ and $(a_2^{EH},a_{2,3})$, those proportional to $C^{*\m}$ cannot be removed.

This shows the inconsistency of the unique $2-2-3$ cubic deformation at second order. Furthermore, we also obtain that the $2-3-3$ deformation and the $2-2-2$ Einstein--Hilbert deformation cannot be part of the same complete theory. As we said in the beginning of the chapter, this indicates that the spin-2 field involved in the $2-3-3$ deformation is not a graviton.

\subsection{Incompatibility of the $1-2-2$ deformation with Einstein-Hilbert theory}

The $2-3-3$ deformation is not the only one that is not compatible with Einstein-Hilbert theory. We can prove that the $1-2-2$ deformation also presents an obstruction. Let us recall the expression of its $a_2$: \begin{eqnarray} a_2=l_{ab}C^*\6_{[\a}C^a_{\b]}\6^{\a}C^{b\b}d^n x\ .\end{eqnarray} The antibracket with $a_2^{EH}$ yields: 
\begin{eqnarray}(a_{2,EH},a_2)&=&\frac{\d a_{2,EH}}{\d C^{*\r}_e}
\frac{\d a_{2}}{\d
C^e_\r}=-2a^e_{bc}l_{ea}C^{b\n}\6_{[\r}C^c_{\n]}\6_\t\left[C^*\6^{[\t}C^{|a|
\r]}\right]\nonumber\\&=&\g(...)+d(...)+2a^e_{bc}l_{ae}C^*\eta^{\s\n}\6_{[\t}
C^b_{\s]}\6_{[\r}C^c_{\n]}\6^{[\t}C^{|a|\r]}\,.
\end{eqnarray}
This cannot be consistent unless $a^e_{(bc}l_{a)e}=0$. One of the main results of the study of the spin-2 deformation \cite{Boulanger:2000rq} is that the coefficients of the Einstein--Hilbert deformation can be chosen diagonal, and thus that spin-2 fields can always be considered as self-interacting.  This condition cannot be released by combining the EH deformation with other $(a_2,a_2)$ expressions, since the antibracket of the $2-2-3$ deformation $a_{2,2}$ with itself does not contain pure spin-2 terms. Since the coefficients of the $1-2-2$ deformation $a_2$ are antisymmetric, we obtain the relation $a^a_{aa}l_{ba}=-2a^a_{ab}l_{aa}=0$. This means that the coefficients of one of the deformations must vanish, and thus we have showed that these deformations are incompatible. This incompatibility means that, though the $1-2-2$ vertex only involves ``lower'' spins, it cannot be considered along with the usual lower spin theories. It is rather a deformation of the ``higher spin'' type.

%%%%%%%%%%%%%%%%%%%%%%%%%%%%%%%%%%%%%%%%%%%%%%%%
\chapter{Conclusions}
%%%%%%%%%%%%%%%%%%%%%%%%%%%%%%%%%%%%%%%%%%%%%%%%
In this thesis, we have investigated the construction of nonabelian consistent deformations involving totally symmetric bosonic massless tensor fields in Minkowski spacetime. We have completed the study of pure spin-3 first order interactions, by achieving the determination of every parity-breaking cubic deformations. We have also determined the complete list of parity-invariant interactions between spin-2 and spin-3 fields. Then, we have extended the study to the $2-s-s$ cubic spin configurations, for which we have proved that there is only one possible nonabelian cubic deformation. Similarly, we have explicitly constructed the unique nonabelian cubic deformation for a $1-s-s$ configuration. We have then proved that some of the known first order deformations, which are all of the form $s-s-s'$, cannot be completed to all orders in the coupling constant, the ones that remain involving a maximal number of derivatives, which is $2s-1$ or $2s-2$. For example, the pure spin-3 deformation that was found by Berends, Burgers and van Dam and that involves three derivatives has been showed to be strongly obstructed, thus invalidating the conjecture that this obstruction could be cured by introducing fields with higher values of the spin. We also found that some of the remaining deformations are incompatible with gravitation. Finally, still in Minkowski spacetime, we established some general rules for finding the possible cubic deformations of the gauge algebra, for any spin configuration $s-s'-s''$, thus allowing one to investigate the exhaustive computation of nonabelian deformations. We found that the highest spin cannot be greater than the sum of the two others, and we found that the maximal number of derivatives in the vertices is related to the middle spin of the configuration: $N\leqslant 2s'-1$. We also obtained a constraint on the repartition of the derivatives in the gauge algebra deformations.

The second main subject of the thesis was to study consistent deformations in a de Sitter or Anti de Sitter spacetime, and to relate the results to those obtained in Minkowski spacetime. The fields considered in $(A)dS$ are totally symmetric gauge fields similar to the Minkowski totally symmetric massless fields. These fields can sometimes be viewed as massive fields, whose mass depends on the cosmological constant, which is a curvature parameter of $(A)dS$. In dimension 4 and 5, some consistent cubic deformations have been found, for any configuration of spins, by Fradkin and Vasiliev. In particular, they have proposed a nonabelian coupling of the spin-$s$ fields with a spin-2 field, that is based on the minimal deformation of the free spin-$s$ Lagrangian, but involves terms with more than two derivatives multiplied by negative powers of the cosmological constant $\L$. We have written explicitly this first order interaction in the case of a $2-3-3$ configuration, and have noticed that its terms containing the highest number of derivatives coincide with the unique Minkowski cubic $2-3-3$ deformation. That is why we have developed an argument to relate them. We have showed that a special flat limit can be established for each cubic $(A)dS$ consistent deformation. This limit does not alter the nonabelian nature of the deformation, hence it provides an injective correspondence between the nonabelian cubic deformations in $(A)dS$ and those in Minkowski spacetime. This argument automatically shows that the Fradkin--Vasiliev quasi-minimal deformation scheme is in fact the unique $(A)dS$ deformation for a $2-s-s$ configuration, since there is only one remaining candidate in Minkowski spacetime. However, this kind of limit must be taken in such a way as to compensate the most negative power of $\L$, which increases with the highest number of derivatives in the cubic deformation and thus with the spins in the considered configuration. The complete Fradkin--Vasiliev first order Lagrangian construction involves every spin, with an arbitrary number of derivatives. Under these conditions, it is impossible to establish a flat limit of this action. In fact, it is well known that a full $(A)dS$ higher spin theory must share the same property, because it appears that the presence of a value of the spin greater than two in the algebra implies the presence of higher values. We have thus showed that, though a correspondence can be established for individual cubic vertices, a possible Minkowski theory cannot be the limit of an $(A)dS$ theory. Furthermore, we have confirmed that a Minkowski higher spin theory is incompatible with Einstein--Hilbert gravitation. If it existed, if would violate the equivalence principle and could not be taken as a local approximation of a theory in a curved space. On the other hand, the $(A)dS$ theory in fact involves particles with a ``mass'', which provides a minimum value of the energy and thus an infrared cutoff on the higher spin processes. Furthermore, the fact that $2-s-s$ $(A)dS$ deformations involve minimal deformation terms allows one to preserve the equivalence principle. The possible Minkowski theory could be the tensionless limit of some string field theories, but does not seem to have a relevance in our current universe. On the other hand, nothing prevents an $(A)dS$ theory to involve classical, macroscopic effects. The nonlocal nature of this kind of theory is still to be made more precise, but we can conjecture that such effects are rather measurable on large scales, probably the cosmological scale. Let us note that Vasiliev's complete theory is an extension of Einstein's theory of gravitation.

The higher spin domain of research is still full of questions that remain to be addressed. Directly in contact with the object of this thesis, an accessible problem is to fully determine the possible deformations of the gauge algebra for bosonic massless fields in Minkowski spacetime, with the help of the antifield formalism. Then, it would be important to find general arguments in favor of the existence of consistent vertices related to these deformations of the gauge algebra. It is also possible to further investigate the deformations at second order in the coupling constant, since it might bring an answer to the question of the possibility of building a Minkowski higher spin theory. Furthermore, the classification of first order deformations could be used to prove the uniqueness of the whole Fradkin--Vasiliev construction. In $(A)dS$, an ambitious problem would be to build a full Lagrangian admitting Vasiliev's equation as equations of motion. Such a Lagrangian would of course be nonlocal. With such an action principle, the even more ambitious problem of quantizing Vasiliev's theory could then be addressed. A preliminary work would be to show that Vasiliev's equations coincide with the FV cubic construction at lowest interacting order. Finally, some progress is still to be made about the precise meaning of a full $(A)dS$ higher spin theory. The computation of some exact solutions already provides some clues about the interaction mechanisms. It remains to find some interpretation of the nonlocal nature of these interactions and to propose some detectable mechanisms, most probably through large scale observations.

%%%%%%%%%%%%%%%%%%%%%%%%%%%%%%%%%%%%%%%%%%%%%%%%

\begin{appendix}
\chapter{Young diagrams\label{app:YD}}

We present here a short review of the Young diagrams and their use in representation theory. We recommend references \cite{Hamermesch,FultonHarris} for more details.

A Young diagram with $d$ boxes is a graphic representation of an ordered partition $\l$ of $d$: 
\begin{eqnarray} \l=(\l_1,...,\l_k)\ |\ \left\{\begin{array}{l}\forall\ i<k\ :\ 1\leqslant\l_{i+1}\leqslant\l_i\\ \sum_{i=1}^k \l_i=d\end{array}\right.\ .\end{eqnarray} The associated Young diagram consists of $k$ rows of identical squares, placed side by side, the length of the $i$th row being $\l_i$. The partition $\l$ is conjugate to another partition $\l'=(\m_1,...,\m_r)$, such that the columns of the Young diagram associated to $\l$ are of length $\m_j$. This can be illustrated by the following generic example:

\begin{picture}(10,95)(0,-55)
\put(1,6){$\l_1$}
\put(1,-4.5){$\l_2$}
\put(1,-24){\vdots}
\put(1,-35){$\l_{k-2}$}
\put(1,-45.5){$\l_{k-1}$}
\put(1,-56){$\l_k$}
\put(31,20){$\m_1$}
\put(42,20){$\m_2$}
\put(59.5,20){\ldots}
%\put(82.5,20){$\m_{\tiny{r-1}}$}
\put(93,20){$\m_r$}
\multiframe(31,5)(10.5,0){2}(10,10){}{}
\multiframe(52,5)(30.5,0){1}(30,10){\ldots}
\multiframe(82.5,5)(10.5,0){2}(10,10){}{}
\multiframe(31,-5.5)(10.5,0){2}(10,10){}{}
\multiframe(52,-5.5)(30.5,0){1}(30,10){\ldots}
\multiframe(82.5,-5.5)(10.5,0){1}(10,10){}
\multiframe(31,-26)(10.5,0){2}(10,20){\vdots}{\vdots}
\multiframe(52,-26)(30.5,0){1}(30,20){}
\multiframe(31,-36)(10.5,0){2}(10,10){}{}
\multiframe(31,-46.5)(10.5,0){2}(10,10){}{}
\multiframe(31,-57)(10.5,0){1}(10,10){}
\end{picture}

\noindent Then, let us provide some particular examples: The partitions $\l=(d)$ and $\l'=(1,...,1)$ are conjugate and correspond to Young diagrams with one row (or one column) of length $d$. For $d=3$, the partition $\l=(2,1)$ is self-conjugate and correspond to the Young diagram \begin{picture}(30,10)(0,0)
\multiframe(1,5)(10.5,0){2}(10,10){}{}
\multiframe(1,-5.5)(10.5,0){1}(10,10){}
\end{picture}
. Finally, a more complicated example for $d=8$: the partition $\l=(4,3,1)$ and its conjugate $\l'=(3,2,2,1)$ correspond to the Young diagram:
\begin{eqnarray}\nonumber
\begin{picture}(50,0)(0,0)
\multiframe(1,5)(10.5,0){4}(10,10){}{}{}{}
\multiframe(1,-5.5)(10.5,0){3}(10,10){}{}{}
\multiframe(1,-16)(10.5,0){1}(10,10){}
\end{picture}\quad.
\end{eqnarray}
\newpage
A Young diagram whose boxes are filled with the integer numbers from $1$ to $d$ is called a Young tableau. A standard Young tableau is a Young tableau such that the numbers increase when going right or down in the diagram. For example, there are two standard Young tableaux for the diagram $(2,1)$:
\begin{eqnarray}
\begin{picture}(60,10)(0,0)
\put(31,0){,}
\multiframe(1,5)(10.5,0){2}(10,10){1}{2}
\multiframe(1,-5.5)(10.5,0){1}(10,10){3}
\multiframe(41,5)(10.5,0){2}(10,10){1}{3}
\multiframe(41,-5.5)(10.5,0){1}(10,10){2}
\end{picture}\quad.
\end{eqnarray}

\section{Representations of $\mathfrak{S}_d$}

A permutation operator, called a Young symmetrizer, is associated to each Young tableau. Two subgroups of the symmetric group $\mathfrak{S}_d$ can be defined: $P=\left\{\s\in \mathfrak{S}_d\ |\ \s \textrm{ preserves each row }\right\}$ and $Q=\left\{\t\in \mathfrak{S}_d\ |\ \t \textrm{ preserves each column }\right\}$. Elements of $P$ (resp $Q$) are products of permutations of the numbers written in the rows (resp. columns). Then the symmetriser\footnote{We consider here sums of permutations. To be more precise, one should rather consider elements $v_\s$ that constitute a basis of the (complex) vector space $\mathbb{C}\mathfrak{S}_d$ underlying the regular representation of $\mathfrak{S}_d$. This vector space being an algebra with the product rule $v_\s.v_\t=v_{\s\t}$.} can be written: \begin{eqnarray}Y=\left[\sum_{\t\in Q} (-1)^{\epsilon_\t} \t\right]\left[\sum_{\s\in P} \s\right]\ ,\end{eqnarray} where $\epsilon_\t$ is the parity of the permutation $\t$. The sum over $Q$ is an antisymmetrizing operator while the sum over $P$ is a symmetrizing operator. Let us note that the operator $\widetilde{Y}$ obtained by first summing over $Q$ then over $P$ appears to be equivalent. For example, the operator associated to \begin{picture}(30,10)(0,0)
\multiframe(1,5)(10.5,0){2}(10,10){1}{2}
\multiframe(1,-5.5)(10.5,0){1}(10,10){3}
\end{picture} 
is: \begin{eqnarray}Y_{\begin{picture}(10,10)(0,0)
\multiframe(1,5.5)(5.5,0){2}(5,5){$\,{}^{}_{{}^1}$}{$\,{}^{}_{{}^2}$}
\multiframe(1,0)(5.5,0){1}(5,5){$\,{}^{}_{{}^3}$}
\end{picture}}=\left[e-(13)\right]\left[e+(12)\right]\ .\end{eqnarray}
Each Young diagram provides an irreducible representation of $\mathfrak{S}_d$, defined through the symmetrizer of one of its standard tableaux: The set $\mathfrak{S}_nY$ is composed of linear combinations of the symmetrizers of the different standard tableaux of $\l$, which can be chosen as the basis of a vectorial space $V_\l$. Then, the matrices that transform this basis correctly are the wanted representation $T_\l$. The dimension of the representation is equal to the number of standard tableaux. Let us define the hook length of the box $(i,j)$ of a Young diagram: $l_{ij}=\l_i+\m_j-i-j+1$. On the diagram, it amounts to counting the number of boxes to the right or below the box $(i,j)$, including the box itself. For example, in the following diagram, the hook length of the dotted box is 4:
\begin{center}
\begin{picture}(80,30)(0,-15)
\multiframe(0,0)(10.5,0){1}(10,10){}
\multiframe(10.5,0)(10.5,0){1}(10,10){}
\multiframe(21,0)(10.5,0){1}(10,10){}
\multiframe(31.5,0)(10.5,0){1}(10,10){}
\multiframe(0,-10.5)(10.5,0){1}(10,10){}
\multiframe(10.5,-10.5)(10.5,0){1}(10,10){}
\multiframe(21,-10.5)(10.5,0){1}(10,10){}
\multiframe(0,-21)(10.5,0){1}(10,10){}
\put(13,-5){$\downarrow$}\put(19,3){$\longrightarrow$}\put(15,3){$-$}
\put(14.3,-3){$|$} \put(50,-8.5){.} \put(13,3){$\bullet$}
\end{picture}
\end{center}
In the following diagram, the hook length of each box has been written: 
\begin{center}
\begin{picture}(80,30)(0,-15)
\multiframe(0,0)(10.5,0){1}(10,10){6}
\multiframe(10.5,0)(10.5,0){1}(10,10){4}
\multiframe(21,0)(10.5,0){1}(10,10){3}
\multiframe(31.5,0)(10.5,0){1}(10,10){1}
\multiframe(0,-10.5)(10.5,0){1}(10,10){4}
\multiframe(10.5,-10.5)(10.5,0){1}(10,10){2}
\multiframe(21,-10.5)(10.5,0){1}(10,10){1}
\multiframe(0,-21)(10.5,0){1}(10,10){1}
\end{picture}
\end{center}
It can be showed that the dimension of $T_\l$ is: \begin{eqnarray}dim\ T_\l = \frac{d!}{\displaystyle\prod_{(i,j)} l_{ij}}\ .\end{eqnarray}
Two different diagrams are related to inequivalent representations, and a diagram can be associated to each inequivalent representation. Thus, the classification of Young diagrams is equivalent to the classification of the representations of $\mathfrak{S}_d$.

\section{Young tableaux and representations of $GL(n,\mathbb{R})$ and $O_g(n)$ \label{as:gln} }

The Young diagrams are also a powerful tool to obtain inequivalent irreducible representations of $GL(n,\mathbb{R})$. $GL(n,\mathbb{R})$ is the group of general transformations of (co)vector components $v_\a$ of $\mathbb{R}^n$. Then, we can consider tensors of $(\mathbb{R}^n)^{\otimes d}$, whose components bear $d$ indices: $T_{\a_1...\a_d}$. These tensors transform under a reducible representation of $GL(n,\mathbb{R})$. This representation can be decomposed into irreducible representations, that act non trivially on tensors with definite symmetries, directly related to the symmetrizers of the Young tableaux with $d$ indices. For example, the two components of a tensor with two indices are its symmetric and antisymmetric parts: $T_{\a_1\a_2}=S_{\a_1\a_2}+A_{\a_1\a_2}$. The symmetric part is provided by the action of $Y_{\begin{picture}(15,10)(0,0)\multiframe(1,5.5)(5.5,0){2}(5,5){$\,{}^{}_{{}^{1}}$}{$\,{}^{}_{{}^{2}}$}\end{picture}}$ in the following way: \begin{eqnarray}S_{\a_1\a_2}=\frac{1}{2}Y_{\begin{picture}(15,10)(0,0)\multiframe(1,5.5)(5.5,0){2}(5,5){$\,{}^{}_{{}^{1}}$}{$\,{}^{}_{{}^{2}}$}\end{picture}}T_{\a_1\a_2}=\frac{1}{2}(T_{\a_1\a_2}+T_{\a_2\a_1})\ .\end{eqnarray} On the other hand, the antisymmetric part is provided by the other Young diagram: \begin{eqnarray}A_{\a_1\a_2}=\frac{1}{2}Y_{\begin{picture}(15,10)(0,0)\multiframe(1,5.5)(5.5,0){1}(5,5){$\,{}^{}_{{}^{1}}$}\multiframe(1,0)(5.5,0){1}(5,5){$\,{}^{}_{{}^{2}}$}\end{picture}}T_{\a_1\a_2}=\frac{1}{2}(T_{\a_1\a_2}-T_{\a_2\a_1})\ .\end{eqnarray} A tensor with three indices can be decomposed into four components: one completely antisymmetric, one completely symmetric and two with a ``hook'' symmetry \begin{picture}(15,0)(0,0)\multiframe(0,5.5)(5.5,0){2}(5,5){}{}\multiframe(0,0)(5.5,0){1}(5,5){}\end{picture}. Let us emphasize that each standard Young tableau provides a different tensor, a tensor can have several inequivalent parts with the same symmetries. On the other hand, the non standard Young tableaux provide a tensor part that is not linearly independent and can then be expressed in function of the other parts. For example: $H^{(1)}_{\a_1\a_2|\a_3}=Y_{\begin{picture}(15,10)(0,0)\multiframe(0,5.5)(5.5,0){2}(5,5){$\,{}^{}_{{}^{1}}$}{$\,{}^{}_{{}^{3}}$}\multiframe(0,0)(5.5,0){1}(5,5){$\,{}^{}_{{}^{2}}$}\end{picture}}A_{\a_1\a_2\a_3}$ and $H^{(2)}_{\a_1\a_3|\a_2}=Y_{\begin{picture}(15,10)(0,0)\multiframe(0,5.5)(5.5,0){2}(5,5){$\,{}^{}_{{}^{1}}$}{$\,{}^{}_{{}^{2}}$}\multiframe(0,0)(5.5,0){1}(5,5){$\,{}^{}_{{}^{3}}$}\end{picture}}A_{\a_1\a_2\a_3}$ are distinct tensors.
More generally, a tensor with $d$ indices and no symmetries can be decomposed into independent parts related to the standard Young tableaux with $d$ boxes, and that transform under inequivalent irreducible representations of $GL(n,\mathbb{R})$. The dimension of these representations, which is equal to the number of components of the associated tensor part, is given by: \begin{eqnarray}dim\ T_\l^{GL(n,\mathbb{R})}=\prod_{(i,j)}\frac{n+j-i}{l_{ij}}\ ,\end{eqnarray} where, once again, $i$ is the row index, $j$ is the column index and $l_{ij}$ is the hook length, the product being made over the boxes $(i,j)$ of the diagram. 

The tensor parts obtained by applying $Y$ are tensors whose indices decompose into $r=\l_1$ antisymmetric groups, of lengths $\m_j\ ,\ j=1,...,k$. Furthermore, the antisymmetrization of all the indices of a group and the first index of the next one gives identically zero. This is what we call the antisymmetric notation of the tensor parts.

{\bf{Remarks:}} 
\begin{itemize}\item The symmetrizers $Y$ are proportional to projectors $P$ (such that $P^2=P$), the ratio being the product of the hook lengths: $Y=\displaystyle\big(\prod_{i,j}l_{ij}\big)P$. 
\item The tensor parts can also be written in the symmetric notation, which is obtained by applying the operators $\widetilde{Y}$ (i.e. by first antisymmetrizing indices then by symmetrizing). In that notation, the indices decompose into $k$ groups of lengths $\l_i$, such that the symmetrization of a group and the first index of the next group gives identically zero. The tensors provided by the two Young symmetrizers of the same Young tableau are equivalent.
\item This construction of independent tensor parts and associated representations in fact provides all the finite-dimensional irreducible representations of $GL(n,\mathbb{R})$.
\item We will sometimes denote the indices of a Young tableau by Greek letters instead of natural numbers in the case where the indices of the tensors are distinct Greek letters instead of subindexed letters (most of the times in the case of a small $d$).
\end{itemize}

Another very interesting feature of Young diagrams is that they ease the determination of the components of a product of tensors. The ``product'' of two young diagrams $\l_1$ and $\l_2$ can be written as follows: First, the indices of the simpler diagram are added one by one to the other in every possible way such that the result is a Young diagram. The symmetries of the simpler diagram must be preserved, hence two boxes belonging to the same column cannot appear in the same row of the result, the boxes of a column cannot all appear in the same column of the result together with a box of the next column, etc. Then, the product is written as the sum of the obtained diagrams, that can appear with a multiplicity. As a first example, let us consider the product of any diagram with a 1-box diagram (which corresponds to a vector): 
\begin{eqnarray}\begin{picture}(50,20)(0,-10)
\multiframe(0,0)(10.5,0){4}(10,10){}{}{}{}\multiframe(0,-10.5)(10.5,0){3}(10,10){}{}{}\multiframe(0,-21)(10.5,0){1}(10,10){}\end{picture}\otimes \begin{picture}(20,20)(-10,1)\multiframe(0,0)(10.5,0){1}(10,10){}\end{picture}=
\begin{picture}(50,20)(0,-10)\multiframe(0,0)(10.5,0){5}(10,10){}{}{}{}{}\multiframe(0,-10.5)(10.5,0){3}(10,10){}{}{}\multiframe(0,-21)(10.5,0){1}(10,10){}\end{picture}\oplus
\begin{picture}(50,20)(0,-10)\multiframe(0,0)(10.5,0){4}(10,10){}{}{}{}\multiframe(0,-10.5)(10.5,0){4}(10,10){}{}{}{}\multiframe(0,-21)(10.5,0){1}(10,10){}\end{picture}\oplus
\begin{picture}(50,20)(0,-10)\multiframe(0,0)(10.5,0){4}(10,10){}{}{}{}\multiframe(0,-10.5)(10.5,0){3}(10,10){}{}{}\multiframe(0,-21)(10.5,0){2}(10,10){}{}\end{picture}\oplus
\begin{picture}(50,20)(0,-10)\multiframe(0,0)(10.5,0){4}(10,10){}{}{}{}\multiframe(0,-10.5)(10.5,0){3}(10,10){}{}{}{}\multiframe(0,-21)(10.5,0){1}(10,10){}\multiframe(0,-31.5)(10.5,0){1}(10,10){}\end{picture}\ .
\end{eqnarray}
Then, the product of symmetric diagrams is interesting to write, for example:
\begin{eqnarray}\begin{picture}(30,20)(0,1)\multiframe(0,0)(10.5,0){3}(10,10){}{}{}\end{picture}\otimes\begin{picture}(20,20)(0,1)\multiframe(0,0)(10.5,0){2}(10,10){}{}\end{picture}=\begin{picture}(50,20)(0,1)\multiframe(0,0)(10.5,0){5}(10,10){}{}{}{}{}\end{picture}\oplus
\begin{picture}(40,20)(0,1)\multiframe(0,0)(10.5,0){4}(10,10){}{}{}{}\multiframe(0,-10.5)(10.5,0){1}(10,10){}\end{picture}\oplus
\begin{picture}(30,20)(0,1)\multiframe(0,0)(10.5,0){3}(10,10){}{}{}\multiframe(0,-10.5)(10.5,0){2}(10,10){}{}\end{picture}\ .
\end{eqnarray} This is the kind of decomposition that arises when one considers the $k$-th derivatives of a symmetric tensor with $s$ indices. In terms of Young diagrams, it is quite obvious that the diagrams in the result have at most two rows. This product properly generates a sum of independent representations, for example of $GL(n,\mathbb{R})$, and thus determines with certainty the independent parts of the direct product of the related tensors. For the sake of argument, let us notice that the numbers of components of the tensors are conserved. In dimension 4, the dimensions for the second example are $20\times10=56+84+60$, using the hook length rule exposed above.

To conclude this appendix, we can now consider tensors that transform under representations of the orthogonal group related to the metric, say $O_g(n)$. In the particular case of the Minkowski metric, this group is $O(n-1,1)$. The independent tensor parts and irreducible representations can be obtained from the $GL(n,\mathbb{R})$ case by separating the tensors into a sum of traceless parts. For example, a symmetric tensor with $s$ indices can be decomposed as a sum of traceless tensors with $s$, $s-2$, $s-4$,... indices, the sum ending with a vector or a scalar depending on whether $s$ is odd or even. A scalar is denoted by a dot in Young diagram notation, and a traceless tensor is denoted with a hat. So, for example: \begin{eqnarray}\begin{picture}(20,0)(0,0)\multiframe(0,0)(10.5,0){2}(10,10){}{}\end{picture}=\begin{picture}(20,0)(0,0)\put(5,6){$\widehat{\phantom{\ldots}}$}\multiframe(0,0)(10.5,0){2}(10,10){}{}\end{picture}\oplus\ \bullet. \end{eqnarray}
The case of a $(2,2)$ symmetry, which is the symmetry of the Riemann tensor, is also interesting:
\begin{eqnarray}\begin{picture}(20,0)(0,0)\multiframe(0,0)(10.5,0){2}(10,10){}{}\multiframe(0,-10.5)(10.5,0){2}(10,10){}{}\end{picture}=
\begin{picture}(20,0)(0,0)\put(5,6){$\widehat{\phantom{\ldots}}$}\multiframe(0,0)(10.5,0){2}(10,10){}{}\multiframe(0,-10.5)(10.5,0){2}(10,10){}{}\end{picture}\oplus\begin{picture}(20,0)(0,0)\put(5,6){$\widehat{\phantom{\ldots}}$}\multiframe(0,0)(10.5,0){2}(10,10){}{}\end{picture}\oplus\ \bullet\ .\end{eqnarray} In the case of the Riemann tensor, the traceless $(2,2)$ tensor is the Weyl tensor, the second term is the traceless part of the Ricci tensor and the scalar is the Gauss scalar curvature. 

%\begin{picture}(80,70)(0,0)
%\multiframe(1,4)(0,-10.5){2}(10,10){}{}
%\multiframe(11.5,4)(55.5,0){1}(55,10){$\stackrel{s''-2}{\ldots}$}
%\multiframe(67.5,4)(10.5,0){1}(10,10){}
%\multiframe(11.5,-6.5)(25.5,0){1}(25,10){$\stackrel{t}{\ldots}$}
%\multiframe(37.5,-6.5)(10.5,0){1}(10,10){}
%\end{picture}

\end{appendix}

%\bibliographystyle{utphys}
%\bibliography{bibthese2}

\providecommand{\href}[2]{#2}\begingroup\raggedright\endgroup

%That's all folks!%

\end{document}